\documentclass[twocolumn,superscriptaddress,showpacs,preprintnumbers,showkeys,amsmath,amssymb,aps,prx,floatfix,includegraphics,longbibliography]{revtex4-2}
\usepackage{amsmath}
\usepackage{xcolor}
\usepackage{graphicx}
\usepackage{bm}
\usepackage[colorlinks=true, citecolor=blue, urlcolor=blue, linkcolor=blue]{hyperref}
\usepackage[normalem]{ulem}
\usepackage{soul}
\usepackage{gensymb}
\usepackage{siunitx}
\usepackage{braket}
\usepackage{blkarray}

\newcommand{\nC}{$n$CCOC}
\newcommand{\nH}{$n$HBCCO}
\newcommand{\dxxyy}{$d_{x^2-y^2}$}
\newcommand{\dzz}{$d_{z^2}$}
\newcommand{\p}{^{\prime}}

\renewcommand{\vr}{\textbf{r}}
\newcommand{\vk}{\textbf{k}}
\newcommand{\vtk}{\mathbf{\Tilde{k}}}
\newcommand{\mypar}[1]{\bigskip{\centering\small \emph{#1}\\\medskip}}
\newcommand{\msc}{m_{\mathrm{SC}}}
\DeclareMathOperator{\Tr}{Tr}

\usepackage{comment}
\usepackage{versions}
\includecomment{versionA}
\includecomment{versionB}
\includecomment{versionC}
\excludeversion{versionAMT}
\includeversion{versionBBL}

\begin{document}

\title{Towards an \emph{ab initio} theory of high-temperature superconductors: a study of multilayer cuprates. 
}

\author{Benjamin Bacq-Labreuil}
\email{benjamin.bacq-labreuil@ipcms.unistra.fr}
\affiliation{D\'epartement de physique, Regroupement qu\'eb\'ecois sur les mat\'eriaux de pointe $\&$ Institut quantique, Universit\'e de Sherbrooke, 2500 Boul. Universit\'e, Sherbrooke, Qu\'ebec J1K2R1, Canada}
\affiliation{Universit\'e de Strasbourg, CNRS, Institut de Physique et Chimie des Mat\'eriaux de Strasbourg, UMR 7504, F-67000 Strasbourg, France}

\author{Benjamin Lacasse}
\affiliation{D\'epartement de physique, Regroupement qu\'eb\'ecois sur les mat\'eriaux de pointe $\&$ Institut quantique, Universit\'e de Sherbrooke, 2500 Boul. Universit\'e, Sherbrooke, Qu\'ebec J1K2R1, Canada}

\author{A.-M. S. Tremblay}
\affiliation{D\'epartement de physique, Regroupement qu\'eb\'ecois sur les mat\'eriaux de pointe $\&$ Institut quantique, Universit\'e de Sherbrooke, 2500 Boul. Universit\'e, Sherbrooke, Qu\'ebec J1K2R1, Canada}

\author{David S\'en\'echal}
\affiliation{D\'epartement de physique, Regroupement qu\'eb\'ecois sur les mat\'eriaux de pointe $\&$ Institut quantique, Universit\'e de Sherbrooke, 2500 Boul. Universit\'e, Sherbrooke, Qu\'ebec J1K2R1, Canada}

\author{Kristjan Haule}
\affiliation{Center for Materials Theory, Department of Physics $\&$ Astronomy, Rutgers University, Piscataway, New Jersey 08854, USA}

\date{\today}

\begin{abstract}

Significant progress towards a theory of high-temperature superconductivity in cuprates has been achieved via the study of effective one- and three-band Hubbard models. 
Nevertheless, material-specific predictions, while essential for constructing a comprehensive theory, remain challenging due to the complex relationship between real materials and the parameters of the effective models. 
By combining cluster dynamical mean-field theory and density functional theory in a charge-self-consistent manner, here we show that the goal of material-specific predictions for high-temperature superconductors from first principles is within reach.
To demonstrate the capabilities of our approach, we take on the challenge of explaining the remarkable physics of multilayer cuprates by focusing on the two representative Ca$_{(1+n)}$Cu$_{n}$O$_{2n}$Cl$_2$ and  HgBa$_2$Ca$_{(n-1)}$Cu$_n$O$_{(2n+2)}$ families.
We shed light on the microscopic origin of many salient features of multilayer cuprates, in particular the $n$-dependence of their superconducting properties. 
The growth of $T_c$ from the single- to the tri-layer compounds is here explained by the reduction of the charge transfer gap and consequently the growth of superexchange $J$ as $n$ increases. 
The origin of both is traced to the appearance of low-energy conduction bands reminiscent of standing wave modes confined within the stack of CuO$_2$ planes. 
We interpret the ultimate drop of $T_c$ for $n\geq 4$ as a consequence of the inhomogeneous doping between the CuO$_{2}$ planes, which prevents the emergence of superconductivity in the inner planes due to their insufficient effective hole doping,  as we also highlight the existence of a minimal doping (4\%) required for superconductivity to appear in one of the planes.
We explain material-specific properties such as the larger propensity  of HgBa$_2$Ca$_{(n-1)}$Cu$_n$O$_{(2n+2)}$ to superconduct compared with Ca$_{(1+n)}$Cu$_{n}$O$_{2n}$Cl$_2$. 
We also find the coexistence of arcs and pockets observed with photoemission, the charge redistribution between copper and oxygen,
and the link to the pseudogap. 
Our work establishes a framework
for comprehensive studies of high-temperature superconducting cuprates, enables detailed comparisons with experiment, and, through its \emph{ab initio} settings, unlocks opportunities for theoretical material design of high-temperature superconductors.
\end{abstract}

\maketitle

\section{Introduction}
\label{sec:intro}
In recent years, significant progress has been made in the description of high-temperature superconductivity. 
Cutting-edge experimental and theoretical works have established clear relations between superconductivity and a few key physical quantities.
The three most important are the oxygen hole-content~\cite{andersen1995b,tranquada1987,fujimori1987,zheng1995,haase2004,jurkutat2014,rybicki2016,kowalski2021,jurkutat2023}, the charge-transfer gap (CTG)~\cite{weber2012,kowalski2021,wang2022,omahony2022a,wang2023a} and the superexchange interaction $J$~\cite{wang2022,kowalski2021}.
Descriptors defined as combinations of physical parameters have also been proposed~\cite{vucivevic2024}, but a central question remains open: how are these quantities related to the details of the high-temperature superconductors' chemical composition and structure?

At the heart of most theoretical investigations so far are effective models designed to reduce the number of degrees of freedom to a well-chosen limited set. 
This strategy has shown to be very successful in describing the physics of high-temperature superconductors well beyond the sole superconducting phase. 
In particular, the one- and three-band Hubbard models~\cite{hubbard1963,gutzwiller1963,kanamori1963,emery1987,varma1987} are able to accurately describe many correlated electronic phenomena such as the Mott transition~\cite{imada1998,park2008,sordi2011,sordi2012,vuvicevic2013,walsh2019}, the pseudogap~\cite{civelli2005,kyung2006,sakai2009,sordi2012,macridin2006,ferrero2009,werner2009,gull2010,merino2014,krien2022}, the high-energy spectral anomalies~\cite{martinez1991a,macridin2007a,manousakis2007a,wang2015a,bacq2023}, the interplay of charge and spin stripes~\cite{zheng2017}, to name a few. 

Yet, the realistic parameterization of these models remains an open issue and difficult task in general, often making the connection to experimentally studied real materials tenuous. 
A representative example is the putative correlation between the superexchange $J$ and superconductivity in cuprates.
On the one hand, calculations based on the three-band Hubbard model predict a direct relation between $J$ and the superconducting order parameter~\cite{kowalski2021}.
On the other hand, experiments show that this correlation seems confined to only a subset of materials, and several outliers are clearly outside the trend (see Fig.~4(d) of Ref.~\onlinecite{wang2022}). 
In short, at this stage the most advanced model calculations cannot explain the existence of materials sharing similar $J$ but displaying markedly different critical temperature $T_c$. 
Therefore, it is crucial to develop novel methods able to perform \emph{material-specific} estimations of the superconducting properties and the aforementioned key quantities.

Density functional theory (DFT) is pivotal to address this challenge since it allows to account, from first principles but approximately, for all electronic degrees of freedom in a given compound. 
Unfortunately, DFT alone is inherently limited in the calculation of response functions and superconducting properties of strongly correlated systems.
It must be complemented by many-body techniques improving the treatment of the few correlated degrees of freedom.

We may distinguish two main philosophies in the recent developments. 
The first consists in using DFT as a base to parameterize effective single- or three-band models~\cite{hirayama2013,hirayama2018b,moree2022a,schmid2023a,held2022,kitatani2023,cataldo2024}.
The many-body tools developed in the context of models can thus be used, such as dynamical mean-field theory (DMFT)~\cite{metzner1989,georges1992,jarrell1992,georges1996} eventually supplemented with diagrammatic expansions around the DMFT solution~\cite{Biermann_Aryasetiawan_Georges_2003,Lan_Shee_Li_Gull_Zgid_2017,stepanov2019a,vandelli2022b,held2022,kitatani2023,cataldo2024}.
For instance, the use of DMFT combined with the dynamical vertex approximation~\cite{held2022,kitatani2023,cataldo2024} leads to accurate predictions for the critical temperature of a nickelate compound~\cite{held2022}. 
Such an approach is very promising, but it is computationally expensive and may not be easily generalized to a broader range of materials. 
Additionally, it lacks the feedback effects on the itinerant degrees of freedom due to the presence of correlated states.
Specifically, strong correlations on the Cu orbitals significantly redistribute charge between the Cu and oxygen orbitals, as well as states in the spacer layers; this shifts the relative positions of these orbitals in the final band structure.

The second type of approaches avoids the use of effective models and  instead retains all degrees of freedom explicitly. 
This has been achieved recently by combining DFT with the density-matrix embedding theory (DMET)~\cite{cui2023a}.
The framework is an \emph{ab initio} approach that treats the correlations local to a given cluster at a higher level than DFT, but the local correlations are still treated approximately.
This is closely related to the treatment of the quantum impurity model by the slave boson-mean field method~\cite{PhysRevB.96.235139}, but applied to a large set of orbitals local to a cluster in space. 
The main drawbacks are its restriction to the ground-state, the high computational cost of including correlations across many different orbitals, and the strong dependence of the DMET procedure on the DFT starting point, which leads to overly persistent antiferromagnetism as a function of hole-doping~\cite{cui2023a}.

%

The complex physics of high-temperature superconductors makes it desirable to develop alternative theoretical methods that offer a complementary perspective  to the aforementioned approaches. 
%
%
Hereby, we present the implementation of a DFT + \emph{cluster} DMFT (CDMFT) framework that offers the following advantages to the existing approaches:
(i) it treats the correlations local to the cluster exactly through solving cluster impurity models, (ii) it provides a simultaneous estimation of the superconducting, spectral, magnetic and electronic properties of a given high-temperature superconductor, (iii) it includes the feedback of strong correlations onto the DFT charge density, (iv) it is computationally cheap enough so as to pave the way towards material-design applications. 

We extend the \emph{embedded} DMFT (eDMFT) formalism of Refs.~\onlinecite{haule2010,HauleJPSJ} to treat cluster impurity problems that are solved using the exact diagonalization (ED) solver of the PyQCM library~\cite{dionne2023a,dionne2023b}. 
We consider $2\times2$ plaquette clusters to estimate the superconducting order parameter of a given material, while having a natural access via DFT+CDMFT to the spectral, magnetic and electronic properties (criteria (i),(ii)).
Taking advantage of the  eDMFT formalism, we are able to perform charge self-consistent DFT+CDMFT calculations (criterion (iii)).
Finally, our choice of impurity solver allows a significant computational speedup (criterion (iv)), opening the door to the study of complex strongly-correlated materials.  

To demonstrate the new capabilities of the method, we perform an exhaustive study of multilayer cuprates, a class of high-temperature superconductors which remained, to date, hardly accessible to theoretical approaches. 
Multilayer cuprates are composed of $n$-CuO$_2$-plane stacks, separated from each other by charge reservoir layers (see Fig.~\ref{fig:fig1}(a)). 
It is essentially the latter that are  material-specific.

The hallmark feature of multilayer cuprates is the universal $n$-dependence of the superconducting critical temperature $T_c$: it is often maximum for the tri-layer ($n=3$) compounds~\cite{mukuda2012}. 
The tri-layer \nH~even yields the record of $T_c$ at ambient pressure among all known superconductors~\cite{dai1995,loret2019a}.
Moreover, the inner CuO$_2$ planes are protected from inhomogeneity and disorder~\cite{mukuda2012,kunisada2020,kurokawa2023}, allowing the direct observation of Fermi pockets and the opening of the superconducting gap at low doping~\cite{kunisada2020,kurokawa2023}.
In other words, the inner planes may provide an experimental access to the intrinsic properties of \emph{clean} CuO$_{2}$ planes. 
Yet, conceiving effective models capturing the essence of these materials is an extremely difficult task, since the model needs to take into account the inter-plane hopping amplitudes, as well as the inhomogeneous hole-doping between the inner and the outer planes. 
Hence, although the multilayer cuprates host a number of mysteries that may crucially impact our understanding of high-temperature superconductivity, there is a real lack of theoretical insight on the microscopic origins of these phenomena. 

We present a detailed theoretical analysis of the physics of multilayer cuprates, focusing on the single- ($n=1$) to the five-layer ($n=5$) compounds of two different families, namely, Ca$_{(n+1)}$Cu$_{n}$O$_{2n}$Cl$_2$ (\nC), and  HgBa$_2$Ca$_{(n-1)}$Cu$_n$O$_{(2n+2)}$ (\nH). Their  crystal structures are illustrated in Fig.~\ref{fig:fig1}(a).
We provide a comprehensive description of the spectral, magnetic, and electronic properties of both undoped and hole-doped materials, examining both the normal and superconducting phases.

For the undoped compounds, we show that the $n$-dependence of the charge transfer gap (CTG) observed experimentally~\cite{wang2023a} can be accurately accounted for by our approach, and we identify its microscopic origin. 
We show that the CTG in multilayer cuprates is reduced by the first low-energy conduction states, appearing for $n\geq2$, which are confined within the CuO$_2$ planes, and can be understood as standing waves in the $n$-CuO$_2$ well potential.
Such confinement induces a smaller CTG in the inner planes in comparison to the outer planes. 
The $n$-dependence of the CTG directly affects the value of the superexchange $J$ and thus explains why antiferromagnetism is stronger in the inner planes~\cite{mukuda2012,oliviero2022a,kunisada2020}. 
We emphasize that this is intrinsically rooted in the physics of the \emph{clean, undoped} multilayer compounds. 

Furthermore, by emulating hole-doping using the virtual crystal approximation (VCA)~\cite{bellaiche2000}, we track the evolution of the materials' properties from the under-doped to the over-doped regime.
In agreement with experiments~\cite{vanveenendaal1994,damascelli2003,hu2021}, we show that the transformation of the spectral function is twofold. 
At very low doping there is essentially only a chemical potential shift, which then combines with a spectral weight reconfiguration as more holes are introduced. 
We show that the redistribution of charge carriers in the structure upon hole-doping is very non-trivial, and that the effective hole-content of the inner CuO$_2$ planes may saturate due to the interplay with other degrees of freedom. 

Most importantly, we estimate the superconducting order parameter of each compound at all values of doping, and show that superconductivity \emph{does not} emerge in the inner planes for $n\geq4$ until the outer planes are overdoped. 
We find a strong link between the normal state spectral properties and the superconducting order parameter. 
Superconductivity disappears in the under-doped and over-doped regions when the normal-state spectral function looses its double-peak, pseudogap-like feature in the vicinity of the Fermi level. 
This explains why the superconducting order parameter vanishes in the inner planes for $n\geq4$: the effective normal-state hole-doping remains below a critical threshold of approximately $4\%$, in remarkable agreement with experiments~\cite{kurokawa2023}.
The tri-layer ($n=3$) compounds thus appear to be the best compromise between (i) hosting protected inner planes, and (ii) reaching sufficient doping for superconductivity to emerge in every plane.    

This work presents a comprehensive theoretical description of both the universal and material-specific characteristics of multilayer cuprates. 
The use of an ED solver is essential, as it significantly reduces the computational cost associated with solving the CDMFT impurity problem. 
%
%
Thus, our work provides \emph{ab initio} predictions of high-temperature superconductivity, paving the way for the theoretical design of superconducting materials.

\subsection*{Organization of the paper}
\label{sec:howto}

The paper is organized as follows. 
In Sec.~\ref{sec:met}, we present our methodology. 
We then discuss our results concerning the undoped multilayer cuprates in Sec.~\ref{sec:res_undop}. 
The evolution of the physics upon hole-doping is presented in Sec.~\ref{sec:res_dop}. 
Finally, we conclude our work in Sec.~\ref{sec:conclusion}.

For a quick overview of the main results, we recommend reading the introduction (Sec.~\ref{sec:intro}), the brief results summaries in Sec.~\ref{sec:res_undop_sum} (undoped compounds) and Sec.\ref{sec:res_dop_sum} (hole-doped compounds), as well as the conclusion (Sec.~\ref{sec:conclusion}). 
Moreover, to ease the reading of the results sections, we divided each subsection into thematic groups of paragraphs that may be read independently. 

\section{Methods}
\label{sec:met}

In the following, we first describe the crystal structures used in this study, as well as how hole-doping is implemented (Sec.~\ref{sec:met_crys}).
We then detail the generalization of the charge self-consistent DFT+eDMFT approach to clusters (Sec.~\ref{sec:met_edmft}), and the use of the ED solver (Sec.~\ref{sec:met_ed}).
Finally, we present our approach to estimate the superconducting properties (Sec.~\ref{sec:met_sc}), and summarize the computational details (Sec.~\ref{sec:met_comp}).

\begin{figure*}
    \centering
    \includegraphics[width=\linewidth]{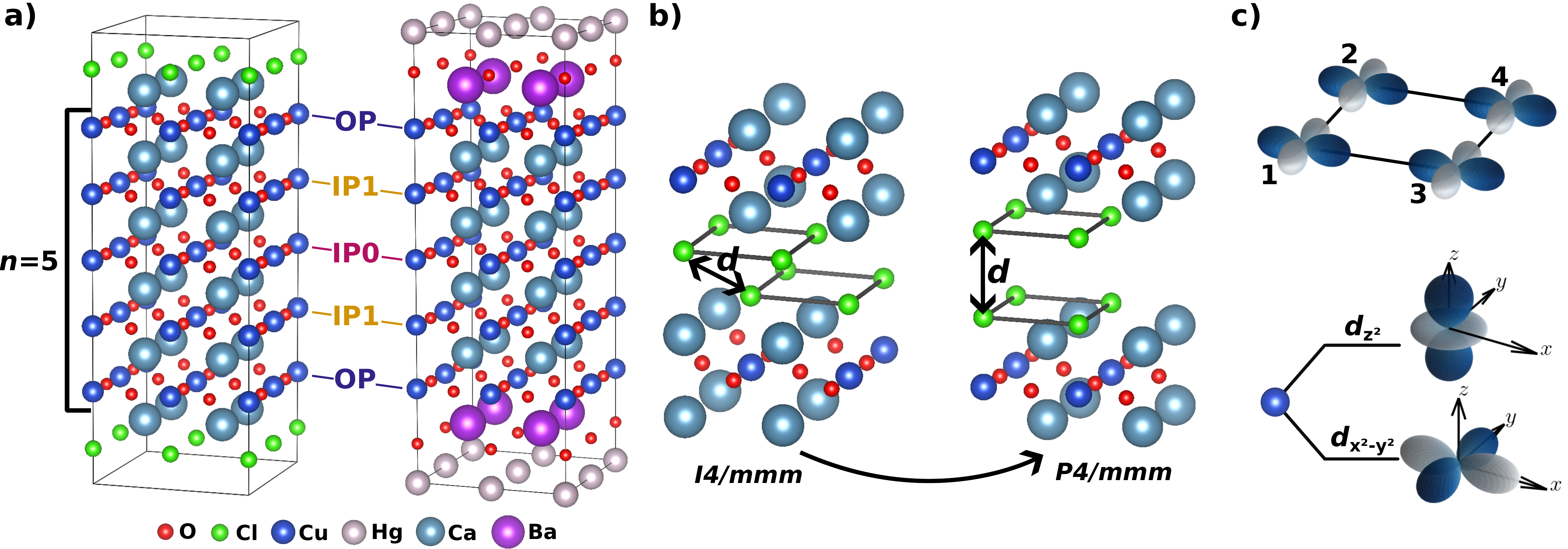}
    \caption{(a) Representative crystal structures for the multilayer cuprate families \nC~(\emph{left}) and \nH~(\emph{right}), for $n=5$. 
    The outer planes (OP), and the two possible types of inner planes (IP1, IP0) are highlighted. 
    (b) Simplification of the \nC~crystal structures, transforming the staggered Cl-$n$CaCuO-Cl stacking ($I4/mmm$) into a vertical stacking ($P4/mmm$), keeping the Cl-Cl distance $d$ fixed.
    (c) Illustration of the two impurity problems solved for each non-equivalent CuO$_2$ plane: a $2\times2$ plaquette of Cu-\dxxyy~orbitals, and a two-orbital single-site including the Cu-\dxxyy~and \dzz~orbitals (\emph{see text for further details}).}
    \label{fig:fig1}
\end{figure*}

\subsection{Crystal structures, hole-doping}
\label{sec:met_crys}

In this study we focus on two multilayer cuprate families: Ca$_{n+1}$Cu$_{n}$O$_{2n}$Cl$_2$ (\nC), and  HgBa$_2$Ca$_{(n-1)}$Cu$_n$O$_{(2n+2)}$ (\nH), see Fig.~\ref{fig:fig1}(a).
The Hg-based compounds have been synthesized and studied experimentally up to $n=16$~\cite{hunter1994,antipov2002,iyo2006,iyo2007}. 
For our calculations, we used the $n=1,2,3$ experimental structure parameters of Ref.~\onlinecite{hunter1994} to be consistent with previous theoretical investigations~\cite{pavarini2001,weber2012}.
The $n=4,5$ ones were determined using the linear scaling of the lattice parameter $c$ \emph{vs.} $n$~\cite{iyo2006,iyo2007}, while the $a$ and $b$ parameters remain constant for $n\geq3$~\cite{mukuda2012}. 
To the best of our knowledge, the \nC~family could be synthesized up to $n=2$~\cite{hiroi1994,kohsaka2002,yamada2005,sowa1990,ruan2016a}. 
We used the experimental structures for the $n=1,2$, and constructed the $n=3,4,5$ structures based on the simple linear scaling $c$ \emph{vs.} $n$ and the distance between the two CuO$_2$ planes in the bi-layer compound. 

We performed DFT+CDMFT calculations with $2\times2$ supercells. 
The $P4/mmm$ space group of the \nH~compounds is not affected by the supercell construction. 
This is however different for the \nC~family, whose $I4/mmm$ space-group reduces to $P4/mmm$ in the supercell. 
From the practical point of view, this means that the number of atoms to be considered in the calculation is multiplied by 8 instead of 4. 
The number of atoms would thus reach $\sim180$ for $n=5$, while our current DFT+CDMFT implementation can reasonably be carried on unit cells with up to $\sim100$ atoms.
To prevent this additional factor of 2, we simplified the \nC~structures as sketched in Fig.~\ref{fig:fig1}(b): the staggered stacking of the Cl-Ca-$nC$uO-Ca-Cl blocks is transformed in the original cell to obtain a vertical stacking $P4/mmm$ structure, keeping the $Cl-Cl$ distance constant. 
Although this may appear as a crude approximation, we show in App.~\ref{ap:struct} that the low-energy properties in a range of $\pm3$~eV around the Fermi level is unaffected by the structural approximation at the level of DFT and beyond.  

The two multilayer families require hole-doping to become superconducting. 
\nC~is usually doped by substituting Ca by Na or vacancies~\cite{hiroi1994,kohsaka2002,yamada2005,sowa1990}. 
\nH, in contrast, is doped by oxygen insertion within the Hg planes~\cite{hunter1994,antipov2002,iyo2006,iyo2007}.
Treating explicitly these atomic substitutions/insertions is expensive from the theoretical point of view since large relaxed supercells must be considered, which is beyond the scope of this study. 
We adopt a simplified procedure by using the Virtual Crystal Approximation (VCA)~\cite{bellaiche2000}: holes are introduced in the structure by lowering the atomic number $Z$ of specific elements to a non-integer value.
For the \nH~compounds we apply the VCA to the Hg atoms, as if O atoms were inserted in the Hg planes. 
The \nC~structures are doped by applying the VCA to the \emph{outer} Ca atoms, to mimic the usual doping of multilayer cuprates, which is restricted to the charge reservoir layers. 
This last point is key: the maximum number of holes that can be introduced in the structure is the same for all $n$ since it only depends on the doping capacity of the charge reservoir layers. 
We thus always introduce the same number of holes $\delta_{\rm tot}=(10,20,30,40,50)\%$ independently of $n$, and let them redistribute over the different CuO$_2$ planes, as it happens in the real compounds. 
We detail in App.~\ref{ap:estim_dop} how we estimate the portion of these holes that end up populating each CuO$_2$ plane.

\subsection{Embedded cluster dynamical mean-field theory}
\label{sec:met_edmft}

Within DFT, the full many-body crystal Hamiltonian is mapped to the effective one-body Kohn-Sham (KS) Hamiltonian which mimics the charge density of the former~\cite{hohenberg1964,kohn1965}.
The mapping consists in representing the two-body interactions via the so-called exchange correlation energy functional $E_{\rm{xc}}$, and is in principle exact for the ground state properties. 
In practice, the exact $E_{\rm{xc}}$ is unknown and approximations are made, most commonly by mapping each point in the material to an auxiliary uniform electron gas problem at the same charge density.
Consequently correlations local to a point in 3D space are accounted for by the local density approximation (LDA)~\cite{kohn1965}, and correlations in the immediate vicinity of each point in 3D space by the generalized gradient approximation (GGA)~\cite{perdew1996}.
Longer range exchange and correlations are ignored. 
Both LDA and GGA face limitations for describing the electronic and structural properties of strongly correlated materials, in which locality to a site in the lattice is required at the minimum. 
%
%
%
These correlated degrees of freedom often live on orbitals spanning a narrow energy range in momentum space, and corresponding to $d$ or $f$ orbitals of an ion in real space. 
Their markedly local character in real space can be efficiently captured by complementing DFT with DMFT~\cite{lichtenstein1998,kotliar2006a,held2007,haule2010,paul2019}, in which exchange and correlations local to a given site are accounted for exactly.

Our approach is based on the DFT+eDMFT implementation of Refs.~\onlinecite{haule2010,HauleJPSJ}, which has been successfully applied to a wide range of transition metal oxides and heavy fermions~\cite{shim2007,haule2010,haule2014,haule2018,mandal2019}.
The key concept is that the electronic structure of strongly correlated materials is better described by the spectral function $A(\mathbf{k},\omega)$ and functional of the spectral function, rather than functional of the charge density alone. 
The spectral function evaluates the probability of adding or removing an electron of energy $\omega$ and momentum $\mathbf{k}$, and is defined in the complex frequency plane as the discontinuity across the real axis of the one-electron Green's function:
\begin{equation}
    \begin{split}
        & A(\mathbf{k},\omega) = -\frac{1}{2\pi i}\left[G(\mathbf{k},\omega+i\eta)-G(\mathbf{k},\omega-i\eta)\right] \\
        & G(\mathbf{k},\omega) = \left[\omega+\mu - H_{\mathbf{k}} - \Sigma(\mathbf{k},\omega)\right]^{-1},
    \end{split}
\end{equation}
where $\eta$ is an infinitesimal positive real number, $\mu$ is the chemical potential, $H_{\mathbf{k}}$ is the non-interacting reference Hamiltonian, and $\Sigma(\mathbf{k},\omega)$ is the self-energy incorporating the effect of electronic correlations.
The latter is \emph{a priori} unknown, and is thus approximated within DFT+DMFT. 
In the following we describe the theoretical formalism necessary to compute the above quantities, from the fundamental functional formalism to the practical implementation used in this work. 

\subsubsection{Functional formalism}

Within DFT+DMFT, the exchange and correlation range is restricted (truncated) to an ion in the solid, and most often only to a given set of $d$ (or $f$) orbitals that form the narrow set of bands near the Fermi level. 
The most appropriate formalism to perform this truncation proceeds from the Luttinger-Ward functional~\cite{kotliar2006a}. 
Similarly to the exchange-correlation functional $E_{xc}$ in DFT, the correlations are here encoded in the Luttinger-Ward functional $\Phi[G]$. 
It can be shown that the exact $\Phi[G]$ contains the sum of all skeleton Feynman diagrams constructed from the single-particle Green's function $G$ and the bare Coulomb interaction. 
If $\Phi[G]$ is simply substituted by 
$$\Phi[G]=E_{XC}[G(\vr,\tau;\vr',\tau')\delta(\vr-\vr')\delta(\tau-\tau')] = E_{XC}[\rho(\vr)],$$
one arrives at the usual DFT equations, but here defined at finite temperature. 
The interpretation of the equations is also different. 
The Kohn-Sham bands are not auxiliary states within the Luttinger-Ward functional approach, but they are physical excitations within the static and space-restricted approximation of $\Phi$.

Within DMFT, only the Feynman diagrams that are restricted to a given ion in the solid are considered in $\Phi[G]$, so that $\Phi[G]=\Phi[G_{\rm{local}}]$, but all topologies and orders of Feynman diagrams are exactly summed up. 
By recognizing that the same quantity $\Phi[G_{\rm{local}}]$ appears also in the exact solution of the corresponding quantum impurity problem, one can compute this quantity without explicitly evaluating the diagrams. 

\subsubsection{Functional formalism for DMFT and CDMFT}
In combining DFT with DMFT one truncates the range of the functional in two steps.
First, all degrees of freedom are treated by the DFT functional $E_{xc}[\rho]$.
Second, for the correlated orbitals the DMFT functional $\Phi[G_{\rm{local}}]$ is added. 
The total $\Phi$ functional is then~\cite{kotliar2006a} 
$$\Phi[G]=E^{V_C}_{xc}[\rho]+\Phi^{U}[G_{\rm{local}}]-E^{U}_{xc}[\rho_{\rm{local}}].$$ 
Here one has to subtract the correlations included in both methods, which is the exchange-correlation energy of the DMFT impurity problem, and is usually called the double-counting.
It is now known exactly~\cite{haule2015}. 
We also marked the functionals with the superscript $V_C$ and $U$, which represents the Coulomb repulsion used in each functional. 
The DFT functional depends on the bare Coulomb repulsion $V_C$ only, while the DMFT functional depends on the screened Coulomb repulsion $U$, because the non-correlated degrees of freedom (core, semicore and other itinerant states) screen the repulsion among the correlated orbitals. 
This screened Coulomb $U$ can be estimated by self-consistent constrained DFT+DMFT calculations~\cite{anisimov1993}.

In the DFT+embedded cluster DMFT method one truncates the range of correlations to a cluster in real space~\cite{lichtenstein2000,kotliar2001,maier2005a,LTP:2006,kotliar2006a,PhysRevB.102.081105}, rather than a single site. 
Since solving a cluster impurity problem with multiple orbitals is numerically very expensive, and because in cuprates only the Cu-$d_{x^2-y^2}$ orbital crosses the Fermi level, we only treat the Cu-$d_{x^2-y^2}$ orbital with cluster-DMFT, while the Cu-$d_{z^2}$ is safely treated within single-site DMFT. 

To properly construct such an approximation we extremize a Kadanoff-Baym functional $\Gamma[G]$ of the Green's function in position space in which the Luttinger-Ward functional $\Phi[G]$ is restricted to certain local quantities defined below:
\begin{eqnarray}
\begin{split}
    \Gamma[G] = & \Tr\log G - \Tr \left(\left(G_0^{-1}-G^{-1}\right)G\right) + E^{V_C}_{H+xc}[\rho] \\
    & + \Phi_{\mathrm{DMFT}}^U[G_{\mathrm{loc}}^{{\rm Cu}(eg)}] - E^{U}_{H+xc}[\rho_{\mathrm{loc}}^{{\rm Cu}(e_g)}] \\
    & + \Phi_{\mathrm{CDMFT}}^U[G_{\mathrm{plaquette}}^{{\rm Cu}(x^2-y^2)}] - \Phi^U_{\rm DMFT}[G_{\textrm{loc}}^{{\rm Cu}(x^2-y^2)}],
\end{split}
\label{LTW}
\end{eqnarray}
where $G$ and $G_0$ are the interacting and non-interacting Green's functions, $E_{H+xc}[\rho]$ is the sum of the Hartree and the exchange-correlation DFT functionals, $\Phi_{\mathrm{DMFT}}[G_{\mathrm{loc}}^{{\rm Cu}(eg)}]$ is the single-site DMFT functional for both the Cu-$e_g$ orbitals, while $\Phi_{\mathrm{CDMFT}}[G_{\mathrm{plaquette}}^{{\rm Cu}(x^2-y^2)}]$ is the cluster-DMFT functional for the Cu-$d_{x^2-y^2}$ orbital. 
These orbitals form the correlated subspace. 
We address the precise way in which they are defined in a following subsection. 

Note that the first line in Eq.~\eqref{LTW} gives the usual DFT equations. 
When the second line is added, we have standard DFT+eDMFT~\cite{haule2010,HauleJPSJ} for the Cu-$e_g$ orbitals, i.e., two-orbital single-site impurity problem. 
Finally, the third line adds the nearest-neighbor and the next-nearest-neighbour correlations on the plaquette, but only for the Cu-$d_{x^2-y^2}$ orbital. 
Note that proper double-counting always has to be subtracted.
In particular, the last term compensates for the fact that the Cu-$d_{x^2-y^2}$ orbital is taken into account in both the single-site and in the $2\times2$ plaquette cluster quantum impurities.


\begin{versionAMT}

\textcolor{red}{\\\\ -----BEGIN AMT Version of method-----}
\subsubsection{Extremization of the functional and Dyson's equation}
In the following, we describe in detail the algorithm used to extremize the above functional (Eq.~\eqref{LTW}).
We need first to define formally the local quantities involved. 
To this end, we recall that $\Gamma[G]$ is a functional of the Green's function
\begin{equation}
\label{eq:G_def}
G(\vr,\tau;\vr',\tau') = -\braket{\mathcal{T}_{\tau}\psi(\mathbf{r,}\tau)\psi^{\dag}(\mathbf{r',}\tau')},
\end{equation}
where $\mathcal{T}_{\tau}$ is the time-ordering operator along imaginary time
$\tau,$ $\psi^{\left(  \dag\right)  }(\mathbf{r,}\tau)$ the creation
(annihilation) operator in an eigenstate of position and the average
$\left\langle {.}\right\rangle $ is a quantum mechanical many-body trace with
a density matrix appropriate for the grand canonical ensemble but that also contains a source field that allows arbitrary variations of the function $G\left(  \mathbf{r,}\tau;\mathbf{r}^{\prime}\mathbf{,}\tau^{\prime}\right)$. 
We do not write spin indices explicitly to simplify the notation.

To functionally differentiate the Luttinger-Ward functionals that depends only on a few correlated orbitals, consider, for example, the following application of the chain rule
\begin{eqnarray}
   && \frac{\delta\Phi^U_{\textrm{DMFT}}[G_{\textrm{loc}}^{{\rm Cu}(eg)}]}{\delta G\left(  \mathbf{r,}\tau;\mathbf{r}^{\prime}\mathbf{,}\tau^{\prime}\right)}  \nonumber \\
   &&=\sum_{\mu \nu} \frac{\delta\Phi^U_{\textrm{DMFT}}[G_{\textrm{loc}}^{{\rm Cu}(eg)}]}{\delta G(\tau_1,\tau_2)_{\mu\nu}}\frac{\delta G(\tau_1,\tau_2)_{\mu\nu}}{\delta G\left(  \mathbf{r,}\tau;\mathbf{r}^{\prime}\mathbf{,}\tau^{\prime}\right)},
\end{eqnarray}
where the Green's function $\delta G(\tau_1,\tau_2)_{\mu\nu}$ is written in a complete set of localized orbitals $\ket{\phi_{\nu}}$ 
\begin{eqnarray}
     &&G(\tau_1,\tau_2)_{\mu\nu} =\nonumber \\ 
     &&\iint d^{3}\mathbf{r}_1d^{3}\mathbf{r}_2\braket{\phi_{\mu}|\vr_1}G(\vr_1,\tau_1;\vr_2,\tau_2)\braket{\vr_2|\phi_{\nu}} \nonumber.
\end{eqnarray}

The functional derivative of the latter Green's function with respect to $G(\vr,\tau;\vr',\tau')$ can easily be computed. The assumption is that 
the Luttinger-Ward functional $\Phi[G]$ is truncated to the correlated local quantities $\Phi_{\mathrm{DMFT}}^U[G_{\mathrm{loc}}^{{\rm Cu}(eg)}]$ and $\Phi_{\mathrm{CDMFT}}^U[G_{\mathrm{plaquette}}^{{\rm Cu}(x^2-y^2)}]$, which means that their derivatives vanish for all localized orbitals, except those of the correlated (active) orbitals $\ket{\phi_{\alpha}}$, that are here centered on the Cu atoms with the compact indices $\alpha$, or $\beta$, or $\gamma$, or $\delta$, standing for the orbital indices $L=\{l,m\}$, the spin index, and the atomic position $\mathbf{R}$ at which the correlated orbitals are centered. 
(Moreover, when used in momentum space, the orbital located at position $\mathbf{R}$ also includes the phase imposed by Bloch's theorem in its definition, namely $e^{i\vk\mathbf{R}}$. 
This phase cancels out for single-site DMFT, while it has to be considered in CDMFT.)

%

Within (C)DMFT these local Green's functions are to be identified with the Green's function of corresponding quantum (cluster) impurity problems, the details of which are presented later.

The Kadanoff-Baym functional of Eq.~\eqref{LTW} can now easily be extremized with respect to $G\left(  \mathbf{r,}\tau;\mathbf{r}^{\prime}\mathbf{,}\tau^{\prime}\right)$, leading to  the following Dyson equation:
\begin{eqnarray}
&& \frac{\delta\Gamma}{\delta G\left(  \mathbf{r}^{\prime}\mathbf{,}\tau^{\prime
}\mathbf{;r,}\tau\right)  } = 0 \nonumber \\
&& = G^{-1}(\vr,\tau;\vr',\tau')-G_0^{-1}(\vr,\tau;\vr',\tau') \nonumber \\&& +V_{H+xc}[\rho]\delta(\vr-\vr')\delta(\tau-\tau')
\nonumber \\
&& +\sum_{\alpha\beta\in\textrm{site}}\braket{\vr|\phi_\alpha}(\Sigma^{\mathrm{DMFT}(eg)}_{\alpha\beta}-V^{DC}_{\alpha\beta})\braket{\phi_\beta|\vr'}
\nonumber\\
&&+\sum_{\gamma\delta\in\textrm{plaquette}}\braket{\vr|\phi_\gamma}\Sigma^{\mathrm{CDMFT}(x^2-y^2)}_{\gamma\delta}\braket{\phi_\delta|\vr'}
\nonumber\\
&&-\sum_{\alpha\beta\in\textrm{site}}\braket{\vr|\phi_\alpha}\Sigma^{\mathrm{DMFT}(x^2-y^2)}_{\alpha\beta}\braket{\phi_\beta|\vr'}
\label{Eq:8}
\end{eqnarray}
where
\begin{eqnarray}
&&V_{H+xc}[\rho]=\frac{\delta E^{V_C}_{H+xc}[\rho]}{\delta\rho}
\label{Eq:9}\\ 
&&\Sigma^{\textrm{DMFT}(eg)}_{\alpha\beta}=\frac{\delta\Phi^U_{\textrm{DMFT}}[G_{\textrm{loc}}^{{\rm Cu}(eg)}]}{\delta (G_{\textrm{loc}}^{{\rm Cu}(eg)})_{\beta\alpha}}
\label{Eq:10}\\ 
&&V^{DC}_{\alpha\beta}=\frac{\delta E^{U}_{H+xc}[\rho_{\textrm{loc}}^{{\rm Cu}(eg)}]}{\delta(\rho_{\textrm{loc}}^{{\rm Cu}(eg)})_{\beta\alpha}}\delta(\vr-\vr')\delta(\tau-\tau')
\label{Eq:11}\\
&&\Sigma^{\textrm{CDMFT}(x^2-y^2)}_{\gamma\delta}=\frac{\delta\Phi^U_{\textrm{CDMFT}}[G_{\textrm{plaquette}}^{{\rm Cu}(x^2-y^2)}]}{\delta (G_{\textrm{plaquette}}^{{\rm Cu}(x^2-y^2)})_{\delta\gamma}}
\label{Eq:12}
\\
&&\Sigma^{\textrm{DMFT}(x^2-y^2)}_{\alpha\beta}=\frac{\delta\Phi^U_{\textrm{DMFT}}[G_{\textrm{loc}}^{{\rm Cu}(x^2-y^2)}]}{\delta (G_{\textrm{loc}}^{{\rm Cu}(x^2-y^2)})_{\beta\alpha}}.
\label{Eq:13}
\end{eqnarray}
The $\tau,\tau'$ dependence of the self-energies has not been written explicitly. 
The first line Eq.~\eqref{Eq:9} is the DFT Hartree and exchange-correlation potential, Eq.~\eqref{Eq:10} is the single-site DMFT self-energy when both $e_g$ orbitals are considered in the quantum impurity model, Eq.~\eqref{Eq:11} is the double-counting for the single-site DMFT calculation with $e_g$ orbitals, Eq.~\eqref{Eq:12} is the cluster-DMFT self-energy on the plaquette for the Cu-$d_{x^2-y^2}$ orbital, and Eq.~\eqref{Eq:13} is the single-site DMFT self-energy for Cu-$d_{x^2-y^2}$ only.

\subsubsection{Dyson's equation in the Kohn-Sham basis}

At the stationary point, translational invariance in imaginary time and invariance under translation by a Bravais lattice vector $\mathbf{R}$ can be assumed, and the Dyson equation~\eqref{Eq:8} can be solved in any complete basis. 
It is convenient to choose the orthonormal Kohn-Sham basis, which is related to the position basis by a unitary transformation
\begin{equation}
\label{eq:ks_bas_def}
\psi^{\dag}(\mathbf{r,}\tau)=\sum_{\mathbf{k,}i}c_{\mathbf{k,}i}^{\dag
}(\tau)\left\langle \psi_{\mathbf{k},i}\right.  \left\vert \mathbf{r}\right\rangle,
\end{equation}
where $c_{\mathbf{k,}i}^{\left(\dag\right)}(\tau)$ is the creation (annihilation) operator for the Kohn-Sham state $i$ at pseudo-momentum $\mathbf{k}$ and imaginary time $\tau$.
Inserting the above into the equation for the Green's function~\eqref{eq:G_def}, one can write it down in the Kohn-Sham basis:
\begin{eqnarray}
  &&G(\vr,\tau;\vr',\tau')= \sum\limits_{\mathbf{k},i,j} \braket{\mathbf{r}|\psi_{\mathbf{k},i}} G_{ij}(\mathbf{k},\tau-\tau') \braket{\psi_{\mathbf{k},j}|\mathbf{r'}}~~~~~~ \\
  &&G_{ij}(\mathbf{k},\tau-\tau') = -\braket{\mathcal{T}_{\tau}c^{\phantom{\dag}}_{\mathbf{k},i}(\tau)c^{\dag  }_{\mathbf{k},j}(\tau')} \nonumber \\
  && = \iint d^{3}\mathbf{r}d^{3}\mathbf{r'}\braket{\psi_{\mathbf{k},i}|\mathbf{r}}G(\vr,\tau;\vr',\tau')\braket{\mathbf{r'}|\psi_{\mathbf{k},j}}
\end{eqnarray}

As mentioned above, within our approach the Luttinger-Ward functional $\Phi[G]$ is truncated to the local quantities $\Phi_{\mathrm{DMFT}}^U[G_{\mathrm{loc}}^{{\rm Cu}(eg)}]$ and $\Phi_{\mathrm{CDMFT}}^U[G_{\mathrm{plaquette}}^{{\rm Cu}(x^2-y^2)}]$, defined by the set of active correlated orbitals $\ket{\phi_{\alpha}}$ mentioned above. 
%
%
%
%
%
In this basis the Dyson equation can be evaluated by taking the matrix elements of Eq.~\eqref{Eq:8} to get 
\begin{multline}
(G^{-1}(\vk,\omega))_{ij}=(\omega+\mu-\varepsilon_{\vk i})\delta_{ij}
\\
-\sum_{\gamma\delta\in\textrm{plaquette}}
\braket{\psi_{\vk,i}|\phi_\gamma}\Sigma^{\textrm{CDMFT}(x^2-y^2)}_{\gamma\delta}(\omega)\braket{\phi_\delta|\psi_{\vk,j}}\\
-\sum_{\alpha\beta\in\textrm{site}}\braket{\psi_{\vk,i}|\phi_\alpha}
(\Sigma^{\textrm{DMFT}(eg)}_{\alpha\beta}(\omega)-V^{DC}_{\alpha\beta})\braket{\phi_\beta|\psi_{\vk,j}}\\
+\sum_{\alpha\beta\in\textrm{site}}\braket{\psi_{\vk,i}|\phi_\alpha}\Sigma^{\textrm{DMFT}(x^2-y^2)}_{\alpha\beta}(\omega)
\braket{\phi_\beta|\psi_{\vk,j}},
\label{eq:Dyson_k}
\end{multline}
where
$(G^{-1}(\vk))_{ij}=\braket{\psi_{\vk,i}|G^{-1}|\psi_{\vk,j}}$, and $\varepsilon_{\vk i}$ is the Kohn-Sham eigenvalue.
The latter solves the Kohn-Sham problem $(-\nabla^2+V_{ext}+V_{H+xc})\psi_{\vk,i}=\varepsilon_{\vk i}\psi_{\vk,i}$ where $V_{ext}$ is the ion potential. 
Note that $G_0^{-1}=\omega+\mu+\nabla^2-V_{ext}$. 
In this form of the Dyson equation, we transition from real to momentum space and from imaginary time to the real frequency domain. 

\subsubsection{The quantum impurity problem}
Using lattice translational invariance in the Dyson's equation for the lattice Green's function in momentum space~\eqref{eq:Dyson_k}, it is clear that the calculation of the single site copper self-energy, or cluster of copper self-energy, is identical for all copper atoms or copper clusters. 

The precise definition of the local orbitals is given later, and we rather focus here on the definition of the local Green's functions $G_{\mathrm{loc}}^{{\rm Cu}(eg)}$ and $G_{\mathrm{plaquette}}^{{\rm Cu}(x^2-y^2)}$ that enter the site or cluster self-consistency equation. 
They can be obtained by performing a projection onto the local orbitals as follows:
\begin{eqnarray}
\label{eq:proj_def}
    && \left(G_{X}(\omega)\right)_{\alpha\beta} \equiv  \mathbb{P}(G) = \sum\limits_{\mathbf{k}}G_{\alpha\beta}(\mathbf{k},\omega) \nonumber \\
    && =  \sum\limits_{\mathbf{k},i,j} \braket{\phi_{\alpha}|\psi_{\mathbf{k,}i}}G_{ij}(\mathbf{k},\omega)\braket{\psi_{\mathbf{k,}j}|\phi_{\beta}} 
    \label{eq:proj_def_2}
\end{eqnarray}
The projection $\mathbb{P}$ is performed either on a single-site ($X=\mathrm{loc}$) or cluster of atoms ($X=\mathrm{plaquette}$), depending on the restriction in space applied to the compound indices $(\alpha,\beta)$.
Here, $(\alpha,\beta)$ refer to a single copper atom or cluster in the impurity problem. 
The sum over wave vectors is over the whole Brillouin zone in the case of the single-site problem, and over the reduced Brillouin zone in the case of the plaquette.
As mentioned above, within (C)DMFT these local Green's functions are to be identified with the Green's function of corresponding quantum (cluster) impurity problems, the details of which are presented later in this section.


The Dyson's equation for the lattice Green's function in momentum space~\eqref{eq:Dyson_k} shows that quantities defined in the local basis, and computed by solving the quantum impurity or quantum cluster problem, need to be \textit{embedded} back into the complete Kohn-Sham basis using
\begin{equation}
\label{eq:embed_def}
\Sigma_{ij}(\vk,\omega) = \sum_{\alpha,\beta} 
\braket{\psi_{\vk,i}|\phi_{\alpha}}\Sigma_{\alpha\beta}(\omega)
\braket{\phi_{{\beta}}|\psi_{\vk,j}} \equiv \mathbb{E}(\Sigma).
\end{equation}
%
%
$\mathbb{E}$ defines the embedding operation while indices $i,j$ run over the Kohn-Sham states, and Greek letters $\alpha,\beta$ over local orbitals.
Note that only the \emph{self-energy} is embedded, not the kinetic energy nor the Hartree and exchange-correlation potentials, as only the dynamic correlations are approximated as being local to the site or to the cluster of sites.

\subsubsection{Construction of the local orbitals}

The orbitals $\braket{\vr|\phi_{\vk,\alpha}}$ that are built from correlated copper orbitals that satisfy Bloch's theorem are chosen to be the quasi-atomic orbitals, centered on the atom located at $\mathbf{R}$ and with orbital indices $L=\{l,m\}$. 
They are called quasi-atomic since they retain much of the character of atomic orbitals such as shape and angular momentum characteristics, and they are localized around an atom or atomic center (copper in our case).
Yet, in contrast to purely atomic orbitals, they are influenced by their surroundings and are cut off at the muffin-tin radius. 
The orbitals $\braket{\vr|\phi_{\vk,\alpha}}$ are constructed from the radial solution of the Schr\"odinger equation $u_l(r)$ 
within a given muffin-tin sphere,  with energy chosen to be the Fermi energy. 
By adding the cubic harmonics $Y_L$, we arrive at the following functions that satisfy Bloch's theorem:
$$\braket{\vr|\widetilde{\phi}_{\vk,\alpha}}=u_l(|\vr-\mathbf{R}|)Y_{L}(\vr-\mathbf{R}) e^{i\vk\mathbf{R}}.$$
Note again that $\alpha$ is combined index for $L$, $\textbf{R}$, spin, and we multiplied by the proper phase-factor.
From these functions, we construct our orthonormal correlated orbitals as
\begin{eqnarray}
\ket{\phi_{\vk,\alpha}}=\sum_{\alpha\p}\ket{\widetilde{\phi}_{\vk,\alpha'}}
\left(\frac{1}{\sqrt{O}}\right)_{\alpha\p,\alpha}
\end{eqnarray}
where the overlap is local
\begin{eqnarray}
O_{\alpha,\alpha\p}=\sum_{i,\vk}\braket{\widetilde{\phi}_{\vk,\alpha}|\psi_{\vk,i}}\braket{\psi_{\vk,i}|\widetilde{\phi}_{{\vk,\alpha\p}}}.
\end{eqnarray}
Here, $\ket{\psi_{\vk,i}}$ is the complete set of Kohn-Sham orbitals with band index $i$. 

Note that the overlap matrix $O$ is unity if we sum over the entire range of bands that have finite overlap with the chosen correlated orbitals $\braket{\vr|\phi_{\vk,\alpha}}$, because $\ket{\psi_{\vk,i}}$ is a complete basis, and because the atomic orbitals are orthogonal. 
However, in practice the sum over $i$ is always truncated. 
We choose a very large window of bands ($-15$eV,$15$eV), so that very little weight of the relevant orbitals falls outside, and  
hence $O\approx 1$ and $\widetilde{\phi}(\textbf{r})\approx \phi(\textbf{r})$. 
Nevertheless, we need to properly normalize the correlated orbitals $\braket{\vr|\phi_{\vk,\alpha}}$ so that the local basis is exactly orthonormal. 

Also note that embedding in eDMFT is very different from Wannier orbital construction, because in the later, the renormalization by overlap is done at each momentum point separately.
Therefore, at some momentum points such overlap may be singular, requiring singular value decomposition of the overlap. 
In eDMFT, the locality of the overlap ensures that it is never singular, and always rather close to unity.

\textcolor{red}{\centering ----END AMT Version of method----}
\end{versionAMT}

\begin{versionBBL}

\subsubsection{Extremization of the functional and Dyson's equation}

In the following, we describe in detail the algorithm used to extremize the above functional (Eq.~\eqref{LTW}).
We first explain how the extremization can be applied to the restricted Luttinger-Ward functional $\Phi[G_{\rm loc/plaquette}^{\mathrm{Cu}(eg/x^2-y^2)}]$. 
To this end, we recall that $\Gamma[G]$ is a functional of the Green's function
\begin{equation}
\label{eq:G_def}
G(\vr,\tau;\vr',\tau') = -\braket{\mathcal{T}_{\tau}\psi(\mathbf{r,}\tau)\psi^{\dag}(\mathbf{r',}\tau')},
\end{equation}
where $\mathcal{T}_{\tau}$ is the time-ordering operator along imaginary time
$\tau,$ $\psi^{\left(  \dag\right)  }(\mathbf{r,}\tau)$ the creation
(annihilation) operator in an eigenstate of position and the average
$\left\langle {.}\right\rangle $ is a quantum mechanical many-body trace with
a density matrix appropriate for the grand canonical ensemble but that also contains a source field that allows arbitrary variations of the function $G\left(  \mathbf{r,}\tau;\mathbf{r}^{\prime}\mathbf{,}\tau^{\prime}\right)$. 
We do not write spin indices explicitly to simplify the notation.

In order to truncate the Luttinger-Ward functional, one needs to pick a set of local orbitals $\ket{\phi_{\alpha}}$, which are here all centered on Cu atoms.
The compact index $\alpha$ (or $\beta$, or $\gamma$, or $\delta$) stands for the orbital index $L=\{l,m\}$, the spin index, and the atomic position $\mathbf{R}$ at which the orbital is centered.

The detailed definition of these orbitals in practice is provided later, and we focus here on the definition of the local Green's function to which is restricted the (C)DMFT Luttinger-Ward functional: 
\begin{eqnarray}
    && G^{Y}_{X}(\tau_1,\tau_2)_{\alpha,\beta} \equiv \mathbb{P}(G)  \nonumber \\
    && = \iint d^3\vr_1d^3\vr_2 \braket{\phi_{\alpha}|\vr_1}G(\vr_1,\tau_1;\vr_2,\tau_2)\braket{\vr_2|\phi_{\beta}}.
    \label{eq:def_gloc}
\end{eqnarray}
The projection $\mathbb{P}$ is performed either on a single-site ($X=\mathrm{loc}$) or cluster of atoms ($X=\mathrm{plaquette}$), and on the Cu-$eg$ orbitals ($Y=eg$) or Cu-\dxxyy~($Y=x^2-y^2$), depending on the restriction in space and orbital applied to the compound indices $(\alpha,\beta)$.

Based on this definition (Eq.~\eqref{eq:def_gloc}), we may differentiate the restricted Luttinger-Ward functionals by considering the following chain rule
\begin{eqnarray}
   && \frac{\delta\Phi^U_{\textrm{(C)DMFT}}[G_X^Y]}{\delta G\left(  \vr,^{\prime}\tau^{\prime};\vr,\tau\right)}  \nonumber \\
   &&=\sum_{\alpha \beta} \frac{\delta\Phi^U_{\textrm{(C)DMFT}}[G_X^Y]}{\delta G_X^Y(\tau_1,\tau_2)_{\beta\alpha}}\frac{\delta G_X^Y(\tau_1,\tau_2)_{\beta\alpha}}{\delta G\left(  \vr^{\prime},\tau^{\prime};\vr,\tau\right)} \nonumber \\
   && = \sum_{\alpha \beta} \braket{\vr|\phi_\alpha} \frac{\delta\Phi^U_{\textrm{(C)DMFT}}[G_X^Y]}{\delta G_X^Y(\tau^{\prime},\tau)_{\beta\alpha}} \braket{\phi_\beta|\vr^{\prime}}.
   \label{eq:phi_diff}
\end{eqnarray}
In principle, the above chain rule should involve the sum over a complete basis of local orbitals.
The (C)DMFT restriction enforces that the derivative of $\Phi^U_{\textrm{(C)DMFT}}[G_X^Y]$ is non-zero only for the active correlated subspace. 

The Kadanoff-Baym functional of Eq.~\eqref{LTW} can now easily be extremized with respect to $G\left( \vr^{\prime}\tau^{\prime};\vr,\tau\right)$, leading to the following Dyson equation:
\begin{eqnarray}
&& \frac{\delta\Gamma}{\delta G\left(  \mathbf{r}^{\prime}\mathbf{,}\tau^{\prime
}\mathbf{;r,}\tau\right)  } = 0 \nonumber \\
&& = G^{-1}(\vr,\tau;\vr',\tau')-G_0^{-1}(\vr,\tau;\vr',\tau') \nonumber \\&& +V_{H+xc}[\rho]\delta(\vr-\vr')\delta(\tau-\tau')
\nonumber \\
&& +\sum_{\alpha\beta\in\textrm{sites}}\braket{\vr|\phi_\alpha}(\Sigma^{\mathrm{DMFT}(eg)}_{\alpha\beta}-V^{DC}_{\alpha\beta})\braket{\phi_\beta|\vr'}
\nonumber\\
&&+\sum_{\gamma\delta\in\textrm{plaquettes}}\braket{\vr|\phi_\gamma}\Sigma^{\mathrm{CDMFT}(x^2-y^2)}_{\gamma\delta}\braket{\phi_\delta|\vr'}
\nonumber\\
&&-\sum_{\alpha\beta\in\textrm{sites}}\braket{\vr|\phi_\alpha}\Sigma^{\mathrm{DMFT}(x^2-y^2)}_{\alpha\beta}\braket{\phi_\beta|\vr'}
\label{Eq:8}
\end{eqnarray}
where
\begin{eqnarray}
&&V_{H+xc}[\rho]=\frac{\delta E^{V_C}_{H+xc}[\rho]}{\delta\rho}
\label{Eq:9}\\ 
&&\Sigma^{\textrm{DMFT}(eg)}_{\alpha\beta}=\frac{\delta\Phi^U_{\textrm{DMFT}}[G_{\textrm{loc}}^{{\rm Cu}(eg)}]}{\delta (G_{\textrm{loc}}^{{\rm Cu}(eg)})_{\beta\alpha}}
\label{Eq:10}\\ 
&&V^{DC}_{\alpha\beta}=\frac{\delta E^{U}_{H+xc}[\rho_{\textrm{loc}}^{{\rm Cu}(eg)}]}{\delta(\rho_{\textrm{loc}}^{{\rm Cu}(eg)})_{\beta\alpha}}\delta(\vr-\vr')\delta(\tau-\tau')
\label{Eq:11}\\
&&\Sigma^{\textrm{CDMFT}(x^2-y^2)}_{\gamma\delta}=\frac{\delta\Phi^U_{\textrm{CDMFT}}[G_{\textrm{plaquette}}^{{\rm Cu}(x^2-y^2)}]}{\delta (G_{\textrm{plaquette}}^{{\rm Cu}(x^2-y^2)})_{\delta\gamma}}
\label{Eq:12}
\\
&&\Sigma^{\textrm{DMFT}(x^2-y^2)}_{\alpha\beta}=\frac{\delta\Phi^U_{\textrm{DMFT}}[G_{\textrm{loc}}^{{\rm Cu}(x^2-y^2)}]}{\delta (G_{\textrm{loc}}^{{\rm Cu}(x^2-y^2)})_{\beta\alpha}}.
\label{Eq:13}
\end{eqnarray}
The $\tau,\tau'$ dependence of the self-energies has not been written explicitly. 
The first line Eq.~\eqref{Eq:9} is the DFT Hartree and exchange-correlation potential, Eq.~\eqref{Eq:10} is the single-site DMFT self-energy when both $e_g$ orbitals are considered in the quantum impurity model, Eq.~\eqref{Eq:11} is the double-counting for the single-site DMFT calculation with $e_g$ orbitals, Eq.~\eqref{Eq:12} is the cluster-DMFT self-energy on the plaquette for the Cu-$d_{x^2-y^2}$ orbital, and Eq.~\eqref{Eq:13} is the single-site DMFT self-energy for Cu-$d_{x^2-y^2}$ only.

\subsubsection{Dyson's equation in the Kohn-Sham basis}

At the stationary point, translational invariance in imaginary time and invariance under translation by a Bravais lattice vector $\mathbf{R}$ can be assumed.
The Dyson equation~\eqref{Eq:8} can be solved in any complete basis.
It is convenient to choose the orthonormal Kohn-Sham basis, which is related to the position basis by a unitary transformation
\begin{equation}
\label{eq:ks_bas_def}
\psi^{\dag}(\mathbf{r,}\tau)=\sum_{\mathbf{k,}i}c_{\mathbf{k,}i}^{\dag
}(\tau)\left\langle \psi_{\mathbf{k},i}\right.  \left\vert \mathbf{r}\right\rangle,
\end{equation}
where $c_{\mathbf{k,}i}^{\left(\dag\right)}(\tau)$ is the creation (annihilation) operator for the Kohn-Sham state $i$ at pseudo-momentum $\mathbf{k}$ and imaginary time $\tau$.
Inserting the above into the equation for the Green's function~\eqref{eq:G_def}, one can write it down in the Kohn-Sham basis:
\begin{eqnarray}
  &&G(\vr,\tau;\vr',\tau')= \sum\limits_{\mathbf{k},i,j} \braket{\mathbf{r}|\psi_{\mathbf{k},i}} G_{ij}(\mathbf{k},\tau-\tau') \braket{\psi_{\mathbf{k},j}|\mathbf{r'}}~~~~~~ \\
  &&G_{ij}(\mathbf{k},\tau-\tau') = -\braket{\mathcal{T}_{\tau}c^{\phantom{\dag}}_{\mathbf{k},i}(\tau)c^{\dag  }_{\mathbf{k},j}(\tau')} \nonumber \\
  && = \iint d^{3}\mathbf{r}d^{3}\mathbf{r'}\braket{\psi_{\mathbf{k},i}|\mathbf{r}}G(\vr,\tau;\vr',\tau')\braket{\mathbf{r'}|\psi_{\mathbf{k},j}}
  \label{eq:g_ks}
\end{eqnarray}

The Dyson equation can be evaluated by inserting~\eqref{eq:g_ks} into~\eqref{Eq:8} to get 
\begin{multline}
(G^{-1}(\vk,\omega))_{ij}=(\omega+\mu-\varepsilon_{\vk i})\delta_{ij}
\\
-\sum_{\gamma\delta\in\textrm{plaquette}}
\braket{\psi_{\vk,i}|\phi_\gamma}\Sigma^{\textrm{CDMFT}(x^2-y^2)}_{\gamma\delta}(\omega)\braket{\phi_\delta|\psi_{\vk,j}}\\
-\sum_{\alpha\beta\in\textrm{site}}\braket{\psi_{\vk,i}|\phi_\alpha}
(\Sigma^{\textrm{DMFT}(eg)}_{\alpha\beta}(\omega)-V^{DC}_{\alpha\beta})\braket{\phi_\beta|\psi_{\vk,j}}\\
+\sum_{\alpha\beta\in\textrm{site}}\braket{\psi_{\vk,i}|\phi_\alpha}\Sigma^{\textrm{DMFT}(x^2-y^2)}_{\alpha\beta}(\omega)
\braket{\phi_\beta|\psi_{\vk,j}},
\label{eq:Dyson_k}
\end{multline}
where $\varepsilon_{\vk i}$ are the Kohn-Sham eigenvalues.
They solve the Kohn-Sham problem $(-\nabla^2+V_{ext}+V_{H+xc})\psi_{\vk,i}=\varepsilon_{\vk i}\psi_{\vk,i}$ where $V_{ext}$ is the ion potential. 
Note that $G_0^{-1}=\omega+\mu+\nabla^2-V_{ext}$. 
In this form of the Dyson equation, we transition from real to momentum space and from imaginary time to the real frequency domain.

Solving the above Dyson's equation~\eqref{eq:Dyson_k} requires the knowledge of the self-energies, which are found from the solution of effective quantum (cluster) impurity problems, the details of which are provided later.
Eq.~\eqref{eq:Dyson_k} defines how these local self-energies need to be \textit{embedded} back into the complete Kohn-Sham basis using
\begin{equation}
\label{eq:embed_def}
\Sigma_{ij}(\vk,\omega) = \sum_{\alpha,\beta} 
\braket{\psi_{\vk,i}|\phi_{\alpha}}\Sigma_{\alpha\beta}(\omega)
\braket{\phi_{{\beta}}|\psi_{\vk,j}} \equiv \mathbb{E}(\Sigma).
\end{equation}
$\mathbb{E}$ defines the embedding operation while indices $i,j$ run over the Kohn-Sham states, and Greek letters $\alpha,\beta$ over local orbitals.
Note that only the \emph{self-energy} is embedded, not the kinetic energy nor the Hartree and exchange-correlation potentials, as only the dynamic correlations are approximated as being local to the site or to the cluster of sites.

The projection operation $\mathbb{P}$, defined in Eq.~\eqref{eq:def_gloc}, can also be written within the Kohn-Sham basis as
\begin{eqnarray}
    G^{Y}_{X}(\omega)_{\alpha,\beta} && = \mathbb{P}(G) \nonumber \\
    && = \sum_{\vk}G^{Y}_{X}(\mathbf{k},\omega)_{\alpha,\beta} \nonumber\\
    && = \sum_{\vk,i,j} \braket{\phi_\alpha|\psi_{\vk,i}}G_{ij}(\vk,\omega)\braket{\psi_{\vk,j}|\phi_\beta}. ~~~
    \label{eq:proj_def_2}
\end{eqnarray}

\subsubsection{Construction of the local orbitals}

We now detail the construction of the set of active correlated orbitals $\ket{\phi_\alpha}$.  
They are chosen to be quasi-atomic orbitals and constructed from the radial solution of the Schr\"odinger equation $u_l(r)$ 
within a given muffin-tin sphere centered at $\mathbf{R}$, with energy chosen to be the Fermi energy. 
They are called quasi-atomic since they retain much of the character of atomic orbitals such as shape and angular momentum characteristics, and they are localized around an atom or atomic center (copper in our case).
Yet, in contrast to purely atomic orbitals, they are influenced by their surroundings and are cut at the muffin-tin radius. 
By adding the cubic harmonics $Y_L$, we arrive at the following functions:
$$\braket{\vr|\widetilde{\phi}_{\alpha}}=u_l(|\vr-\mathbf{R}|)Y_{L}(\vr-\mathbf{R}).$$
Note again that $\alpha$ is combined index for $L$, $\textbf{R}$, and spin.
%
%

From these functions, we construct our orthonormal correlated orbitals as
\begin{eqnarray}
\ket{\phi_{\alpha}}=\sum_{\alpha\p}\ket{\widetilde{\phi}_{\alpha'}}
\left(\frac{1}{\sqrt{O}}\right)_{\alpha\p,\alpha}
\end{eqnarray}
where the overlap is local
\begin{eqnarray}
O_{\alpha,\alpha\p}=\sum_{i,\vk}\braket{\widetilde{\phi}_{\alpha}|\psi_{\vk,i}}\braket{\psi_{\vk,i}|\widetilde{\phi}_{{\alpha\p}}}.
\end{eqnarray}
Here, $\ket{\psi_{\vk,i}}$ is the complete set of Kohn-Sham orbitals with band index $i$, but the sum over bands is restricted to a given energy window. 

Note that the overlap matrix $O$ is unity if we sum over the entire range of bands that have finite overlap with the chosen correlated orbitals $\ket{\widetilde{\phi}_{\alpha}}$, because $\ket{\psi_{\vk,i}}$ is a complete basis, and because the atomic orbitals are orthonormal. 
However, in practice the sum over $i$ is always truncated. 
We choose a very large window of bands ($-15$eV,$15$eV), so that very little weight of the relevant orbitals falls outside, and  
hence $O\approx 1$ and $\widetilde{\phi}(\textbf{r})\approx \phi(\textbf{r})$. 
Nevertheless, we need to properly normalize the correlated orbitals $\braket{\vr|\phi_{\vk,\alpha}}$ so that the local basis is exactly orthonormal, otherwise the high-frequency tails are inaccurate. 

\begin{figure*}
    \centering
    \includegraphics[width=0.75\linewidth]{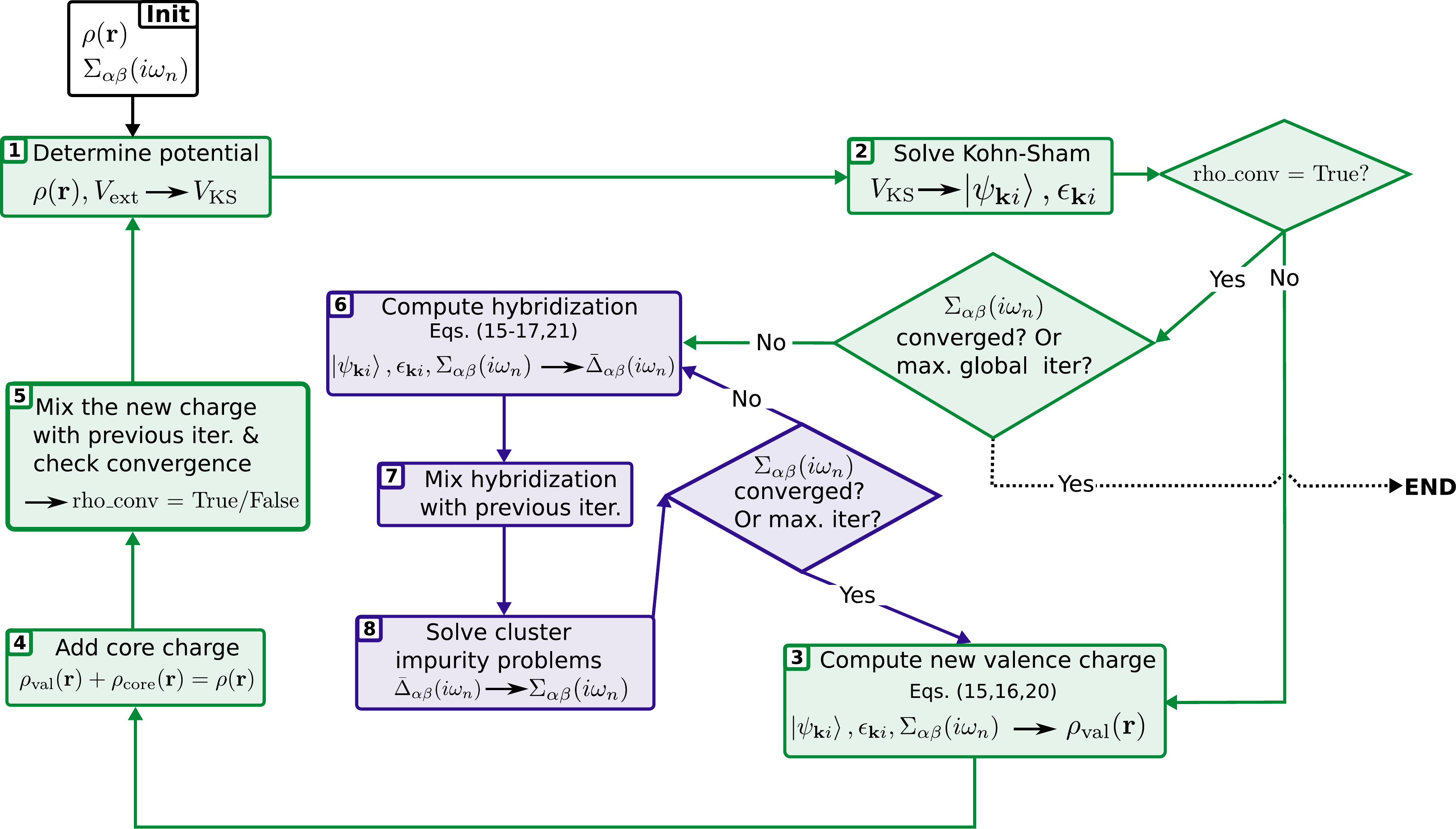}
    \caption{Flowchart of the charge self-consistent DFT+CDMFT algorithm.
    Self-consistency is reached when both $\rho(\vr)$ and $\Sigma_{\alpha,\beta}(i\omega_n)$ are converged. 
    The numbering follows the steps described in the text. 
    The DFT (CDMFT) subloop is highlighted in green (purple). 
    Note that superconductivity is treated in a separate post-processing procedure (see Sec.~\ref{sec:met_sc}).  
    }
    \label{fig:fig_flowchart}
\end{figure*}

Note also that embedding in eDMFT is very different from Wannier orbital construction, because in the latter the renormalization by overlap is done at each momentum point separately.
Since this overlap may be singluar at some momentum points, a singular value  decomposition is required.
By contrast, in eDMFT, the locality of the overlap ensures that it is never singular, and always rather close to unity.
Moreover, within eDMFT, the Hilbert space in which the kinetic energy is represented is spanned by the original Kohn-Sham basis, i.e., it is not truncated as in the Wannier construction. 
Hence optimizing Wannier orbitals to faithfully represent bands is not needed. 
Consequently, the Coulomb interaction is applied to local orbitals, which are more localized than maximally localized Wannier orbitals.

\end{versionBBL}


\subsubsection{Iterative procedure and hybridization function}
\label{sec:iter_proc}

The iterative procedure to extremize the Luttinger-Ward functional is illustrated in Fig.~\ref{fig:fig_flowchart}. 
It proceeds in the following way:
\\
\\
\emph{Starting point}
\begin{enumerate}
    \setcounter{enumi}{-1}
    \item Start with an approximate self-energy of the quantum impurity and quantum cluster problems, and an approximate (often DFT-only) charge density of the solid. 
\end{enumerate}
\emph{DFT subloop: fixed self-energy (green boxes)}
\begin{enumerate}
    \item Determine the Hartree and exchange-correlation potential from the new charge density, and obtain the Kohn-Sham potential via
    $$
    V_{\rm KS} = V_{H+xc} + V_{\rm ext},
    $$
    where $V_{\rm ext}$ contains the ionic potential. 
    Compute also the eDMFT double-counting potential $V_{\alpha\beta}^{DC}$ by integrating the screened exchange-correlation potential of the eDMFT charge density over the unit cell, and project it to the relevant orbitals. 
    \item Solve the Kohn-Sham problem for its eigen functions $\ket{\psi_{\vk,i}}$ and eigen energies $\varepsilon_{\vk i}$:
    $$
    \left[-\nabla^{2}+V_{\rm KS}\right]\ket{\psi_{\vk,i}} = \varepsilon_{\vk i}\ket{\psi_{\vk,i}},
    $$ which determine the computational basis.
\end{enumerate}

At this stage, if the charge density is not converged, proceed to step 3. 
If it is converged there are two options: (i) if the self-energy is not converged, jump to step 6; (iii) if the self-energy is converged, or the maximum number of global iterations is reached, end the calculation.
%
\\
\\

\begin{enumerate}
\setcounter{enumi}{2}
    \item Embed the self-energy following Eq.~\eqref{eq:embed_def} and solve the Dyson Eq.~\eqref{eq:Dyson_k} to obtain the Green's function $G_{ij}(\vk,\omega)$, expressed in complete Kohn-Sham basis. 
    Then compute the new valence charge density $\rho_{\rm val}(\vr)$ of the solid via: 
    \begin{equation}
    \rho_{\rm val}(\vr)=\frac{1}{\beta}\sum_{i\omega}\sum_{\vk,ij}    \braket{\vr|\psi_{\vk,i}}G_{ij}    (\vk,i\omega)\braket{\psi_{\vk,j}|\vr}.
    \label{Eq:charge}
    \end{equation}
    The valence contribution to the total energy and the total free energy of the solid is also computed. 
    \item Get the total charge density by adding the core contribution
    $$
    \rho(\vr) = \rho_{\rm val}(\vr) + \rho_{\rm core}(\vr)
    $$
    The core contribution to the total energy and total free energy is added.
    \item Mix the new charge density with that of previous iterations, compute the total energy and free energy of the solid.
    Go back to step 1. 
    
\end{enumerate}

\emph{CDMFT subloop: fixed charge density (purple boxes)}
\begin{enumerate}
\setcounter{enumi}{5}
    
    \item Embed the self-energy following Eq.~\eqref{eq:embed_def}, compute the local Green's function with Eq.~\eqref{eq:proj_def_2}, and solve the Dyson Eq.~\eqref{eq:Dyson_k} to obtain the impurity hybridization functions $\Bar{\Delta}$ and the cluster impurity levels $\epsilon$ of each impurity problem via:
    \begin{equation}
    \Bar{\Delta}_{\alpha\beta}(\omega) + \epsilon_{\alpha\beta} = \omega - \Sigma_{\alpha\beta}(\omega) - \left(G^{-1}_{\textrm{loc}}(\omega)\right)_{\alpha\beta}.
\end{equation}

    \item  Mix the new hybridization function with the existing one.

    \item Solve the quantum impurity problems to obtain new single-site and plaquette self-energies $\Sigma^{\mathrm{DMFT}}_{\alpha\beta}(\omega)$ and $\Sigma^{\mathrm{CDMFT}}_{\gamma\delta}(\omega)$
\end{enumerate}
If the number of CDMFT steps for a given charge density is reached, return to step 3.
Else, go back to step 6. 
\\
\\

The charge self-consistency is stopped when the charge density and self-energy are converged. 
For the sake of readability, we skip here some technical details such as the method used to find the chemical potential, to integrate over momentum, etc. 
The interested reader can find all these information in the exhaustive descriptions of Refs.~\onlinecite{haule2010,haule2018}.
We finally draw readers' attention to the fact that the charge self-consistency is performed on the Matsubara axis. 
Since the ED solver effectively works at zero temperature, we set a fictitious inverse temperature $\beta$ only to define a mesh of Matsubara frequencies. 

\subsubsection{Choice of correlated subspace}

In the context of our study of multilayer cuprates, it is sufficient to treat as correlated degrees of freedom the Cu-$e_g$ orbitals, namely \dxxyy~and \dzz. 
The Cu-\dzz~orbital is usually neglected in single- and three-band effective models. 
However, in ab-initio calculations, it appears relatively close to the Fermi level. 
As a result in hole doped system it can be pushed very close to EF. 
Therefore, we decided to include it in the correlated subspace.
The $t_{2g}$ orbitals are located further due to  strong crystal field splitting and can thus be treated solely at the DFT level.

Each CuO$_2$ plane is associated with two impurity problems as depicted in Fig.~\ref{fig:fig1}(c): (i) a $2\times2$ cluster of single Cu-\dxxyy~orbitals (providing $\Phi^U_{\rm CDMFT}[G_{\rm plaquette}^{{\rm Cu}(x^2-y^2)}]$ in Eq.~\eqref{LTW}), and (ii) a two-orbital single-site impurity problem involving both \dxxyy~and \dzz~(providing $\Phi^U_{\rm DMFT}[G_{\rm loc}^{{\rm Cu}(eg)}]$ in Eq.~\eqref{LTW}).
The latter allows a better treatment of the \dzz~orbital by taking into account the Coulomb interaction with \dxxyy. 

We use the Slater form of the Coulomb repulsion as follows:
\begin{equation}
\begin{split}
    \hat{U} = & \frac{1}{2}\bigg[
    \sum_{\substack{m_\alpha,m_\beta \\ s,s'}} U_{\alpha\beta\beta\alpha}\phi^{\dagger}_{m_\alpha,s}\phi^{\dagger}_{m_\beta,s'}\phi^{\phantom{\dagger}}_{m_\beta,s'}\phi^{\phantom{\dagger}}_{m_\alpha,s} \\
    & + \sum_{\substack{m_\alpha,m_\beta \\ m_\alpha\neq m_\beta \\ s,s'}} U_{\alpha\beta\alpha\beta}\phi^{\dagger}_{m_\alpha,s}\phi^{\dagger}_{m_\beta,s'}\phi^{\phantom{\dagger}}_{m_\alpha,s'}\phi^{\phantom{\dagger}}_{m_\beta,s}
    \bigg],
\end{split}
\end{equation}
where $m_{\alpha/\beta}$, $s/s'$ denote explicitly the orbital and spin indices. 
The Coulomb matrix elements are defined by: 
\begin{equation}
    U_{\alpha\beta\gamma\delta} = \sum_{m,k} \frac{4\pi \mathrm{F}^{k}}{2k+1}\braket{Y_{lm_\alpha}|Y_{km}|Y_{lm_\delta}} \braket{Y_{lm_\beta}|Y^{*}_{km}|Y_{lm_\gamma}},
\nonumber
\end{equation}
and the Slater integrals $\mathrm{F}^{k}$ are related to more commonly quoted parameters $U$ and $J_{\rm Hund}$ by: 
\begin{equation}
\begin{split}
    & \mathrm{F}^{0} = U, \\
    & \mathrm{F}^{2} = \frac{112}{13}J_{\rm Hund}, \\
    & \mathrm{F}^{4} = \frac{70}{13}J_{\rm Hund}. 
\end{split}
\end{equation}
Note that the cluster impurity problem contains only a single orbital per site, therefore terms beyond Hubbard repulsion do not play an important role in this step. 

To avoid the double counting of the local \dxxyy~self-energy, we need to subtract the single-site DMFT self-energy of the \dxxyy~orbital (see Eq.~\eqref{eq:Dyson_k}), keeping only the cluster impurity contribution for this orbital which incorporates the the non-local (inter-site) correlations on the plaquette. 
For each CuO$_2$ plane, the self-energy can thus be written as: 
\begin{equation}
\begin{split}
    &\Sigma = \\
    &\stackrel{d_{x^{2}-y^{2}}^{1} ~~~ d_{x^{2}-y^{2}}^{2} ~~~ d_{x^{2}-y^{2}}^{3} ~~~ d_{x^{2}-y^{2}}^{4} ~~~ d_{z^{2}}^{1} ~~~ d_{z^{2}}^{2} ~~~ d_{z^{2}}^{3} ~~~ d_{z^{2}}^{4}}{
    \begin{pmatrix}
        \begin{blockarray}{cccccccc}
        \begin{block}{cccc|cccc}
         \Sigma_{x^2-y^2}^{(a)} & \Sigma_{x^2-y^2}^{(b)} & \Sigma_{x^2-y^2}^{(b)} & \Sigma_{x^2-y^2}^{(c)} & 0 & 0 & 0 & 0  \\
         \Sigma_{x^2-y^2}^{(b)} & \Sigma_{x^2-y^2}^{(a)} & \Sigma_{x^2-y^2}^{(c)} & \Sigma_{x^2-y^2}^{(b)} & 0 & 0 & 0 & 0 \\
         \Sigma_{x^2-y^2}^{(b)} & \Sigma_{x^2-y^2}^{(c)} & \Sigma_{x^2-y^2}^{(a)} & \Sigma_{x^2-y^2}^{(b)} & 0 & 0 & 0 & 0 \\
         \Sigma_{x^2-y^2}^{(c)} & \Sigma_{x^2-y^2}^{(b)} & \Sigma_{x^2-y^2}^{(b)} & \Sigma_{x^2-y^2}^{(a)} & 0 & 0 & 0 & 0 \\
        \end{block}
         \hline
         \begin{block}{cccc|cccc}
         0 & 0 & 0 & 0 & \Sigma_{z^2}^{(a)} & 0 & 0 & 0 \\
         0 & 0 & 0 & 0  & 0 & \Sigma_{z^2}^{(a)} & 0 & 0 \\
         0 & 0 & 0 & 0& 0 & 0 & \Sigma_{z^2}^{(a)} & 0  \\
         0 & 0 & 0 & 0 & 0 & 0 & 0 & \Sigma_{z^2}^{(a)}  \\
       \end{block}
        \end{blockarray}
    \end{pmatrix}
    }
\end{split}
\end{equation}
in the $\left\{d^{i}_{L}\right\}$ basis  where $i=1,\dots,4$ are the cluster sites (see the numbering convention in Fig.~\ref{fig:fig1}(c)), and $L$ denotes the orbital. 
The $\Sigma_{L}^{(a)}$, $\Sigma_{L}^{(b)}$ and $\Sigma_{L}^{(c)}$ are respectively the local, nearest-neighbour and next-nearest-neighbour components.

\subsubsection{Double counting}
The double-counting $V^{DC}(\vr)$ is the Hartree and exchange-correlation potential of the impurity problem~\cite{haule2015}, and is local to a point in 3D space.
Since the local orbitals $\phi_\alpha(\vr)$ are localized within the muffin-tin sphere of the correlated atom centered at $\mathbf{R}$, and since we only consider the local interaction $U$, the double-counting remains the same as in DFT+eDMFT and has no inter-site terms. 
Thereby, the double-counting per CuO$_2$ plane may be written as: 
\begin{equation}
\begin{split}
    &V^{DC} = \\
   & \hspace*{-0.4cm}\begin{pmatrix}
        \begin{blockarray}{cccccccc}
        \begin{block}{cccc|cccc}
         V^{DC}_{x^2-y^2} & 0 & 0 & 0 & 0 & 0 & 0 & 0\\
         0 & V^{DC}_{x^2-y^2} & 0 & 0 & 0 & 0 & 0 & 0 \\
         0 & 0 & V^{DC}_{x^2-y^2} & 0 & 0 & 0 & 0 & 0 \\
         0 & 0 & 0 & V^{DC}_{x^2-y^2} & 0 & 0 & 0 & 0 \\
        \end{block}
         \hline
         \begin{block}{cccc|cccc}
         0 & 0 & 0 & 0 & V^{DC}_{z^2} & 0 & 0 & 0 \\
         0 & 0 & 0 & 0 & 0 & V^{DC}_{z^2} & 0 & 0 \\
         0 & 0 & 0 & 0 & 0 & 0 & V^{DC}_{z^2} & 0  \\
         0 & 0 & 0 & 0 & 0 & 0 & 0 & V^{DC}_{z^2}   \\
       \end{block}
        \end{blockarray}
    \end{pmatrix},
\end{split}
\end{equation}
where the individual $V^{DC}_{L}$ components are evaluated using the exact double-counting scheme of Ref.~\onlinecite{haule2015}.
This method involves the evaluation of the exchange-correlation energy of the homogeneous electron gas with a screened interaction $V_{\mathrm{DMFT}}(\mathbf{r},\mathbf{r}\p)$, representing the screened local Hubbard $U$ and Hunds interaction $J_{\mathrm{Hund}}$ used within the impurity problems. 
$V_{\mathrm{DMFT}}(\mathbf{r},\mathbf{r}\p)$ can be modeled in various ways, and is implemented in practice as a mixture of Yukawa and dielectric screening~\cite{haule2015}:
\begin{equation}
    V_{\mathrm{DMFT}}(\mathbf{r},\mathbf{r}\p) = \frac{\mathrm{e}^{-\lambda\lvert \mathbf{r} - \mathbf{r}\p \rvert}}{\epsilon\lvert\mathbf{r}-\mathbf{r}\p\rvert},
\end{equation}
where $\lambda$ and $\epsilon$ are constants that can be uniquely determined from the value of $U$ and $J_{\mathrm{Hund}}$. 
Since the undoped \nC~compounds are insulating the Yukawa screening should be negligible, and we hence model it by dielectric screening only (set $\lambda=0$).
%
In all other calculations in which the system is metallic we use the mixed dielectric and Yukawa screening representation.

\subsubsection{Periodization}

A well-known drawback of the real-space cluster approach is that it breaks translational symmetry~\cite{maier2005a, LTP:2006, kotliar2006a}.
However, during charge self-consistent calculations, this does not pose a problem, as we work directly within the superlattice Brillouin zone. 
Once the DFT+CDMFT procedure has converged, accessing momentum-resolved observables in the original Brillouin zone requires a periodization procedure.

In our approach we avoid the complications arising from unfolding the full band structure, as is done, e.g., in the studies of alloys~\cite{rubel2014}. 
Here we only periodize the Green's function projected on the Cu-\dxxyy~cluster. 
The projection allows to use the standard Green's function periodization developed in the context of effective models~\cite{senechal2000,maier2005a,LTP:2006,stanescu2006,verret2019}:
\begin{equation}
    G_{\mathrm{per}}(\vtk,\omega) = \frac{1}{N_{\mathrm{C}}}\sum_{\mathbf{R}\mathbf{R\p}}\mathrm{e}^{-i\vtk(\mathbf{R}-\mathbf{R\p})}G^{x^2-y^2}_{\mathbf{R}\mathbf{R\p}}(\vk,\omega),
\end{equation}
where $N_{\mathrm{C}}$ is the number of cluster sites and $\vtk$ is the momentum in the primitive BZ, while $\vk$ is momentum in the BZ corresponding to the supercell. 
$G_{\mathrm{per}}(\vtk,z)$ is the periodized quantity expressed in the primitive BZ.
The projected Green's function $G^{x^2-y^2}_{\mathbf{R}\mathbf{R\p}}(\vk,\omega)$ in the supercell BZ may be obtained from: 
\begin{equation}
    G^{x^2-y^2}_{\mathbf{R}\mathbf{R\p}}(\vk,\omega) = \sum_{i,j}\braket{\phi^{x^2-y^2}_{\mathbf{R}}|\psi_{\vk,i}} G_{ij}(\vk,\omega)  \braket{\psi_{\vk,j}|\phi^{x^2-y^2}_{\mathbf{R\p}}}.
\end{equation}

We chose to periodize the Green's function instead of the cumulant or the self-energy since we are interested in the momentum-resolved spectral function (i.e. the imaginary part of the Green's function). 
In this way, the periodization approximation is carried after the necessary matrix inversions so as to avoid the emergence of intractable errors~\cite{verret2019}.
Moreover, since the Cu-\dxxyy~orbital is the leading contribution to the spectral weight close to the Fermi level, the additional projection is not a limitation. 
The strong Cu-O covalency ensures that the spectral weight of the oxygen closely follows that of copper. 

%
%
%
%

\subsection{Exact diagonalization impurity solver}
\label{sec:met_ed}

We have integrated the ED solver of the PyQCM library~\cite{dionne2023a,dionne2023b} into the DFT+CDMFT implementation described above. 
As input, the solver receives the impurity hopping matrix $\hat{T}$, the interaction $\hat{V}$ tensor, and the impurity hybridization function $\Bar{\Delta}_{\alpha\beta}(i\omega_n)$. 
A set of discrete bath orbitals is optimized by minimizing the distance between the impurity and \emph{discretized} hybridization $\Delta_{\alpha\beta}$:
\begin{equation}
\begin{split}
    & d = \sum_{n}W(i\omega_n)\lvert\Bar{\Delta}_{\alpha\beta}(i\omega_n)-\Delta_{\alpha\beta}(i\omega_n)\rvert^2 \\
    & \Delta_{\alpha\beta}(i\omega_n) = \sum_{b} \frac{\theta^{\phantom{*}}_{\alpha b}\theta^{*}_{\beta b}}{i\omega_n - \epsilon_{b}},
\end{split}
\end{equation}
where $\left\{\theta,\epsilon\right\}$ is the set of bath hoppings and energies parameters, and $W(i\omega_n)$ is a weight function.
The optimization is carried out using a Matsubara frequency mesh at a given fictitious inverse temperature $\beta$, which is also used throughout the DFT+CDMFT calculation. 
We emphasize that the ED solver nevertheless works at zero-temperature.

Different types of weight functions have been proposed~\cite{bolech2003,kyung2006,stanescu2006,kancharla2008b}, but we found that different variations did not improve significantly our results. 
We thus adopted a simple cutoff version:
\begin{equation}
    W(i\omega_n)=
     \begin{cases}
       1 \quad \text{if} \quad \omega_n < \omega_c \\
       0 \quad \text{else}
     \end{cases}.
\end{equation}
The cutoff frequency is typically chosen to be $\omega_c\sim10-15$~eV.

In order to accelerate the calculations of the quantum cluster impurity problem, the bath geometry is designed to exploit the symmetries of the $2\times2$ cluster. 
We followed the prescriptions of Ref.~\cite{foley2019a} in which a general 8-orbital bath geometry was proposed (2 bath orbitals per irreducible representation of the cluster's point group). 
The advantages of this geometry are twofold: not only does it accelerate the computations, but it also improves the estimation of the superconducting properties~\cite{foley2019a,dash2019,kowalski2021}. 

The bath optimization results in the following impurity Hamiltonian:
\begin{equation}
    \hat{H}_{\mathrm{imp}} = \hat{T} + \hat{V} + \hat{H}_{\mathrm{hyb}}\left[\theta,\epsilon\right].
\end{equation}
The ground-state is determined via the Lanczos algorithm, and then used to compute the impurity Green's function using the band Lanczos approach~\cite{bai2000}. 
The impurity self-energy needed to carry the DFT+CDMFT self-consistency is then obtained via the Dyson equation. 
Details about how these steps are implemented in practice within PyQCM may be found in Ref.~\onlinecite{dionne2023a}.

The impurity function $\bar{\Delta}_{\alpha\beta}(\omega)$ contains the hybridization amplitude between the correlated orbitals and \emph{all} other degrees of freedom in the crystal. 
One may thus wonder to which extent a discrete set of 8 bath orbitals may faithfully account for the physics at play. 
Yet, in all our calculations the discrepancy between the \emph{lattice} and \emph{impurity} occupation of the correlated subspace per Cu atom is of the order $10^{-3}$ once convergence is achieved. 
This level of agreement is attainable because, as noted in  Sec.~\ref{sec:iter_proc}, the charge self-consistency is carried out entirely in Matsubara space. 
Hence, the discretized hybridization is accurate enough to ensure a good convergence. 
Similar accuracy was also demonstrated for the single-band Hubbard model away from half-filling solved with 4-site CDMFT, where an accuracy better than 1\% can be reached already at 2 bath-sites per cluster sites~\cite{koch2008}. 
To further support our approach, we provide in App.~\ref{ap:ed} a comparison between DFT + single-site DMFT calculations carried out with ED and a continuous-time Monte Carlo (CTQMC) solver~\cite{haule2007}.

A parallel can be drawn with recent advances showing that noisy Green's function on the imaginary axis can be accurately represented using a limited set of poles~\cite{kaye2022}.
The information lost due to the bath discretization would act, to some extent, similarly as the information lost in the noise.

We conclude our discussion of the impurity solver by explaining our decision to use ED rather than the well-establish CTQMC solvers~\cite{gull2011}. 
While CTQMC would enable direct treatment of the infinite bath and prediction of transition temperatures - unlike zero temperature ED - it is hindered by a severe sign problem in the so-called covalent regime of three-band models~\cite{kowalski2021}. 
This regime typically corresponds to the conditions encountered in our first-principles calculations, making CTQMC unsuitable for this work. 
A further consideration is the computational efficiency required for solving the impurity problem.
The unmatched speed of ED is essential for studying large supercells, such as those present in the $n=4,5$ \nC~and \nH~compounds. 


\subsection{Effective model for superconductivity}
\label{sec:met_sc}

The superconducting gap in cuprates is usually of the order of $\sim 50$~meV~\cite{kurokawa2023,luo2023a}.
Hence the typical energy scale associated with superconductivity is much smaller than any other energy scale involved in our calculations. 
In particular, the opening of such a small gap in the spectral function should not significantly affect the charge density. 
Therefore, we estimate the superconducting properties in a post-processing CDMFT calculation where all degrees of freedom but the \dxxyy~orbitals are being frozen, i.e., are kept in the normal state.

We reduce the computational burden by fixing the projector between the Kohn-Sham and the local basis. 
As such, one may construct a generalized single-band model in which the non-interacting dispersion $\varepsilon_\vk$ is replaced by a \emph{retarded} dispersion $\Gamma(\vk,\omega)$ which encodes the effect of all other degrees of freedom in the crystal. 
This is a direct generalization of the usual procedure used in the CDMFT studies of three-band models, in which the non-interacting oxygen degrees of freedom are integrated out~\cite{dash2019,kowalski2021}.

The retarded dispersion $\Gamma(\vk,z)$ is defined via the following ansatz:
\begin{equation}
\begin{split}
    G_{\alpha\beta}(\vk,\omega) & = \sum_{i,j}
    \braket{\phi_\alpha|\psi_{\vk,i}}
    G_{ij}(\vk,\omega)
    \braket{\psi_{\vk,j}|\phi_{\beta}}\\
    & \equiv \left[\omega+\mu-\Gamma(\vk,\omega)-\Sigma(\omega)\right]^{-1}_{\alpha\beta},
\end{split}
\end{equation}
where $\Sigma(\omega)$ is the DFT+CDMFT self-energy. 
By inverting the above equation, one may obtain:
\begin{equation}
    \Gamma_{\alpha\beta}(\vk,\omega) =  \omega + \mu - \Sigma_{\alpha\beta}(\omega) - \left[G^{-1}\right]_{\alpha\beta}(\vk,\omega).
\end{equation}
In this way, the local Green's function can be computed without having to explicitly construct the projector's matrix elements which are all encoded in $\Gamma_{\alpha\beta}(\vk,\omega)$. 

In practice, $\Gamma_{\alpha\beta}(\vk,\omega)$ is computed for the Cu-\dxxyy~orbitals of the $2\times2$ cluster, from the self-energy $\Sigma_{\alpha\beta}$ and Green's function $G_{\alpha\beta}$ converged in the normal state.
To incorporate superconductivity in the post-processing CDMFT calculation, we use the Nambu formalism as implemented within the PyQCM package~\cite{dionne2023a,dionne2023b}.
It amounts to performing a particle-hole transformation of the spin-down operators:
\begin{equation}
    \left\{
    \begin{split}
        & c^{\phantom{\dagger}}_{\alpha} = c^{\phantom{\dagger}}_{\alpha,\uparrow} \\
        &d^{\phantom{\dagger}}_{\alpha} = c^{\dagger}_{\alpha,\downarrow}
    \end{split}
    \right.
\end{equation}
Within the $\{c,d\}$ basis, the self-energy reads:
\begin{equation}
\label{eq:self_nambu}
    \Sigma_{\alpha\beta}(\omega) = \begin{pmatrix}
        \Sigma^{n}_{\alpha\beta}(\omega) & \Sigma^{an}_{\alpha\beta}(\omega) \\
        \Sigma^{an}_{\alpha\beta}(\omega) & -\Sigma^{n}_{\alpha\beta}(\omega)^{\dagger}
    \end{pmatrix},
\end{equation}
where we distinguish the normal and anomalous part which have to be found self-consistently. 
Since all degrees of freedom but the \dxxyy~orbitals are fixed in the normal state, the effective dispersion $\Gamma_{\alpha\beta}(\vk,\omega)$ only contains normal components:
\begin{equation}
    \Gamma_{\alpha\beta}(\vk,\omega) = \begin{pmatrix}
        \Gamma^{n}_{\alpha\beta}(\vk,\omega) & 0 \\
        0 & -\Gamma^{n}_{\alpha\beta}(\vk,\omega)^{\dagger}
    \end{pmatrix}.
\end{equation}
The latter remains constant during the self-consistency. 
However, breaking the $U(1)$ symmetry is allowed in the bath in a self-consistent way.

The key output of this last CDMFT calculation is the value of the superconducting order parameter $\msc$.
It is computed as the expectation of the following operator: 
\begin{multline}
    \hat{D}_{\mathrm{SC}} = \frac{1}{2}\Biggl[\sum_{\braket{ij}_{x}}\left(d^{\dagger}_{i\uparrow}d^{\dagger}_{j\downarrow}-d^{\dagger}_{i\downarrow}d^{\dagger}_{j\uparrow}\right)\\
    -\sum_{\braket{ij}_{y}}\left(d^{\dagger}_{i\uparrow}d^{\dagger}_{j\downarrow}-d^{\dagger}_{i\downarrow}d^{\dagger}_{j\uparrow}\right)
    +~\mathrm{H.c.}\Biggr],
\end{multline}
where $\braket{ij}_{x}$ ($\braket{ij}_{y}$) denotes nearest-neighbour Cu sites along the $\mathbf{x}$ ($\mathbf{y}$) direction, and $d^{\dagger}_{i\sigma}$ is the electron creation operator on the \dxxyy~orbital on site $i$ with spin $\sigma$. 

In this article, we compute $\msc$ for every non-equivalent CuO$_2$ plane in the compound \emph{separately}, and thus ignore proximity effects between neighbouring planes. 
We checked that such effects are small and do not affect our conclusions, as described in App.~\ref{ap:prox}.
To ensure the robustness of the $\msc$ prediction, we perform two post-processing procedures: (i) a two-step process in which we first converge the CDMFT calculation with a 100~meV static pairing field, and then re-converge after turning it off; (ii) a single-step process in which no pairing field is applied and superconductivity emerges self-consistently from the normal state. 
In most cases, at values of doping where superconductivity is either well established or undoubtedly vanishing, the two procedures lead to the same $\msc$ (the mismatch is smaller than $10^{-4}$). 
In the under-doped (over-doped) doping region where superconductivity is at the verge of emerging (vanishing), the two estimations may differ markedly. 
In such cases, the order parameter is taken from another two-step procedure in which we first apply 1~meV static pairing field.
This uncertainty in the transition regions is related to the discretized bath used in the ED solver, and can even be traced back to instabilities in the normal state solution at these values of doping. 

\subsection{Computational details}
\label{sec:met_comp}

The DFT+CDMFT calculations in the normal state, which involve the DFT package Wien2k~\cite{blaha2020} and eDMFT package~\cite{haule2010,HauleJPSJ}, are carried with $12\times12\times8$ $\vk$-grid, using a Matsubara frequency mesh at a fictitious temperature $\beta^{-1}=0.02$~eV. 
For all compounds, we set the local Hubbard interaction to $U=12$~eV and chose the Perdew–Burke–Ernzerhof exchange-correlation functional~\cite{perdew1996}. 
In App.~\ref{ap:u}, we provide detailed arguments motivating this choice, as well as a proof that our conclusions are robust against variations of $U$ ranging from 10~eV to 14~eV.
Within the two-orbital single-site impurity problem, we also use the Hund's term $J_{\mathrm{Hund}}=1$~eV.

We compute observables on the real frequency axis only when charge self-consistency is reached. 
Since the ED solver evaluates the impurity Green's function from the Lehmann representation, we have a direct access to real-frequency observables without relying on analytic continuation. 
Real frequency spectral functions and density of states (DOS) are computed with an additional broadening of 0.1~eV. 
The broadening is used only for data visualization and does not affect our conclusions in any way.
Finally, for the final post-processing estimation of the superconducting order parameter we compute $\Gamma_{\alpha\beta}(\vk,\omega)$ on the same Matsubara frequency mesh and on a $16\times16\times4$ $\vk$-grid. 
We checked that changing the fictitious temperature and the number of momentum points did not affect the value of the order parameter. 

Preparation of the crystal structures and visualisation of the charge density differences were carried out with VESTA~\cite{momma2011} software.

\section{Results: Undoped multilayer cuprates}
\label{sec:res_undop}

Key quantities such as the CTG and the effective superexchange $J$, which have been shown to be related to superconductivity~\cite{weber2012,kowalski2021,wang2022,omahony2022a,wang2023a}, can be accessed from the electronic structure of the parent compounds. 
It has been shown experimentally that the CTG in the Bi-based multilayer cuprates has a very similar $n$-dependence as $T_c$~\cite{wang2023a}, and signs of this dependence have also been observed in the \nC~family~\cite{ruan2016a}. 
Yet, to this date, the universality of the CTG \emph{vs.} $n$ relation is not fully established and, most importantly, the physical origin remains elusive. 
Another pivotal question deals with the value of the superexchange $J$, which may be modulated with $n$ since it is expected to be inversely proportional to the CTG. 

The $n$-dependent physics of multilayer cuprates is usually conceived to be caused by the uneven hole-doping and the related inhomogeneities between non-equivalent CuO$_2$ outer and inner planes.
These arguments are mainly hand-wavy, and would especially not hold when considering the undoped compounds. 
Therefore, we first study the undoped (stoichiometric) \nC~and \nH.
We will show that this $n$-dependence is rooted in the intrinsic physics of the clean parent compounds.
The comparative study of two families allows moreover to distinguish between the \emph{universal} and \emph{material-specific} trends in multilayer cuprates.

We start this section by a short summary of the main results (Sec.~\ref{sec:res_undop_sum}).
Then, we detail the analysis of \nC, whose simplicity enables us to describe the origin of the $n$-dependent physics (Sec.~\ref{sec:res_undop_nc}).
We then present in Sec.~\ref{sec:res_undop_nh} the \nH~family, a more complex example due to the self-doping induced by the Hg and apical O atoms. 

\subsection{Brief summary of the undoped results}
\label{sec:res_undop_sum}

We find several hallmarks of universality in the physics of the multilayer cuprates with respect to the number of consecutive CuO$_2$ planes, $n$: i) the CTG is decreasing from $n=1$ to $n=3$, and is saturated at $n=5$. It is smaller (larger) in the inner (outer) planes (Figs.~\ref{fig:fig3}(a),\ref{fig:fig10}(a)); ii) the superexchange $J$ is increasing from $n=1$ to $n=3$, and it is also saturated at $n=5$. In accordance with CTG, $J$ is larger (smaller) in the inner (outer) planes (Figs.~\ref{fig:fig3}(b),\ref{fig:fig10}(b)); iii) the carrier distribution between the Cu-$d$ and O-$p$ orbitals is also changing with $n$, and larger number of CuO$_2$ planes typically leads to more holes on oxygen (more mixed valent Cu ion), and the inner planes are most mixed valent (this last point is detailed in App.~\ref{ap:pcharges}, Figs.~\ref{fig:fig4},\ref{fig:fig11}).
Similar ideas were phenomenologically drawn from understanding the trends in $T_c$ across different multilayer families~\cite{mukuda2012,wang2023a}. 

We identify the origin of this strong $n$-dependent physics as the appearance of standing wave modes, confined in real space between the $n$ CuO$_2$ planes, and in energy space as the low-energy conduction bands (Figs.~\ref{fig:fig5},\ref{fig:fig6},\ref{fig:fig7}).
The confinement is directly visible in the DFT-only band structure of the \nC~family, and less pronounced in the \nH~compounds due to the presence of the self-doping Hg-O bands. 
The effect of the Hg-O states is mostly independent of $n$, and is mainly responsible for the quantitative differences observed between the two families. 
Hence, our study unambiguously shows that the key parameters relevant for superconductivity such as the CTG, the superexchange $J$ and the oxygen occupancy~\cite{kowalski2021}, are influenced by the properties of the first conduction states (other than the upper Hubbard band). 
Moreover, based on the evidence that the CTG is a key descriptor for the superconducting order parameter~\cite{kowalski2021,omahony2022a}, or directly $T_c$~\cite{wang2023a}, our results are reminiscent of the experimental evidence that different multilayer cuprate families can share the same universal $T_c$ \emph{vs.} $n$ trend, while quantitatively differing in the absolute value of $T_c$~\cite{mukuda2012,wang2023a}.

Another important conclusion is that the differentiation between outer and inner CuO planes in multilayer cuprates is \emph{intrinsic}.
It exists prior to the effects of doping and disorder which may enhance the differentiation between the different planes. 
Moreover, we surprisingly observe in the undoped tri-layer \nH~ that the inner plane (IP1) captures more holes than the outer plane (OP) as a consequence of self-doping. 
This may provide a route towards a more homogeneous doping of multilayer compounds.
Indeed, we show in Sec.~\ref{sec:res_dop} that at low dopings the inner plane of the tri-layer \nH~is more efficiently doped than its counterpart in the tri-layer \nC.

It is remarkable that the states participating in the $n$-dependence of the multilayer cuprates' physics display salient features of a confined two-dimensional electron gas.
In other words, we find evidence of states in a \emph{bulk} system that behave as in a thin film. 
This suggests that model Hamiltonian approaches would require the Hubbard $U$ to be a strong function of $n$, as it is well known for thin films that the electronic screening depends strongly on the system's thickness. 
One may also \emph{engineer} the first conduction states by using thin films and heterostructures to tune the CTG and $J$ to potentially improve the superconducting properties. 
We leave these ideas for future investigations as they are beyond the scope of the present work.

\subsection{nCCOC: a pedagogical example}
\label{sec:res_undop_nc}

We start our detailed analysis with the \nC~family since its electronic structure is simpler, and therefore easier to understand than \nH.
We first present the $n$-dependence of the CTG and the superexchange $J$ (Sec.~\ref{sec:res_undop_nc_ctg}), and then discuss its physical origin (Sec.~\ref{sec:res_undop_nc_cond}).

\subsubsection{Charge-transfer gap and superexchange}
\label{sec:res_undop_nc_ctg}
\begin{figure*}
    \centering
    \includegraphics[width=\linewidth]{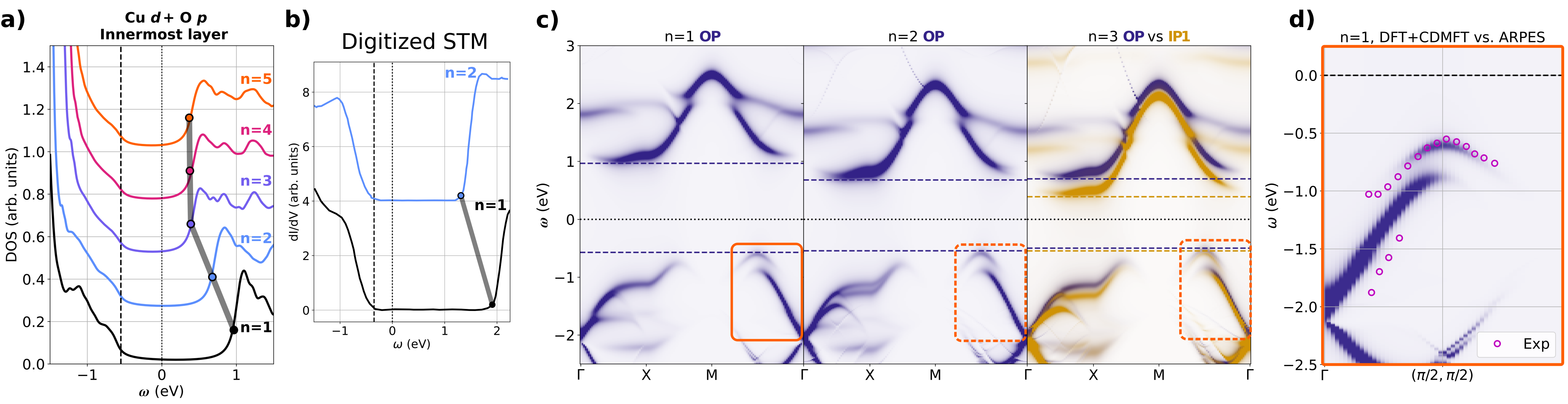}
    \caption{(a) Projected DOS on the Cu-$d$ and O-$p$ orbitals of the \emph{innermost} CuO$_2$ plane of \nC~($n$=1\textendash5).
    The chemical potential was shifted in the gap so to align the DOS in the occupied region, as marked by the black dashed line. 
    The colored dots and gray line in the unoccupied part highlight the variations of the CTG. 
    (b) Digitized STM measurements from Ref.~\onlinecite{ruan2016a}, the chemical potentials were aligned. 
    (c) Momentum-resolved spectral function projected on the Cu-\dxxyy~orbital for $n$=1\textendash3, along $\Gamma$\textendash X\textendash M\textendash$\Gamma$. 
    The purple and orange dashed lines highlight the CTG for the OP and IP1 planes, respectively.
    The orange boxes highlight the spin-polaron feature (\emph{see text}). 
    (d) Comparison between our calculations and ARPES data from Ref.~\onlinecite{ronning2005}, for the $n=1$ \nC~compound along the $\Gamma-M$ path. 
    }
    \label{fig:fig2}
\end{figure*}

The CTG is the first key quantity that has a strong effect on superconductivity, and is directly accessible within DFT+CDMFT. 
It can also be measured via optical spectroscopy~\cite{perkins1993,falck1994,perkins1998,waku2004}, electron energy loss spectroscopy (EELS)~\cite{schuster2009,wang2023a}, and scanning tunneling miscroscopy (STM)~\cite{ruan2016a,omahony2022a}. 
The CTG can be understood as the energy necessary to transfer an electron from the ligand atom to the transition metal, i.e., in the case of cuprates from the O-$p$ to the Cu-$d$ orbitals. 
In the undoped compounds, and in absence of self-doping, the strong electronic correlations open a Mott gap in the spectra, which is the CTG. 

\mypar{Spectral function analysis}

In Fig.~\ref{fig:fig2}(a), we show the DOS of \nC~($n$=1\textendash5), projected on the Cu-$d$ and O-$p$ orbitals of the \emph{innermost} CuO$_2$ plane for a better visibility. 
For all multilayer compounds ($n$), the DOS is insulating, and we observe a clear reduction of the CTG from the single-layer ($\Delta^{n=1}_{\rm CTG}\approx1.5\;$eV) to the tri-layer case ($\Delta^{n=3}_{\rm CTG}\approx0.9\;$eV), which is then stabilized for $n\geq3$. 
Optical conductivity~\cite{waku2004} and STM measurements~\cite{ruan2016a} converge to an estimate of $\Delta^{n=1}_{\rm CTG}\approx2.0\;$eV, which is reduced to 1.5~eV for the bi-layer compound~\cite{ruan2016a} (see Fig.~\ref{fig:fig2}(b)).
The absolute value of these measured gaps should be interpreted with caution, since non-zero (but small) spectral weight in the optical conductivity spectrum can be observed up to 1.6~eV~\cite{waku2004}, and STM measurements of insulators can suffer from the band bending phenomenon~\cite{feenstra1987,weimer1989}.
The drop of the experimentally measured CTG by 0.5~eV from the single- to the bi-layer compound (see Fig.~\ref{fig:fig2}(b)) is nevertheless meaningful, and is well captured by our calculation, which predicts a reduction of 0.3~eV. 
Our simulations also predict a further reduction of the CTG by another 0.3~eV for the tri-layer compound.
This is also in remarkable agreement with recent STM measurements on insulating $n=1,2,3$ Bi$_2$Sr$_2$Ca$_{n-1}$Cu$_{n}$O$_{2n+4+\delta}$ samples~\cite{wang2023a} showing a similar 0.3~eV drop from the single- to the bi-layer case, and a further 0.3~eV drop in the tri-layer compound. 
To the best of our knowledge, insulating samples with four or five layers ($n=4,5$) have not been synthetized yet, which prevents further comparison to experimental data. 

\begin{figure}
    \centering
    \includegraphics[width=0.8\linewidth]{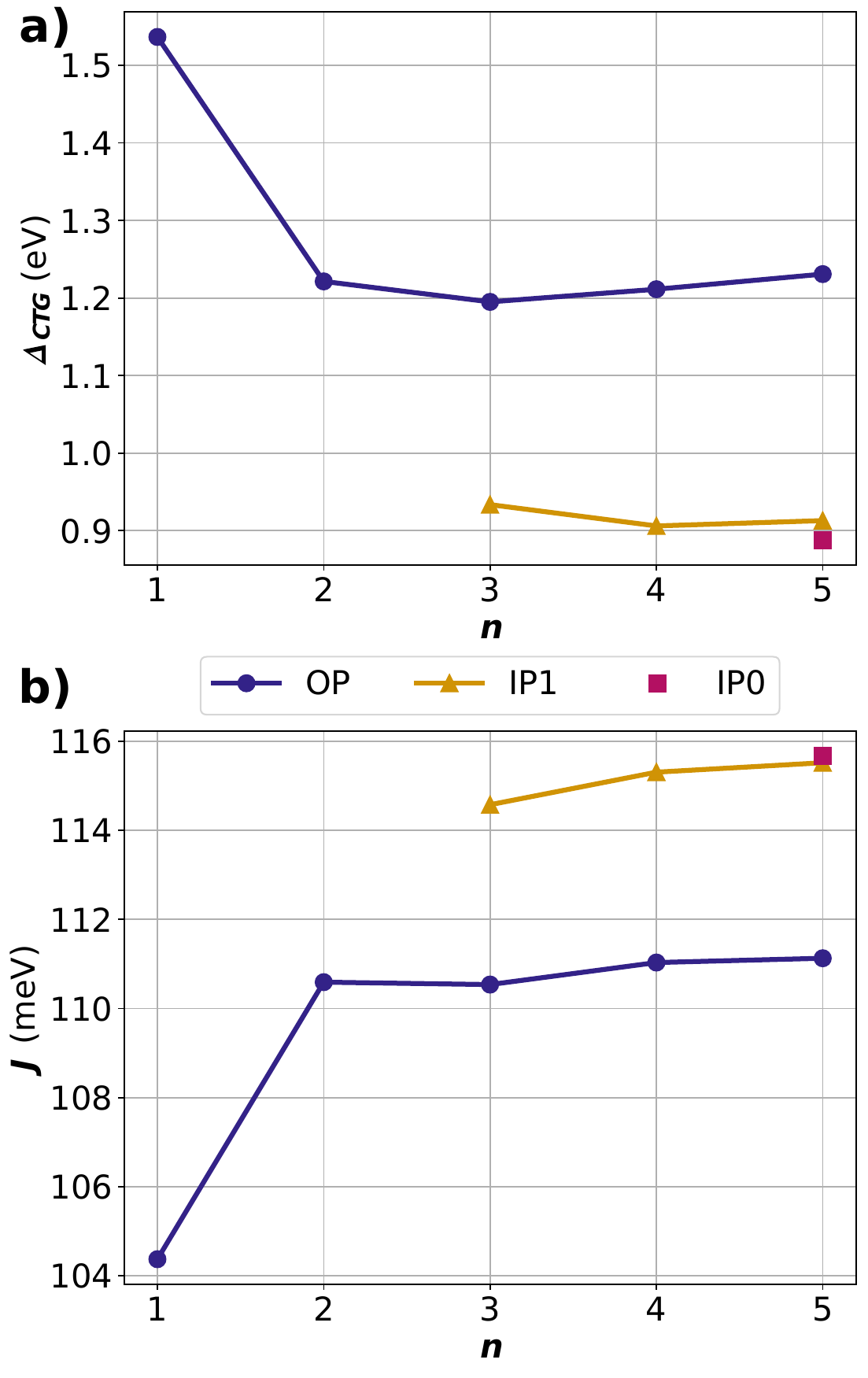}
    \caption{(a) CTG and (b) effective superexchange $J$ estimated for the OP, IP1 and IP0 of \nC~($n$=1\textendash5).}
    \label{fig:fig3}
\end{figure}

We present in Fig.~\ref{fig:fig2}(c) the momentum-resolved spectral function, projected onto the Cu-\dxxyy~orbital, along the high symmetry momentum path between $\Gamma=(0,0)$, $\mathrm{X}=(\pi,0)$, and $\mathrm{M}=(\pi,\pi)$ points, for $n=1,2,3$. 
%
%
A typical signature of the universal \emph{waterfall} and \emph{kink} features in cuprates~\cite{martinez1991a,macridin2007a,manousakis2007a,wang2015a,bacq2023} can be seen at the \emph{nodal} $(\pi/2,\pi/2)$ point in the $\Gamma$-M path (highlighted by orange boxes). 
The low-energy electronic dispersion is strongly renormalized and backfolded around the nodal point, which was explained recently by the formation of spin-polaron, i.e. electrons heavily dressed by magnons~\cite{bacq2023}. 
This peculiar shape of the spin-polaron dispersion is responsible for the emergence of Fermi pockets at low dopings (see Sec.~\ref{sec:res_dop_spec}). 
To further support our results we present in Fig.~\ref{fig:fig2}(d) a comparison between the calculated spectral function and angle-resolved photoemission spectroscopy (ARPES) measurements from Ref.~\onlinecite{ronning2005}, for the single-layer \nC~compound along the $\Gamma-M$ path. 
We find an excellent agreement of the renormalized spin-polaron dispersion.  
This highlights that our cluster impurity allows to capture precisely the low energy electronic properties of cuprate superconductors. 
To improve the correspondence with experiment for the high-energy dispersion between -1.5 and -2.0~eV, larger impurity clusters should be used~\cite{bacq2023}.

\mypar{Charge-transfer gap vs. $n$}

It can be clearly seen in Fig.~\ref{fig:fig2}(c) that the CTG is a strong function of the number consecutive CuO$_2$ planes ($n$), and that different planes experience different CTG. 
In the OP, the gap decreases from $n=1$ to $n=2$, and nearly saturates beyond $n\geq 2$. 
However, for $n\geq 2$, the inner planes appear and yield an even smaller CTG.  
In other words, the effective electronic interactions felt by the \dxxyy~electrons is reduced with increasing $n$.  

These results are gathered in Fig.~\ref{fig:fig3}(a), where is shown the CTG ($\Delta^{n}_{\rm CTG}$) in all CuO$_2$ planes for all \nC~compounds. 
The CTG does not change appreciably anymore beyond $n=3$, both in the outer and inner planes. 

%

\mypar{Superexchange $J$ vs. $n$}

The CTG is not the only descriptor that strongly affects superconductivity~\cite{kowalski2021}.
Another important parameter is the effective superexchange $J$, which encapsulates the interaction between spins localized on neighbouring Cu atoms.
In the localized limit, $J$ is related to the CTG ($\Delta_{\rm CTG}$) by $J\approx t_{\rm eff}^{2}/\Delta_{\rm CTG}$, where $t_{\rm eff}$ is an effective hopping amplitude between the neighbouring Cu atoms. 
The direct correlation between $J$ and the superconducting order parameter predicted theoretically~\cite{kowalski2021} is partly confirmed by experimental data relating $J$ and $T_c$~\cite{wang2022}.
Despite a few outliers, in most multilayer compounds $J$ correlates with $T_c$.
Hence it is an important quantity to estimate.

Here we compute $J$ from the $2\times2$ cluster susceptibility at the cluster momentum  $(\pi,\pi)$~\cite{kyung2009,kowalski2021}.
We emphasize that this is only a very approximate estimation of $J$ that tends to underestimate it, because we extract it from the momentum averaged susceptibility, as captured by the small cluster.
To calculate the susceptibility precisely at $(\pi,\pi)$ point, one would need the Bethe-Salpether equation which requires the two particle vertex function~\cite{musshoff2021a}, a task beyond the current work.
We note that the \emph{ab initio} DFT+DMET formalism, using $2\times2$ clusters of Ref.~\onlinecite{cui2023a}, suffers from the same limitations and accordingly predict values of $J$ which are close to ours.  

Fig.~\ref{fig:fig3}(b) shows our estimated $J$ for all planes and all \nC~compounds. 
From the single-layer to the bi-layer compound, $J$ increases by 6 meV ($\approx50$~K). 
In the three-layer compound, the OP has almost the same $J$ as the bi-layer, but the IP1 has a $J$ that is 4 meV larger than the OP. 
We observe only a slight increase in $J$ in both the outer and inner planes from $n=3$ to $n=4$ and $n=5$.
Note that our estimation of $J$ is somewhat smaller than the values measured experimentally, where for the $n=1$ compound, $J$ was found to be 150 meV~\cite{lebert2017a,lebert2023a}, hence 1.5 times larger that our estimation. 
However, we emphasize that the variations of the estimated $J$ across the different planes and different compounds are meaningful, and unambiguously show that intrinsically stronger antiferromagnetic fluctuations in the inner planes can be expected, without any influence from inhomogeneous hole-doping or structural distortions.
We also mention that, in agreement with our calculations, there is experimental evidence that the inner planes host particularly strong antiferromagnetic correlations~\cite{mukuda2012,kunisada2017,kurokawa2023}.

\mypar{Orbital occupations vs. $n$}

In App.~\ref{ap:pcharges}, we provide an additional analysis of the $n$-dependence of the orbital occupations. 
It is remarkably similar to the behavior of the CTG. 
The smaller the CTG, the more mixed-valent is the occupation of the Cu-\dxxyy/\dzz~orbitals.
This confirms the interpretation of effectively smaller local charge correlations ($U$) to the benefit of stronger non-local spin correlations ($J$) when the CTG is reduced. 

\subsubsection{Quantum confinement of conduction states}
\label{sec:res_undop_nc_cond}

\begin{figure*}
    \centering
    \includegraphics[width=\linewidth]{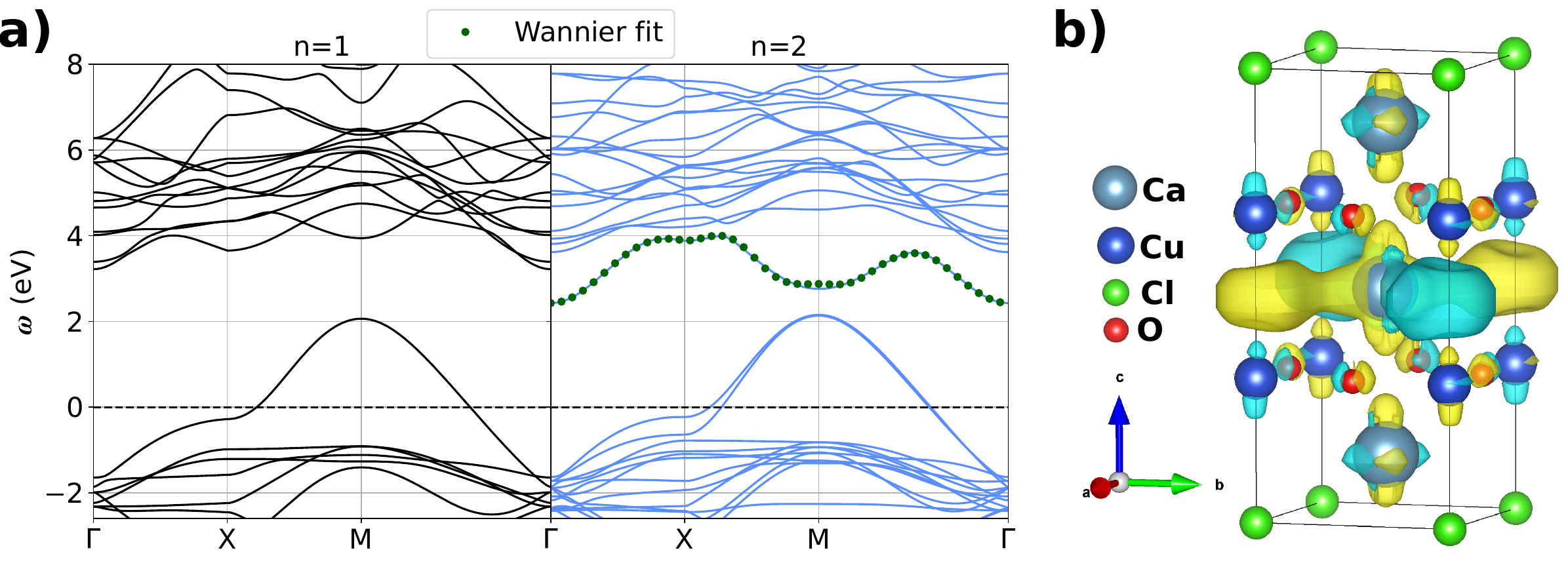}
    \caption{(a) \emph{DFT-only} band structures for the $n=1,2$ \nC. 
    The isolated first conduction state of the $n=2$ compound is fitted with a maximally localized Wannier orbital shown in (b).}
    \label{fig:fig5}
\end{figure*}

Up to now, we have analyzed how the normal state properties of the undoped nCCOC multilayer cuprates depend on the number of consecutive CuO$_{2}$ planes ($n$).
We have shown that the CTG is reduced with an increasing number of planes and that the inner ones have the smallest CTG.
These inner planes also have larger superexchange $J$, they are more mixed-valent, and the number of holes in the $p_x$ orbitals (pointing towards Cu, see App.~\ref{ap:pcharges}) is increased. 
All these properties are known to be favorable for superconductivity, but we have not yet presented their microscopic origin.

In the following, we argue that the $n$-dependence is mostly due to the appearance of additional conduction bands, which are confined in-between the CuO$_2$ planes, and limit the size of the charge gap in the unoccupied part of the spectrum. 
This physics can be understood and extracted even from the DFT band structure.

\mypar{Single- vs. bi-layer compounds}

We first focus on the reduction of the CTG from the single- to the bi-layer  \nC~compounds. 
They display similar DFT band structures, as shown in Fig.~\ref{fig:fig5}(a). 
The main difference is the presence of an additional isolated conduction band for the $n=2$ compound, located approximately 1~eV below the bulk of the other conduction bands. 
Taking advantage of its isolation, we performed a fit using maximally localized Wannier orbitals with wannier90~\cite{mostofi2008,mostofi2014}, which resulted in the orbital shown in Fig.~\ref{fig:fig5}(b). 
The latter is confined in the interstitial space between the two CuO$_2$ planes, with a maximum weight located in between the Cu atoms. 
Note that this orbital is very extended in real space, with most of its weight \emph{in between} the CuO$_2$ planes.
Hence correlations on such orbital are weak and should not be affected much by beyond-DFT treatment.
%
%
This isolated band reduces the CTG by providing an additional source of hybridization, thereby enabling more efficient electron hopping, and by pinning the upper Hubbard band at lower energies. 
We note that these effects are beyond previous model Hamiltonian approaches, such as solving the realistic three-band Hubbard model, which would neglect such conduction states. 

\begin{figure*}
    \centering
    \includegraphics[width=\linewidth]{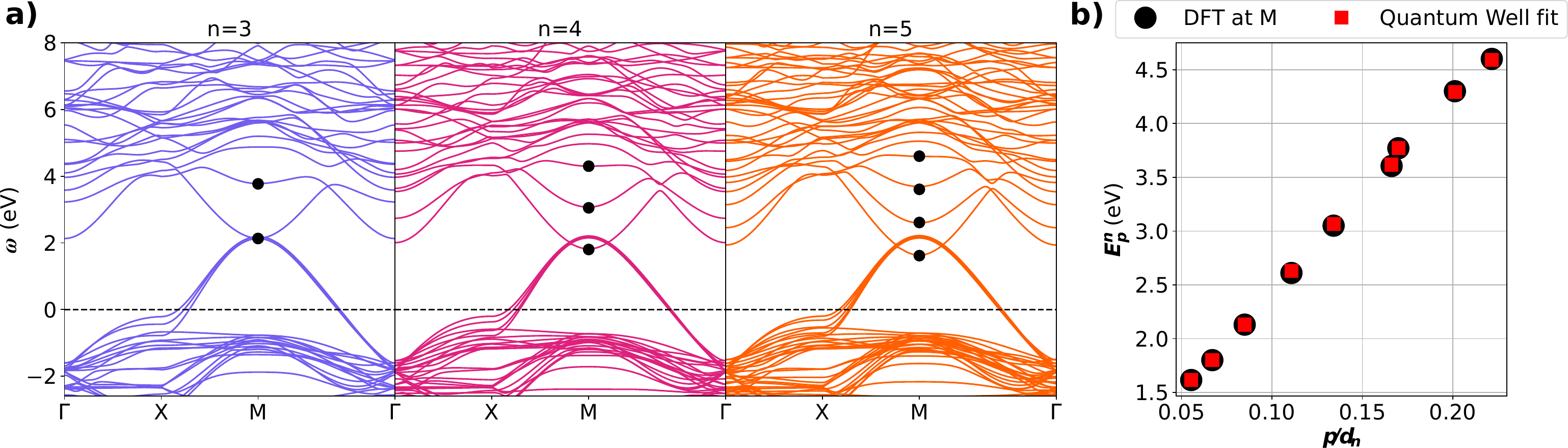}
    \caption{(a) \emph{DFT-only} band structures for the $n=3,4, 5$ \nC. 
    Black circles correspond to the energies at M estimated from a simple linear relation with $n$ (Eq.~\eqref{eq:deltaE}, \emph{see text}). 
    These energies are plotted in (b) and compared to a fit based on a quantum well toy model.}
    \label{fig:fig6}
\end{figure*}

\mypar{Generalization to all $n$}

The above analysis can be generalized to the $n\geq3$ compounds.
As shown in Fig.~\ref{fig:fig6}(a), in analogy with the $n=2$ case we observe $n-1$ conduction states detached from the bulk of conduction bands.
They are mostly visible at the M point, where they strikingly resemble quantum confined states in thin films~\cite{gauthier2020,chiang2000,kawakami1999,paggel1999}. 
Remarkably, at M the $n-1$ states are separated by a constant energy shift $\Delta E_{n}$: 
\begin{equation}
\label{eq:deltaE}
    E^{n}_{p+1} = E^{n}_{p} + \Delta E_{n},
\end{equation}
where $p\in[1,n-1]$ is the index of the mode.
The energy shifts $\Delta E_{n=3}\approx1.64\;$eV, $\Delta E_{n=4}\approx1.25\;$eV, $\Delta E_{n=5}\approx0.995\;$eV follow the simple relation:
\begin{equation}
\label{eq:deltaE_rel}
\frac{\Delta E^{n+1}}{\Delta E^{n}} \approx \frac{n}{n+1} \approx \frac{d_n}{d_{n+1}},
\end{equation}
where $d_{n}$ is the thickness of the $n$-CuO$_2$-plane stack (taken as the distance between the surrounding Cl atoms).  
In analogy with quantum confinement in thin films, the above relation could be understood from the phase-accumulation model which describes the necessary condition for the formation of a standing wave in a quantum well of thickness $d_{n}$~\cite{chiang2000}:
\begin{equation}
\label{eq:phase_acc}
    2d_{n}k_{z}(E^{n}_{p}) + \Theta(E^{n}_{p}) = 2p\pi
\end{equation}
where $k_{z}$ is the original dispersion along $z$ if there was no confinement along this direction, and $\Theta(E^{n}_{p})$ a correcting phase corresponding to the leakage of the standing wave outside of the well, which is \emph{a priori} unknown. 
If the dispersion relation $k_{z}(E^{n}_{p})$ is linear, then the energies of the modes are inversely proportional to the thickness $d_{n}$, i.e., $E^{n}_{p}\propto1/d_n$, from which Eq.~\eqref{eq:deltaE_rel} would directly follow. 

\mypar{Physical interpretation of the first conduction states}

The first $n-1$ conduction states may therefore be seen as \textit{standing wave modes} confined within the stack of $n$ CuO$_2$ planes. 
To assert this interpretation, we cannot use Eq.~\eqref{eq:phase_acc} directly since $\Theta(E^{n}_{p})$ is unknown, and the usual empirical formulas used for thin films are not applicable in our context. 
Instead, we consider a simple uni-dimensional quantum well with an energy barrier $V_{0}=8\;$eV, corresponding to the energy of the first conduction states of the surrounding Cl atoms, which form the barrier of the well. 
We assume a linear dispersion relation $k_{z}(E)=\alpha E$, where $\alpha$ is the only free parameter that is adjusted so to reproduce the energies $E^{n,\rm DFT}_{p}$. 
For each $n$, $\alpha$ is optimized by minimizing the distance 
$$\Delta E^{n}=\sum_{p=1}^{n-1}\left|E^{n,\rm DFT}_{p}-E^{n,\rm QW}_{p}\right|,$$
where $E^{n,\rm QW}_{p}$ are the quantum well energies. 
We obtain $\alpha^{n}\in[1.125,1.151]$, i.e., a relatively stable value given the simplicity of our toy model. 
As is usually done in the thin film community, we then reconstruct the dispersion $k_{z}(E)$ by plotting all the modes' energies $E^{n}_{p}$ against $p/d_{n}$.
Remarkably, all the $E^{n,\rm QW}_{p}$ are in excellent agreement with the DFT predicted bands $E^{n,\rm DFT}_{p}$, and form a linear relationship, as shown in Fig.~\ref{fig:fig6}(b).

This analysis of the \emph{nature} of the first conduction states of multilayer cuprates allows to understand the dependence of the CTG and $J$ with respect to the number of CuO$_{2}$ planes and their position in the structure.  
Since these standing wave modes are confined within the stack of $n$ CuO$_2$ planes, the hybridization with the outer planes is significant only on one side of the plane (towards the interior), in contrast to the inner planes for which the hybridization is strong on both sides. 
Hence, similarly to a surface phenomenon, but here found in a \emph{bulk} system, the inner planes hybridize twice as strong as the outer planes with these standing wave modes, which leads to a reduction of the CTG. 
We emphasize the generality and importance of this interpretation for unconventional superconductors in general, since the formation of such conduction states can also be noticed in multilayer nickelates~\cite{labollita2022}.

\subsection{nHBCCO: contrast from self-doping}
\label{sec:res_undop_nh}

\begin{figure*}
    \centering
    \includegraphics[width=\linewidth]{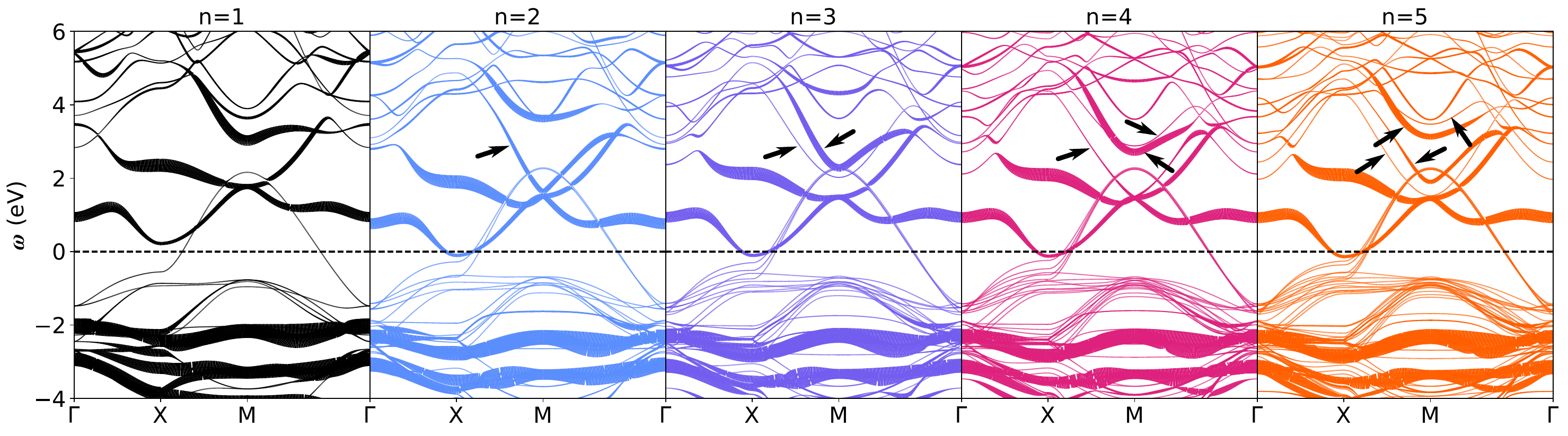}
    \caption{(a) \emph{DFT-only} band structures for the $n$=1\textendash5 \nH. 
    The relative linewidth denotes the contribution from the Hg and apical O. 
    The black arrows highlight the additional bands that, in analogy with the \nC~case, are expected to contribute to the $n$-dependent behavior.}
    \label{fig:fig7}
\end{figure*}

We now turn to the \nH~family, and start by presenting the DFT band structure in Fig.~\ref{fig:fig7}. 
In contrast to \nC, we observe two bands just above the Fermi level for all $n$, which stem from Hg and the apical O~\cite{singh1993,singh1994}, and are marked by thick lines.
For $n\geq 2$, these bands induce a self-doping effect similar to that observed for the Bi-based multilayer compounds~\cite{chan1991,lin2006,nokelainen2020}.
In addition, we notice the emergence of $n-1$ low-energy conduction bands (marked by black arrows), which are repelled at energies slightly above the self-doping bands. 
Based on the arguments developed previously, we argue that the Hg-O bands cause a reduction of the CTG in all compounds, including the single layer compound. In addition, the $n$ dependence is weaker than in \nC, because Hg-O bands push the standing wave modes higher up in energy.

\begin{figure*}
    \centering
    \includegraphics[width=\linewidth]{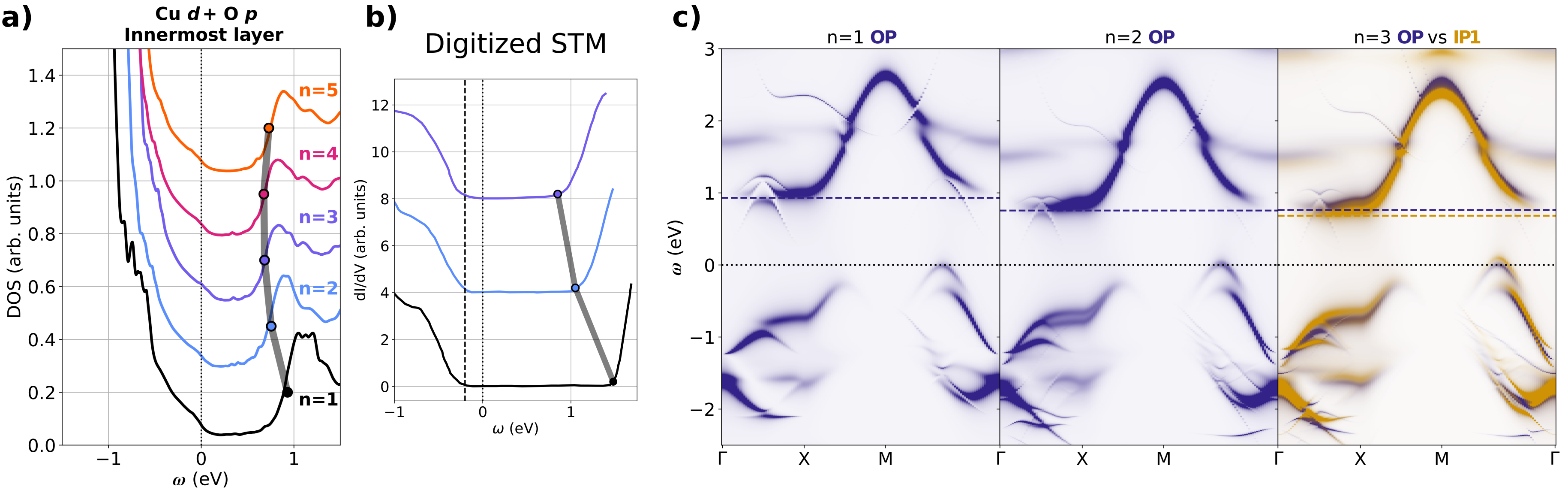}
    \caption{(a) DFT+CDMFT projected DOS on the Cu-$d$ and O-$p$ orbitals of the \emph{innermost} CuO$_2$ plane of \nH~($n$=1\textendash5).
    The colored dots and gray line in the unoccupied part highlight the variations of the CTG.
    To ease the comparison, we align the bottom of the CTG with the Fermi level for the innermost planes of $n=1$ and $n=5$ (which otherwise would be below the Fermi level) by applying a rigid shift in energy.
    (b) STM measurements digitized from Ref.~\onlinecite{wang2023a}, performed on $n=(1,2,3)$ Bi$_2$Sr$_2$Ca$_{n-1}$Cu$_{n}$O$_{2n+4+\delta}$ insulating samples. 
    (c) Momentum-resolved spectral function of the Cu-\dxxyy~orbital for $n=1,2,3$, along $\Gamma$\textendash X \textendash M\textendash$\Gamma$. 
    The purple and orange dashed lines highlight the CTG for the OP and IP1 planes, respectively.}
    \label{fig:fig8}
\end{figure*}

\mypar{Spectral function analysis}

The DFT+CDMFT DOS projected onto the Cu-$d$ and O-$p$ orbitals of the innermost CuO$_2$ plane of \nH~is shown in Fig.~\ref{fig:fig8}(a). 
Only the single layer compound is strictly insulating, while the other ones remain metallic. 
The Hg-O bands induce a self-doping effect even when correlations in CuO$_2$ planes are considered. 
Surprisingly, the charge gap in the innermost planes remains finite for all compounds, as seen in Fig.~\ref{fig:fig8}(a), which means that the CuO$_2$ planes are in a selective Mott state in the parent compound, making multilayer compounds very bad metals.
Note that to ease the visual comparison, we applied a rigid shift to the $n=1$ and $n=5$ DOS in Fig.~\ref{fig:fig8}(a) to cancel self-doping effects. 
Overall, the $n$-dependence of the DOS is very similar to the STM measurements of Ref.~\onlinecite{wang2023a} (shown in Fig.~\ref{fig:fig8}(b)), which were performed on insulating  Bi$_2$Sr$_2$Ca$_{n-1}$Cu$_{n}$O$_{2n+4+\delta}$ samples.
%
%

The CTG is evaluated (without the shift) as the distance between the Fermi level and the bottom of the upper Hubbard band (see Fig.~\ref{fig:fig8}(c)). 
The DFT+CDMFT calculations confirm our anticipation: the CTG is overall smaller than in the \nC~compounds, and the dependence on $n$ is less pronounced. 
The mechanism that leads to the reduction of the CTG is however the same, as can be seen from the spectral functions displayed in Fig.~\ref{fig:fig8}(c). 
In the outer plane, the CTG is reduced from $n=1$ ($\Delta_{\rm CTG}^{n=1}\approx0.92\;$eV) to $n=2$ ($\Delta_{\rm CTG}^{n=2}\approx0.75\;$eV) and then saturates, and is smaller in the inner planes for $n\geq3$ ($\Delta_{\rm CTG}^{n\geq3,\rm IP1}\approx0.68\;$eV).

\begin{figure}
    \centering
    \includegraphics[width=0.8\linewidth]{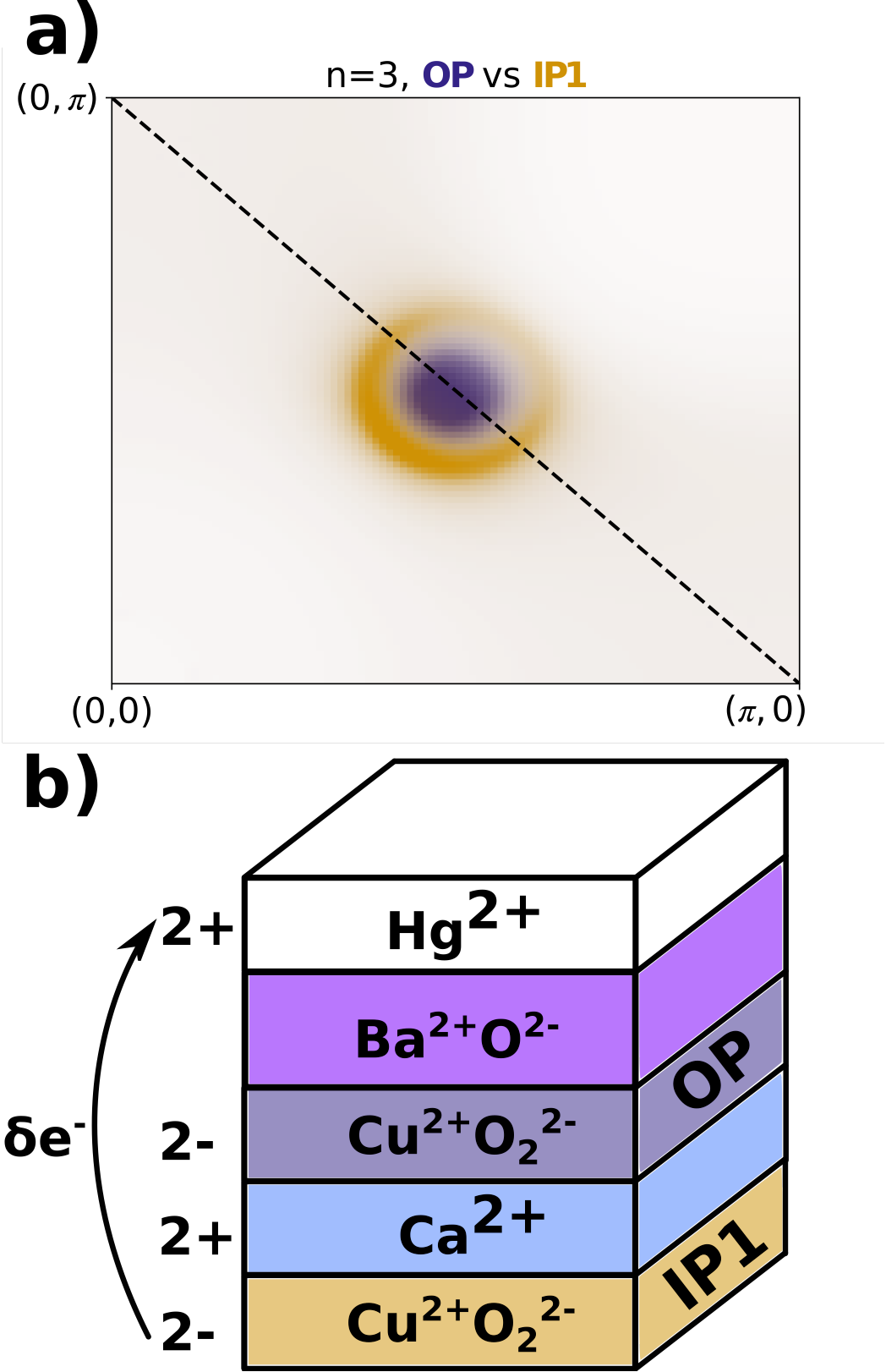}
    \caption{(a) FS for the $n=3$ \nH.
    (b) Simplified sketch of the charged layers composing the system. 
    It is favorable to transfer a portion of electrons $\delta e^{-}$ from IP1 rather than OP to minimize the electrostatic energy.}
    \label{fig:fig9}
\end{figure}

\mypar{Effect of self-doping}

We also noticed a subtle difference between the tri-layer spectral functions of \nC~and \nH. 
The first valence band of the inner plane in \nH~lies closer to the Fermi level than the outer plane, while in \nC~it is the opposite. 
This is related to the self-doping, which favors hole-doping of the inner planes rather than the outer ones.  
We confirm it by showing the Fermi surface (FS) in Fig.~\ref{fig:fig9}(a): the IP1 pocket is indeed larger than that of the OP. 
The reader interested in the fact that we obtain Fermi pockets, and not arcs, at low doping levels will find a detailed discussion in Sec.~\ref{sec:res_dop_spec}.
As an additional check, we also estimated the doping in each layer using the procedure described in App.~\ref{ap:estim_dop}.
We obtained $\delta^{\rm OP}_{\rm CuO_2}\simeq1.5\%$ and $\delta^{\rm IP1}_{\rm CuO_2}\simeq1.9\%$. 

This result is surprising since one may expect that the self-doping due to the apical O and Hg atoms would introduce more holes in the nearest CuO$_2$ plane, the OP. 
We propose a simple electrostatic picture, shown in Fig.~\ref{fig:fig9}(b), from which we argue that the system can minimize its energy by creating more holes in the IP1 instead of OP. 
Due to self-doping, a portion of electrons $2\delta e^{-}$ has to be transferred to the surrounding Hg$^{2+}$ planes from (one of) the (Cu$^{2+}$O$^{-2}_2$)$^{2-}$ planes.
The two possibilities lead to different energy costs, that we denote $E_{\textrm{OP}}$ and $E_{\textrm{IP1}}$. 
Considering the layers as a collection of point charges, one may find that the difference between the two energies, at the first order in $\delta$, will scale as
\begin{equation}
    E_{\textrm{OP}} - E_{\textrm{IP1}} \propto 4e^2\delta\left(\frac{1}{r_{\rm OP}}-\frac{1}{r_{\rm IP1}}\right) > 0
\end{equation}
where $r_{\rm OP/IP1}$ is the distance between the Hg$^{2+}$ plane and the outer/inner (Cu$^{2+}$O$^{-2}_2$)$^{2-}$ ones. 

The energy difference is positive: from a purely electrostatic point of view, it is advantageous to screen the positive charge of the Hg$^{2+}$ plane by maximizing the number of electrons in its vicinity. 
In other words, it is energetically favorable to take the portion $2\delta$ of electrons from the IP1 which is located further away from the Hg$^{2+}$ plane.

We will show in Sec.~\ref{sec:res_dop} that this self-doping effect is important for doping and enhancing superconductivity. 
Namely, hole-doping in the charge reservoir layers easily increases the hole concentration in the OP, but much less efficiently in the inner planes.
The low concentration of holes in the inner planes is a strong limiting factor for superconductivity.
In contrast to \nC, the intrinsic self-doping effect in \nH~seems to create more favorable conditions for superconductivity to emerge intrinsically in the inner planes. 

\begin{figure}
    \centering
    \includegraphics[width=0.8\linewidth]{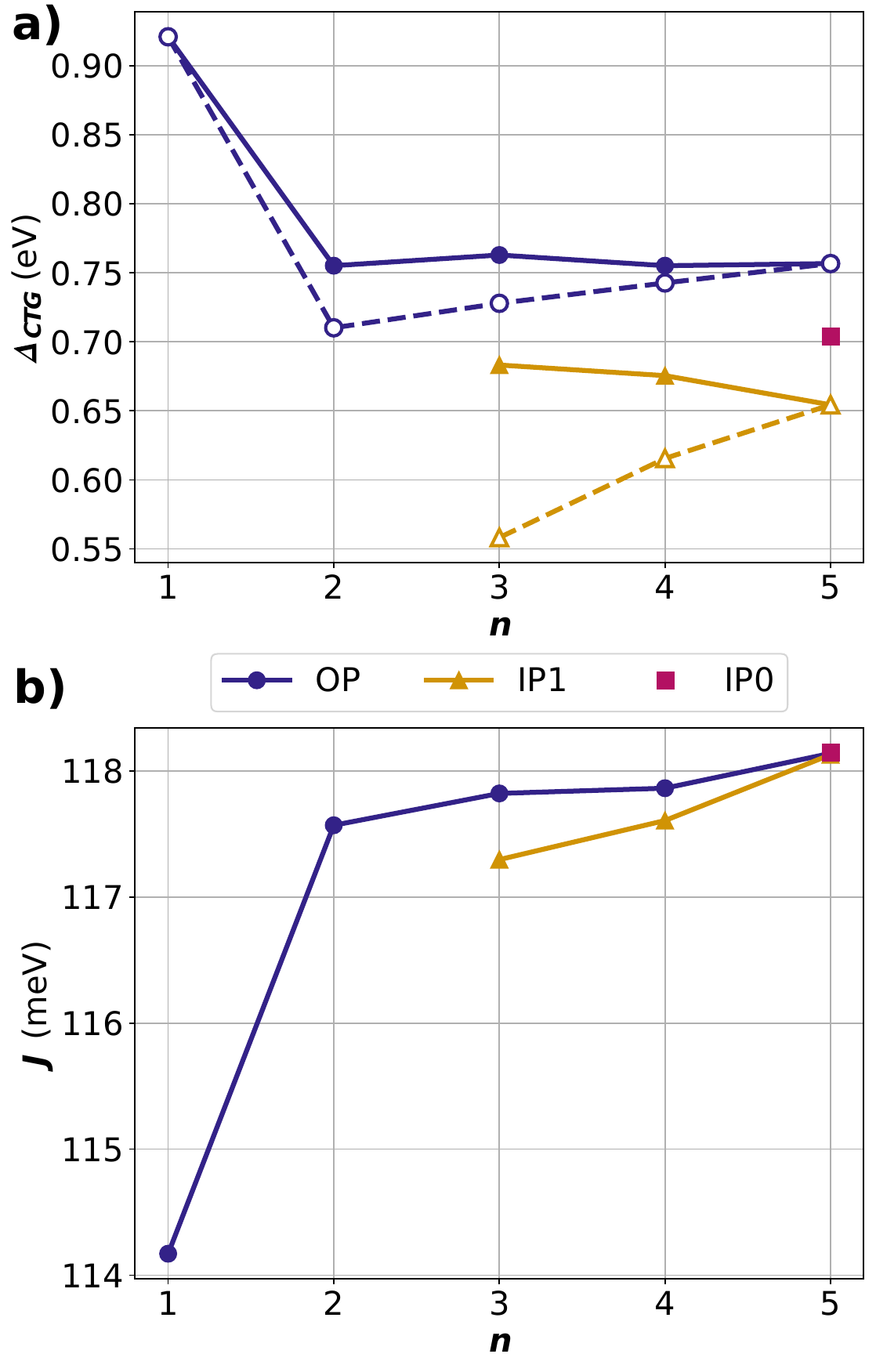}
    \caption{(a) CTG and (b) effective superexchange $J$ estimated for the OP, IP1 and IP0 for \nH~($n$=1\textendash5).
    The dashed curves with open symbols in (a) correspond to the case where the bottom of the CTG is considered to be the top of the Cu-$d$ and O-$p$ Zhang-Rice band instead of the Fermi level (solid lines full symbols, \emph{see text}). 
    }
    \label{fig:fig10}
\end{figure}

\mypar{Charge transfer gap vs. $n$}

We summarize in Fig.~\ref{fig:fig10}(a) the values of the CTG in \nH~for each type of CuO$_2$ planes with respect to $n$, as evaluated from the Fermi level (solid lines full symbols). 
Overall the CTG is smaller than in \nC, and the $n$-dependence is similar but less pronounced.
The evolution of the CTG from $n=1$ to $n=3$ agrees remarkably well with the STM measurements of Ref.~\onlinecite{wang2023a} (see Fig.~\ref{fig:fig8}(b)), performed on Bi-based multilayer cuprates.
These were \emph{insulating} parent compounds, obtained by performing chemical substitution to prevent self-doping~\cite{wang2023a}. 
We can anticipate the CTG for the case where the self-doping is removed, by evaluating it from the top of the ZRS band, and not from the Fermi level.
The CTG obtained in this way is shown in Fig.~\ref{fig:fig10}(a) (dashed lines empty symbols), and suggests that the dependence on $n$ could be stronger in such insulating materials.
In particular, the tri-layer compound shows a significantly smaller gap, yielding a CTG \emph{vs.} $n$ profile strikingly reminiscent of the $n$-dependence of $T_{c}$~\cite{mukuda2012,wang2023a}.
We note that this crude prediction may, however, slightly overestimate the reduction of the CTG since the Hg-O self-doping bands would be shifted away from the Fermi level. 

\mypar{Superexchange $J$ vs. $n$}

We now turn to the superexchange $J$, which is shown in Fig.~\ref{fig:fig10}(b) with respect to the number and type of consecutive CuO$_2$ planes.
First, $J$ in \nH~ is overall larger than in \nC~compounds, which is consistent with the smaller CTG. 
Second, it shows a clear growth with $n$, as in \nC, consistent with the reduction of the CTG. 
In contrast to \nC, however, the inner planes show smaller $J$ than outer planes, which seems inconsistent with CTG. 
This is likely due to limitation of our method to extract parameter $J$, which is simply given by $(\pi,\pi)$ susceptibility for insulators as explained in Sec.~\ref{sec:res_undop_nc_ctg}. 
Our method thus most likely underestimate the superexchange in the metallic $n\ge 2$ compounds. 

The absolute values of $J$ are of correct order of magnitude, but not directly comparable to experiments. 
Our estimated values of $J=114\;$meV for $n=1$ and $J=118\;$meV for $n=2$ are not very far from experimental values of $J\sim 135\;$meV ($n=1$) and $J\sim 180\;$meV ($n=2$)~\cite{wang2022}. 
Larger discrepancy for $n=2$ as compared to $n=1$ is likely due to the fact that $n=2$ compound is metallic and our extracted value of $J$ is already screened. 
Lastly, as mentioned in Sec.~\ref{sec:res_undop_nc} similar \emph{ab initio} approach based on DFT+DMET applied to $2\times2$ clusters predicts very similar values of $J$ for the single- and bi-layer \nH~compounds~\cite{cui2023a}.

\section{Results: hole-doped multilayer cuprates}
\label{sec:res_dop}

\begin{figure*}
    \centering
    \includegraphics[width=0.8\linewidth]{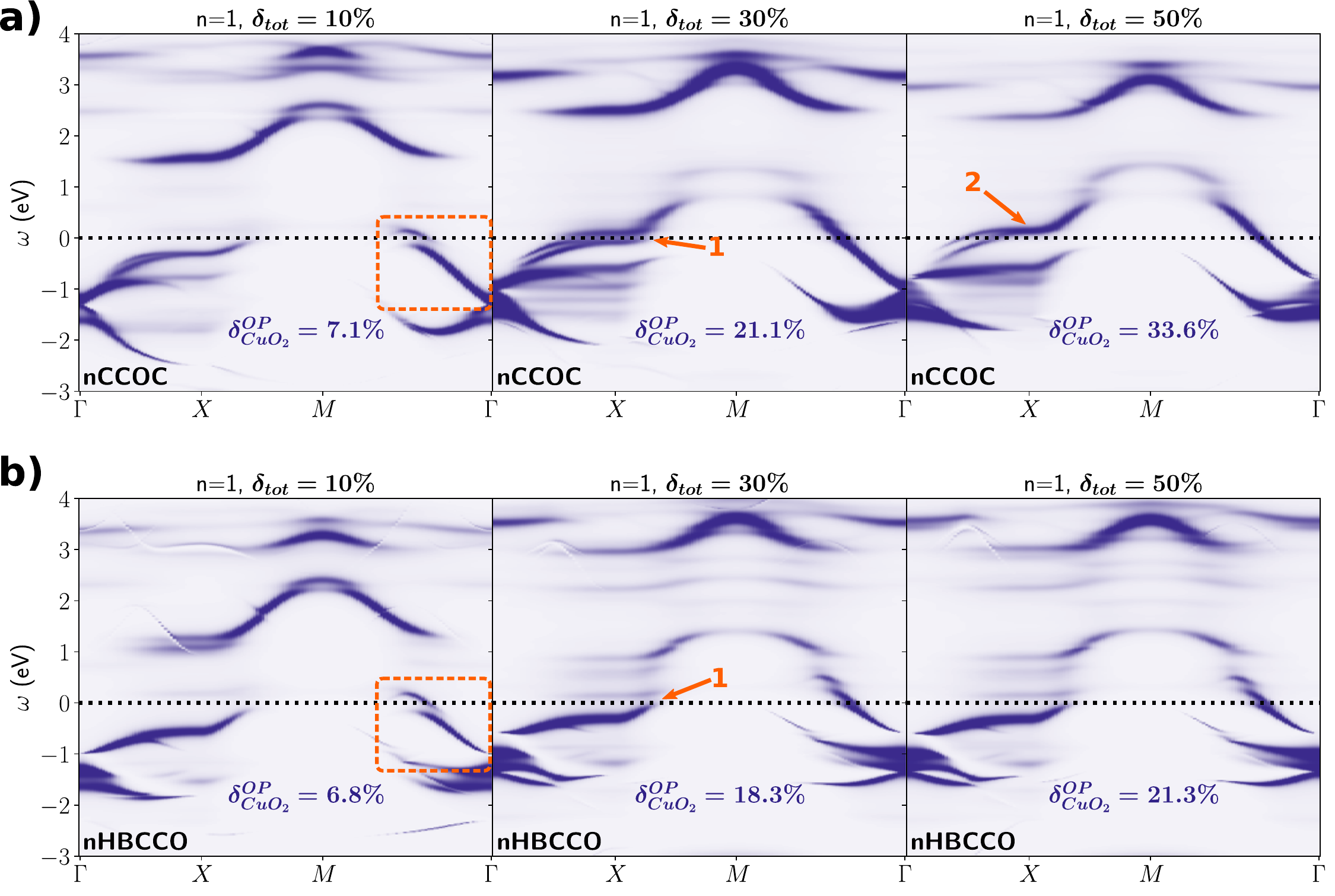}
    \caption{Momentum-resolved spectral function projected on the Cu-\dxxyy~orbital for three values of total doping $\delta_{\rm tot}$=[$10\%,30\%,50\%$], computed for the single-layer ($n=1$) (a) \nC~and (b) \nH.
    The estimated doping in the OP $\delta^{\rm OP}_{\rm CuO_2}$ is specified on each panel.
    The orange boxes and numbered arrows highlight several features discussed in the text. }
    \label{fig:fig12}
\end{figure*}

We now turn to the analysis of the hole-doped multilayer cuprates. 
They are to be thought of as doped Mott insulators, hence their properties should strongly inherit from the physics of the parent compounds~\cite{lee2006,kancharla2008b,sordi2010}.
This question of inheritance has been mostly studied with effective models. 
Therefore, it is important to quantify how each material specificity may influence the evolution of the electronic properties upon hole-doping. 

Another key aspect of the hole-doped multilayer cuprates is the inhomogeneous doping of the CuO$_2$ planes~\cite{mukuda2012,kunisada2020,oliviero2022a,kurokawa2023}. 
It is at the source of the observation of Fermi pockets coexisting with arcs in the FS~\cite{kunisada2020,kurokawa2023}, and it strongly influences the superconducting properties~\cite{mukuda2012,kurokawa2023}. 
In particular, although the inner planes present favorable conditions for superconductivity in terms of CTG~\cite{wang2023a}, $J$ and homogeneity~\cite{mukuda2012}, the small amount of holes that may reach them prevents the observation of large critical temperatures in these planes~\cite{mukuda2012}. 
Theoretical insight are crucially needed to understand better how the doped holes redistribute in the structure, to establish whether there is a doping threshold for the inner planes or if there are ways to enhance their capacity to host holes, and finally to study how superconductivity could be stimulated. 

In our first principles approach, the doping of each plane is determined self-consistently. 
We can control only the total doping of the material. 
We use the same VCA values for {\it total} doping $\delta_{tot} = 10,~20,~30,~40\text{ and }50\%$ for all compounds ($n=1,\dots,5$). 
This allows doping in the different planes to range from the usual underdoped to optimally-doped and overdoped regimes.

In the following, we start with a short summary of the main results for the hole-doped multilayer cuprates (Sec.~\ref{sec:res_dop_sum}).
Then, we investigate the evolution of the spectral properties upon hole-doping (Sec.~\ref{sec:res_dop_spec}), and we analyze how the insertion of charge carriers rearranges the electron and hole distribution (Sec.~\ref{sec:res_dop_hd}). 
Finally, in Sec.~\ref{sec:res_dop_sc} we examine the superconducting properties and their relation to the normal state described previously. 

\subsection{Summary of the hole-doped results}
\label{sec:res_dop_sum}

We track the evolution of the spectral properties from the under-doped to the over-doped regimes. 
In agreement with experimental evidence~\cite{vanveenendaal1994,damascelli2003,hu2021}, we show that the transformation is twofold (Figs.~\ref{fig:fig12},\ref{fig:fig13}): (i) at very low dopings, there is essentially a chemical potential shift; (ii) above roughly $4\%$ of effective doping, the shift is combined with a substantial spectral weight transfer from the high-energy Hubbard bands to the low-energy features. 
The experimental observation of Fermi pockets coexisting with arcs in multilayer cuprates~\cite{kunisada2020,kurokawa2023} is naturally explained by our calculations (Fig.~\ref{fig:fig14}). 
The inhomogeneous doping is accurately taken into account, and lead to situations where the inner planes are in the very under-doped regime, while the outer planes are close to the optimal doping. 
Hence, the spectral function of the inner planes only undergo the chemical potential shift which promotes the nodal backfolded spin-polaron dispersion (see Sec.~\ref{sec:res_undop_nc_ctg}) to the Fermi level, thus forming pockets, while the optimally-doped outer planes display an arc.

We find that these transformations are universal to the two families studied, but our method also allows to identify material-specific trends. 
Namely, the effective doping of the CuO$_2$ plane in the single-layer \nH~saturates close to optimal doping, while its \nC~counterpart clearly crosses the over-doped regime. 
We emphasize that our method enables to theoretically address \emph{ab initio} the key 
elements of the HTSC spectral function, such as 
the pseudogap~\cite{alloul1989,warren1989,ito1993,damascelli2003,timusk1999,Huscroft2001,civelli2005,kyung2006,sakai2009,sordi2012,macridin2006,ferrero2009,werner2009,gull2010,merino2014,krien2022}, and the high-energy anomalies~\cite{martinez1991a,macridin2007a,manousakis2007a,wang2015a,bacq2023,kohsaka2003,graf2007a,graf2007,kordyuk2006,borisenko2006a,borisenko2006b,valla2007,xie2007,moritz2009,zhang2008,damascelli2003,basak2009,zhou2010}.

Not only can we track the doping evolution through the spectral properties, we can also directly map out the carrier redistribution in real space, by virtue of the charge self-consistency. 
This allows to visualize and understand precisely how the holes redistribute in the crystal structure (Figs.~\ref{fig:fig15},\ref{fig:fig16}). 
In particular, we identify two doping regimes in which the different orbitals involved behave differently. 
While at low and intermediate doping the Cu-\dxxyy~orbital gain holes upon doping, its hole-content saturates at large doping while other Cu-$d$ orbitals start to host holes. 
Moreover, we show that increasing the hole-content on a given orbital is always compensated by an additional electronic cloud in the surrounding orbitals, leading to a non-trivial charge carrier redistribution on the Cu and O atoms. 
This may provide important information for the experimental determination of the effective hole-content, using for instance nuclear magnetic resonance~\cite{haase2004,jurkutat2014,jurkutat2023}.

We estimate the superconducting properties of the multilayer cuprates by computing the zero-temperature superconducting order parameter $\msc$ (Figs.~\ref{fig:fig17},\ref{fig:fig18}). 
We show that $\msc$ remains zero until a hole-doping threshold of $\sim4\%$, in remarkable agreement with experiments~\cite{kurokawa2023} (see Fig.~\ref{fig:fig18}). 
The existence of this doping threshold explains one of our most important finding: the inner planes of the $n\geq4$ compounds \emph{do not} host superconductivity because their effective doping remains too small (Fig.~\ref{fig:fig18}). 
This highlights an apparent competition between the favorable conditions of the inner planes (smaller CTG, larger $J$, protection from inhomogeneities), and the unfavorable lack of holes.
In addition to these universal trends, our method can also reproduce the expected material-specific trends since $\msc$ is larger in the \nH~compounds in comparison to \nC~(Fig.~\ref{fig:fig17}).

Finally, we are able to provide a combined analysis of the normal state and superconducting properties (Figs.~\ref{fig:fig19},\ref{fig:fig20}). 
We show that for superconductivity to emerge, the normal state DOS should display a nearly particle-hole symmetric feature in the vicinity ($\pm200\;$meV) of the Fermi level reminescent of a pseudogap. 
Both the emergence and disappearance of superconductivity in the under-doped and over-doped regimes can be tracked down to this pseudogap-like feature (Fig.~\ref{fig:fig19}). 
Moreover, the superconducting trends may be influenced by material-specific characteristics, such as the relative energy difference between the Cu-\dzz~and \dxxyy~orbitals, in agreement with previous works~\cite{sakakibara2010,sakakibara2012} (Fig.~\ref{fig:fig20}). 
This highlights the need for \emph{ab initio} methods for constructing a more complete theory of high-temperature superconductivity.

\subsection{Evolution of the spectral properties}
\label{sec:res_dop_spec}

Upon doping, the spectral function of high-temperature superconductors undergoes major transformations related to key aspects of the electronic properties~\cite{meinders1993,vanveenendaal1994,damascelli2003,wang2020,hu2021,bacq2023,kurokawa2023,sakai2023}. 
ARPES measurements carried on multilayer cuprates have revealed the inhomogeneous doping between the outer and inner planes~\cite{kunisada2020,kurokawa2023,luo2023}, confirming the interpretation of nuclear magnetic resonance and quantum oscillations measurements~\cite{mukuda2012,oliviero2022a}.
Fermi pockets have been observed for the first time in hole-doped multilayer compounds, revealing at the same time that arbitrarily small doping turns the inner CuO$_2$ planes metallic~\cite{kurokawa2023}. 
All these intriguing spectral properties of multilayer cuprates have been seen experimentally, but have not been understood theoretically.
While much progress has been done by studying the doping-dependent properties of the one-band and three-band effective Hubbard models, these can hardly account for a markedly inhomogeneous distribution of holes between adjacent CuO$_2$ planes.

As described in Sec.~\ref{sec:met_crys}, we emulate the hole-doping within VCA, with extra charge added to the outer Ca and Hg atoms, ranging from $\delta_{\rm tot}=10\%$ to $\delta_{\rm tot}=50\%$, which is sufficient to describe the main trends observed experimentally. 
We emphasize that in this study we do not consider the questions related to the thermodynamics of dopants and defects~\cite{zunger2021,lany2007}.
Such considerations, and a more realistic treatment of doping, are kept for future works. 

\mypar{Spectral function of the single-layer compounds}

\begin{figure*}
	\centering
	\includegraphics[width=0.8\linewidth]{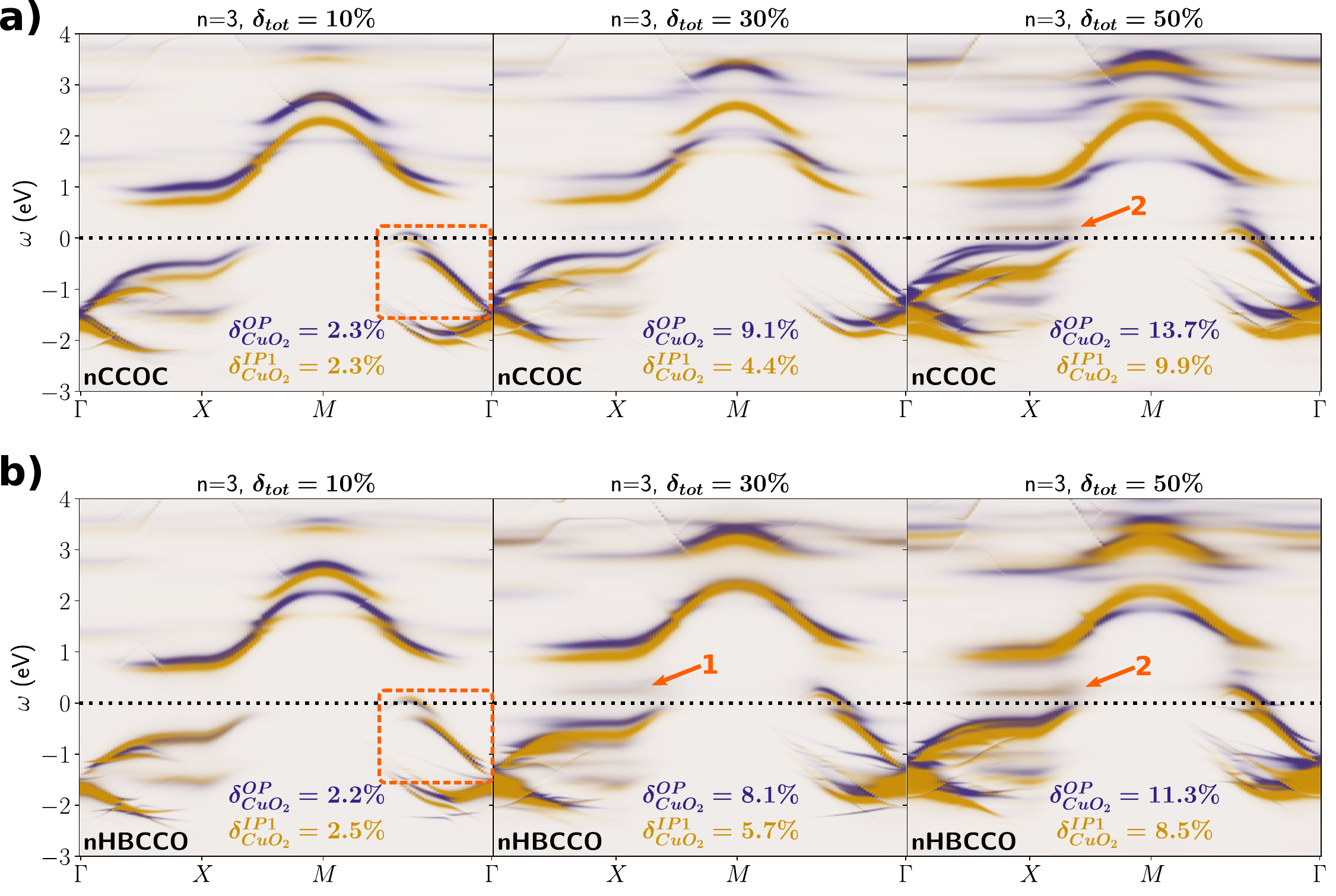}
	\caption{Momentum-resolved spectral function projected on the Cu-\dxxyy~orbital for three values of total doping $\delta_{\rm tot}$=[$10\%,30\%,50\%$], computed for the tri-layer ($n=3$) (a) \nC~and (b) \nH.
		The estimated dopings in the OP and IP1 $\delta^{\rm OP/IP1}_{\rm CuO_2}$ are specified on each panel.
		The orange boxes and numbered arrows highlight several features discussed in the text.}
	\label{fig:fig13}
\end{figure*}

We start with the analysis of the momentum-resolved spectral functions of the single-layer \nC~and \nH~compounds, projected onto the Cu-\dxxyy~orbital, as shown in Fig.~\ref{fig:fig12}.
Note first that approximately $70\%$ of all the inserted holes migrate to the CuO$_2$ plane, as shown by the estimated doping $\delta^{\rm OP}_{\rm CuO_2}$ (see App.~\ref{ap:estim_dop} for details about the estimation method).
This is in agreement with previous \emph{ab initio} studies using the VCA and the rigid band approximation~\cite{cui2023a}.
We however observe a stark deviation from this tendency for the $\delta_{\rm tot}=50\%$ \nH~compound: the effective hole-content saturates at a much lower value than its \nC~counterpart.
Consequently, the spectral function remains unchanged beyond $\delta_{\rm tot}=30\%$. 
The apparent pinning of the Fermi level is due to a substantial shift of the occupied Hg and apical O bands, which starts to cross the Fermi level at $\delta_{\rm tot}=30\%$, consistently with previous findings~\cite{singh1994}.
Clear evidence of this phenomenon is provided in App.~\ref{ap:estim_dop}.  
We shall come back to this point later in Sec.~\ref{sec:res_dop_hd}, while we now focus on the transformations of the spectral function.

At low doping $\delta_{\rm tot}=10\%$, the spectral function resembles that of the undoped compounds (see Figs~\ref{fig:fig2},\ref{fig:fig8}) to which a rigid chemical potential shift was applied. 
The backfolded spin-polaron feature at the nodal point $(\pi/2,\pi/2)$ (in the path $\Gamma$-M, highlighted by the orange boxes), is promoted to the Fermi level without major transformation. 
Moreover, no substantial spectral weight transfer is noticed above the Fermi level at the anti-nodal $(\pi,0)$ point.
The transfer is visible at larger values of dopings, such as $\delta_{\rm tot}=30\%$, at which the pseudogap is opened in the anti-nodal region (orange arrows n°1). 
%
%
At $\delta_{\rm tot}=50\%$, the pseudogap in \nC~closes (but not in \nH) leaving the Van Hove singularity above the Fermi level (orange arrow n°2). 

In short, the spectral function analysis of the $n=1$ compounds shows that our method effectively captures the hole-doping evolution observed experimentally in cuprates.
Indeed, the chemical potential first shifts towards the top of the valence band, and upon further doping the spectral weight is substantially reconfigured while opening the pseudogap~\cite{vanveenendaal1994,damascelli2003,hu2021}. 
Similar spectral evolution has been seen before in simulating effective models for cuprates~\cite{meinders1993,civelli2005,wang2020,sakai2023}.
Yet, our method reveals that even these simple single-layer cuprates may display unexpected material-specific features: the \nH~compound cannot be easily over-doped and remains in the pseudogap phase even at high levels of doping due to the Fermi level pinning. 

\begin{figure*}
	\centering
	\includegraphics[width=0.75\linewidth]{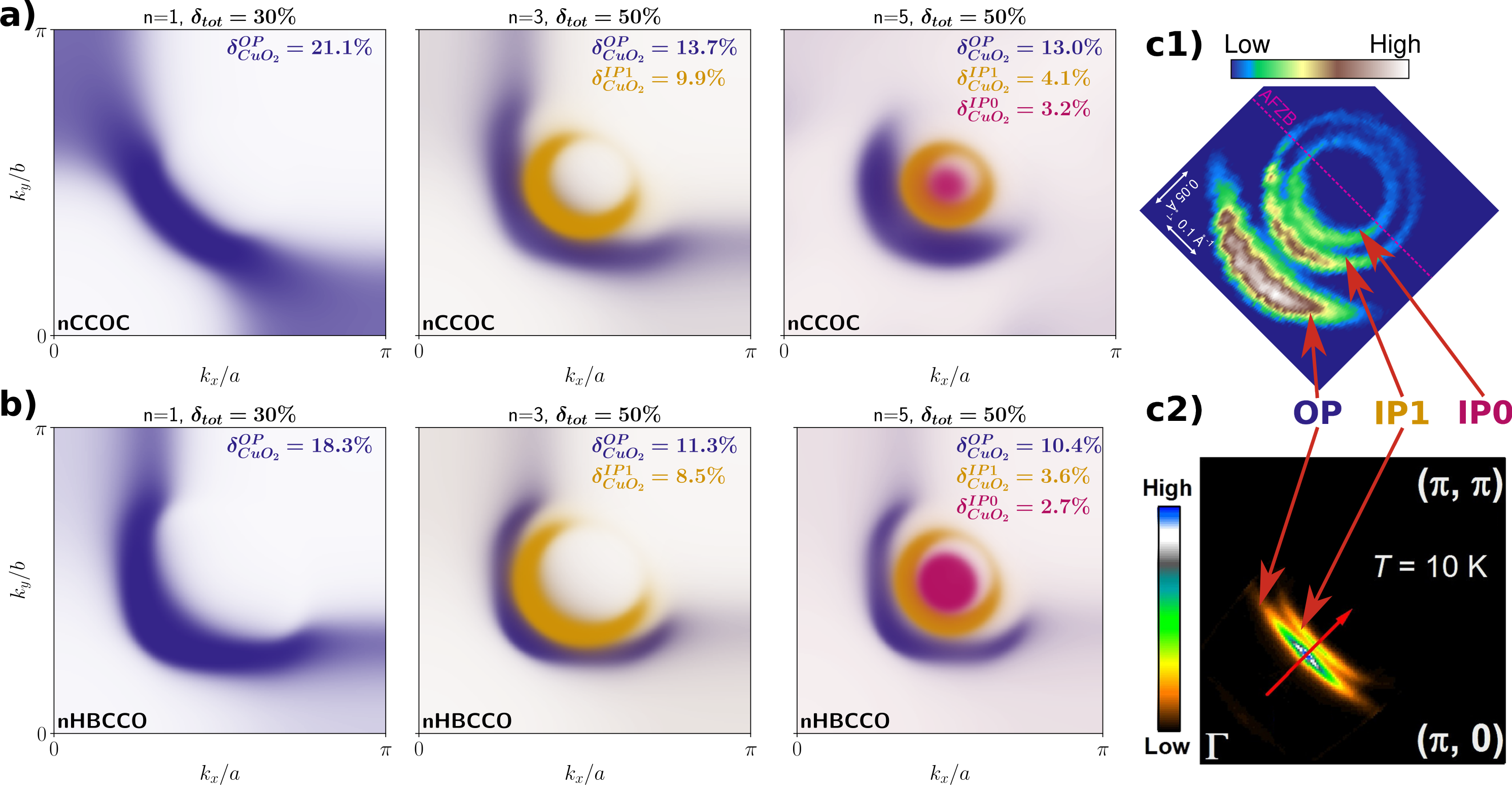}
	\caption{FS projected on the Cu-\dxxyy~orbital for the $n=1,3,5$ (a) \nC~and (b) \nH.
		The total VCA doping is $\delta_{\rm tot}=30\%$ for the single-layer compound, and $\delta_{\rm tot}=50\%$ for the $n=3,5$ ones.
		The estimated dopings in the OP, IP1 and IP0 $\delta^{\rm OP,IP1,IP0}_{\rm CuO_2}$ are specified on each panel.
		(c1,2) Fermi surfaces measured via ARPES on, respectively, $n=5$ Ba$_2$Ca$_4$Cu$_5$O$_{10}$(F,O)$_2$ and $n=3$ Bi$_2$Sr$_2$Ca$_2$Cu$_3$O$_{10+\delta}$ samples. 
		The two panels were adapted with permission from Refs.~\onlinecite{kunisada2020,ideta2010}.
	}
	\label{fig:fig14}
\end{figure*}

\mypar{Spectral function of the tri-layer compounds}

We now turn to the tri-layer compounds ($n=3$) in Fig~\ref{fig:fig13}.
We inserted the same number of holes $\delta_{\rm tot}$ in the charge-reservoir layers as before.
Since the holes are shared between the three CuO$_2$ planes, the effective doping of each plane is expected to be much smaller here, but one would still expect their sum to be around 70\% of nominal doping. 
Summing over the estimated dopings $\delta^{\rm OP/IP1}_{\rm CuO_2}$ however shows that, \emph{in total}, more holes are absorbed within the CuO$_2$ planes for $n=3$, in comparison to $n=1$. 

At low doping ($\delta_{\rm tot}=10\%$), the holes are distributed rather uniformly between the planes.
The spin-polaron feature is rigidly shifted across the Fermi level (orange boxes), and no spectral weight transfer is visible at the anti-nodal point. 
Many of the key features discussed in the undoped compounds (see Sec.~\ref{sec:res_undop_sum}) are recognizable: (i) the CTG,  measured by the onset of the UHB, is smaller for $n=3$ as compared to $n=1$, and is minimized in the inner plane, and (ii) the inner plane in the \nH~compound contains slightly more holes than the outer plane, in contrast to \nC. 

Upon further doping ($\delta_{\rm tot}=30\%$), the inhomogeneous distribution of holes identified in experiments~\cite{mukuda2012,oliviero2022a,kunisada2020,kurokawa2023,luo2023} is clearly accounted for in our calculations.
\nC~does not show significant spectral weight transfer above the Fermi level at the anti-node (at the node it is simply a consequence of the chemical potential shift), while a weak feature can be identified in all planes of \nH~ (orange arrow n°1). 
The smaller CTG in \nH~may thus be favorable to the spectral weight transfer upon hole-doping.

Finally, at substantial doping ($\delta_{\rm tot}=50\%$), where CuO$_2$ planes are still under-doped, we identify a spectral weight transfer above the Fermi level for all planes in both \nC~and \nH, with a clear pseudogap feature in the anti-nodal region (orange arrows n°2). 
A key difference between the two families is the difference between the energies of the ZRS band of the IP1 and of the OP planes: the energy distance is much larger in \nC~than in \nH.
This translates into a larger difference between the doping levels in the \nC~(3.8\%) as compared to \nH~(2.8\%). 
We note that from the spectral functions alone, we would expect smaller doping level in the inner plane of \nC~compound as compared to the one of \nH~compound. 
Our estimations, however, are $\delta^{\rm IP1}_{\textrm{nCCOC}}=9.9\%$ and $\delta^{\rm IP1}_{\textrm{nHBCCO}}=8.5\%$, respectively. 
This apparent discrepancy is related to the method used to estimate the doping, \emph{a posteriori}, which is perturbed by the shift of the Hg and apical O occupied bands at the Fermi level in \nH~at large values of doping (see App.~\ref{ap:estim_dop}).
Hence we emphasize that this is not a limitation of our DFT+CDMFT calculations. 
Indeed, our estimates, the first theoretical ones, for the doping differentiation between the inner and the outer planes agree qualitatively with the experimental estimates from nuclear magnetic resonance and quantum oscillations~\cite{mukuda2012,oliviero2022a}.

\begin{figure*}
	\centering
	\includegraphics[width=0.85\linewidth]{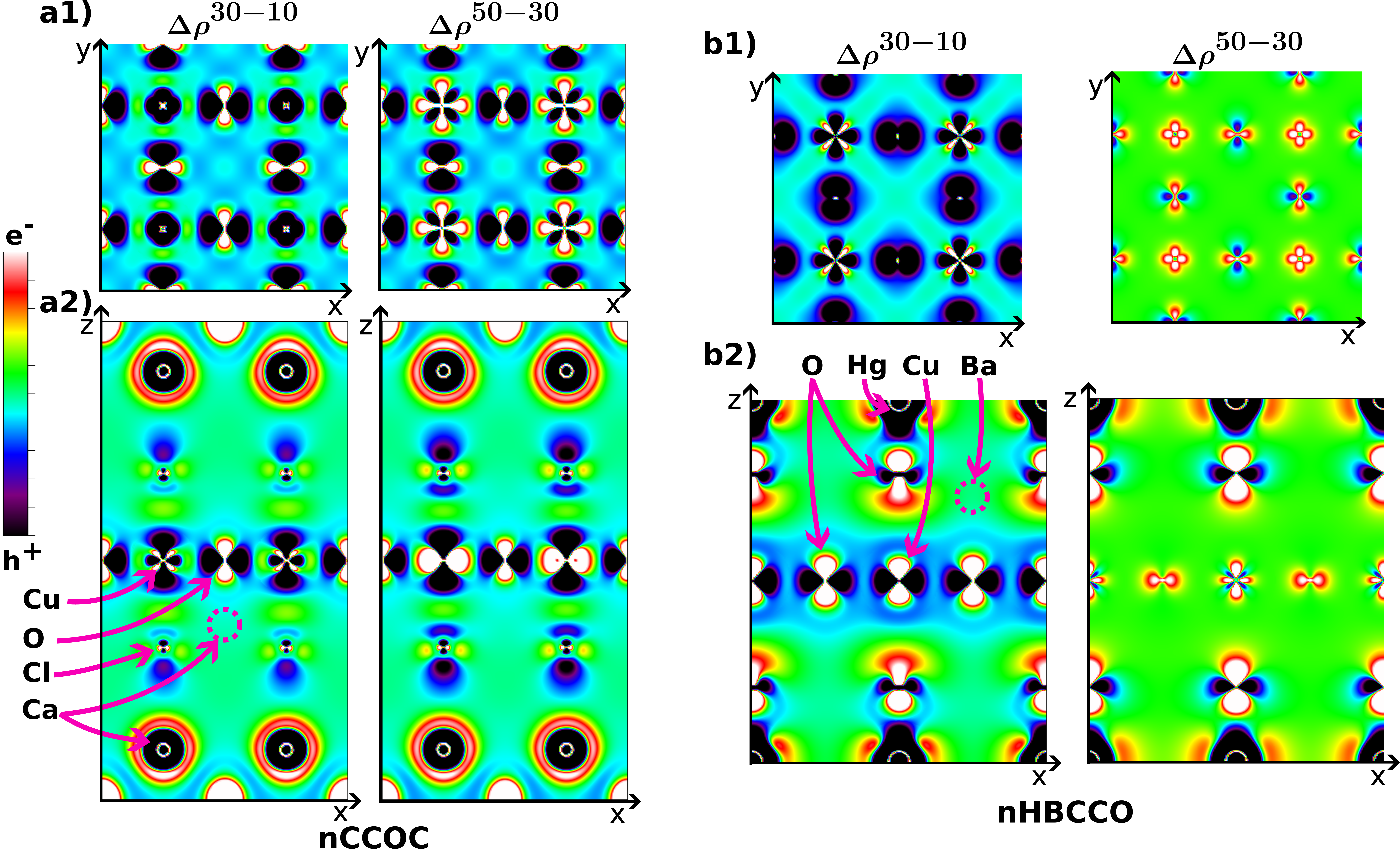}
	\caption{Charge density difference $\Delta\rho(\mathbf{r})$ between two different total VCA doping values for the single-layer ($n=1$) (a1-a2) \nC~and (b1-b2) \nH. 
		Panels (a1),(b1) are top views of the CuO$_2$ OP, while (a2),(b2) are side views of the whole structure.}
	\label{fig:fig15}
\end{figure*}

\mypar{Coexistence of Fermi arcs and pockets}

A striking consequence of the inhomogeneous doping across different planes is the coexistence of different FS: arcs and pockets of different sizes, which were recently identified with ARPES measurements~\cite{ideta2010,kunisada2020,kurokawa2023} (see Fig.~\ref{fig:fig14}(c1,2)). 
In Fig.~\ref{fig:fig14}(a-b) we present the FS obtained from the momentum-resolved spectral function projected onto the Cu-\dxxyy~orbital, for $n=1,3,5$ \nC~and \nH~compounds. 
As expected, the single-layer systems yield a single Fermi arc, which shows a pseudogap at the antinode. 
We choose a representative doping of $\delta_{\rm tot}=30\%$, at which the CuO$_2$ plane is slightly over-doped. 
For three-layer ($n=3$) we choose $\delta_{\rm tot}=50\%$, which makes the outer planes close to optimal doping, and inner planes  under-doped. 
We see a coexistence of a small and a large arc for the outer and inner planes, respectively.
The calculated Fermi surfaces are in excellent agreement with ARPES measurements performed on $n=3$ Bi$_2$Sr$_2$Ca$_2$Cu$_3$O$_{10+\delta}$ samples~\cite{ideta2010} and reproduced in Fig.~\ref{fig:fig14}(c2). 
As discussed above, the inner arc of \nH~is wider than the one of \nC, although our estimates for the relative dopings would suggest otherwise. 
For $n=5$, we choose $\delta_{\rm tot}=50\%$, which again puts the outer plane close to optimum doping, while the inner planes are heavily under-doped. 
We observe a coexistence of Fermi arcs (OP) and pockets (IP1 and IP0), in striking resemblance to the ARPES measurements~\cite{kunisada2020,kurokawa2023}, as shown in Fig.~\ref{fig:fig14}(c1) in which we reproduce the Fermi surface of $n=5$ Ba$_2$Ca$_4$Cu$_5$O$_{10}$(F,O)$_2$ from Ref.~\onlinecite{kunisada2020}.

Moreover, we can identify the mechanism behind the transformation from pockets to arcs.
In agreement with the interpretation of Refs.~\onlinecite{kunisada2020,kurokawa2023}, we find that the Fermi pocket is a consequence of the strong antiferromagnetic correlations.
At very low doping only the chemical potential shift occurs without major transformation of the spectral function. 
Hence, the backfolded spin-polaron dispersion at the nodal point across the antiferromagnetic zone boundary (most clearly visible in the undoped spectra in Fig.~\ref{fig:fig2},\ref{fig:fig8}) survives and yields a closed Fermi pocket. 
As doping increases, the bending of the dispersion disappears and it is pushed away from the Fermi level, thus transforming the pocket to an arc. 
At the anti-node, some weak spectral weight appears due to the incremental filling of the pseudogap.

\begin{figure*}
	\centering
	\includegraphics[width=0.9\linewidth]{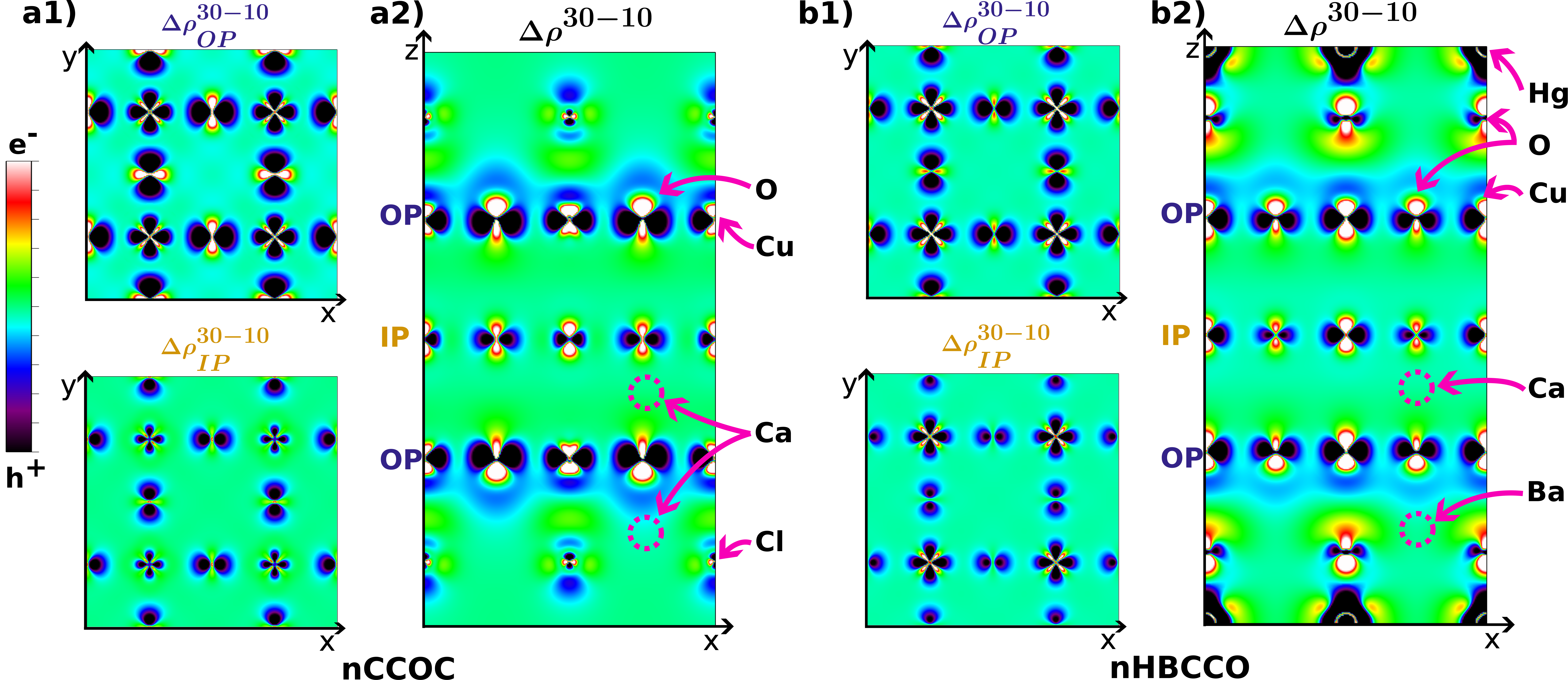}
	\caption{Charge density difference $\Delta\rho(\mathbf{r})$ between two different total VCA doping values for the tri-layer ($n=3$) (a1-a2) \nC~and (b1-b2) \nH. 
		Panels (a1),(b1) are top views of the CuO$_2$ OP and IP, while (a2),(b2) are side views of the whole structure.}
	\label{fig:fig16}
\end{figure*}

\subsection{Carrier redistribution}
\label{sec:res_dop_hd}

Until now we presented the momentum and energy dependence of doping holes in cuprates. 
However, the real space reorganization of the holes is also important~\cite{haase2004,jurkutat2014,rybicki2016,jurkutat2023}, especially since there is a lack of theoretical insights in this area.
In the following, we discuss the redistribution of charges (both holes and electrons) as a whole in the multilayer cuprates. 
We show that doping yields non-trivial phenomena which highlight the complex nature of doping quantum materials~\cite{zunger2021}.

\mypar{Charge redistribution in the single-layer compounds}

In order to understand the doping-dependent redistribution of charges, we compute the charge density difference between two DFT+CDMFT simulations at different doping levels $\Delta\rho^{x-y}(\vr) = \rho^{x\%}(\vr)-\rho^{y\%}(\vr)$.
This quantity is displayed in Fig.~\ref{fig:fig15} for the single-layer \nC~and \nH~compounds at two different levels of doping. 
At each point in space $\mathbf{r}$, a blue-purple-black (yellow-red-white) color indicates that there is effectively more holes (electrons) at doping $x\%$ as compared to $y\%$, i.e., $\rho^{x\%}(\vr)\leq\rho^{y\%}(\vr)$ ($\rho^{x\%}(\vr)\geq\rho^{y\%}(\vr)$).

We first focus on the real space $xy$ cut through the CuO$_2$ plane and $xz$ cut through the structure of \nC, see Figs.~\ref{fig:fig15}(a1) and (a2).   
In both doping regimes (\emph{left} and \emph{right} panels) the charge carriers redistribute in a non-trivial way, some orbitals effectively gaining holes while other gain electrons.
Within the CuO$_2$ plane at intermediate doping $\Delta\rho^{30-10}$, the extra holes populate the Cu-\dxxyy~ and \dzz~orbitals, as well as the O-$p_x$ orbitals (pointing towards the Cu atoms). 
The O-$p_y$ and $p_z$ orbitals (pointing away from Cu) gain electrons instead, so that the O atoms remain close to the O$^{2-}$ oxidation state. 

At larger values of doping $\Delta\rho^{50-30}$ the Cu-\dxxyy~orbital seems to stop accepting additional holes.
They instead start to populate the Cu-$d_{xy}$ orbital.
However, the behavior of the Cu-\dzz~and the O-$p$ orbitals remains similar to the intermediate doping regime, where $p_x$ gains extra holes while $p_y$, $p_z$ gain electrons.  
This observation is consistent with the transformations of the spectral function shown in Fig.~\ref{fig:fig12}. 
At $\delta_{\rm tot}=50\%$ the Fermi level has been shifted below the Van Hove singularity, and $t_{2g}$ valence states, such as Cu-$d_{xy}$, start to get populated with holes.  

The charge redistribution in the $n=1$ \nH~compound at low doping ($\Delta\rho^{30-10}$ is shown in Fig.~\ref{fig:fig15}(b1).
The strongest electronic cloud is now in proximity to Hg atom (to which VCA is applied). 
In similarity with \nC, the holes in the CuO$_2$ plane occupy the Cu-\dxxyy~and O-$p_x$ orbitals, but the compensating electrons are here found only on the O-$p_z$ orbital, not $p_y$.
This is linked with the absence of interstitial Ca atoms (see below the discussion of the $n=3$ compounds).  
Another difference is that the Cu-\dzz~absorbs electron rather than holes upon hole-doping. 
This may be related to the larger crystal-field splitting between the Cu-\dxxyy~and \dzz~orbitals in \nH~compounds, discussed in Sec.~\ref{sec:res_dop_sc}.
At large values of doping ($\Delta\rho^{50-30}$) the CuO$_2$ plane becomes almost inert, and the vast majority of the holes screen electrons close to the Hg and apical O atoms. This is consistent with the momentum-resolved spectral functions in Fig.~\ref{fig:fig12}, which remains almost unchanged beyond $\delta_{\rm tot}=30\%$.

\mypar{Charge redistribution in the tri-layer compounds}

We now turn to the charge redistribution in the tri-layer compounds. 
Since the CuO$_2$ planes remain in the intermediate regime even at $\delta_{\rm tot}=50\%$, both $\Delta\rho^{30-10}$ and $\Delta\rho^{50-30}$ are similar, and 
we only show the former in Fig.~\ref{fig:fig16}.
Both the inner (IP1) and the two outer planes (OP) in \nC~compound (Fig.~\ref{fig:fig16}(a1)) are similar to the single-layer compound, namely holes go on the Cu-\dxxyy~orbital and O-$p_x$, while electrons go on $p_y$ and $p_z$. 
There is a substantial difference in the intensity between the two planes:  the IP1 is less hole-doped than the OP, which is consistent with the spectral functions discussed in Sec.~\ref{sec:res_dop_spec}. 

This is also evident in Fig.~\ref{fig:fig16}(a2), which shows the density difference along the $xz$ plane. 
The redistribution of holes in the outer plane O-$p_z$ orbital is very peculiar: the outer part, pointing towards the charge-reservoir layers is confined, while in inner part merges with an extended electronic cloud between the OP and Ca atoms.
This behavior is reminiscent of orbitals at the surface of a material, and thus connects with our analysis of the undoped compounds: the physics of the $n$-CuO$_2$-plane stack is similar to that of a thin film when $n\geq2$ (see Sec.~\ref{sec:res_undop_nc_cond}). 

In Fig.~\ref{fig:fig16}(b1,b2) we show the charge density difference $\Delta\rho^{30-10}$ for the tri-layer \nH~compound. 
Overall, it is very similar to its \nC~counterpart: the inner plane is again less hole-doped than the outer planes, and the same orbitals gain holes of electrons.  
We notice that in the OP the O-$p$ orbitals pointing away from Cu tend to gain electrons (see Fig.~\ref{fig:fig16}(b1)), which was not the case in the $n=1$ \nH~compound (see Fig.~\ref{fig:fig15}(b1)).
Such difference may be related to the presence of additional Ca atoms separating the CuO$_2$ planes for the $n\geq2$ \nH~structures. 

\subsection{Superconductivity}
\label{sec:res_dop_sc}

Having analyzed the normal state properties of the hole-doped compounds in Sec.~\ref{sec:res_dop_spec},\ref{sec:res_dop_hd}, we now turn to superconductivity. 
As detailed in Sec.~\ref{sec:met_edmft} we compute the zero-temperature superconducting $d$-wave order parameter $\msc=\braket{\hat{D}_{\mathrm{SC}}}$ for each plane separately.
We thus neglect the proximity effect in the superconducting state between adjacent CuO$_2$ planes for $n\geq2$, as justified in App.~\ref{ap:prox}.

\begin{figure}
    \centering
    \includegraphics[width=\linewidth]{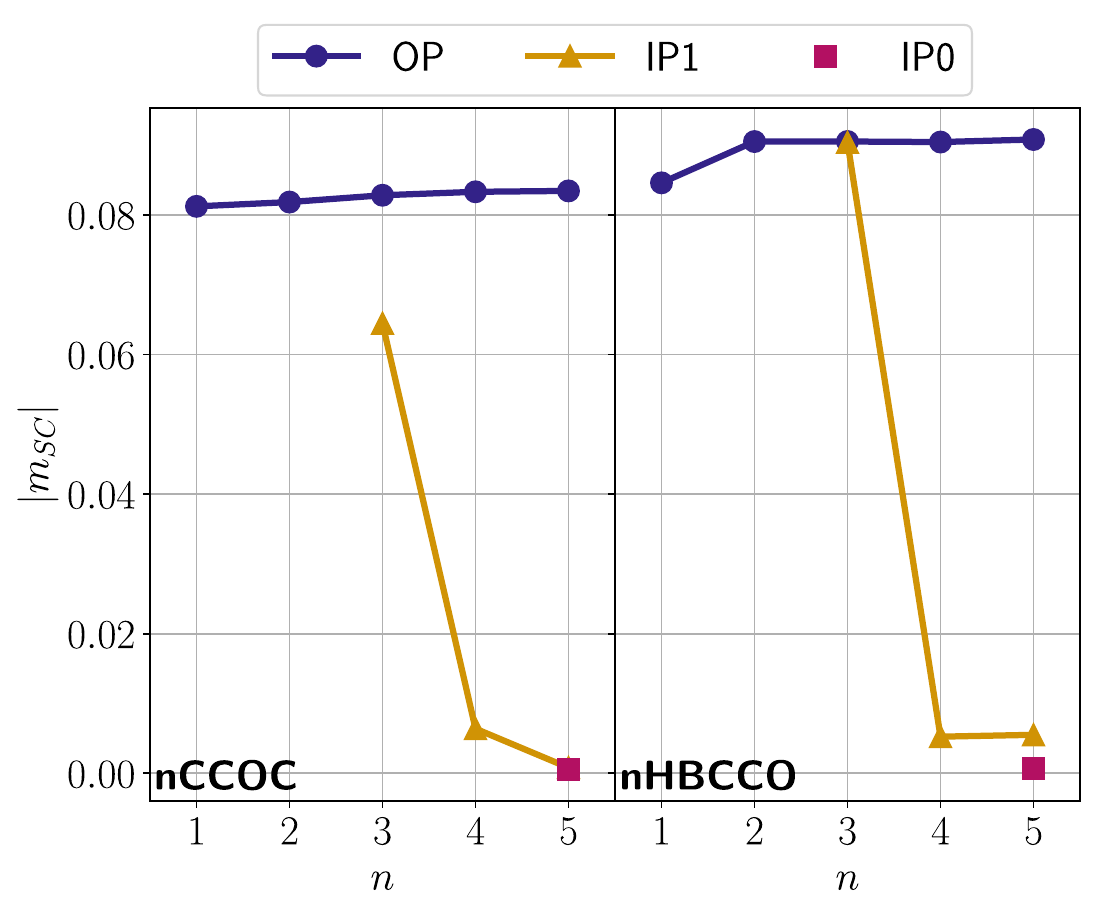}
    \caption{Optimal superconducting order parameter $\msc$ for the $n=1-5$ \nC~(\emph{left}) and \nH~(\emph{right}), in each type of CuO$_2$ planes.}
    \label{fig:fig17}
\end{figure}

\mypar{Optimal superconducting order parameter}

In Fig.~\ref{fig:fig17}, we show the \emph{optimal} superconducting order parameter $\msc$ in \nC~and \nH, with respect to the number of consecutive CuO$_{2}$ planes and their position in the structure. 
The \emph{optimal} $\msc$ is obtained by taking the largest value among all dopings $\delta_{\rm tot}=10-50\%$.
We observe that the order parameter is overall larger in the \nH~compounds, consistently with expectations from our analysis of the normal state properties in Sec.~\ref{sec:res_undop}, as well as from experimentally known critical temperatures. 
In the outer planes of \nH, $\msc$ sharply increases from $n=1$ to $n=2$ and then saturates, consistent with the $n$-dependence of the CTG and $J$ (see Fig.~\ref{fig:fig10}).
This is contrasted with the behavior in the outer planes of \nC, in which the order parameter is only slowly and monotonically increasing with $n$ with no sharp jump (compare with Fig.~\ref{fig:fig3} for the evolution of the CTG and $J$).

The most important result in Fig.~\ref{fig:fig17} is the non-trivial dependence of the superconducting order parameter in the inner planes. 
In particular, the disappearance of superconductivity in IP0 and IP1 for $n\geq4$ is striking.
We checked that this disappearance resists to the inclusion of the proximity effect, see App.~\ref{ap:prox}.
The $n=3$ compounds are the only one hosting significant superconductivity in the inner planes (IP1), and \nH~has much stronger superconductivity in IP1 as compared to \nC.
Our results are similar to the experimental work of Ref.~\onlinecite{mukuda2012}, in which it was found that the estimated $T_c$ in the inner planes is almost always smaller than in the outer planes. 
In the next paragraph we explain these observations, which were so far not understood theoretically.

\begin{figure*}
    \centering
    \includegraphics[width=\linewidth]{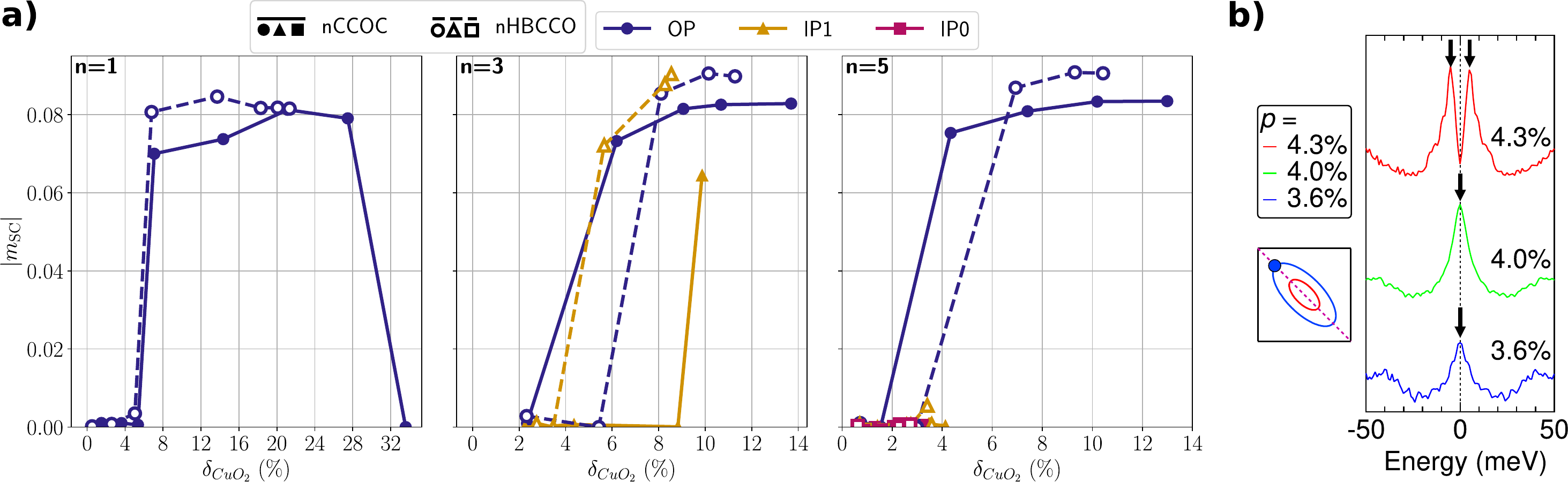}
    \caption{(a) Superconducting order parameter $\msc$ \emph{vs.} estimated doping on the corresponding CuO$_2$ plane for $n=1,3,5$ \nC~and \nH.
    (b) ARPES symmetrized energy distribution curve showing the opening of the superconducting gap in the IP1 plane, adapted with permission from Ref.~\onlinecite{kurokawa2023}. 
    The hole doping $p$ corresponds to our convention $\delta_{CuO_2}$. 
    }
    \label{fig:fig18}
\end{figure*}

\mypar{Superconducting order parameter vs. doping}

We show $\msc$ with respect to the estimated doping for each plane ($\delta^{\rm OP,IP1,IP0}_{\rm CuO_2}$) in Fig.~\ref{fig:fig18}(a).
The filled and the open symbols stand for \nC~and \nH, respectively. 
As observed in Sec.~\ref{sec:res_dop_sc},\ref{sec:res_dop_hd}, the single-layer \nC~can be efficiently doped so that we can find the under-doped ($<16\%$), optimally doped ($\approx 20\%$), and also the over-doped regime, with the fast drop of $\msc$ around $\delta^{\rm OP}_{\rm CuO_2}\simeq33.6\%$. 
In contrast, the effective doping in the single-layer \nH~saturates, so that the strongly over-doped regime is unreachable within this theoretical setup (and definitely hard in experiment). 
The largest estimated doping is 22\%, which is only slightly beyond optimal doping with strong superconductivity.
Note that we performed additional calculations for the single-layer compounds at $\delta_{\rm tot} = 2.5,~5,~\text{and }7.5\%$ to resolve the entire superconducting dome. 

We observe that $\msc$ is always vanishing when the estimated doping is below $\sim4\%$, in remarkable agreement with the opening of the superconducting gap at $4.3\%$ doping observed with ARPES in the $n=6$ multilayer cuprate~\cite{kurokawa2023} and reproduced in Fig.~\ref{fig:fig18}(b).  
It is also consistent with the experimental phase diagram of clean hole-doped cuprates~\cite{stewart2017,alloul2024}.
This observation is very interesting because the same CDMFT approach in the single- and three-band model simulations shows persistent superconducting dome all the way to the infinitesimally low doping~\cite{aichhorn2006,civelli2009,dash2019,sakai2023}.
Hence the finite doping threshold in superconductivity must be related to chemistry of cuprates, and is possibly material dependent\footnote{We have not investigated the competition with antiferromagnetism, that may also be a factor explaining the finite doping threshold~\cite{senechal2005competition,capone2006competition,kancharla2008b,foley2019a}.}. 
Furthermore, the results for $n=5$ explain why the superconducting order parameter is vanishing in the inner planes of the $n\geq4$ compounds: the effective doping never reaches the minimum threshold for superconductivity to \emph{intrinsically} emerge.
Hence, the $n=3$ compounds appear, at the level of our calculations, as the best compromise between (i) hosting protected inner planes, and (ii) reaching sufficient doping for superconductivity to emerge in every plane.  

We note that the inner plane (IP1) of the tri-layer \nH~compound has the largest order parameter, and is still slightly under-doped at $\delta_{\rm tot}=50\%$.
Thus, additional hole doping by other means may even slightly increase the superconductivity in the inner plane, while slightly reducing it in the outer plane. 
It would however be hard to reach doping levels in the five-layer compounds at which superconductivity would be stabilized in all layers.

\begin{figure*}
    \centering
    \includegraphics[width=\linewidth]{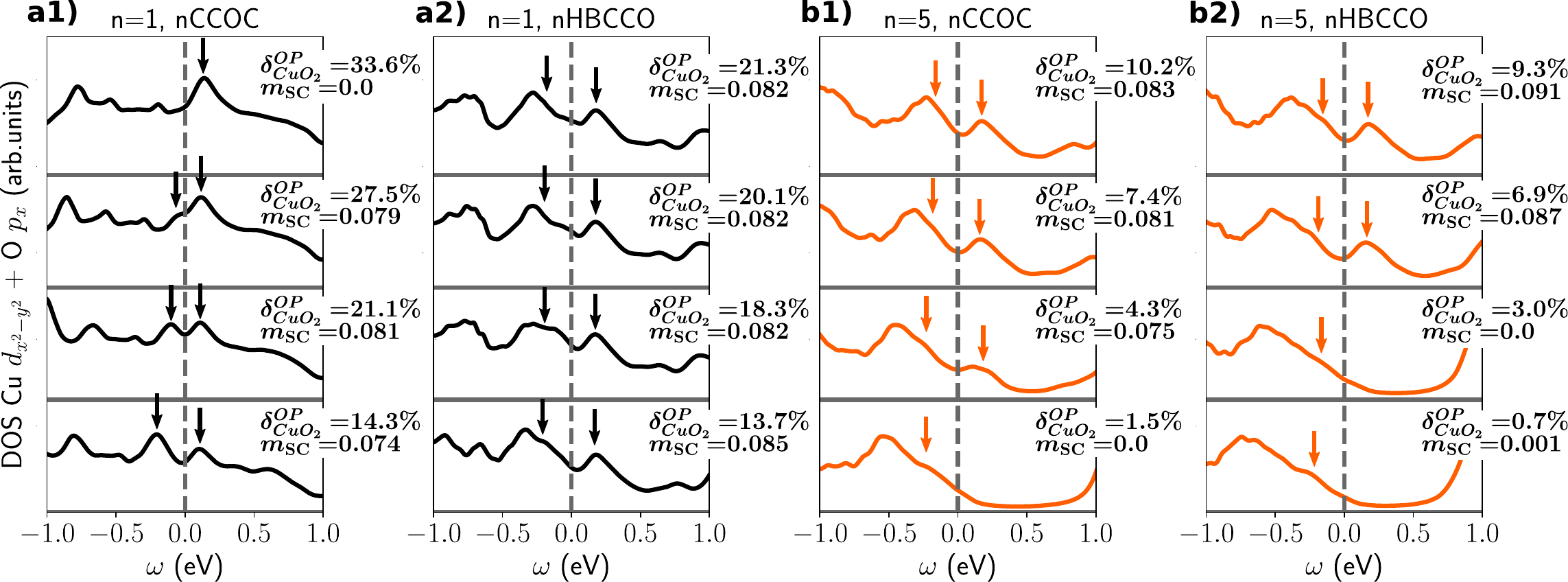}
    \caption{DFT+CDMFT DOS projected on the OP planes of the $n=1,5$ (a1,b1) \nC~and (a2,b2) \nH~compounds at different estimated dopings. 
    The latter, and the corresponding superconducting order parameter $\msc$ are explicitly written alongside each DOS.
    The arrows highlight the presence or absence of the symmetric pseudogap-like features around the Fermi level. }
    \label{fig:fig19}
\end{figure*}

\mypar{Link with the normal state properties}

Overall, our results support the interpretation that the CTG and $J$ control the superconducting order parameter~\cite{kowalski2021,weber2021}, but also show that the latter depends on other material-specific and doping-dependent parameters.
In the following, we draw further connection between $\msc$ and the normal state properties. 
We aim to answer two questions: (i) what drives the under- and over-doped transitions?; (ii) why is the order parameter in the \nH~compounds always larger than in \nC~(for a given doping value)?

We present in Fig.~\ref{fig:fig19} the \emph{normal-state} DOS at different estimated dopings, for $n=1$ and $n=5$, projected onto the Cu-\dxxyy~and the O-$p_x$ orbitals. 
We first inspect the panel (a1): it can be seen that as long as the order parameter $\msc$ is non-zero (three bottom curves), there exist a double-peak structure in a $\pm200\;$meV range around the Fermi level. 
Hereafter, we refer to this feature as a pseudogap. 
The top curve in Fig.~\ref{fig:fig19}(a1) shows that the over-doped transition in \nC, at $\delta^{\rm OP}_{\rm CuO_2}=33.6\%$, is directly linked with the loss of the pseudogap-like feature. 
This is a consequence of the transfer of the Van Hove singularity to the unoccupied part of the spectrum.
In Fig.~\ref{fig:fig19}(a2) are shown the normal-state DOS for $n=1$ \nH. 
As discussed above, the effective doping $\delta^{\rm OP}_{\rm CuO_2}$ saturates: the pseudogap does not close, and thus $\msc$ does not vanish. 

This relation between the normal-state pseudogap-like feature and the superconducting order parameter can also be observed in the underdoped side of the dome. 
As shown in Fig.~\ref{fig:fig19}(b1,b2), the order parameter in both $n=5$ \nC~and \nH~ is zero for small dopings at which the normal-state DOS does not contain spectral weight around $100-200\;$meV above the Fermi level. 
Hence the particle-hole symmetry of the pseudogap-like feature seems essential for superconductivity to emerge, which is consistent with conclusions drawn from previous model studies~\cite{sakai2023}. 

A noteworthy finding in our calculations is that enhanced particle-hole symmetry in the normal-state densities of states appears to be associated with optimal doping for superconductivity. 
In non-interacting models, the thermopower vanishes when the density of states is particle-hole symmetric. 
This correlation between particle-hole symmetry and optimal doping suggests a straightforward theoretical explanation for the observed coincidence of vanishing thermopower and optimal doping in cuprates (See Fig. 2 of~\cite{Honma_Hor_2008}). 
Previous studies using CDMFT~\cite{Chakraborty_Galanakis_Phillips_2010,Haule_Kotliar_Avoided_2007} and DCA~\cite{Vidhyadhiraja:2009} found a dynamically generated particle-hole symmetry similar to the one found here near optimal doping.
Moreover, Ref.~\cite{Chakraborty_Galanakis_Phillips_2010} finds that, when vertex corrections are neglected, the thermopower vanishes precisely at the particle-hole-symmetric optimal doping.

As discussed above, our calculations reproduce the doping threshold of around $4\%$ in the under-doped region of the phase diagram even though the model calculations using CDMFT on a small $2\times2$ cluster find superconductivity at arbitrary small doping~\cite{aichhorn2006,civelli2009,fratino2016a,dash2019,sakai2023}.
It is important to stress that in these calculations the normal state spectral function contains the pseudogap-like feature in its spectrum spectrum even at this infinitesimal dopings~\cite{fratino2016a,sakai2023}.
In contrast, using larger clusters and similar quantum embedding methods, it was found that there is a doping threshold at small doping~\cite{dionne2023a}. 
Interestingly, larger clusters suppress the pseudogap-like feature at these low doping levels~\cite{wang2020}. 
Our results are thus in agreement with these findings, although we use a small $2\times2$ impurity cluster, but more realistic electronic structure. 
Taking into account all other degrees of freedom in the structure, even at the DFT level, therefore allows to recover the right doping-dependence of $\msc$. 
It is as if the impurity problem is ``blind'' to very low doping, the spectral function looks thus similar to the insulating case and we observe only a chemical potential shift to the top of the ZRS band. 

\begin{figure*}
    \centering
    \includegraphics[width=\linewidth]{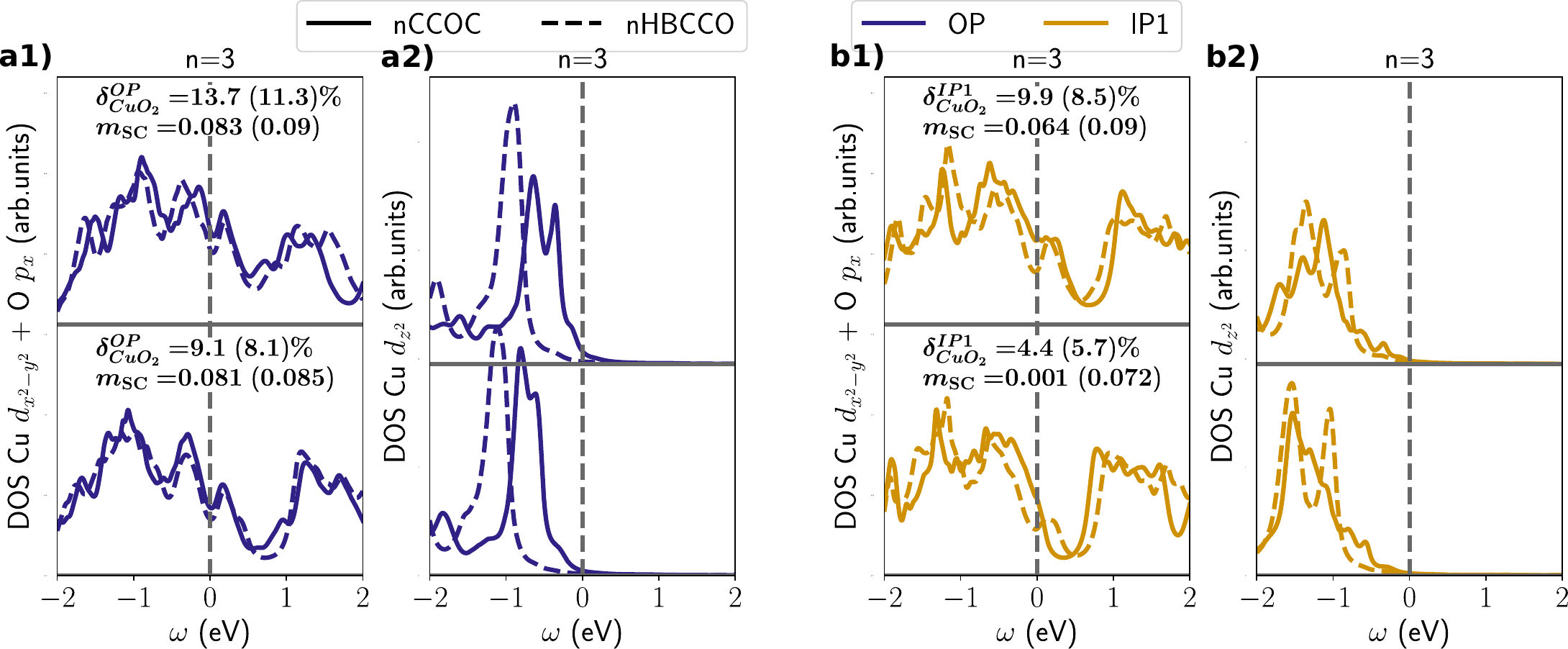}
    \caption{DFT+CDMFT DOS projected on the (a1,a2) OP and the (b1,b2) IP1 planes of the $n=3$ \nC~and \nH~compounds. 
    The estimated doping and superconducting order parameter $\msc$ for each plane and structure are displayed, the values for \nH~are given between parenthesis.}
    \label{fig:fig20}
\end{figure*}

To answer the question on the difference in superconductivity between the \nC~and \nH~compound, we show in Fig.~\ref{fig:fig20} the normal-state DOS of the different CuO$_2$ planes in the two systems, at different dopings. 
Their Cu-\dxxyy~and O-$p_x$ DOS are very similar when projecting to the outer plane (see Fig.~\ref{fig:fig20}(a1)).
The Cu-\dzz~DOS is however very different in the outer planes of the two compounds (see Fig.~\ref{fig:fig20}(a2)). 
The crystal-field splitting is much larger in \nH,
which has been shown to favor superconductivity~\cite{sakakibara2010,sakakibara2012}.

The normal state DOS of the inner plane is shown in Figs.~\ref{fig:fig20}(b1,b2). 
We note that projection onto the Cu-\dxxyy~and O-$p_x$ orbitals at small doping of $4.4\%$ in  \nC~compound does not have the pseudogap-like feature, and the superconductivity is vanishing. 
In contrast, \nH~compound at only slightly larger dopings of $5.7\%$ has this feature, and has finite $\msc$. 
At larger dopings of $9.9\%$ in \nC, the pseudogap-like feature starts to emerge, but it not fully formed in contrast to \nH. 
Consequently, $\msc$ becomes non-zero in \nC~but is weaker than in \nH. 
Moreover, the crystal field splitting between the Cu-\dzz~and \dxxyy~orbitals is equally large in both compounds in the inner planes, and thus does not contribute to differentiate the strength of $\msc$.
Based on this discussion, we conclude that the pseudogap-like particle-hole symmetric feature of the normal state DOS is essential for the emergence superconductivity, and also explains why superconductivity is hindered in the IP1 of \nC.

\section{Discussion and Conclusion}
\label{sec:conclusion}

In this work, we developed a charge self-consistent density functional theory plus \emph{cluster} dynamical mean-field theory framework, aiming towards an \emph{ab initio} theory of high-temperature superconductors.
This approach unlocks the simultaneous prediction of the material-specific superconducting, spectral and antiferromagnetic properties in unconventional superconductors.
To demonstrate these capabilities we investigated two multilayer cuprate families, Ca$_{(1+n)}$Cu$_{n}$O$_{2n}$Cl$_2$, and  HgBa$_2$Ca$_{(n-1)}$Cu$_n$O$_{(2n+2)}$, from the single- ($n=1$) to the five-layer ($n=5$) compounds, shedding light on their still-mysterious and often universal characteristics.  

Before delving deeper into our findings, it is important to note that the zero-temperature superconducting order parameter, $m_{sc}$, and the superconducting critical temperature, $T_c$, are not strictly proportional. 
In the conventional BCS theory, the superconducting gap is proportional to $T_c$, but our case is more complex. 
Here, BCS theory does not apply, and phase fluctuations likely influence $m_{sc}$ and $T_c$ differently. 
As a result, the peak of the $T_c$ dome as a function of doping is determined by finite-temperature long-distance phase fluctuations~\cite{Emery_Kivelson_1995,Kowalski_MSc}.
In the underdoped regime, Kosterlitz-Thouless physics becomes increasingly more important~\cite{Maier_Scalapino_2019}, with the superfluid stiffness decreasing~\cite{uemura1989probing,simard2019}.
Nevertheless, at the level of dynamical mean-field theory and one band model, studies have shown that the maximum $T_c$—in the absence of Kosterlitz-Thouless vortex-antivortex physics—is proportional to the maximum zero-temperature $m_{sc}$~\cite{fratino2016}. 
Thus, we have made this assumption in our analysis, and subsequent paragraphs are based on it. 
Further work will be necessary to compute the actual $T_c$, including Kosterlitz-Thouless phase fluctuations.

The most remarkable feature of the multilayer cuprates is the universal $n$-dependence of the critical temperature: $T_c$ is often maximum for $n=3$~\cite{mukuda2012}.
As was shown in Ref.~\onlinecite{wang2023a} this translates into a similar trend in the CTG, since the latter is expected to be inversely proportional to $T_c$~\cite{omahony2022a,kowalski2021}. 
In particular, the reduction of the CTG could be directly visualized through scanning transmission microscopy in undoped compounds from $n=1$ to $n=3$~\cite{wang2023a}. 
We show that the $n$-dependence of the CTG is indeed universal and that our DFT+CDMFT approach is able to reproduce it accurately. 
Most importantly, we trace its origin to the emergence of low-energy conduction states for $n\geq2$, reminiscent of quantum well states, which provide an additional hybridization channel for the correlated orbitals, and pin the upper Hubbard band closer to the Fermi level. 
These well states explain  the reduction of the CTG from the single- to the bi-layer compounds.
Furthermore, their confinement along the $z$ direction within the $n$-CuO$_2$-plane stack, as if they were trapped in a thin film, induces a reduction of the CTG in the inner planes in comparison to the outer ones, as the former can hybridize on both sides (above \emph{and} below) while the latter can hybridize on one side only (above \emph{or} below).  
Consequently the CTG is further reduced in the inner planes for $n\geq3$. 
This translates into an increase of the superexchange $J$ and relates to the experimental evidence of stronger antiferromagnetism within the inner planes~\cite{mukuda2012,kunisada2020,kurokawa2023}. 
It is remarkable that this physics is \emph{intrinsic} to the multilayer compounds: it exists prior to the inhomogeneous doping and disorder which are expected to increase the differentiation between the outer and inner planes. 

While the $n$-dependence of the CTG (and $J$) is universal to all multilayer families of cuprates, the absolute value of the gap is a material-specific characteristic well captured by our method. 
Namely, the presence of self-doping bands in the \nH~compounds induces a global, i.e., independent of $n$, lowering of the CTG in comparison to \nC. 
This is consistent with the lower $T_c$ of \nC~compounds as compared to \nH, and agrees with the STM measurements of Ref.~\onlinecite{wang2023a} showing that Bi-based compounds have an overall smaller (larger) CTG ($T_\mathrm{C}$) than \nC.
We emphasize that these phenomena involve a subtle interplay between the correlated and non-correlated degrees of freedom that is beyond the reach of usual model Hamiltonians. 
Our study thus underlines the need for \emph{ab initio} methods that include this interplay naturally, to enable material-specific predictions for high-temperature superconductors. 

We studied the hole-doped compounds using the virtual crystal approximation. 
There are numerous experimental evidence of the inhomogeneous doping between the outer and inner planes~\cite{mukuda2012,kunisada2020,oliviero2022a,kurokawa2023}, that our calculations capture accurately.
%
%
This inhomogeneity leads to the emergence of a composite Fermi surface made of Fermi arcs and pockets, in agreement with photoemission spectroscopy data~\cite{kunisada2020,kurokawa2023}. 
Having access to the entire doping evolution, from the undoped to the over-doped regimes, we can support the interpretation of Refs.~\onlinecite{kunisada2020,kurokawa2023} that pockets are a consequence of antiferromagnetic spin fluctuations and originate from the inner planes. 
This is deeply linked with the physics of the undoped compounds, in which spin-polaron quasiparticles emerge close to the Fermi level due to the strong spin fluctuations.
These quasiparticles display a backfolded dispersion across the nodal point along the $\Gamma$-M direction.
In the low-doping region to which the inner planes are confined, the spectral function mainly experiences a chemical potential shift that promotes the unperturbed backfolded feature at the Fermi level, thus leading to the emergence of pockets. 

The redistribution of charges within the material is a complex process. 
However, our charge self-consistent approach offers direct insight into this redistribution, incorporating the corrections to charge density introduced by electronic correlations. 
We identify markedly different behaviors at intermediate doping in contrast to large doping. 
At intermediate doping, the holes primarily occupy the Cu-\dxxyy~and O-$p_{x}$~(pointing towards Cu) orbitals, as well as the VCA-doped atom (Ca, Hg).
These extra holes are always accompanied by additional electrons located on nearby orbitals, which vary depending on the surrounding crystal environment. 
The doping inhomogeneity between the inner and outer planes can be directly witnessed. 
In the strongly overdoped regime, which is reached for the single-layer compounds, the Cu-\dxxyy~orbital both in \nC~and \nH~seems not to accept additional holes, and even slightly gains electrons. 
In fact, the effective hole-doping in \nH~saturates due to the promotion of other degrees of freedom to the Fermi level, which points out the importance of possible doping limitations of certain cuprates.  

Another key element of this work is the \emph{ab initio} estimation of the superconducting order parameter $\msc$ as a function of hole-doping.  
Each CuO$_2$ plane was treated separately after converging the DFT+CDMFT calculation in the normal state.
As discussed in App.~\ref{ap:prox}, the proximity effect is anyway weak and does not limit our interpretation.  
One of the key result is the absence of superconductivity in the inner planes of the $n\geq4$ compounds. 
This is unambiguously explained by the fact that, until a total doping of $\delta_{\mathrm{tot}}=50\%$, these planes do not reach sufficient hole-doping to host superconductivity. 
Indeed, in remarkable agreement with experiments~\cite{kurokawa2023}, our calculations show that below $\delta_{\rm CuO_2}\sim4\%$ hole-doping in the CuO$_2$ plane, superconductivity does not emerge. 
Therefore, the tri-layer compounds ($n=3$) appear as the best compromise between (i) hosting protected inner planes, and (ii) reaching sufficient doping in every plane. 
Moreover, $\msc$ is globally larger in \nH~compounds in comparison to \nC, as expected from the value of $T_c$, $J$ and CTG discussed above. 

\begin{table}
    \centering
    \begin{tabular}{c|c|c|c|c|c}
    \hline
       Family & $n=1$ & $n=2$ & $n=3$ & $n=4$ & $n=5$\\
       \hline
       \hline
       \nH  & 98 K & 127 K & 137 K & 127 K & 111 K \\
       \hline
       \nC &  28 K (43 K) &  49 K & - & - & -      \\ 
    \end{tabular}
    \caption{Critical temperature $T_c$ \emph{vs} $n$ for three families of multilayer cuprates, taken from Refs.~[\onlinecite{mukuda2012},\onlinecite{yamada2005},\onlinecite{zenitani1995}].
    The two values for single-layer \nC~corresponds to, respectively, Na-doped and vacancy-doped (between parenthesis) samples. 
    The two-layer \nC~$T_c$ is measured on Na-doped samples. }
    \label{tab:tc}
\end{table}

As mentioned in the beginning of the conclusion, we infer the superconducting properties via the zero-temperature order parameter $\msc$, which is not strictly proportional to $T_c$ but can in principle be measured experimentally~\cite{omahony2022a}. 
We provide in Table~\ref{tab:tc} the experimentally known $T_c$ for the \nC~and \nH~compounds. 
The variations in critical temperature are overall more pronounced than the variations of the estimated $\msc$.
Yet, the trends we obtain are robust since all the calculations were performed under the exact same approximations. 
Moreover, we emphasize that our predictions allow to understand how three properties of mutlilayer cuprates, apparently unrelated, can be combined together to understand the universal trend in $T_c$ \emph{vs.} $n$ in multilayer cuprates.
First, there exists a doping threshold for the CuO$_2$ planes below which superconductivity cannot emerge, as shown experimentally~\cite{kurokawa2023} and in this work.
Second, hole-doping in multilayer cuprates is strongly inhomogeneous, and doping the inner planes is particularly challenging.
Third, if they were not so difficult to dope, inner planes would be intrinsically more favorable for superconductivity since they display smaller CTG, larger $J$, and in real materials they would be protected from distortions etc. 

Our results also enable linking the \emph{normal state} properties with the evolution of the zero-temperature superconducting order parameter. 
For superconductivity to emerge, the normal-state DOS in a range of $\pm200$~meV around the Fermi level needs to display a pseudogap-like feature having a particle-hole symmetric character. 
Both the under-doped and over-doped superconducting phase transitions are linked with the disappearance of this feature. 
This also explains the difference between the IP1 of the \nC~and \nH~tri-layer compounds. 
The former host a significantly weaker intrinsic order parameter since the normal-state pseudogap-like feature does not become markedly particle-hole symmetric: we observe only a shoulder above the Fermi level, instead of a well-defined second peak. 
In fact, the difference may be traced back to the self-doping from the Hg-O bands which, in the tri-layer \nH~promotes hole-doping in the IP1 rather than the OP. 
This observation might provide a way to favorably tune the hole distribution in multilayer cuprates. 

We believe that our work not only contributes to a better understanding of high-temperature superconductors, but that it also paves the way towards new research perspectives. 
On the one hand, the advancement of \emph{ab initio} predictive approaches may enable a more complete understanding of doping in high-temperature superconductors, which is necessary to address the material-specific tendencies.
For instance, we expect that chemical substitution in \nC~and oxygen insertion in \nH~may lead to significantly different carrier redistributions and structural deformations, which will necessarily affect the superconducting properties. 
This may enable to deepen the variations in predicted $\msc$. 
Similarly, our method will enable studies of the effect of non-magnetic impurities within the CuO$_2$ layers, which affect differently the pseudogap and superconductivity~\cite{alloul2024}.

At the same time, our work motivates further methodological improvements to quantum impurity solvers. 
Being able to afford more bath orbitals within the ED solver would be a great advantage, thus recent developments using matrix-product states to perform the Lanczos algorithm are promising~\cite{paeckel2023}. 
New algorithmic advances in the field of Quantum Monte Carlo solvers~\cite{fernandez2022,erpenbeck2023,haule2023strongcouplingquantumimpurity} may also be beneficial to directly evaluate $T_c$ is these compounds. 

Moreover, the estimation of the superexchange parameter $J$ could be improved by solving the Bethe-Salpeter equation. 
This would open the possibility to study the full spin excitation spectrum \emph{ab initio}. 
At this stage, the link between $J$ and superconducting order parameter is our weakest conclusion since the observed relative change in $T_c$ as a function of the number of layers is much larger than the calculated relative change in $J$. 

Finally, while the superconducting order parameter informs about the tendency of the system to create electron pairs, it is crucial to also estimate the tendency of these pairs to form a coherent state by computing the superfluid stiffness. 
We believe that the latter may be estimated \emph{ab initio} using our approach, thanks to developments enabling computation of two-particle quantities with Lanczos-based exact diagonalization solvers~\cite{tanaka2019}, and the computation of the superfluid stiffness within DMFT~\cite{liang2017}. 

Our work is the first step of a fascinating long-term research program. 
Indeed, further understanding of the problem could be obtained through many additional studies, such as the \emph{ab initio} competition of $d$-wave superconductivity and antiferromagnetism, the pressure dependence of the superconducting order parameter, better estimates for the transition temperature, large throughput exploratory calculations for new superconducting materials etc.

In conclusion, our contribution paves the way towards predictive approaches for high-temperature superconductors which, as exemplified in this work, host a huge potential for theoretical material design.


\begin{acknowledgments}

We acknowledge Nicolas Gauthier and Gabriel Kotliar for fruitful discussions. 
B.B.-L. was supported by a prestige postdoctoral fellowship from Institut quantique, and acknowledge support from the US National Science Foundation (NSF) Grant Number 2201516 under the Accelnet program of Office of International Science and Engineering (OISE).
B.B.-L., B. L., A.-M.S.T. and D.S. acknowledge financial support by the Canada First Research Excellence Fund. 
Numerical calculations were performed on computers provided by Rutgers University, by Calcul Québec and by the Digital Research Alliance of Canada.
Work by K.H. was supported by NSF DMR-2233892 and by a grant from the Simons Foundation (SFI-MPS-NFS-00006741-06, K.H.)

\textit{Author contributions}: B. B.-L. integrated the PyQCM library into the eDMFT software conceived, respectively, by D. S. and K. H.. 
B. B.-L. carried the calculations, as well as the analysis with help from the other authors.
B. L. performed the proximity effect calculations. 
K.H., A.-M.S.T. and D.S. initiated the project and supervised it. 
All authors participated in the writing of the present manuscript.

\end{acknowledgments}

\newpage

\appendix

\section{Benchmark of the structure approximation for \nC}
\label{ap:struct}

To perform calculations for the \nC~compounds from $n=1$ to $n=5$ with $2\times2$ supercells, we had to introduce a structural approximation for $n\geq2$ (see Sec.~\ref{sec:met_crys}).
The $I4/mmm$ space group is transformed to $P4/mmm$ before generating the supercell, which allows to effectively divide the number of atoms considered by two. 

\begin{figure}
    \centering
    \includegraphics[width=\linewidth]{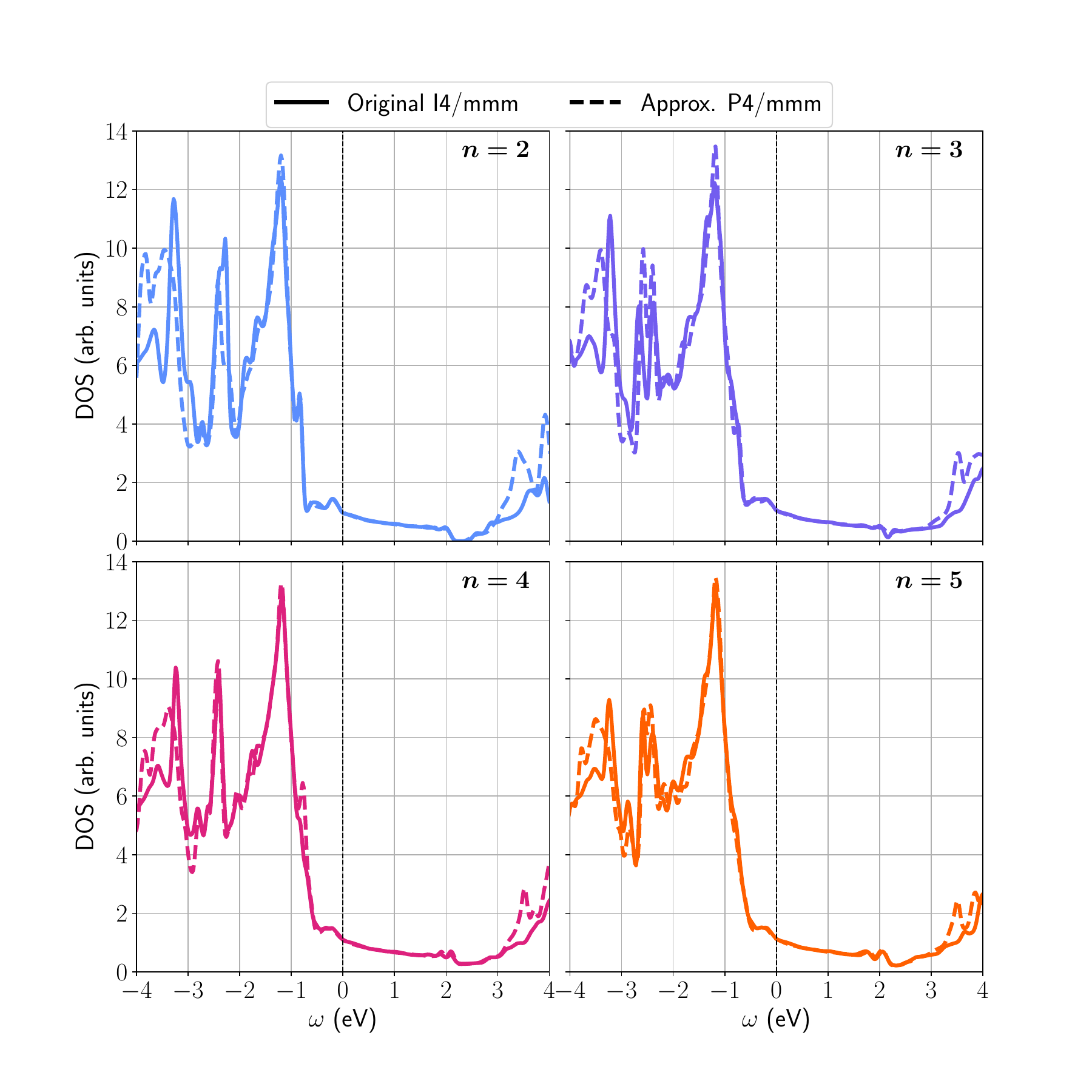}
    \caption{DFT DOS obtained using the original crystal structure ($I4/mmm$) and the approximated one ($P4/mmm$), for all $n\geq2$ \nC~compounds.}
    \label{fig:fig_s5}
\end{figure}

We show in Fig.~\ref{fig:fig_s5} the DFT-only DOS for the $n\geq2$ \nC~compounds with and without structural approximation. 
For all compounds, the DOS are indistinguishable in a window of $\pm3$~eV around the Fermi level. 
Some small discrepancy start to be noticed beyond $\pm3$~eV. 
The low-energy region of the spectrum is dominated by degrees of freedom related to the CuO$_2$ planes, which are unaffected by structural modifications in the Cl-Cl planes. 
It is this region mostly that will be corrected by the CDMFT self-energy. 
Our approximation is therefore justified since it leaves the DFT starting point, over which the DFT+CDMFT is carried out, unchanged. 
\section{Estimation of the hole-content in the CuO$_2$ planes}
\label{ap:estim_dop}

\begin{figure*}
    \centering
    \includegraphics[width=0.9\linewidth]{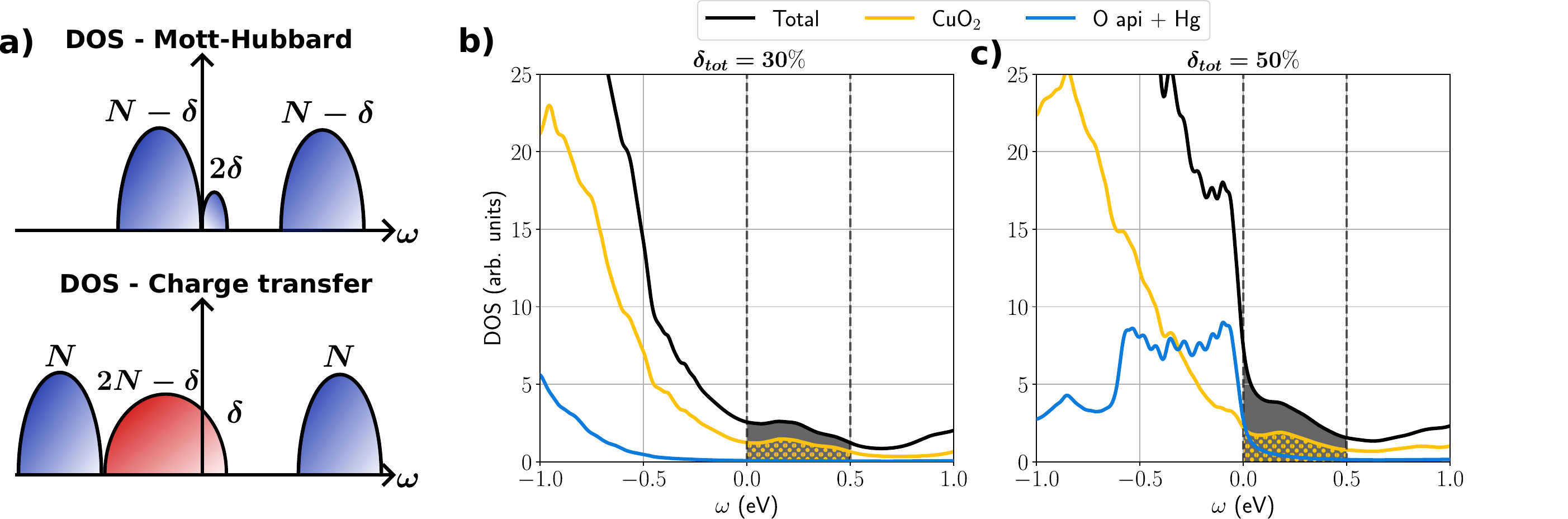}
    \caption{(a) Simplified sketch of the DOS of a hole-doped Mott insulator (\emph{top panel}) and a hole-doped charge-transfer insulator (\emph{bottom panel}). 
    (b)-(c) Representative example of a hole-doped multilayer cuprate DOS, obtained from the DFT+CDMFT calculation at $\delta_{\rm tot}=30,~50\%$. 
    The DOS projected onto the Cu-\dxxyy~and O-$p_x$ orbitals of the outer plane, and onto the apical O and Hg atoms are also displayed. 
    The shaded grey area represents the integrated area for the full DOS, and the dotted region the area for the CuO$_2$ plane. 
    The ratio of the two provides an estimate of the proportion of holes $\delta_{\rm CuO_2}$ plane (\emph{see text}).}
    \label{fig:fig_s1}
\end{figure*}

Similarly to experiment, estimating the amount of effective hole-doping of each CuO$_2$ plane within our realistic calculations is not straightforward.
Because of the exact double-counting scheme, in which the screening model is adapted accoding to the ``metallicity'' of the system (see Sec.~\ref{sec:met_edmft}), the double-counting correction may be different in the parent and hole-doped compound. 
This prevents a straightforward comparison of the orbital occupations. 
Another possible way to estimate the doping is to quantify the density of state (DOS) in the low-energy region above the Fermi level. 
In the following, we detail our estimation procedure. 

The parent cuprate compounds are charge-transfer insulators (say, in absence of self-doping). 
The charge-transfer gap originates from the opening of the Mott gap of the Cu-$d$ band. 
Hence, it is instructive to discuss briefly the behavior of Mott and charge-transfer insulators upon hole-doping~\cite{meinders1993}, as sketched in Fig.~\ref{fig:fig_s1}(a). 
In the pictorial Mott system (made of only the Cu-$d$), the integrated spectral weight just above the Fermi level is expected to be equal to twice the amount of hole-doping $\delta$, since the spectral weight is transferred from both Hubbard bands. 
In contrast, in a charge-transfer system, the Fermi level crosses the O-$p$ bands, and therefore hole-doping is similar to the semi-conductors for which the integrated spectral weight just above the Fermi level is equal to the amount of hole-doping $\delta$. 

In reality, both these pictures are wrong since a Zhang-Rice singlet (ZRS) band forms at low-energy. 
The ZRS has a mixed Cu-$d$ and O-$p$ character.
Thus, for an effective doping $\delta_{\rm CuO_2}$ in the CuO$_2$ planes the integrated spectral weight $\mathcal{I}_{\rm CuO_2}$ just above the Fermi level, projected onto the Cu-$d$ and O-$p$ orbitals, will most probably correspond to $\mathcal{I}_{\rm CuO_2} = x\delta_{\rm CuO_2}$, with $x\in [1,2]$. 
It is therefore not possible to directly estimate $\delta_{\rm CuO_2}$ from the spectral weight just above the Fermi level. 

We however exploit the fact that most of the holes $\delta_{\rm tot}$ are captured by the Cu-$d$ and O-$p$ orbitals. 
Hence, the changes in the \emph{total} low-energy spectral weight above the Fermi level is driven by the behavior of the Cu-$d$ and O-$p$ orbitals. 
In other words, if the projected low-energy spectral weight onto the Cu-$d$ and O-$p$ orbitals amounts to $\mathcal{I}_{\rm CuO_2} =x\delta_{\rm CuO_2}$, then the total one should be very close to $\mathcal{I}_{\rm tot} =x\delta_{\rm tot}$. 
Since $\delta_{\rm tot}$ is known (set by the virtual crystal approximation), the effective doping on the CuO$_2$ planes can be estimated by taking the ratio of the spectral weight integrands close to the Fermi level:
\begin{equation}
    \delta_{\rm CuO_2} = \frac{\mathcal{I}_{\rm CuO_2}}{\mathcal{I}_{\rm tot}}\delta_{\rm tot}.
\end{equation}
This is illustrated in Fig.~\ref{fig:fig_s1}(b), where $\mathcal{I}_{\rm tot}$ correspond to the shaded gray area, and $\mathcal{I}_{\rm CuO_2}$ to the dotted one. 

As mentioned in the text, this approximation works well when the doped holes primarily occupy the CuO$_2$ planes. 
This is the case of all \nC~compounds studied, but it poses a limitation for the \nH~compounds at large doping. 
This can be understood in Fig.~\ref{fig:fig_s1}(c): the Hg and apical O spectral weight starts to cross the Fermi level at large doping (here illustrated at $\delta_{\rm tot}=50\%$). 
This weakens our assumption that the \emph{total} and \emph{projected} areas share the same $x$, and leads to poorer estimations of $\delta_{\rm CuO_2}$. 
We however emphasize that the limitation only concerns the data analysis at large doping, and not our \emph{ab initio} method. 
Hence, it is not detrimental to any conclusion of the present work. 

\section{Benchmark of the exact diagonalisation solver}
\label{ap:ed}

We carried out numerous benchmarks of PyQCM's exact diagonalization (ED) solver~\cite{dionne2023a,dionne2023b} in comparison to the Continuous-time Quantum Monte Carlo (CTQMC) solver implemented within the eDMFT framework~\cite{haule2007,haule2010}. 
We detail in this section one representative example: a DFT + single-site DMFT calculation for the single-layer \nC~compound, in which we treat the Cu-$e_g$ orbitals within a two-orbital impurity. 
We set $U=12$~eV and $J_{\mathrm{Hund}}=1$~eV. 

\begin{figure}
    \centering
    \includegraphics[width=\linewidth]{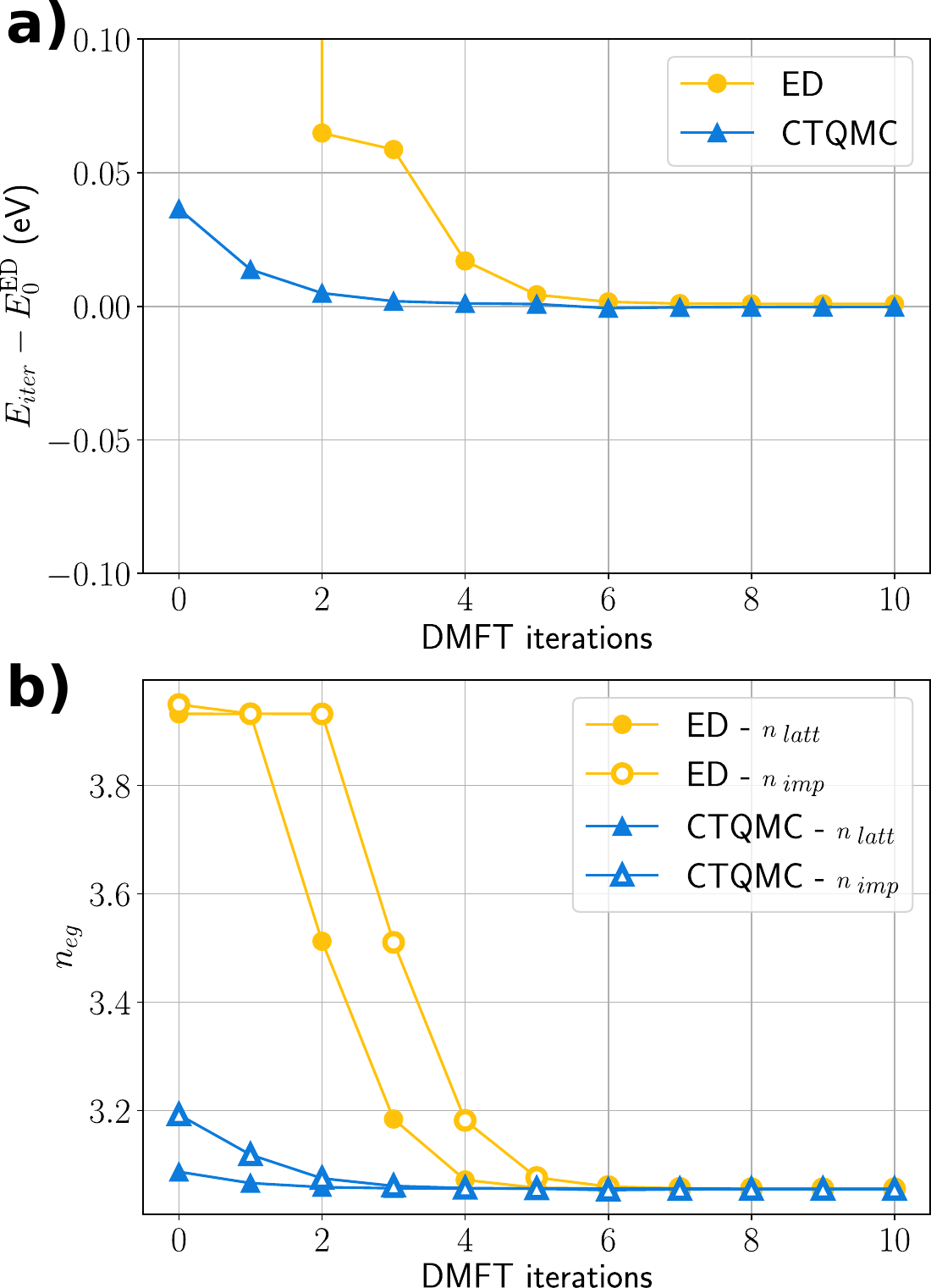}
    \caption{Convergence of (a) the DFT+DMFT total energy difference (taking the converged ED energy as reference), and (b) the lattice ($n_{imp}$) and impurity occupations ($n_{latt}$) as obtained with the CTQMC and ED solvers.}
    \label{fig:fig_s3}
\end{figure}

We show in Fig.~\ref{fig:fig_s3}(a) the evolution of the DFT+DMFT total energy all along the self-consistent calculation. 
Both solvers converge to the same final total energy within a precision of $2\cdot10^{-4}$. 
In Fig.~\ref{fig:fig_s3}(b), we also compare the convergence of the occupation of the correlated Cu-$e_g$ orbitals estimated from the lattice ($n_{latt}$) and from the impurity ($n_{imp}$).
Similarly, the ED results are in excellent agreement with the ones obtained using CTQMC: the occupations differ by less than $10^{-3}$. 
Most importantly, the convergence of $n_{latt}$ and $n_{imp}$ when using the ED solver shows that the discretization of the bath does not hinder the accuracy.  

\begin{figure}
    \centering
    \includegraphics[width=\linewidth]{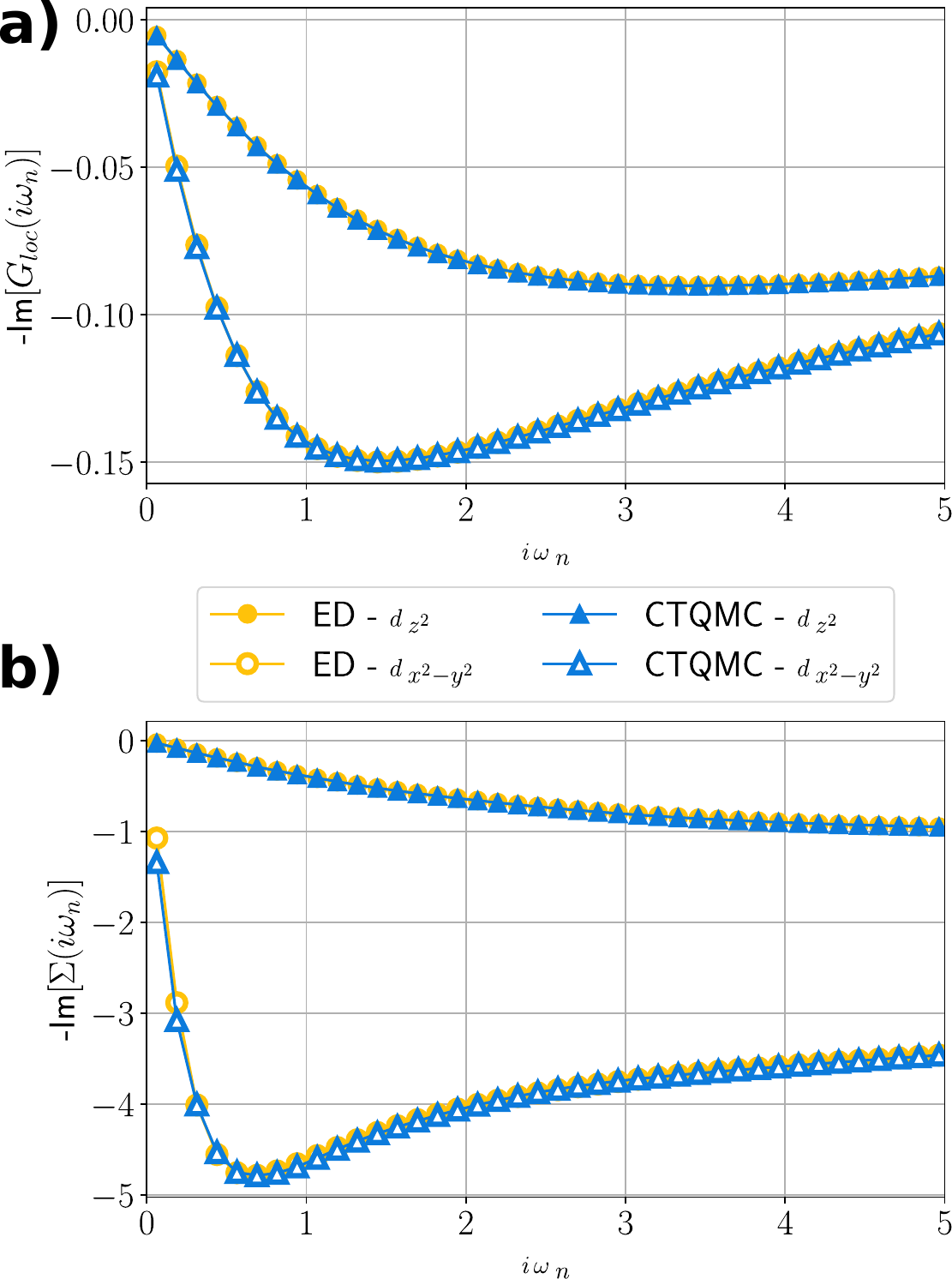}
    \caption{Matsubara (a) Green's function and (b) self-energy projected onto the local subspace, as obtained with the CTQMC and ED solvers.}
    \label{fig:fig_s4}
\end{figure}

The excellent agreement is also witnessed in the resulting Green's function and self-energy projected onto the local subspace, which are shown in Fig.~\ref{fig:fig_s4}(a,b). 
The ED and CTQMC results match almost perfectly down to the lowest Matsubara frequency, which allows an excellent agreement of all static observables as discussed above. 
Hence, as demonstrated in our work and this representative example, exact diagonalization solvers can be successfully and efficiently used within realistic \emph{ab initio} calculations. 

\section{Including the proximity effect}
\label{ap:prox}

All the estimations of the superconducting order parameter $\msc$ presented in the main text were obtained by neglecting the proximity effect between the neighbouring CuO$_2$ planes. 
Given the short distance that separates the outer from the inner planes, one may question the validity of this approximation. 
In particular, we may imagine a scenario in which superconductivity is triggered in originally non-superconducting inner planes (say, because of insufficient doping), due to the proximity with superconducting outer planes. 


\begin{figure*}
    \centering
    \includegraphics[width=0.9\linewidth]{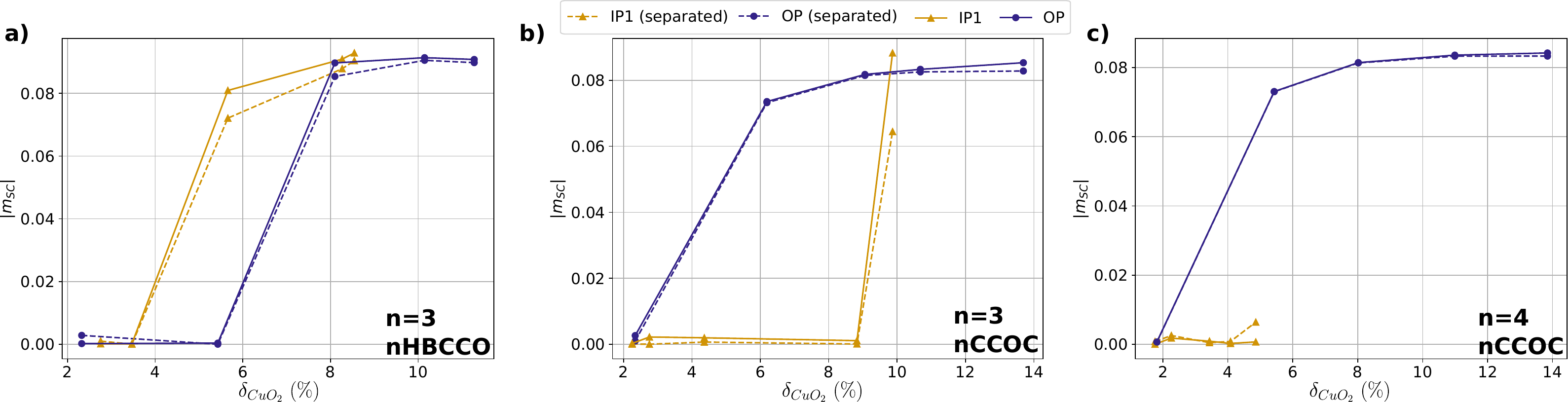}
    \caption{Superconducting order parameter $\msc$ \emph{vs.} estimated doping $\delta_{\rm CuO_2}$ for the (a) tri-layer \nH, (b) tri-layer \nC~and (c) four-layer \nC~compounds. 
    The dashed lines do not include the proximity effect while the solid lines do. 
    The observed proximity effect is small.}
    \label{fig:fig_s9}
\end{figure*}

In the following, we first present a way to include the proximity effect in our calculations, and then discuss its impact on the superconducting order parameter. 
We show that it is a weak effect in multilayer compounds, at least at ambient pressure, and especially that it is not able to trigger superconductivity in the inner planes even in the presence of superconducting outer planes.  
Most importantly, these results strongly reinforce our conclusions regarding the $n\geq4$ multilayer compounds: until $\delta_{\rm tot}=50\%$ doping, even when taking into account the proximity effect, the inner panes remain in the normal phase due to the lack of holes. 

To include the proximity effect, we generalized the scheme described in Sec.~\ref{sec:met_sc} to treat super-clusters made of the $2\times2$ plaquettes of every consecutive CuO$_2$ plane.
Following the super-cluster philosophy~\cite{charlebois2015,pahlevanzadeh2021}, we neglect the inter-cluster self-energy but we include explicitly the coupling between the CuO$_2$ planes \emph{on the lattice}, i.e., at the mean-field level. 
Formally, the self-energy thus reads
\begin{equation}
    \Sigma(\omega) = \begin{pmatrix}
        \Sigma_{\alpha\beta}^{1}(\omega) & 0 & \dots \\
        0 & \Sigma_{\alpha\beta}^{2}(\omega) & 0 & \\
        \vdots & 0 & \ddots &  0\\
        & & 0 & \Sigma_{\alpha\beta}^{n}(\omega)\\
    \end{pmatrix},
\end{equation}
where each block $\Sigma_{\alpha\beta}^{i}(\omega)$ corresponds to the self-energy of the $i^{th}$ ($i\in\{1,n\}$) CuO$_2$ plane (Eq.~\eqref{eq:self_nambu}, Sec.~\ref{sec:met_sc}).
On the lattice, the coupling between the planes is included within the effective dispersion: 
\begin{equation}
\Gamma(\vk,\omega) = \begin{pmatrix}
        \Gamma_{\alpha\beta}^{11}(\vk,\omega) & \Gamma_{\alpha\beta}^{12}(\vk,\omega) & \dots \\
        \Gamma_{\alpha\beta}^{21}(\vk,\omega) & \Gamma_{\alpha\beta}^{22}(\vk,\omega) &  & \\
        \vdots &  & \ddots &  \\
        & &  & \Gamma_{\alpha\beta}^{nn}(\vk,\omega)\\
    \end{pmatrix}
\end{equation}
We emphasize that, as described in Sec.~\ref{sec:met_sc}, the computation of $\Gamma(\vk,\omega)$ does not involve any free parameter. 

We show in Fig.~\ref{fig:fig_s9} a representative example of the impact of the proximity effect on the superconducting order parameter. 
The latter is only weakly affected in the tri-layer \nC~and \nH~compounds, see panels (a) and (b). 
Moreover, superconductivity is not triggered in the inner planes of the $n\geq4$ compounds, as exemplified in the panel (c) for the \nC~family (\nH~behaves similarly).
Hence our conclusions regarding the predominant role of the effective hole-doping of the inner planes in multilayer cuprates are unambiguously confirmed. 

Finally, we note one sizable change due to the proximity effect : $\msc$ in the IP1 of the tri-layer \nC~compounds is enhanced and even exceeds that of the OP. 
This leads to a smaller difference between the IP1 of the \nC~and \nH~compounds.

\section{On the choice of the on-site $U$}
\label{ap:u}

Choosing the value of the Hubbard on-site $U$ is a critical issue when studying effective low energy models of strongly-correlated systems. 
Variations of a few hundreds of meV may drastically change the resulting physics. 
However, this sensitivity to $U$ is substantially reduced within realistic electronic structure methods like ours. 
In our ab initio embedding approach, the relevant $U$ scale is much larger. 
For example, the unscreened Coulomb $U$ is approximately 31$\,$eV for the chosen orbital, and the screened $U$ falls between 10 and 14,eV (as estimated for single-layer CCOC and HBCCO). 
The dominant screening contributions come from other orbitals on Cu site, i.e. the $1s,2s,2p,3s,3p,4s,...$ and $t_{2g}$ orbitals of Cu, which are not explicitly included in our impurity problem. 
Consequently, the $U$ values we use are significantly larger than those typically employed in single-band effective models. 
Most importantly, within our framework, large variations in $U$ have only a minor effect on the physical picture, ensuring that the results remain robust against uncertainties in $U$.

\begin{figure*}
    \centering
    \includegraphics[width=\linewidth]{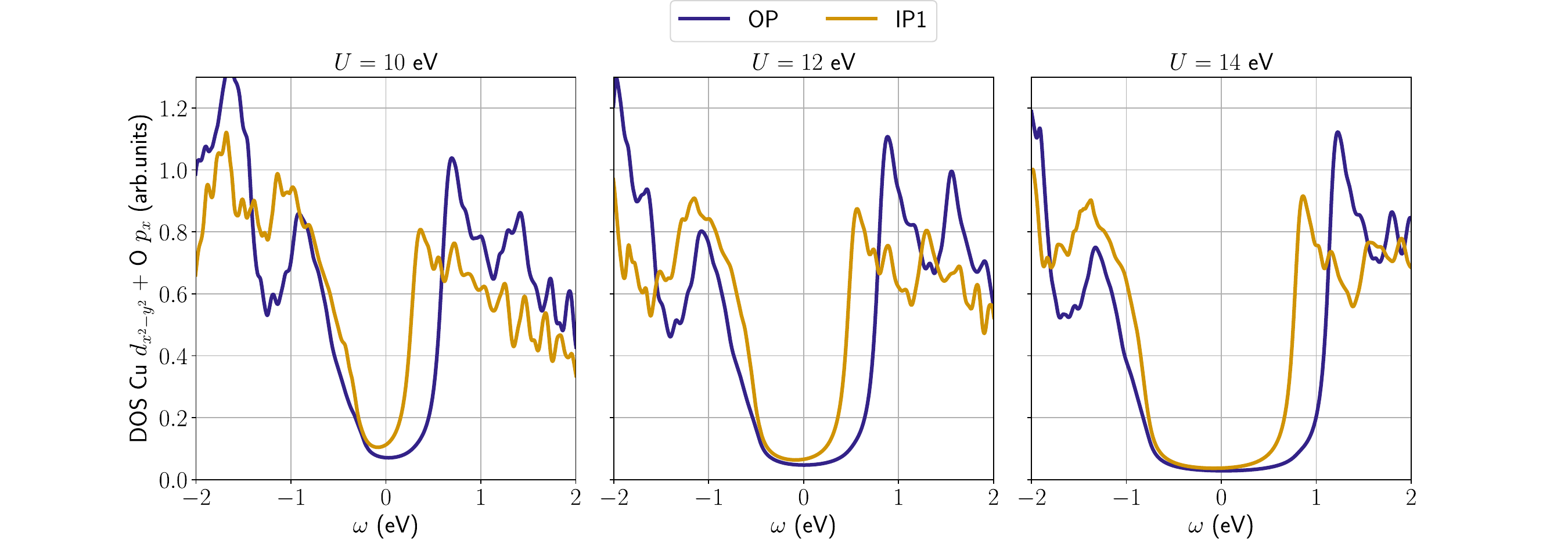}
    \caption{Comparison of the DFT+CDMFT DOS projected onto the Cu-\dxxyy~and O-$p_x$ orbitals, obtained with three different $U$ values for the tri-layer \nC~compound.}
    \label{fig:fig_s7}
\end{figure*}

\begin{figure*}
    \centering
    \includegraphics[width=0.9\linewidth]{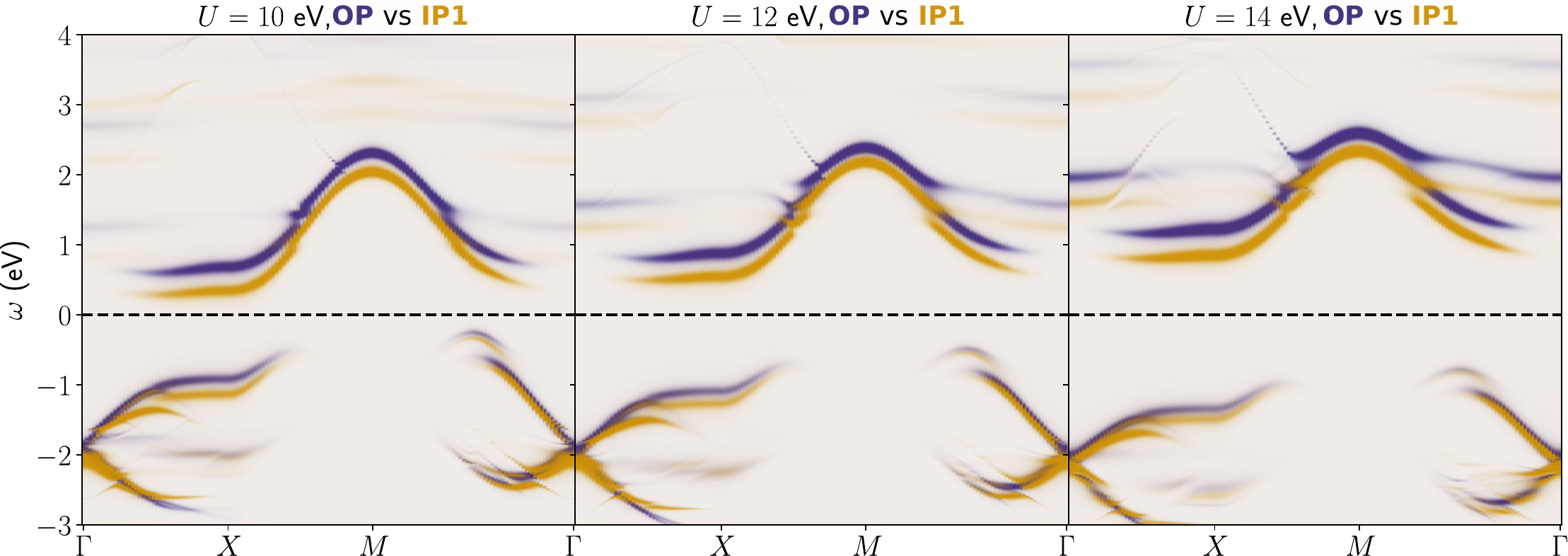}
    \caption{Comparison of the momentum-resolved spectral function projected onto the Cu-\dxxyy~orbital, obtained with three different $U$ values for the tri-layer \nC~compound.}
    \label{fig:fig_s8}
\end{figure*}

We have performed multiple tests to ensure that the conclusions of our work are robust against the choice of $U$, and we detail one representative example in this section. 
We performed DFT+CDMFT calculations on the tri-layer \nC~compound, in the same conditions as for all other computations presented in the main manuscript, simply varying $U$ from 10~eV to 14~eV. 
The resulting DOS are shown in Fig.~\ref{fig:fig_s7}. 
Despite variations of the absolute value of the CTG, at all $U$ considered we observe the same differentiation between outer and inner planes. 
The latter always display a smaller charge gap, which is consistent with our main conclusions. 
Moreover, the overall shape of the DOS remains relatively stable at all $U$ values, which is confirmed by inspecting the momentum-resolved spectral functions displayed in Fig.~\ref{fig:fig_s8}. 
All the salient spectral features discussed in the main manuscript can be unambiguously seen at all values of $U$. 

We emphasize that variations of $U$ of the order of 4~eV in single-band effective models, parameterized for usual cuprates, would lead to a dramatic change of the physics, most likely a metal-insulator transition. 
Therefore, if $U$ remains an adjustable parameter in our current approach, the robustness shown in this representative example strongly supports the \emph{ab initio} character of our method.

For a fully quantitative analysis, predictions of material-specific $U$ is necessary. 
One can estimate $U$ self-consistently via the constrained eDMFT method~\cite{anisimov1993}.
In single-site eDMFT it gives Coulomb $U$ and $J_{\rm Hund}$ around 10~eV and 1~eV, respectively.
For the present study, however, we found that many experimental observations can be reproduced and interpreted with a fixed $U$ value. 
Building on the arguments developed here, we predict that$U$ should decrease from $n=1$ to $n=2$ in the outer planes, and then saturate for $n\geq3$ in the outer planes. 
In the inner planes, $U$ would likely be smaller due to additional screening. 
We anticipate that estimating material-specific $U$ values would reinforce the trends observed in this work, but we defer such investigations to future studies.

%
%
%
%

\section{$n$-dependence of the orbital occupations in the undoped compounds}
\label{ap:pcharges}

\subsection{$n$CCOC}

\begin{figure*}
    \centering
    \includegraphics[width=0.9\linewidth]{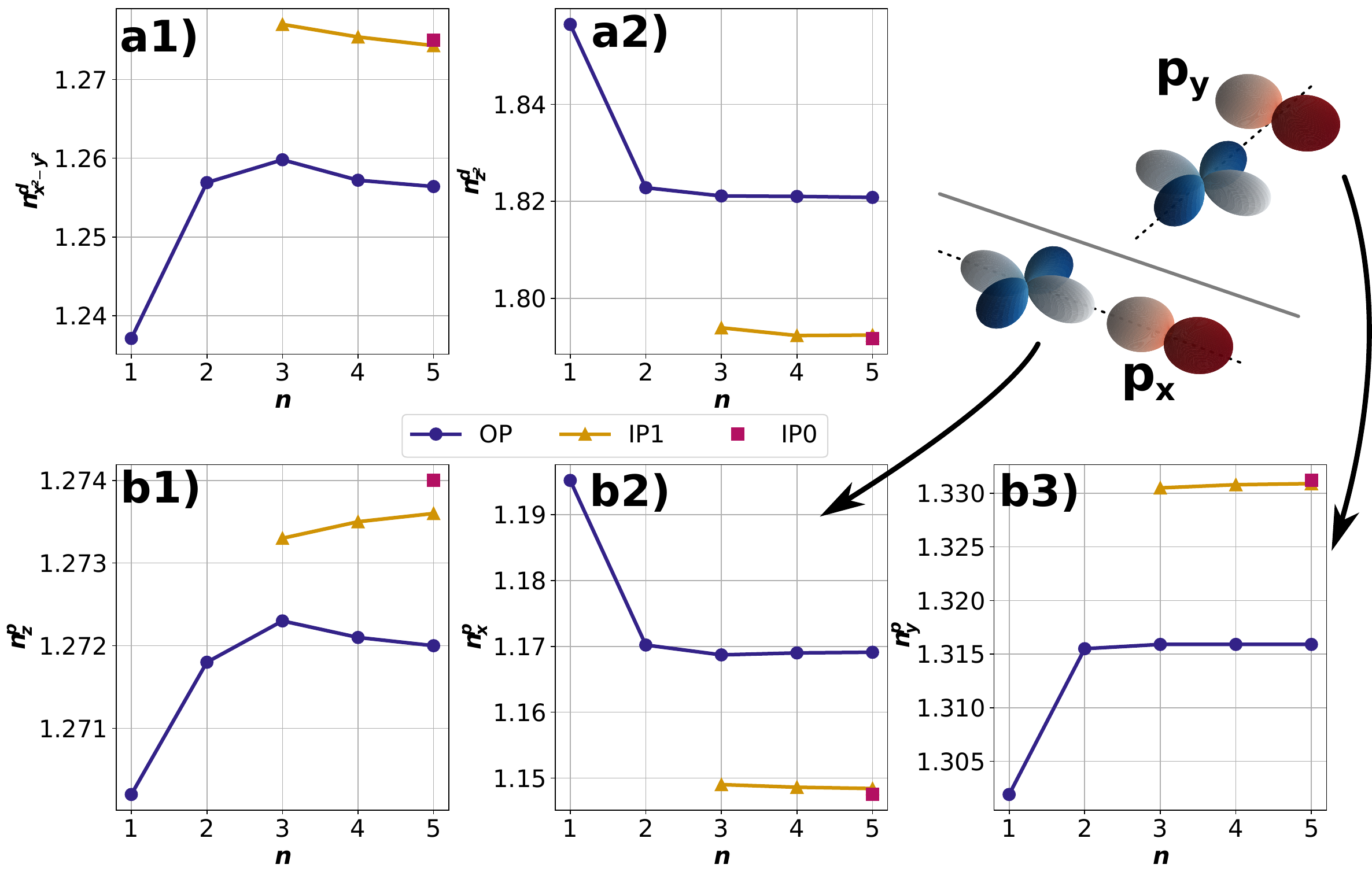}
    \caption{Partial charges of, respectively, (a1-2) the Cu-\dxxyy~and \dzz~orbitals, and (b1-3) the O-$p_z$, $p_x$ and $p_y$ orbitals of \nC~($n$=1\textendash5).
    The O-$p_x$ ($p_y$) orbital points towards (away from) the Cu-\dxxyy, as illustrated in (c). }
    \label{fig:fig4}
\end{figure*}

Another quantity of interest in cuprate superconductors is the oxygen hole content~\cite{andersen1995b,tranquada1987,fujimori1987,zheng1995,haase2004,jurkutat2014,rybicki2016,kowalski2021,jurkutat2023}.
Even in the undoped compounds the distribution of charges on the Cu-\dxxyy/\dzz~and O-$p_x$/$p_y$/$p_z$ orbitals is already instructive.
In Fig.~\ref{fig:fig4}, we show the evolution of the charges on each of these orbitals, with respect to the number of CuO$_{2}$ planes and their position in the structure.   
A striking feature reminiscent of the CTG evolution is that the occupation of the orbitals varies significantly.
Specifically, with increasing $n$ the Cu-$d$ orbitals drift away from their nominal occupancy towards less correlated mixed valency. 
The nominal occupancy of the Cu-\dxxyy~and \dzz~is 1 and 2, respectively. 
This means that electrons in the bi-layer compound experience weaker correlations than in the single-layer one, in agreement with previously found smaller CTG. 
Furthermore, the inner planes are also more mixed-valent, hence explaining the smaller CTG in these planes. 

We note that such strong correlation between CTG and occupancy is an effect that is beyond model Hamiltonian simulations, and requires a first-principles method as developed here. 
Namely, changes of Cu occupancy here lead to substantial changes of the Hartree and exchange-correlation potential, which affects the double-counting potential, and consequently the on-site energy levels of Cu-$d$ orbitals. 
Even more importantly, these occupancies tune the crystal field splitting between the Cu-\dxxyy~and \dzz~orbitals, which plays a crucial role in enhancing superconductivity as discussed in Sec.~\ref{sec:res_dop_sc}.

The occupancy of the O-$p$ orbitals also varies with the number of layers and their position in the structure. 
In particular, when the occupancy of the Cu-\dxxyy~orbital increases (and consequently the CTG decreases), the holes get transferred to the oxygen $p_x$ orbital, which points towards the Cu-\dxxyy~orbital (see Fig.~\ref{fig:fig4}(b2)).
Similar correlation between CTG and $p_x$ orbital occupancy was reported in Ref.~\cite{kowalski2021} on simulating the three-band Hubbard model. 
This introduction of additional holes on oxygen $p_x$ orbital is particularly important as it was shown that higher concentration of holes on oxygen favors superconductivity. 
The occupancy of the $p_y$ orbitals (pointing away from Cu) increases when $p_x$ decreases in our simulation. 
Hence $p_y$ orbital partially  compensates for $p_x$ change of occupancy.
The relation between the relative occupancy of the $p_x$ and $p_y$ orbitals is a precursor of the behavior upon hole-doping, as detailed in Sec.~\ref{sec:res_dop_hd}.

\subsection{$n$HBCCO}

\begin{figure*}
    \centering
    \includegraphics[width=0.9\linewidth]{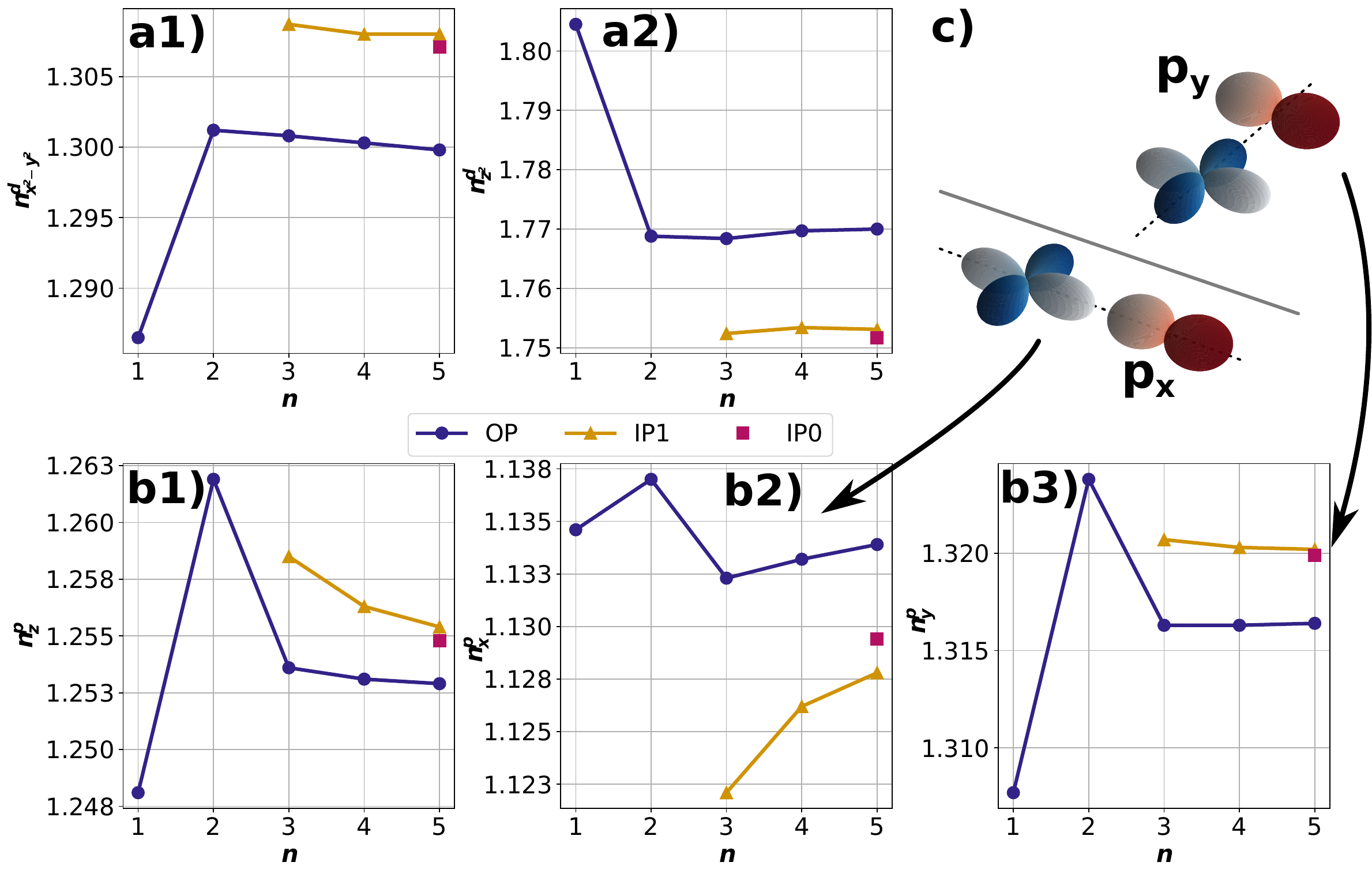}
    \caption{Partial charges of, respectively, (a1-2) the Cu-\dxxyy~and \dzz~orbitals, and (b1-3) the O-$p_z$, $p_x$ and $p_y$ orbitals in \nH~($n$=1\textendash5).
    The O-$p_x$ ($p_y$) orbital points towards (away from) the Cu-\dxxyy, as illustrated in (c).}
    \label{fig:fig11}
\end{figure*}

We also analyze the evolution of the orbital occupancies in \nH, which are displayed in Fig.~\ref{fig:fig11}. 
The Cu-\dxxyy~and \dzz~orbitals show very similar trends as in \nC~although, similarly to the CTG, the layer dependence is weaker. 
For instance the maximum difference $\Delta n^{d}_{x^2-y^2}\sim0.02$ for \nH~(difference taken between the OP at $n=0$ and IP1 at $n=3$), while it is $\Delta n^{d}_{x^2-y^2}\sim0.04$ for \nC. 
The same conclusion applies to the O-$p$ occupancies, to the exception of the $n=2$ compound which shows a slightly peculiar behavior of $n^{p}_{z}$ and $n^{p}_{y}$. 
This may be linked with the fact that for $n=1$ the Hg compounds do not contain any Ca atom, which are introduced at $n=2$.

\bibliography{references}

\begin{thebibliography}{196}%
\makeatletter
\providecommand \@ifxundefined [1]{%
 \@ifx{#1\undefined}
}%
\providecommand \@ifnum [1]{%
 \ifnum #1\expandafter \@firstoftwo
 \else \expandafter \@secondoftwo
 \fi
}%
\providecommand \@ifx [1]{%
 \ifx #1\expandafter \@firstoftwo
 \else \expandafter \@secondoftwo
 \fi
}%
\providecommand \natexlab [1]{#1}%
\providecommand \enquote  [1]{``#1''}%
\providecommand \bibnamefont  [1]{#1}%
\providecommand \bibfnamefont [1]{#1}%
\providecommand \citenamefont [1]{#1}%
\providecommand \href@noop [0]{\@secondoftwo}%
\providecommand \href [0]{\begingroup \@sanitize@url \@href}%
\providecommand \@href[1]{\@@startlink{#1}\@@href}%
\providecommand \@@href[1]{\endgroup#1\@@endlink}%
\providecommand \@sanitize@url [0]{\catcode `\\12\catcode `\$12\catcode
  `\&12\catcode `\#12\catcode `\^12\catcode `\_12\catcode `\%12\relax}%
\providecommand \@@startlink[1]{}%
\providecommand \@@endlink[0]{}%
\providecommand \url  [0]{\begingroup\@sanitize@url \@url }%
\providecommand \@url [1]{\endgroup\@href {#1}{\urlprefix }}%
\providecommand \urlprefix  [0]{URL }%
\providecommand \Eprint [0]{\href }%
\providecommand \doibase [0]{https://doi.org/}%
\providecommand \selectlanguage [0]{\@gobble}%
\providecommand \bibinfo  [0]{\@secondoftwo}%
\providecommand \bibfield  [0]{\@secondoftwo}%
\providecommand \translation [1]{[#1]}%
\providecommand \BibitemOpen [0]{}%
\providecommand \bibitemStop [0]{}%
\providecommand \bibitemNoStop [0]{.\EOS\space}%
\providecommand \EOS [0]{\spacefactor3000\relax}%
\providecommand \BibitemShut  [1]{\csname bibitem#1\endcsname}%
\let\auto@bib@innerbib\@empty
\bibitem [{\citenamefont {Andersen}\ \emph {et~al.}(1995)\citenamefont
  {Andersen}, \citenamefont {Liechtenstein}, \citenamefont {Jepsen},\ and\
  \citenamefont {Paulsen}}]{andersen1995b}%
  \BibitemOpen
  \bibfield  {author} {\bibinfo {author} {\bibfnamefont {O.~K.}\ \bibnamefont
  {Andersen}}, \bibinfo {author} {\bibfnamefont {A.~I.}\ \bibnamefont
  {Liechtenstein}}, \bibinfo {author} {\bibfnamefont {O.}~\bibnamefont
  {Jepsen}},\ and\ \bibinfo {author} {\bibfnamefont {F.}~\bibnamefont
  {Paulsen}},\ }\bibfield  {title} {\bibinfo {title} {{{LDA}} energy bands,
  low-energy hamiltonians, {{$t',~t'',~t_{\perp}(k)$ and $J_{\perp}$}}},\
  }\href {https://doi.org/10.1016/0022-3697(95)00269-3} {\bibfield  {journal}
  {\bibinfo  {journal} {Journal of Physics and Chemistry of Solids}\ }\bibinfo
  {series} {Proceedings of the {{Conference}} on {{Spectroscopies}} in {{Novel
  Superconductors}}},\ \textbf {\bibinfo {volume} {56}},\ \bibinfo {pages}
  {1573} (\bibinfo {year} {1995})}\BibitemShut {NoStop}%
\bibitem [{\citenamefont {Tranquada}\ \emph {et~al.}(1987)\citenamefont
  {Tranquada}, \citenamefont {Heald},\ and\ \citenamefont
  {Moodenbaugh}}]{tranquada1987}%
  \BibitemOpen
  \bibfield  {author} {\bibinfo {author} {\bibfnamefont {J.~M.}\ \bibnamefont
  {Tranquada}}, \bibinfo {author} {\bibfnamefont {S.~M.}\ \bibnamefont
  {Heald}},\ and\ \bibinfo {author} {\bibfnamefont {A.~R.}\ \bibnamefont
  {Moodenbaugh}},\ }\bibfield  {title} {\bibinfo {title} {{{X-ray-absorption
  near-edge-structure study of La$_{2-x}$(Ba,Sr)CuO$_{4-y}$
  superconductors}}},\ }\href {https://doi.org/10.1103/PhysRevB.36.5263}
  {\bibfield  {journal} {\bibinfo  {journal} {Phys. Rev. B}\ }\textbf {\bibinfo
  {volume} {36}},\ \bibinfo {pages} {5263} (\bibinfo {year}
  {1987})}\BibitemShut {NoStop}%
\bibitem [{\citenamefont {Fujimori}\ \emph {et~al.}(1987)\citenamefont
  {Fujimori}, \citenamefont {{Takayama-Muromachi}}, \citenamefont {Uchida},\
  and\ \citenamefont {Okai}}]{fujimori1987}%
  \BibitemOpen
  \bibfield  {author} {\bibinfo {author} {\bibfnamefont {A.}~\bibnamefont
  {Fujimori}}, \bibinfo {author} {\bibfnamefont {E.}~\bibnamefont
  {{Takayama-Muromachi}}}, \bibinfo {author} {\bibfnamefont {Y.}~\bibnamefont
  {Uchida}},\ and\ \bibinfo {author} {\bibfnamefont {B.}~\bibnamefont {Okai}},\
  }\bibfield  {title} {\bibinfo {title} {Spectroscopic evidence for strongly
  correlated electronic states in {{La-Sr-Cu}} and {{Y-Ba-Cu}} oxides},\ }\href
  {https://doi.org/10.1103/PhysRevB.35.8814} {\bibfield  {journal} {\bibinfo
  {journal} {Phys. Rev. B}\ }\textbf {\bibinfo {volume} {35}},\ \bibinfo
  {pages} {8814} (\bibinfo {year} {1987})}\BibitemShut {NoStop}%
\bibitem [{\citenamefont {Zheng}\ \emph {et~al.}(1995)\citenamefont {Zheng},
  \citenamefont {Kitaoka}, \citenamefont {Ishida},\ and\ \citenamefont
  {Asayama}}]{zheng1995}%
  \BibitemOpen
  \bibfield  {author} {\bibinfo {author} {\bibfnamefont {G.-q.}\ \bibnamefont
  {Zheng}}, \bibinfo {author} {\bibfnamefont {Y.}~\bibnamefont {Kitaoka}},
  \bibinfo {author} {\bibfnamefont {K.}~\bibnamefont {Ishida}},\ and\ \bibinfo
  {author} {\bibfnamefont {K.}~\bibnamefont {Asayama}},\ }\bibfield  {title}
  {\bibinfo {title} {Local hole distribution in the {{CuO}}{\textsubscript{2}}
  plane of {{High-T}}{\textsubscript{c}} {{Cu-Oxides}} studied by {{Cu}} and
  {{Oxygen NQR}}/{{NMR}}},\ }\href {https://doi.org/10.1143/jpsj.64.2524}
  {\bibfield  {journal} {\bibinfo  {journal} {Journal of the Physical Society
  of Japan}\ }\textbf {\bibinfo {volume} {64}},\ \bibinfo {pages} {2524}
  (\bibinfo {year} {1995})}\BibitemShut {NoStop}%
\bibitem [{\citenamefont {Haase}\ \emph {et~al.}(2004)\citenamefont {Haase},
  \citenamefont {Sushkov}, \citenamefont {Horsch},\ and\ \citenamefont
  {Williams}}]{haase2004}%
  \BibitemOpen
  \bibfield  {author} {\bibinfo {author} {\bibfnamefont {J.}~\bibnamefont
  {Haase}}, \bibinfo {author} {\bibfnamefont {O.~P.}\ \bibnamefont {Sushkov}},
  \bibinfo {author} {\bibfnamefont {P.}~\bibnamefont {Horsch}},\ and\ \bibinfo
  {author} {\bibfnamefont {G.~V.~M.}\ \bibnamefont {Williams}},\ }\bibfield
  {title} {\bibinfo {title} {{{Planar Cu and O hole densities in high-${T}_{c}$
  cuprates determined with NMR}}},\ }\href
  {https://doi.org/10.1103/PhysRevB.69.094504} {\bibfield  {journal} {\bibinfo
  {journal} {Phys. Rev. B}\ }\textbf {\bibinfo {volume} {69}},\ \bibinfo
  {pages} {094504} (\bibinfo {year} {2004})}\BibitemShut {NoStop}%
\bibitem [{\citenamefont {Jurkutat}\ \emph {et~al.}(2014)\citenamefont
  {Jurkutat}, \citenamefont {Rybicki}, \citenamefont {Sushkov}, \citenamefont
  {Williams}, \citenamefont {Erb},\ and\ \citenamefont {Haase}}]{jurkutat2014}%
  \BibitemOpen
  \bibfield  {author} {\bibinfo {author} {\bibfnamefont {M.}~\bibnamefont
  {Jurkutat}}, \bibinfo {author} {\bibfnamefont {D.}~\bibnamefont {Rybicki}},
  \bibinfo {author} {\bibfnamefont {O.~P.}\ \bibnamefont {Sushkov}}, \bibinfo
  {author} {\bibfnamefont {G.~V.~M.}\ \bibnamefont {Williams}}, \bibinfo
  {author} {\bibfnamefont {A.}~\bibnamefont {Erb}},\ and\ \bibinfo {author}
  {\bibfnamefont {J.}~\bibnamefont {Haase}},\ }\bibfield  {title} {\bibinfo
  {title} {Distribution of electrons and holes in cuprate superconductors as
  determined from ${}^{17}\mathrm{O}$ and $^{63}\mathrm{Cu}$ nuclear magnetic
  resonance},\ }\href {https://doi.org/10.1103/PhysRevB.90.140504} {\bibfield
  {journal} {\bibinfo  {journal} {Phys. Rev. B}\ }\textbf {\bibinfo {volume}
  {90}},\ \bibinfo {pages} {140504} (\bibinfo {year} {2014})}\BibitemShut
  {NoStop}%
\bibitem [{\citenamefont {Rybicki}\ \emph {et~al.}(2016)\citenamefont
  {Rybicki}, \citenamefont {Jurkutat}, \citenamefont {Reichardt}, \citenamefont
  {Kapusta},\ and\ \citenamefont {Haase}}]{rybicki2016}%
  \BibitemOpen
  \bibfield  {author} {\bibinfo {author} {\bibfnamefont {D.}~\bibnamefont
  {Rybicki}}, \bibinfo {author} {\bibfnamefont {M.}~\bibnamefont {Jurkutat}},
  \bibinfo {author} {\bibfnamefont {S.}~\bibnamefont {Reichardt}}, \bibinfo
  {author} {\bibfnamefont {C.}~\bibnamefont {Kapusta}},\ and\ \bibinfo {author}
  {\bibfnamefont {J.}~\bibnamefont {Haase}},\ }\bibfield  {title} {\bibinfo
  {title} {Perspective on the phase diagram of cuprate high-temperature
  superconductors},\ }\href {https://doi.org/10.1038/ncomms11413} {\bibfield
  {journal} {\bibinfo  {journal} {Nat Commun}\ }\textbf {\bibinfo {volume}
  {7}},\ \bibinfo {pages} {11413} (\bibinfo {year} {2016})}\BibitemShut
  {NoStop}%
\bibitem [{\citenamefont {Kowalski}\ \emph {et~al.}(2021)\citenamefont
  {Kowalski}, \citenamefont {Dash}, \citenamefont {S{\'e}mon}, \citenamefont
  {S{\'e}n{\'e}chal},\ and\ \citenamefont {Tremblay}}]{kowalski2021}%
  \BibitemOpen
  \bibfield  {author} {\bibinfo {author} {\bibfnamefont {N.}~\bibnamefont
  {Kowalski}}, \bibinfo {author} {\bibfnamefont {S.~S.}\ \bibnamefont {Dash}},
  \bibinfo {author} {\bibfnamefont {P.}~\bibnamefont {S{\'e}mon}}, \bibinfo
  {author} {\bibfnamefont {D.}~\bibnamefont {S{\'e}n{\'e}chal}},\ and\ \bibinfo
  {author} {\bibfnamefont {A.-M.}\ \bibnamefont {Tremblay}},\ }\bibfield
  {title} {\bibinfo {title} {Oxygen hole content, charge-transfer gap,
  covalency, and cuprate superconductivity},\ }\href
  {https://doi.org/10.1073/pnas.2106476118} {\bibfield  {journal} {\bibinfo
  {journal} {Proc. Natl. Acad. Sci. U.S.A.}\ }\textbf {\bibinfo {volume}
  {118}},\ \bibinfo {pages} {e2106476118} (\bibinfo {year} {2021})}\BibitemShut
  {NoStop}%
\bibitem [{\citenamefont {Jurkutat}\ \emph {et~al.}(2023)\citenamefont
  {Jurkutat}, \citenamefont {Kattinger}, \citenamefont {Tsankov}, \citenamefont
  {Reznicek}, \citenamefont {Erb},\ and\ \citenamefont {Haase}}]{jurkutat2023}%
  \BibitemOpen
  \bibfield  {author} {\bibinfo {author} {\bibfnamefont {M.}~\bibnamefont
  {Jurkutat}}, \bibinfo {author} {\bibfnamefont {C.}~\bibnamefont {Kattinger}},
  \bibinfo {author} {\bibfnamefont {S.}~\bibnamefont {Tsankov}}, \bibinfo
  {author} {\bibfnamefont {R.}~\bibnamefont {Reznicek}}, \bibinfo {author}
  {\bibfnamefont {A.}~\bibnamefont {Erb}},\ and\ \bibinfo {author}
  {\bibfnamefont {J.}~\bibnamefont {Haase}},\ }\bibfield  {title} {\bibinfo
  {title} {How pressure enhances the critical temperature of superconductivity
  in {{YBa$_2$Cu$_3$O$_{6+y}$}}},\ }\href
  {https://doi.org/10.1073/pnas.2215458120} {\bibfield  {journal} {\bibinfo
  {journal} {Proceedings of the National Academy of Sciences}\ }\textbf
  {\bibinfo {volume} {120}},\ \bibinfo {pages} {e2215458120} (\bibinfo {year}
  {2023})}\BibitemShut {NoStop}%
\bibitem [{\citenamefont {Weber}\ \emph {et~al.}(2012)\citenamefont {Weber},
  \citenamefont {Yee}, \citenamefont {Haule},\ and\ \citenamefont
  {Kotliar}}]{weber2012}%
  \BibitemOpen
  \bibfield  {author} {\bibinfo {author} {\bibfnamefont {C.}~\bibnamefont
  {Weber}}, \bibinfo {author} {\bibfnamefont {C.}~\bibnamefont {Yee}}, \bibinfo
  {author} {\bibfnamefont {K.}~\bibnamefont {Haule}},\ and\ \bibinfo {author}
  {\bibfnamefont {G.}~\bibnamefont {Kotliar}},\ }\bibfield  {title} {\bibinfo
  {title} {Scaling of the transition temperature of hole-doped cuprate
  superconductors with the charge-transfer energy},\ }\href
  {https://doi.org/10.1209/0295-5075/100/37001} {\bibfield  {journal} {\bibinfo
   {journal} {EPL}\ }\textbf {\bibinfo {volume} {100}},\ \bibinfo {pages}
  {37001} (\bibinfo {year} {2012})}\BibitemShut {NoStop}%
\bibitem [{\citenamefont {Wang}\ \emph {et~al.}(2022)\citenamefont {Wang},
  \citenamefont {He}, \citenamefont {Yang}, \citenamefont {{Garcia-Fernandez}},
  \citenamefont {Nag}, \citenamefont {Zhou}, \citenamefont {Minola},
  \citenamefont {Tacon}, \citenamefont {Keimer}, \citenamefont {Peng},\ and\
  \citenamefont {Li}}]{wang2022}%
  \BibitemOpen
  \bibfield  {author} {\bibinfo {author} {\bibfnamefont {L.}~\bibnamefont
  {Wang}}, \bibinfo {author} {\bibfnamefont {G.}~\bibnamefont {He}}, \bibinfo
  {author} {\bibfnamefont {Z.}~\bibnamefont {Yang}}, \bibinfo {author}
  {\bibfnamefont {M.}~\bibnamefont {{Garcia-Fernandez}}}, \bibinfo {author}
  {\bibfnamefont {A.}~\bibnamefont {Nag}}, \bibinfo {author} {\bibfnamefont
  {K.}~\bibnamefont {Zhou}}, \bibinfo {author} {\bibfnamefont {M.}~\bibnamefont
  {Minola}}, \bibinfo {author} {\bibfnamefont {M.~L.}\ \bibnamefont {Tacon}},
  \bibinfo {author} {\bibfnamefont {B.}~\bibnamefont {Keimer}}, \bibinfo
  {author} {\bibfnamefont {Y.}~\bibnamefont {Peng}},\ and\ \bibinfo {author}
  {\bibfnamefont {Y.}~\bibnamefont {Li}},\ }\bibfield  {title} {\bibinfo
  {title} {Paramagnons and high-temperature superconductivity in a model family
  of cuprates},\ }\href {https://doi.org/10.1038/s41467-022-30918-z} {\bibfield
   {journal} {\bibinfo  {journal} {Nat Commun}\ }\textbf {\bibinfo {volume}
  {13}},\ \bibinfo {pages} {3163} (\bibinfo {year} {2022})}\BibitemShut
  {NoStop}%
\bibitem [{\citenamefont {O'Mahony}\ \emph {et~al.}(2022)\citenamefont
  {O'Mahony}, \citenamefont {Ren}, \citenamefont {Chen}, \citenamefont {Chong},
  \citenamefont {Liu}, \citenamefont {Eisaki}, \citenamefont {Uchida},
  \citenamefont {Hamidian},\ and\ \citenamefont {Davis}}]{omahony2022a}%
  \BibitemOpen
  \bibfield  {author} {\bibinfo {author} {\bibfnamefont {S.~M.}\ \bibnamefont
  {O'Mahony}}, \bibinfo {author} {\bibfnamefont {W.}~\bibnamefont {Ren}},
  \bibinfo {author} {\bibfnamefont {W.}~\bibnamefont {Chen}}, \bibinfo {author}
  {\bibfnamefont {Y.~X.}\ \bibnamefont {Chong}}, \bibinfo {author}
  {\bibfnamefont {X.}~\bibnamefont {Liu}}, \bibinfo {author} {\bibfnamefont
  {H.}~\bibnamefont {Eisaki}}, \bibinfo {author} {\bibfnamefont
  {S.}~\bibnamefont {Uchida}}, \bibinfo {author} {\bibfnamefont {M.~H.}\
  \bibnamefont {Hamidian}},\ and\ \bibinfo {author} {\bibfnamefont {J.~C.~S.}\
  \bibnamefont {Davis}},\ }\bibfield  {title} {\bibinfo {title} {On the
  electron pairing mechanism of copper-oxide high temperature
  superconductivity},\ }\href {https://doi.org/10.1073/pnas.2207449119}
  {\bibfield  {journal} {\bibinfo  {journal} {Proceedings of the National
  Academy of Sciences}\ }\textbf {\bibinfo {volume} {119}},\ \bibinfo {pages}
  {e2207449119} (\bibinfo {year} {2022})}\BibitemShut {NoStop}%
\bibitem [{\citenamefont {Wang}\ \emph {et~al.}(2023)\citenamefont {Wang},
  \citenamefont {Zou}, \citenamefont {Lin}, \citenamefont {Luo}, \citenamefont
  {Yan}, \citenamefont {Yin}, \citenamefont {Xu}, \citenamefont {Zhou},
  \citenamefont {Wang},\ and\ \citenamefont {Zhu}}]{wang2023a}%
  \BibitemOpen
  \bibfield  {author} {\bibinfo {author} {\bibfnamefont {Z.}~\bibnamefont
  {Wang}}, \bibinfo {author} {\bibfnamefont {C.}~\bibnamefont {Zou}}, \bibinfo
  {author} {\bibfnamefont {C.}~\bibnamefont {Lin}}, \bibinfo {author}
  {\bibfnamefont {X.}~\bibnamefont {Luo}}, \bibinfo {author} {\bibfnamefont
  {H.}~\bibnamefont {Yan}}, \bibinfo {author} {\bibfnamefont {C.}~\bibnamefont
  {Yin}}, \bibinfo {author} {\bibfnamefont {Y.}~\bibnamefont {Xu}}, \bibinfo
  {author} {\bibfnamefont {X.}~\bibnamefont {Zhou}}, \bibinfo {author}
  {\bibfnamefont {Y.}~\bibnamefont {Wang}},\ and\ \bibinfo {author}
  {\bibfnamefont {J.}~\bibnamefont {Zhu}},\ }\bibfield  {title} {\bibinfo
  {title} {Correlating the charge-transfer gap to the maximum transition
  temperature in {{Bi$_2$Sr$_2$Ca$_{n-1}$Cu$_n$O$_{2n+4+\delta}$}}},\ }\href
  {https://doi.org/10.1126/science.add3672} {\bibfield  {journal} {\bibinfo
  {journal} {Science}\ }\textbf {\bibinfo {volume} {381}},\ \bibinfo {pages}
  {227} (\bibinfo {year} {2023})}\BibitemShut {NoStop}%
\bibitem [{\citenamefont {Vu\ifmmode \check{c}\else \v{c}\fi{}i\ifmmode
  \check{c}\else \v{c}\fi{}evi\ifmmode~\acute{c}\else \'{c}\fi{}}\ and\
  \citenamefont {Ferrero}(2024)}]{vucivevic2024}%
  \BibitemOpen
  \bibfield  {author} {\bibinfo {author} {\bibfnamefont {J.~c.~v.}\
  \bibnamefont {Vu\ifmmode \check{c}\else \v{c}\fi{}i\ifmmode \check{c}\else
  \v{c}\fi{}evi\ifmmode~\acute{c}\else \'{c}\fi{}}}\ and\ \bibinfo {author}
  {\bibfnamefont {M.}~\bibnamefont {Ferrero}},\ }\bibfield  {title} {\bibinfo
  {title} {Simple predictors of ${T}_{c}$ in superconducting cuprates and the
  role of interactions between effective wannier orbitals in the
  $d\text{\ensuremath{-}}p$ three-band model},\ }\href
  {https://doi.org/10.1103/PhysRevB.109.L081115} {\bibfield  {journal}
  {\bibinfo  {journal} {Phys. Rev. B}\ }\textbf {\bibinfo {volume} {109}},\
  \bibinfo {pages} {L081115} (\bibinfo {year} {2024})}\BibitemShut {NoStop}%
\bibitem [{\citenamefont {Hubbard}\ and\ \citenamefont
  {Flowers}(1963)}]{hubbard1963}%
  \BibitemOpen
  \bibfield  {author} {\bibinfo {author} {\bibfnamefont {J.}~\bibnamefont
  {Hubbard}}\ and\ \bibinfo {author} {\bibfnamefont {B.~H.}\ \bibnamefont
  {Flowers}},\ }\bibfield  {title} {\bibinfo {title} {Electron correlations in
  narrow energy bands},\ }\href {https://doi.org/10.1098/rspa.1963.0204}
  {\bibfield  {journal} {\bibinfo  {journal} {Proceedings of the Royal Society
  of London. Series A. Mathematical and Physical Sciences}\ }\textbf {\bibinfo
  {volume} {276}},\ \bibinfo {pages} {238} (\bibinfo {year}
  {1963})}\BibitemShut {NoStop}%
\bibitem [{\citenamefont {Gutzwiller}(1963)}]{gutzwiller1963}%
  \BibitemOpen
  \bibfield  {author} {\bibinfo {author} {\bibfnamefont {M.~C.}\ \bibnamefont
  {Gutzwiller}},\ }\bibfield  {title} {\bibinfo {title} {Effect of correlation
  on the ferromagnetism of transition metals},\ }\href
  {https://doi.org/10.1103/PhysRevLett.10.159} {\bibfield  {journal} {\bibinfo
  {journal} {Phys. Rev. Lett.}\ }\textbf {\bibinfo {volume} {10}},\ \bibinfo
  {pages} {159} (\bibinfo {year} {1963})}\BibitemShut {NoStop}%
\bibitem [{\citenamefont {Kanamori}(1963)}]{kanamori1963}%
  \BibitemOpen
  \bibfield  {author} {\bibinfo {author} {\bibfnamefont {J.}~\bibnamefont
  {Kanamori}},\ }\bibfield  {title} {\bibinfo {title} {Electron correlation and
  ferromagnetism of transition metals},\ }\href
  {https://doi.org/10.1143/PTP.30.275} {\bibfield  {journal} {\bibinfo
  {journal} {Progress of Theoretical Physics}\ }\textbf {\bibinfo {volume}
  {30}},\ \bibinfo {pages} {275} (\bibinfo {year} {1963})}\BibitemShut
  {NoStop}%
\bibitem [{\citenamefont {Emery}(1987)}]{emery1987}%
  \BibitemOpen
  \bibfield  {author} {\bibinfo {author} {\bibfnamefont {V.~J.}\ \bibnamefont
  {Emery}},\ }\bibfield  {title} {\bibinfo {title} {Theory of high-{{$T_C$}}
  superconductivity in oxides},\ }\href
  {https://doi.org/10.1103/PhysRevLett.58.2794} {\bibfield  {journal} {\bibinfo
   {journal} {Phys. Rev. Lett.}\ }\textbf {\bibinfo {volume} {58}},\ \bibinfo
  {pages} {2794} (\bibinfo {year} {1987})}\BibitemShut {NoStop}%
\bibitem [{\citenamefont {Varma}\ \emph {et~al.}(1987)\citenamefont {Varma},
  \citenamefont {{Schmitt-Rink}},\ and\ \citenamefont {Abrahams}}]{varma1987}%
  \BibitemOpen
  \bibfield  {author} {\bibinfo {author} {\bibfnamefont {C.~M.}\ \bibnamefont
  {Varma}}, \bibinfo {author} {\bibfnamefont {S.}~\bibnamefont
  {{Schmitt-Rink}}},\ and\ \bibinfo {author} {\bibfnamefont {E.}~\bibnamefont
  {Abrahams}},\ }\bibfield  {title} {\bibinfo {title} {Charge transfer
  excitations and superconductivity in ``ionic'' metals},\ }\href
  {https://doi.org/10.1016/0038-1098(87)90407-8} {\bibfield  {journal}
  {\bibinfo  {journal} {Solid State Communications}\ }\textbf {\bibinfo
  {volume} {62}},\ \bibinfo {pages} {681} (\bibinfo {year} {1987})}\BibitemShut
  {NoStop}%
\bibitem [{\citenamefont {Imada}\ \emph {et~al.}(1998)\citenamefont {Imada},
  \citenamefont {Fujimori},\ and\ \citenamefont {Tokura}}]{imada1998}%
  \BibitemOpen
  \bibfield  {author} {\bibinfo {author} {\bibfnamefont {M.}~\bibnamefont
  {Imada}}, \bibinfo {author} {\bibfnamefont {A.}~\bibnamefont {Fujimori}},\
  and\ \bibinfo {author} {\bibfnamefont {Y.}~\bibnamefont {Tokura}},\
  }\bibfield  {title} {\bibinfo {title} {Metal-insulator transitions},\ }\href
  {https://doi.org/10.1103/RevModPhys.70.1039} {\bibfield  {journal} {\bibinfo
  {journal} {Rev. Mod. Phys.}\ }\textbf {\bibinfo {volume} {70}},\ \bibinfo
  {pages} {1039} (\bibinfo {year} {1998})}\BibitemShut {NoStop}%
\bibitem [{\citenamefont {Park}\ \emph {et~al.}(2008)\citenamefont {Park},
  \citenamefont {Haule},\ and\ \citenamefont {Kotliar}}]{park2008}%
  \BibitemOpen
  \bibfield  {author} {\bibinfo {author} {\bibfnamefont {H.}~\bibnamefont
  {Park}}, \bibinfo {author} {\bibfnamefont {K.}~\bibnamefont {Haule}},\ and\
  \bibinfo {author} {\bibfnamefont {G.}~\bibnamefont {Kotliar}},\ }\bibfield
  {title} {\bibinfo {title} {Cluster dynamical mean field theory of the
  {{Mott}} transition},\ }\href
  {https://doi.org/10.1103/PhysRevLett.101.186403} {\bibfield  {journal}
  {\bibinfo  {journal} {Phys. Rev. Lett.}\ }\textbf {\bibinfo {volume} {101}},\
  \bibinfo {pages} {186403} (\bibinfo {year} {2008})}\BibitemShut {NoStop}%
\bibitem [{\citenamefont {Sordi}\ \emph {et~al.}(2011)\citenamefont {Sordi},
  \citenamefont {Haule},\ and\ \citenamefont {Tremblay}}]{sordi2011}%
  \BibitemOpen
  \bibfield  {author} {\bibinfo {author} {\bibfnamefont {G.}~\bibnamefont
  {Sordi}}, \bibinfo {author} {\bibfnamefont {K.}~\bibnamefont {Haule}},\ and\
  \bibinfo {author} {\bibfnamefont {A.-M.~S.}\ \bibnamefont {Tremblay}},\
  }\bibfield  {title} {\bibinfo {title} {Mott physics and first-order
  transition between two metals in the normal-state phase diagram of the
  two-dimensional {{Hubbard}} model},\ }\href
  {https://doi.org/10.1103/PhysRevB.84.075161} {\bibfield  {journal} {\bibinfo
  {journal} {Phys. Rev. B}\ }\textbf {\bibinfo {volume} {84}},\ \bibinfo
  {pages} {075161} (\bibinfo {year} {2011})}\BibitemShut {NoStop}%
\bibitem [{\citenamefont {Sordi}\ \emph {et~al.}(2012)\citenamefont {Sordi},
  \citenamefont {S{\'e}mon}, \citenamefont {Haule},\ and\ \citenamefont
  {Tremblay}}]{sordi2012}%
  \BibitemOpen
  \bibfield  {author} {\bibinfo {author} {\bibfnamefont {G.}~\bibnamefont
  {Sordi}}, \bibinfo {author} {\bibfnamefont {P.}~\bibnamefont {S{\'e}mon}},
  \bibinfo {author} {\bibfnamefont {K.}~\bibnamefont {Haule}},\ and\ \bibinfo
  {author} {\bibfnamefont {A.-M.~S.}\ \bibnamefont {Tremblay}},\ }\bibfield
  {title} {\bibinfo {title} {Pseudogap temperature as a {{Widom}} line in doped
  {{Mott}} insulators},\ }\href {https://doi.org/10.1038/srep00547} {\bibfield
  {journal} {\bibinfo  {journal} {Sci Rep}\ }\textbf {\bibinfo {volume} {2}},\
  \bibinfo {pages} {547} (\bibinfo {year} {2012})}\BibitemShut {NoStop}%
\bibitem [{\citenamefont {Vu\ifmmode \check{c}\else \v{c}\fi{}i\ifmmode
  \check{c}\else \v{c}\fi{}evi\ifmmode~\acute{c}\else \'{c}\fi{}}\ \emph
  {et~al.}(2013)\citenamefont {Vu\ifmmode \check{c}\else \v{c}\fi{}i\ifmmode
  \check{c}\else \v{c}\fi{}evi\ifmmode~\acute{c}\else \'{c}\fi{}},
  \citenamefont {Terletska}, \citenamefont {Tanaskovi\ifmmode~\acute{c}\else
  \'{c}\fi{}},\ and\ \citenamefont {Dobrosavljevi\ifmmode~\acute{c}\else
  \'{c}\fi{}}}]{vuvicevic2013}%
  \BibitemOpen
  \bibfield  {author} {\bibinfo {author} {\bibfnamefont {J.}~\bibnamefont
  {Vu\ifmmode \check{c}\else \v{c}\fi{}i\ifmmode \check{c}\else
  \v{c}\fi{}evi\ifmmode~\acute{c}\else \'{c}\fi{}}}, \bibinfo {author}
  {\bibfnamefont {H.}~\bibnamefont {Terletska}}, \bibinfo {author}
  {\bibfnamefont {D.}~\bibnamefont {Tanaskovi\ifmmode~\acute{c}\else
  \'{c}\fi{}}},\ and\ \bibinfo {author} {\bibfnamefont {V.}~\bibnamefont
  {Dobrosavljevi\ifmmode~\acute{c}\else \'{c}\fi{}}},\ }\bibfield  {title}
  {\bibinfo {title} {Finite-temperature crossover and the quantum {{Widom line
  near the Mott}} transition},\ }\href
  {https://doi.org/10.1103/PhysRevB.88.075143} {\bibfield  {journal} {\bibinfo
  {journal} {Phys. Rev. B}\ }\textbf {\bibinfo {volume} {88}},\ \bibinfo
  {pages} {075143} (\bibinfo {year} {2013})}\BibitemShut {NoStop}%
\bibitem [{\citenamefont {Walsh}\ \emph {et~al.}(2019)\citenamefont {Walsh},
  \citenamefont {S\'emon}, \citenamefont {Poulin}, \citenamefont {Sordi},\ and\
  \citenamefont {Tremblay}}]{walsh2019}%
  \BibitemOpen
  \bibfield  {author} {\bibinfo {author} {\bibfnamefont {C.}~\bibnamefont
  {Walsh}}, \bibinfo {author} {\bibfnamefont {P.}~\bibnamefont {S\'emon}},
  \bibinfo {author} {\bibfnamefont {D.}~\bibnamefont {Poulin}}, \bibinfo
  {author} {\bibfnamefont {G.}~\bibnamefont {Sordi}},\ and\ \bibinfo {author}
  {\bibfnamefont {A.-M.~S.}\ \bibnamefont {Tremblay}},\ }\bibfield  {title}
  {\bibinfo {title} {Thermodynamic and information-theoretic description of the
  {{Mott}} transition in the two-dimensional {{Hubbard}} model},\ }\href
  {https://doi.org/10.1103/PhysRevB.99.075122} {\bibfield  {journal} {\bibinfo
  {journal} {Phys. Rev. B}\ }\textbf {\bibinfo {volume} {99}},\ \bibinfo
  {pages} {075122} (\bibinfo {year} {2019})}\BibitemShut {NoStop}%
\bibitem [{\citenamefont {Civelli}\ \emph {et~al.}(2005)\citenamefont
  {Civelli}, \citenamefont {Capone}, \citenamefont {Kancharla}, \citenamefont
  {Parcollet},\ and\ \citenamefont {Kotliar}}]{civelli2005}%
  \BibitemOpen
  \bibfield  {author} {\bibinfo {author} {\bibfnamefont {M.}~\bibnamefont
  {Civelli}}, \bibinfo {author} {\bibfnamefont {M.}~\bibnamefont {Capone}},
  \bibinfo {author} {\bibfnamefont {S.~S.}\ \bibnamefont {Kancharla}}, \bibinfo
  {author} {\bibfnamefont {O.}~\bibnamefont {Parcollet}},\ and\ \bibinfo
  {author} {\bibfnamefont {G.}~\bibnamefont {Kotliar}},\ }\bibfield  {title}
  {\bibinfo {title} {Dynamical breakup of the {{Fermi}} surface in a doped
  {{Mott}} insulator},\ }\href {https://doi.org/10.1103/PhysRevLett.95.106402}
  {\bibfield  {journal} {\bibinfo  {journal} {Phys. Rev. Lett.}\ }\textbf
  {\bibinfo {volume} {95}},\ \bibinfo {pages} {106402} (\bibinfo {year}
  {2005})}\BibitemShut {NoStop}%
\bibitem [{\citenamefont {Kyung}\ \emph {et~al.}(2006)\citenamefont {Kyung},
  \citenamefont {Kancharla}, \citenamefont {S{\'e}n{\'e}chal}, \citenamefont
  {Tremblay}, \citenamefont {Civelli},\ and\ \citenamefont
  {Kotliar}}]{kyung2006}%
  \BibitemOpen
  \bibfield  {author} {\bibinfo {author} {\bibfnamefont {B.}~\bibnamefont
  {Kyung}}, \bibinfo {author} {\bibfnamefont {S.~S.}\ \bibnamefont
  {Kancharla}}, \bibinfo {author} {\bibfnamefont {D.}~\bibnamefont
  {S{\'e}n{\'e}chal}}, \bibinfo {author} {\bibfnamefont {A.-M.~S.}\
  \bibnamefont {Tremblay}}, \bibinfo {author} {\bibfnamefont {M.}~\bibnamefont
  {Civelli}},\ and\ \bibinfo {author} {\bibfnamefont {G.}~\bibnamefont
  {Kotliar}},\ }\bibfield  {title} {\bibinfo {title} {Pseudogap induced by
  short-range spin correlations in a doped {{Mott}} insulator},\ }\href
  {https://doi.org/10.1103/PhysRevB.73.165114} {\bibfield  {journal} {\bibinfo
  {journal} {Phys. Rev. B}\ }\textbf {\bibinfo {volume} {73}},\ \bibinfo
  {pages} {165114} (\bibinfo {year} {2006})}\BibitemShut {NoStop}%
\bibitem [{\citenamefont {Sakai}\ \emph {et~al.}(2009)\citenamefont {Sakai},
  \citenamefont {Motome},\ and\ \citenamefont {Imada}}]{sakai2009}%
  \BibitemOpen
  \bibfield  {author} {\bibinfo {author} {\bibfnamefont {S.}~\bibnamefont
  {Sakai}}, \bibinfo {author} {\bibfnamefont {Y.}~\bibnamefont {Motome}},\ and\
  \bibinfo {author} {\bibfnamefont {M.}~\bibnamefont {Imada}},\ }\bibfield
  {title} {\bibinfo {title} {Evolution of electronic structure of doped
  {{Mott}} insulators: Reconstruction of poles and zeros of {{Green}}'s
  function},\ }\href {https://doi.org/10.1103/PhysRevLett.102.056404}
  {\bibfield  {journal} {\bibinfo  {journal} {Phys. Rev. Lett.}\ }\textbf
  {\bibinfo {volume} {102}},\ \bibinfo {pages} {056404} (\bibinfo {year}
  {2009})}\BibitemShut {NoStop}%
\bibitem [{\citenamefont {Macridin}\ \emph {et~al.}(2006)\citenamefont
  {Macridin}, \citenamefont {Jarrell}, \citenamefont {Maier}, \citenamefont
  {Kent},\ and\ \citenamefont {D'Azevedo}}]{macridin2006}%
  \BibitemOpen
  \bibfield  {author} {\bibinfo {author} {\bibfnamefont {A.}~\bibnamefont
  {Macridin}}, \bibinfo {author} {\bibfnamefont {M.}~\bibnamefont {Jarrell}},
  \bibinfo {author} {\bibfnamefont {T.}~\bibnamefont {Maier}}, \bibinfo
  {author} {\bibfnamefont {P.~R.~C.}\ \bibnamefont {Kent}},\ and\ \bibinfo
  {author} {\bibfnamefont {E.}~\bibnamefont {D'Azevedo}},\ }\bibfield  {title}
  {\bibinfo {title} {Pseudogap and antiferromagnetic correlations in the
  {{Hubbard}} model},\ }\href {https://doi.org/10.1103/PhysRevLett.97.036401}
  {\bibfield  {journal} {\bibinfo  {journal} {Phys. Rev. Lett.}\ }\textbf
  {\bibinfo {volume} {97}},\ \bibinfo {pages} {036401} (\bibinfo {year}
  {2006})}\BibitemShut {NoStop}%
\bibitem [{\citenamefont {Ferrero}\ \emph {et~al.}(2009)\citenamefont
  {Ferrero}, \citenamefont {Cornaglia}, \citenamefont {De~Leo}, \citenamefont
  {Parcollet}, \citenamefont {Kotliar},\ and\ \citenamefont
  {Georges}}]{ferrero2009}%
  \BibitemOpen
  \bibfield  {author} {\bibinfo {author} {\bibfnamefont {M.}~\bibnamefont
  {Ferrero}}, \bibinfo {author} {\bibfnamefont {P.~S.}\ \bibnamefont
  {Cornaglia}}, \bibinfo {author} {\bibfnamefont {L.}~\bibnamefont {De~Leo}},
  \bibinfo {author} {\bibfnamefont {O.}~\bibnamefont {Parcollet}}, \bibinfo
  {author} {\bibfnamefont {G.}~\bibnamefont {Kotliar}},\ and\ \bibinfo {author}
  {\bibfnamefont {A.}~\bibnamefont {Georges}},\ }\bibfield  {title} {\bibinfo
  {title} {Pseudogap opening and formation of {{Fermi}} arcs as an
  orbital-selective {{Mott}} transition in momentum space},\ }\href
  {https://doi.org/10.1103/PhysRevB.80.064501} {\bibfield  {journal} {\bibinfo
  {journal} {Phys. Rev. B}\ }\textbf {\bibinfo {volume} {80}},\ \bibinfo
  {pages} {064501} (\bibinfo {year} {2009})}\BibitemShut {NoStop}%
\bibitem [{\citenamefont {Werner}\ \emph {et~al.}(2009)\citenamefont {Werner},
  \citenamefont {Gull}, \citenamefont {Parcollet},\ and\ \citenamefont
  {Millis}}]{werner2009}%
  \BibitemOpen
  \bibfield  {author} {\bibinfo {author} {\bibfnamefont {P.}~\bibnamefont
  {Werner}}, \bibinfo {author} {\bibfnamefont {E.}~\bibnamefont {Gull}},
  \bibinfo {author} {\bibfnamefont {O.}~\bibnamefont {Parcollet}},\ and\
  \bibinfo {author} {\bibfnamefont {A.~J.}\ \bibnamefont {Millis}},\ }\bibfield
   {title} {\bibinfo {title} {Momentum-selective metal-insulator transition in
  the two-dimensional {{Hubbard}} model: {{An}} 8-site dynamical cluster
  approximation study},\ }\href {https://doi.org/10.1103/PhysRevB.80.045120}
  {\bibfield  {journal} {\bibinfo  {journal} {Phys. Rev. B}\ }\textbf {\bibinfo
  {volume} {80}},\ \bibinfo {pages} {045120} (\bibinfo {year}
  {2009})}\BibitemShut {NoStop}%
\bibitem [{\citenamefont {Gull}\ \emph {et~al.}(2010)\citenamefont {Gull},
  \citenamefont {Ferrero}, \citenamefont {Parcollet}, \citenamefont {Georges},\
  and\ \citenamefont {Millis}}]{gull2010}%
  \BibitemOpen
  \bibfield  {author} {\bibinfo {author} {\bibfnamefont {E.}~\bibnamefont
  {Gull}}, \bibinfo {author} {\bibfnamefont {M.}~\bibnamefont {Ferrero}},
  \bibinfo {author} {\bibfnamefont {O.}~\bibnamefont {Parcollet}}, \bibinfo
  {author} {\bibfnamefont {A.}~\bibnamefont {Georges}},\ and\ \bibinfo {author}
  {\bibfnamefont {A.~J.}\ \bibnamefont {Millis}},\ }\bibfield  {title}
  {\bibinfo {title} {Momentum-space anisotropy and pseudogaps: {{A}}
  comparative cluster dynamical mean-field analysis of the doping-driven
  metal-insulator transition in the two-dimensional {{Hubbard}} model},\ }\href
  {https://doi.org/10.1103/PhysRevB.82.155101} {\bibfield  {journal} {\bibinfo
  {journal} {Phys. Rev. B}\ }\textbf {\bibinfo {volume} {82}},\ \bibinfo
  {pages} {155101} (\bibinfo {year} {2010})}\BibitemShut {NoStop}%
\bibitem [{\citenamefont {Merino}\ and\ \citenamefont
  {Gunnarsson}(2014)}]{merino2014}%
  \BibitemOpen
  \bibfield  {author} {\bibinfo {author} {\bibfnamefont {J.}~\bibnamefont
  {Merino}}\ and\ \bibinfo {author} {\bibfnamefont {O.}~\bibnamefont
  {Gunnarsson}},\ }\bibfield  {title} {\bibinfo {title} {Pseudogap and singlet
  formation in organic and cuprate superconductors},\ }\href
  {https://doi.org/10.1103/PhysRevB.89.245130} {\bibfield  {journal} {\bibinfo
  {journal} {Phys. Rev. B}\ }\textbf {\bibinfo {volume} {89}},\ \bibinfo
  {pages} {245130} (\bibinfo {year} {2014})}\BibitemShut {NoStop}%
\bibitem [{\citenamefont {Krien}\ \emph {et~al.}(2022)\citenamefont {Krien},
  \citenamefont {Worm}, \citenamefont {Chalupa-Gantner}, \citenamefont
  {Toschi},\ and\ \citenamefont {Held}}]{krien2022}%
  \BibitemOpen
  \bibfield  {author} {\bibinfo {author} {\bibfnamefont {F.}~\bibnamefont
  {Krien}}, \bibinfo {author} {\bibfnamefont {P.}~\bibnamefont {Worm}},
  \bibinfo {author} {\bibfnamefont {P.}~\bibnamefont {Chalupa-Gantner}},
  \bibinfo {author} {\bibfnamefont {A.}~\bibnamefont {Toschi}},\ and\ \bibinfo
  {author} {\bibfnamefont {K.}~\bibnamefont {Held}},\ }\bibfield  {title}
  {\bibinfo {title} {Explaining the pseudogap through damping and antidamping
  on the {{Fermi}} surface by imaginary spin scattering},\ }\href
  {https://doi.org/https://doi.org/10.1038/s42005-022-01117-5} {\bibfield
  {journal} {\bibinfo  {journal} {Communications Physics}\ }\textbf {\bibinfo
  {volume} {5}},\ \bibinfo {pages} {336} (\bibinfo {year} {2022})}\BibitemShut
  {NoStop}%
\bibitem [{\citenamefont {Martinez}\ and\ \citenamefont
  {Horsch}(1991)}]{martinez1991a}%
  \BibitemOpen
  \bibfield  {author} {\bibinfo {author} {\bibfnamefont {G.}~\bibnamefont
  {Martinez}}\ and\ \bibinfo {author} {\bibfnamefont {P.}~\bibnamefont
  {Horsch}},\ }\bibfield  {title} {\bibinfo {title} {Spin polarons in the
  {{$t-J$}} model},\ }\href {https://doi.org/10.1103/PhysRevB.44.317}
  {\bibfield  {journal} {\bibinfo  {journal} {Phys. Rev. B}\ }\textbf {\bibinfo
  {volume} {44}},\ \bibinfo {pages} {317} (\bibinfo {year} {1991})}\BibitemShut
  {NoStop}%
\bibitem [{\citenamefont {Macridin}\ \emph {et~al.}(2007)\citenamefont
  {Macridin}, \citenamefont {Jarrell}, \citenamefont {Maier},\ and\
  \citenamefont {Scalapino}}]{macridin2007a}%
  \BibitemOpen
  \bibfield  {author} {\bibinfo {author} {\bibfnamefont {A.}~\bibnamefont
  {Macridin}}, \bibinfo {author} {\bibfnamefont {M.}~\bibnamefont {Jarrell}},
  \bibinfo {author} {\bibfnamefont {T.}~\bibnamefont {Maier}},\ and\ \bibinfo
  {author} {\bibfnamefont {D.~J.}\ \bibnamefont {Scalapino}},\ }\bibfield
  {title} {\bibinfo {title} {High-energy kink in the single-particle spectra of
  the two-dimensional {{Hubbard}} model},\ }\href
  {https://doi.org/10.1103/PhysRevLett.99.237001} {\bibfield  {journal}
  {\bibinfo  {journal} {Phys. Rev. Lett.}\ }\textbf {\bibinfo {volume} {99}},\
  \bibinfo {pages} {237001} (\bibinfo {year} {2007})}\BibitemShut {NoStop}%
\bibitem [{\citenamefont {Manousakis}(2007)}]{manousakis2007a}%
  \BibitemOpen
  \bibfield  {author} {\bibinfo {author} {\bibfnamefont {E.}~\bibnamefont
  {Manousakis}},\ }\bibfield  {title} {\bibinfo {title} {String excitations of
  a hole in a quantum antiferromagnet and photoelectron spectroscopy},\ }\href
  {https://doi.org/10.1103/PhysRevB.75.035106} {\bibfield  {journal} {\bibinfo
  {journal} {Phys. Rev. B}\ }\textbf {\bibinfo {volume} {75}},\ \bibinfo
  {pages} {035106} (\bibinfo {year} {2007})}\BibitemShut {NoStop}%
\bibitem [{\citenamefont {Wang}\ \emph {et~al.}(2015)\citenamefont {Wang},
  \citenamefont {Wohlfeld}, \citenamefont {Moritz}, \citenamefont {Jia},
  \citenamefont {{van Veenendaal}}, \citenamefont {Wu}, \citenamefont {Chen},\
  and\ \citenamefont {Devereaux}}]{wang2015a}%
  \BibitemOpen
  \bibfield  {author} {\bibinfo {author} {\bibfnamefont {Y.}~\bibnamefont
  {Wang}}, \bibinfo {author} {\bibfnamefont {K.}~\bibnamefont {Wohlfeld}},
  \bibinfo {author} {\bibfnamefont {B.}~\bibnamefont {Moritz}}, \bibinfo
  {author} {\bibfnamefont {C.~J.}\ \bibnamefont {Jia}}, \bibinfo {author}
  {\bibfnamefont {M.}~\bibnamefont {{van Veenendaal}}}, \bibinfo {author}
  {\bibfnamefont {K.}~\bibnamefont {Wu}}, \bibinfo {author} {\bibfnamefont
  {C.-C.}\ \bibnamefont {Chen}},\ and\ \bibinfo {author} {\bibfnamefont
  {T.~P.}\ \bibnamefont {Devereaux}},\ }\bibfield  {title} {\bibinfo {title}
  {Origin of strong dispersion in {{Hubbard}} insulators},\ }\href
  {https://doi.org/10.1103/PhysRevB.92.075119} {\bibfield  {journal} {\bibinfo
  {journal} {Phys. Rev. B}\ }\textbf {\bibinfo {volume} {92}},\ \bibinfo
  {pages} {075119} (\bibinfo {year} {2015})}\BibitemShut {NoStop}%
\bibitem [{\citenamefont {Bacq-Labreuil}\ \emph {et~al.}(2025)\citenamefont
  {Bacq-Labreuil}, \citenamefont {Fawaz}, \citenamefont {Okazaki},
  \citenamefont {Obata}, \citenamefont {Cercellier}, \citenamefont
  {Le~F\`evre}, \citenamefont {Bertran}, \citenamefont {Santos-Cottin},
  \citenamefont {Yamamoto}, \citenamefont {Yamada}, \citenamefont {Azuma},
  \citenamefont {Horiba}, \citenamefont {Kumigashira}, \citenamefont
  {d'Astuto}, \citenamefont {Biermann},\ and\ \citenamefont {Lenz}}]{bacq2023}%
  \BibitemOpen
  \bibfield  {author} {\bibinfo {author} {\bibfnamefont {B.}~\bibnamefont
  {Bacq-Labreuil}}, \bibinfo {author} {\bibfnamefont {C.}~\bibnamefont
  {Fawaz}}, \bibinfo {author} {\bibfnamefont {Y.}~\bibnamefont {Okazaki}},
  \bibinfo {author} {\bibfnamefont {Y.}~\bibnamefont {Obata}}, \bibinfo
  {author} {\bibfnamefont {H.}~\bibnamefont {Cercellier}}, \bibinfo {author}
  {\bibfnamefont {P.}~\bibnamefont {Le~F\`evre}}, \bibinfo {author}
  {\bibfnamefont {F.~m.~c.}\ \bibnamefont {Bertran}}, \bibinfo {author}
  {\bibfnamefont {D.}~\bibnamefont {Santos-Cottin}}, \bibinfo {author}
  {\bibfnamefont {H.}~\bibnamefont {Yamamoto}}, \bibinfo {author}
  {\bibfnamefont {I.}~\bibnamefont {Yamada}}, \bibinfo {author} {\bibfnamefont
  {M.}~\bibnamefont {Azuma}}, \bibinfo {author} {\bibfnamefont
  {K.}~\bibnamefont {Horiba}}, \bibinfo {author} {\bibfnamefont
  {H.}~\bibnamefont {Kumigashira}}, \bibinfo {author} {\bibfnamefont
  {M.}~\bibnamefont {d'Astuto}}, \bibinfo {author} {\bibfnamefont
  {S.}~\bibnamefont {Biermann}},\ and\ \bibinfo {author} {\bibfnamefont
  {B.}~\bibnamefont {Lenz}},\ }\bibfield  {title} {\bibinfo {title} {Universal
  waterfall feature in cuprate superconductors: Evidence of a momentum-driven
  crossover},\ }\href {https://doi.org/10.1103/PhysRevLett.134.016502}
  {\bibfield  {journal} {\bibinfo  {journal} {Phys. Rev. Lett.}\ }\textbf
  {\bibinfo {volume} {134}},\ \bibinfo {pages} {016502} (\bibinfo {year}
  {2025})}\BibitemShut {NoStop}%
\bibitem [{\citenamefont {Zheng}\ \emph {et~al.}(2017)\citenamefont {Zheng},
  \citenamefont {Chung}, \citenamefont {Corboz}, \citenamefont {Ehlers},
  \citenamefont {Qin}, \citenamefont {Noack}, \citenamefont {Shi},
  \citenamefont {White}, \citenamefont {Zhang},\ and\ \citenamefont
  {Chan}}]{zheng2017}%
  \BibitemOpen
  \bibfield  {author} {\bibinfo {author} {\bibfnamefont {B.-X.}\ \bibnamefont
  {Zheng}}, \bibinfo {author} {\bibfnamefont {C.-M.}\ \bibnamefont {Chung}},
  \bibinfo {author} {\bibfnamefont {P.}~\bibnamefont {Corboz}}, \bibinfo
  {author} {\bibfnamefont {G.}~\bibnamefont {Ehlers}}, \bibinfo {author}
  {\bibfnamefont {M.-P.}\ \bibnamefont {Qin}}, \bibinfo {author} {\bibfnamefont
  {R.~M.}\ \bibnamefont {Noack}}, \bibinfo {author} {\bibfnamefont
  {H.}~\bibnamefont {Shi}}, \bibinfo {author} {\bibfnamefont {S.~R.}\
  \bibnamefont {White}}, \bibinfo {author} {\bibfnamefont {S.}~\bibnamefont
  {Zhang}},\ and\ \bibinfo {author} {\bibfnamefont {G.~K.-L.}\ \bibnamefont
  {Chan}},\ }\bibfield  {title} {\bibinfo {title} {Stripe order in the
  underdoped region of the two-dimensional {{Hubbard}} model},\ }\href
  {https://doi.org/10.1126/science.aam7127} {\bibfield  {journal} {\bibinfo
  {journal} {Science}\ }\textbf {\bibinfo {volume} {358}},\ \bibinfo {pages}
  {1155} (\bibinfo {year} {2017})}\BibitemShut {NoStop}%
\bibitem [{\citenamefont {Hirayama}\ \emph {et~al.}(2013)\citenamefont
  {Hirayama}, \citenamefont {Miyake},\ and\ \citenamefont
  {Imada}}]{hirayama2013}%
  \BibitemOpen
  \bibfield  {author} {\bibinfo {author} {\bibfnamefont {M.}~\bibnamefont
  {Hirayama}}, \bibinfo {author} {\bibfnamefont {T.}~\bibnamefont {Miyake}},\
  and\ \bibinfo {author} {\bibfnamefont {M.}~\bibnamefont {Imada}},\ }\bibfield
   {title} {\bibinfo {title} {Derivation of static low-energy effective models
  by an {\emph{ab initio}} downfolding method without double counting of
  {{Coulomb}} correlations: {{Application}} to {{SrVO$_3$}}, {{FeSe}}, and
  {{FeTe}}},\ }\href {https://doi.org/10.1103/PhysRevB.87.195144} {\bibfield
  {journal} {\bibinfo  {journal} {Phys. Rev. B}\ }\textbf {\bibinfo {volume}
  {87}},\ \bibinfo {pages} {195144} (\bibinfo {year} {2013})}\BibitemShut
  {NoStop}%
\bibitem [{\citenamefont {Hirayama}\ \emph {et~al.}(2018)\citenamefont
  {Hirayama}, \citenamefont {Yamaji}, \citenamefont {Misawa},\ and\
  \citenamefont {Imada}}]{hirayama2018b}%
  \BibitemOpen
  \bibfield  {author} {\bibinfo {author} {\bibfnamefont {M.}~\bibnamefont
  {Hirayama}}, \bibinfo {author} {\bibfnamefont {Y.}~\bibnamefont {Yamaji}},
  \bibinfo {author} {\bibfnamefont {T.}~\bibnamefont {Misawa}},\ and\ \bibinfo
  {author} {\bibfnamefont {M.}~\bibnamefont {Imada}},\ }\bibfield  {title}
  {\bibinfo {title} {Ab initio effective {{Hamiltonians}} for cuprate
  superconductors},\ }\href {https://doi.org/10.1103/PhysRevB.98.134501}
  {\bibfield  {journal} {\bibinfo  {journal} {Phys. Rev. B}\ }\textbf {\bibinfo
  {volume} {98}},\ \bibinfo {pages} {134501} (\bibinfo {year}
  {2018})}\BibitemShut {NoStop}%
\bibitem [{\citenamefont {Mor{\'e}e}\ \emph {et~al.}(2022)\citenamefont
  {Mor{\'e}e}, \citenamefont {Hirayama}, \citenamefont {Schmid}, \citenamefont
  {Yamaji},\ and\ \citenamefont {Imada}}]{moree2022a}%
  \BibitemOpen
  \bibfield  {author} {\bibinfo {author} {\bibfnamefont {J.-B.}\ \bibnamefont
  {Mor{\'e}e}}, \bibinfo {author} {\bibfnamefont {M.}~\bibnamefont {Hirayama}},
  \bibinfo {author} {\bibfnamefont {M.~T.}\ \bibnamefont {Schmid}}, \bibinfo
  {author} {\bibfnamefont {Y.}~\bibnamefont {Yamaji}},\ and\ \bibinfo {author}
  {\bibfnamefont {M.}~\bibnamefont {Imada}},\ }\bibfield  {title} {\bibinfo
  {title} {Ab initio low-energy effective hamiltonians for the high-temperature
  superconducting cuprates
  {{${\mathrm{Bi}}_{2}{\mathrm{Sr}}_{2}{\mathrm{CuO}}_{6},$
  ${\mathrm{Bi}}_{2}{\mathrm{Sr}}_{2}{\mathrm{CaCu}}_{2}{\mathrm{O}}_{8},$
  ${\mathrm{HgBa}}_{2}{\mathrm{CuO}}_{4},$ and ${\mathrm{CaCuO}}_{2}$}}},\
  }\href {https://doi.org/10.1103/PhysRevB.106.235150} {\bibfield  {journal}
  {\bibinfo  {journal} {Phys. Rev. B}\ }\textbf {\bibinfo {volume} {106}},\
  \bibinfo {pages} {235150} (\bibinfo {year} {2022})}\BibitemShut {NoStop}%
\bibitem [{\citenamefont {Schmid}\ \emph {et~al.}(2023)\citenamefont {Schmid},
  \citenamefont {Mor\'ee}, \citenamefont {Kaneko}, \citenamefont {Yamaji},\
  and\ \citenamefont {Imada}}]{schmid2023a}%
  \BibitemOpen
  \bibfield  {author} {\bibinfo {author} {\bibfnamefont {M.~T.}\ \bibnamefont
  {Schmid}}, \bibinfo {author} {\bibfnamefont {J.-B.}\ \bibnamefont {Mor\'ee}},
  \bibinfo {author} {\bibfnamefont {R.}~\bibnamefont {Kaneko}}, \bibinfo
  {author} {\bibfnamefont {Y.}~\bibnamefont {Yamaji}},\ and\ \bibinfo {author}
  {\bibfnamefont {M.}~\bibnamefont {Imada}},\ }\bibfield  {title} {\bibinfo
  {title} {Superconductivity studied by solving ab initio low-energy effective
  hamiltonians for carrier doped {{ ${\mathrm{CaCuO}}_{2}$,
  ${\mathrm{Bi}}_{2}{\mathrm{Sr}}_{2}{\mathrm{CuO}}_{6}$,
  ${\mathrm{Bi}}_{2}{\mathrm{Sr}}_{2}{\mathrm{CaCu}}_{2}{\mathrm{O}}_{8}$, and
  ${\mathrm{HgBa}}_{2}{\mathrm{CuO}}_{4}$}}},\ }\href
  {https://doi.org/10.1103/PhysRevX.13.041036} {\bibfield  {journal} {\bibinfo
  {journal} {Phys. Rev. X}\ }\textbf {\bibinfo {volume} {13}},\ \bibinfo
  {pages} {041036} (\bibinfo {year} {2023})}\BibitemShut {NoStop}%
\bibitem [{\citenamefont {Held}\ \emph {et~al.}(2022)\citenamefont {Held},
  \citenamefont {Si}, \citenamefont {Worm}, \citenamefont {Janson},
  \citenamefont {Arita}, \citenamefont {Zhong}, \citenamefont {Tomczak},\ and\
  \citenamefont {Kitatani}}]{held2022}%
  \BibitemOpen
  \bibfield  {author} {\bibinfo {author} {\bibfnamefont {K.}~\bibnamefont
  {Held}}, \bibinfo {author} {\bibfnamefont {L.}~\bibnamefont {Si}}, \bibinfo
  {author} {\bibfnamefont {P.}~\bibnamefont {Worm}}, \bibinfo {author}
  {\bibfnamefont {O.}~\bibnamefont {Janson}}, \bibinfo {author} {\bibfnamefont
  {R.}~\bibnamefont {Arita}}, \bibinfo {author} {\bibfnamefont
  {Z.}~\bibnamefont {Zhong}}, \bibinfo {author} {\bibfnamefont {J.~M.}\
  \bibnamefont {Tomczak}},\ and\ \bibinfo {author} {\bibfnamefont
  {M.}~\bibnamefont {Kitatani}},\ }\bibfield  {title} {\bibinfo {title} {Phase
  diagram of nickelate superconductors calculated by dynamical vertex
  approximation},\ }\bibfield  {journal} {\bibinfo  {journal} {Frontiers in
  Physics}\ }\textbf {\bibinfo {volume} {9}},\ \href
  {https://doi.org/10.3389/fphy.2021.810394} {10.3389/fphy.2021.810394}
  (\bibinfo {year} {2022})\BibitemShut {NoStop}%
\bibitem [{\citenamefont {Kitatani}\ \emph {et~al.}(2023)\citenamefont
  {Kitatani}, \citenamefont {Si}, \citenamefont {Worm}, \citenamefont
  {Tomczak}, \citenamefont {Arita},\ and\ \citenamefont {Held}}]{kitatani2023}%
  \BibitemOpen
  \bibfield  {author} {\bibinfo {author} {\bibfnamefont {M.}~\bibnamefont
  {Kitatani}}, \bibinfo {author} {\bibfnamefont {L.}~\bibnamefont {Si}},
  \bibinfo {author} {\bibfnamefont {P.}~\bibnamefont {Worm}}, \bibinfo {author}
  {\bibfnamefont {J.~M.}\ \bibnamefont {Tomczak}}, \bibinfo {author}
  {\bibfnamefont {R.}~\bibnamefont {Arita}},\ and\ \bibinfo {author}
  {\bibfnamefont {K.}~\bibnamefont {Held}},\ }\bibfield  {title} {\bibinfo
  {title} {Optimizing superconductivity: {{From}} cuprates via nickelates to
  palladates},\ }\href {https://doi.org/10.1103/PhysRevLett.130.166002}
  {\bibfield  {journal} {\bibinfo  {journal} {Phys. Rev. Lett.}\ }\textbf
  {\bibinfo {volume} {130}},\ \bibinfo {pages} {166002} (\bibinfo {year}
  {2023})}\BibitemShut {NoStop}%
\bibitem [{\citenamefont {Di~Cataldo}\ \emph {et~al.}(2024)\citenamefont
  {Di~Cataldo}, \citenamefont {Worm}, \citenamefont {Tomczak}, \citenamefont
  {Si},\ and\ \citenamefont {Held}}]{cataldo2024}%
  \BibitemOpen
  \bibfield  {author} {\bibinfo {author} {\bibfnamefont {S.}~\bibnamefont
  {Di~Cataldo}}, \bibinfo {author} {\bibfnamefont {P.}~\bibnamefont {Worm}},
  \bibinfo {author} {\bibfnamefont {J.~M.}\ \bibnamefont {Tomczak}}, \bibinfo
  {author} {\bibfnamefont {L.}~\bibnamefont {Si}},\ and\ \bibinfo {author}
  {\bibfnamefont {K.}~\bibnamefont {Held}},\ }\bibfield  {title} {\bibinfo
  {title} {Unconventional superconductivity without doping in infinite-layer
  nickelates under pressure},\ }\href
  {https://doi.org/https://doi.org/10.1038/s41467-024-48169-5} {\bibfield
  {journal} {\bibinfo  {journal} {Nature Communications}\ }\textbf {\bibinfo
  {volume} {15}},\ \bibinfo {pages} {3952} (\bibinfo {year}
  {2024})}\BibitemShut {NoStop}%
\bibitem [{\citenamefont {Metzner}\ and\ \citenamefont
  {Vollhardt}(1989)}]{metzner1989}%
  \BibitemOpen
  \bibfield  {author} {\bibinfo {author} {\bibfnamefont {W.}~\bibnamefont
  {Metzner}}\ and\ \bibinfo {author} {\bibfnamefont {D.}~\bibnamefont
  {Vollhardt}},\ }\bibfield  {title} {\bibinfo {title} {Correlated lattice
  fermions in $d=\ensuremath{\infty}$ dimensions},\ }\href
  {https://doi.org/10.1103/PhysRevLett.62.324} {\bibfield  {journal} {\bibinfo
  {journal} {Phys. Rev. Lett.}\ }\textbf {\bibinfo {volume} {62}},\ \bibinfo
  {pages} {324} (\bibinfo {year} {1989})}\BibitemShut {NoStop}%
\bibitem [{\citenamefont {Georges}\ and\ \citenamefont
  {Kotliar}(1992)}]{georges1992}%
  \BibitemOpen
  \bibfield  {author} {\bibinfo {author} {\bibfnamefont {A.}~\bibnamefont
  {Georges}}\ and\ \bibinfo {author} {\bibfnamefont {G.}~\bibnamefont
  {Kotliar}},\ }\bibfield  {title} {\bibinfo {title} {Hubbard model in infinite
  dimensions},\ }\href {https://doi.org/10.1103/PhysRevB.45.6479} {\bibfield
  {journal} {\bibinfo  {journal} {Phys. Rev. B}\ }\textbf {\bibinfo {volume}
  {45}},\ \bibinfo {pages} {6479} (\bibinfo {year} {1992})}\BibitemShut
  {NoStop}%
\bibitem [{\citenamefont {Jarrell}(1992)}]{jarrell1992}%
  \BibitemOpen
  \bibfield  {author} {\bibinfo {author} {\bibfnamefont {M.}~\bibnamefont
  {Jarrell}},\ }\bibfield  {title} {\bibinfo {title} {Hubbard model in infinite
  dimensions: A quantum monte carlo study},\ }\href
  {https://doi.org/10.1103/PhysRevLett.69.168} {\bibfield  {journal} {\bibinfo
  {journal} {Phys. Rev. Lett.}\ }\textbf {\bibinfo {volume} {69}},\ \bibinfo
  {pages} {168} (\bibinfo {year} {1992})}\BibitemShut {NoStop}%
\bibitem [{\citenamefont {Georges}\ \emph {et~al.}(1996)\citenamefont
  {Georges}, \citenamefont {Kotliar}, \citenamefont {Krauth},\ and\
  \citenamefont {Rozenberg}}]{georges1996}%
  \BibitemOpen
  \bibfield  {author} {\bibinfo {author} {\bibfnamefont {A.}~\bibnamefont
  {Georges}}, \bibinfo {author} {\bibfnamefont {G.}~\bibnamefont {Kotliar}},
  \bibinfo {author} {\bibfnamefont {W.}~\bibnamefont {Krauth}},\ and\ \bibinfo
  {author} {\bibfnamefont {M.~J.}\ \bibnamefont {Rozenberg}},\ }\bibfield
  {title} {\bibinfo {title} {Dynamical mean-field theory of strongly correlated
  fermion systems and the limit of infinite dimensions},\ }\href
  {https://doi.org/10.1103/RevModPhys.68.13} {\bibfield  {journal} {\bibinfo
  {journal} {Rev. Mod. Phys.}\ }\textbf {\bibinfo {volume} {68}},\ \bibinfo
  {pages} {13} (\bibinfo {year} {1996})}\BibitemShut {NoStop}%
\bibitem [{\citenamefont {Biermann}\ \emph {et~al.}(2003)\citenamefont
  {Biermann}, \citenamefont {Aryasetiawan},\ and\ \citenamefont
  {Georges}}]{Biermann_Aryasetiawan_Georges_2003}%
  \BibitemOpen
  \bibfield  {author} {\bibinfo {author} {\bibfnamefont {S.}~\bibnamefont
  {Biermann}}, \bibinfo {author} {\bibfnamefont {F.}~\bibnamefont
  {Aryasetiawan}},\ and\ \bibinfo {author} {\bibfnamefont {A.}~\bibnamefont
  {Georges}},\ }\bibfield  {title} {\bibinfo {title} {First-principles approach
  to the electronic structure of strongly correlated systems: Combining the
  $gw$ approximation and dynamical mean-field theory},\ }\href
  {https://doi.org/10.1103/PhysRevLett.90.086402} {\bibfield  {journal}
  {\bibinfo  {journal} {Physical Review Letters}\ }\textbf {\bibinfo {volume}
  {90}},\ \bibinfo {pages} {086402} (\bibinfo {year} {2003})}\BibitemShut
  {NoStop}%
\bibitem [{\citenamefont {Lan}\ \emph {et~al.}(2017)\citenamefont {Lan},
  \citenamefont {Shee}, \citenamefont {Li}, \citenamefont {Gull},\ and\
  \citenamefont {Zgid}}]{Lan_Shee_Li_Gull_Zgid_2017}%
  \BibitemOpen
  \bibfield  {author} {\bibinfo {author} {\bibfnamefont {T.~N.}\ \bibnamefont
  {Lan}}, \bibinfo {author} {\bibfnamefont {A.}~\bibnamefont {Shee}}, \bibinfo
  {author} {\bibfnamefont {J.}~\bibnamefont {Li}}, \bibinfo {author}
  {\bibfnamefont {E.}~\bibnamefont {Gull}},\ and\ \bibinfo {author}
  {\bibfnamefont {D.}~\bibnamefont {Zgid}},\ }\bibfield  {title} {\bibinfo
  {title} {Testing self-energy embedding theory in combination with gw},\
  }\href {https://doi.org/10.1103/PhysRevB.96.155106} {\bibfield  {journal}
  {\bibinfo  {journal} {Physical Review B}\ }\textbf {\bibinfo {volume} {96}},\
  \bibinfo {pages} {155106} (\bibinfo {year} {2017})}\BibitemShut {NoStop}%
\bibitem [{\citenamefont {Stepanov}\ \emph {et~al.}(2019)\citenamefont
  {Stepanov}, \citenamefont {Harkov},\ and\ \citenamefont
  {Lichtenstein}}]{stepanov2019a}%
  \BibitemOpen
  \bibfield  {author} {\bibinfo {author} {\bibfnamefont {E.~A.}\ \bibnamefont
  {Stepanov}}, \bibinfo {author} {\bibfnamefont {V.}~\bibnamefont {Harkov}},\
  and\ \bibinfo {author} {\bibfnamefont {A.~I.}\ \bibnamefont {Lichtenstein}},\
  }\bibfield  {title} {\bibinfo {title} {Consistent partial bosonization of the
  extended {{Hubbard}} model},\ }\href
  {https://doi.org/10.1103/PhysRevB.100.205115} {\bibfield  {journal} {\bibinfo
   {journal} {Phys. Rev. B}\ }\textbf {\bibinfo {volume} {100}},\ \bibinfo
  {pages} {205115} (\bibinfo {year} {2019})}\BibitemShut {NoStop}%
\bibitem [{\citenamefont {Vandelli}\ \emph {et~al.}(2022)\citenamefont
  {Vandelli}, \citenamefont {Kaufmann}, \citenamefont {{El-Nabulsi}},
  \citenamefont {Harkov}, \citenamefont {Lichtenstein},\ and\ \citenamefont
  {Stepanov}}]{vandelli2022b}%
  \BibitemOpen
  \bibfield  {author} {\bibinfo {author} {\bibfnamefont {M.}~\bibnamefont
  {Vandelli}}, \bibinfo {author} {\bibfnamefont {J.}~\bibnamefont {Kaufmann}},
  \bibinfo {author} {\bibfnamefont {M.}~\bibnamefont {{El-Nabulsi}}}, \bibinfo
  {author} {\bibfnamefont {V.}~\bibnamefont {Harkov}}, \bibinfo {author}
  {\bibfnamefont {A.}~\bibnamefont {Lichtenstein}},\ and\ \bibinfo {author}
  {\bibfnamefont {E.}~\bibnamefont {Stepanov}},\ }\bibfield  {title} {\bibinfo
  {title} {Multi-band {{D-TRILEX}} approach to materials with strong electronic
  correlations},\ }\href {https://doi.org/10.21468/SciPostPhys.13.2.036}
  {\bibfield  {journal} {\bibinfo  {journal} {SciPost Physics}\ }\textbf
  {\bibinfo {volume} {13}},\ \bibinfo {pages} {036} (\bibinfo {year}
  {2022})}\BibitemShut {NoStop}%
\bibitem [{\citenamefont {Cui}\ \emph {et~al.}(2025)\citenamefont {Cui},
  \citenamefont {Yang}, \citenamefont {T{\"o}lle}, \citenamefont {Ye},
  \citenamefont {Yuan}, \citenamefont {Zhai}, \citenamefont {Park},
  \citenamefont {Kim}, \citenamefont {Zhang}, \citenamefont {Lin} \emph
  {et~al.}}]{cui2023a}%
  \BibitemOpen
  \bibfield  {author} {\bibinfo {author} {\bibfnamefont {Z.-H.}\ \bibnamefont
  {Cui}}, \bibinfo {author} {\bibfnamefont {J.}~\bibnamefont {Yang}}, \bibinfo
  {author} {\bibfnamefont {J.}~\bibnamefont {T{\"o}lle}}, \bibinfo {author}
  {\bibfnamefont {H.-Z.}\ \bibnamefont {Ye}}, \bibinfo {author} {\bibfnamefont
  {S.}~\bibnamefont {Yuan}}, \bibinfo {author} {\bibfnamefont {H.}~\bibnamefont
  {Zhai}}, \bibinfo {author} {\bibfnamefont {G.}~\bibnamefont {Park}}, \bibinfo
  {author} {\bibfnamefont {R.}~\bibnamefont {Kim}}, \bibinfo {author}
  {\bibfnamefont {X.}~\bibnamefont {Zhang}}, \bibinfo {author} {\bibfnamefont
  {L.}~\bibnamefont {Lin}}, \emph {et~al.},\ }\bibfield  {title} {\bibinfo
  {title} {Ab initio quantum many-body description of superconducting trends in
  the cuprates},\ }\href
  {https://doi.org/https://doi.org/10.1038/s41467-025-56883-x} {\bibfield
  {journal} {\bibinfo  {journal} {Nature Communications}\ }\textbf {\bibinfo
  {volume} {16}},\ \bibinfo {pages} {1845} (\bibinfo {year}
  {2025})}\BibitemShut {NoStop}%
\bibitem [{\citenamefont {Ayral}\ \emph {et~al.}(2017)\citenamefont {Ayral},
  \citenamefont {Lee},\ and\ \citenamefont {Kotliar}}]{PhysRevB.96.235139}%
  \BibitemOpen
  \bibfield  {author} {\bibinfo {author} {\bibfnamefont {T.}~\bibnamefont
  {Ayral}}, \bibinfo {author} {\bibfnamefont {T.-H.}\ \bibnamefont {Lee}},\
  and\ \bibinfo {author} {\bibfnamefont {G.}~\bibnamefont {Kotliar}},\
  }\bibfield  {title} {\bibinfo {title} {Dynamical mean-field theory,
  density-matrix embedding theory, and rotationally invariant slave bosons: A
  unified perspective},\ }\href {https://doi.org/10.1103/PhysRevB.96.235139}
  {\bibfield  {journal} {\bibinfo  {journal} {Phys. Rev. B}\ }\textbf {\bibinfo
  {volume} {96}},\ \bibinfo {pages} {235139} (\bibinfo {year}
  {2017})}\BibitemShut {NoStop}%
\bibitem [{\citenamefont {Haule}\ \emph {et~al.}(2010)\citenamefont {Haule},
  \citenamefont {Yee},\ and\ \citenamefont {Kim}}]{haule2010}%
  \BibitemOpen
  \bibfield  {author} {\bibinfo {author} {\bibfnamefont {K.}~\bibnamefont
  {Haule}}, \bibinfo {author} {\bibfnamefont {C.-H.}\ \bibnamefont {Yee}},\
  and\ \bibinfo {author} {\bibfnamefont {K.}~\bibnamefont {Kim}},\ }\bibfield
  {title} {\bibinfo {title} {Dynamical mean-field theory within the
  full-potential methods: {{Electronic}} structure of {{CeIrIn$_5$, CeCoIn$_5$,
  and CeRhIn$_5$}}},\ }\href {https://doi.org/10.1103/PhysRevB.81.195107}
  {\bibfield  {journal} {\bibinfo  {journal} {Phys. Rev. B}\ }\textbf {\bibinfo
  {volume} {81}},\ \bibinfo {pages} {195107} (\bibinfo {year}
  {2010})}\BibitemShut {NoStop}%
\bibitem [{\citenamefont {Haule}(2018{\natexlab{a}})}]{HauleJPSJ}%
  \BibitemOpen
  \bibfield  {author} {\bibinfo {author} {\bibfnamefont {K.}~\bibnamefont
  {Haule}},\ }\bibfield  {title} {\bibinfo {title} {Structural predictions for
  correlated electron materials using the functional dynamical mean field
  theory approach},\ }\href {https://doi.org/10.7566/JPSJ.87.041005} {\bibfield
   {journal} {\bibinfo  {journal} {Journal of the Physical Society of Japan}\
  }\textbf {\bibinfo {volume} {87}},\ \bibinfo {pages} {041005} (\bibinfo
  {year} {2018}{\natexlab{a}})}\BibitemShut {NoStop}%
\bibitem [{\citenamefont {Dionne}\ \emph
  {et~al.}(2023{\natexlab{a}})\citenamefont {Dionne}, \citenamefont {Foley},
  \citenamefont {Rousseau},\ and\ \citenamefont {Sénéchal}}]{dionne2023a}%
  \BibitemOpen
  \bibfield  {author} {\bibinfo {author} {\bibfnamefont {T.~N.}\ \bibnamefont
  {Dionne}}, \bibinfo {author} {\bibfnamefont {A.}~\bibnamefont {Foley}},
  \bibinfo {author} {\bibfnamefont {M.}~\bibnamefont {Rousseau}},\ and\
  \bibinfo {author} {\bibfnamefont {D.}~\bibnamefont {Sénéchal}},\ }\bibfield
   {title} {\bibinfo {title} {{Pyqcm: An open-source Python library for quantum
  cluster methods}},\ }\href {https://doi.org/10.21468/SciPostPhysCodeb.23}
  {\bibfield  {journal} {\bibinfo  {journal} {SciPost Phys. Codebases}\ ,\
  \bibinfo {pages} {23}} (\bibinfo {year} {2023}{\natexlab{a}})}\BibitemShut
  {NoStop}%
\bibitem [{\citenamefont {Dionne}\ \emph
  {et~al.}(2023{\natexlab{b}})\citenamefont {Dionne}, \citenamefont {Foley},
  \citenamefont {Rousseau},\ and\ \citenamefont {Sénéchal}}]{dionne2023b}%
  \BibitemOpen
  \bibfield  {author} {\bibinfo {author} {\bibfnamefont {T.~N.}\ \bibnamefont
  {Dionne}}, \bibinfo {author} {\bibfnamefont {A.}~\bibnamefont {Foley}},
  \bibinfo {author} {\bibfnamefont {M.}~\bibnamefont {Rousseau}},\ and\
  \bibinfo {author} {\bibfnamefont {D.}~\bibnamefont {Sénéchal}},\ }\bibfield
   {title} {\bibinfo {title} {{Codebase release 2.2 for Pyqcm}},\ }\href
  {https://doi.org/10.21468/SciPostPhysCodeb.23-r2.2} {\bibfield  {journal}
  {\bibinfo  {journal} {SciPost Phys. Codebases}\ ,\ \bibinfo {pages} {23}}
  (\bibinfo {year} {2023}{\natexlab{b}})}\BibitemShut {NoStop}%
\bibitem [{\citenamefont {Mukuda}\ \emph {et~al.}(2012)\citenamefont {Mukuda},
  \citenamefont {Shimizu}, \citenamefont {Iyo},\ and\ \citenamefont
  {Kitaoka}}]{mukuda2012}%
  \BibitemOpen
  \bibfield  {author} {\bibinfo {author} {\bibfnamefont {H.}~\bibnamefont
  {Mukuda}}, \bibinfo {author} {\bibfnamefont {S.}~\bibnamefont {Shimizu}},
  \bibinfo {author} {\bibfnamefont {A.}~\bibnamefont {Iyo}},\ and\ \bibinfo
  {author} {\bibfnamefont {Y.}~\bibnamefont {Kitaoka}},\ }\bibfield  {title}
  {\bibinfo {title} {High-{{$T_C$}} superconductivity and antiferromagnetism in
  multilayered copper oxides - {{A}} new paradigm of superconducting
  mechanism},\ }\href {https://doi.org/10.1143/JPSJ.81.011008} {\bibfield
  {journal} {\bibinfo  {journal} {J. Phys. Soc. Jpn.}\ }\textbf {\bibinfo
  {volume} {81}},\ \bibinfo {pages} {011008} (\bibinfo {year}
  {2012})}\BibitemShut {NoStop}%
\bibitem [{\citenamefont {Dai}\ \emph {et~al.}(1995)\citenamefont {Dai},
  \citenamefont {Chakoumakos}, \citenamefont {Sun}, \citenamefont {Wong},
  \citenamefont {Xin},\ and\ \citenamefont {Lu}}]{dai1995}%
  \BibitemOpen
  \bibfield  {author} {\bibinfo {author} {\bibfnamefont {P.}~\bibnamefont
  {Dai}}, \bibinfo {author} {\bibfnamefont {B.}~\bibnamefont {Chakoumakos}},
  \bibinfo {author} {\bibfnamefont {G.}~\bibnamefont {Sun}}, \bibinfo {author}
  {\bibfnamefont {K.}~\bibnamefont {Wong}}, \bibinfo {author} {\bibfnamefont
  {Y.}~\bibnamefont {Xin}},\ and\ \bibinfo {author} {\bibfnamefont
  {D.}~\bibnamefont {Lu}},\ }\bibfield  {title} {\bibinfo {title} {Synthesis
  and neutron powder diffraction study of the superconductor
  {{HgBa$_2$Ca$_2$Cu$_3$O$_{8+\delta}$}} by {{Tl}} substitution},\ }\href
  {https://doi.org/https://doi.org/10.1016/0921-4534(94)02461-8} {\bibfield
  {journal} {\bibinfo  {journal} {Physica C: Superconductivity}\ }\textbf
  {\bibinfo {volume} {243}},\ \bibinfo {pages} {201} (\bibinfo {year}
  {1995})}\BibitemShut {NoStop}%
\bibitem [{\citenamefont {Loret}\ \emph {et~al.}(2019)\citenamefont {Loret},
  \citenamefont {Auvray}, \citenamefont {Gallais}, \citenamefont {Cazayous},
  \citenamefont {Forget}, \citenamefont {Colson}, \citenamefont {Julien},
  \citenamefont {Paul}, \citenamefont {Civelli},\ and\ \citenamefont
  {Sacuto}}]{loret2019a}%
  \BibitemOpen
  \bibfield  {author} {\bibinfo {author} {\bibfnamefont {B.}~\bibnamefont
  {Loret}}, \bibinfo {author} {\bibfnamefont {N.}~\bibnamefont {Auvray}},
  \bibinfo {author} {\bibfnamefont {Y.}~\bibnamefont {Gallais}}, \bibinfo
  {author} {\bibfnamefont {M.}~\bibnamefont {Cazayous}}, \bibinfo {author}
  {\bibfnamefont {A.}~\bibnamefont {Forget}}, \bibinfo {author} {\bibfnamefont
  {D.}~\bibnamefont {Colson}}, \bibinfo {author} {\bibfnamefont {M.-H.}\
  \bibnamefont {Julien}}, \bibinfo {author} {\bibfnamefont {I.}~\bibnamefont
  {Paul}}, \bibinfo {author} {\bibfnamefont {M.}~\bibnamefont {Civelli}},\ and\
  \bibinfo {author} {\bibfnamefont {A.}~\bibnamefont {Sacuto}},\ }\bibfield
  {title} {\bibinfo {title} {Intimate link between charge density wave,
  pseudogap and superconducting energy scales in cuprates},\ }\href
  {https://doi.org/10.1038/s41567-019-0509-5} {\bibfield  {journal} {\bibinfo
  {journal} {Nat. Phys.}\ }\textbf {\bibinfo {volume} {15}},\ \bibinfo {pages}
  {771} (\bibinfo {year} {2019})}\BibitemShut {NoStop}%
\bibitem [{\citenamefont {Kunisada}\ \emph {et~al.}(2020)\citenamefont
  {Kunisada}, \citenamefont {Isono}, \citenamefont {Kohama}, \citenamefont
  {Sakai}, \citenamefont {Bareille}, \citenamefont {Sakuragi}, \citenamefont
  {Noguchi}, \citenamefont {Kurokawa}, \citenamefont {Kuroda}, \citenamefont
  {Ishida}, \citenamefont {Adachi}, \citenamefont {Sekine}, \citenamefont
  {Kim}, \citenamefont {Cacho}, \citenamefont {Shin}, \citenamefont {Tohyama},
  \citenamefont {Tokiwa},\ and\ \citenamefont {Kondo}}]{kunisada2020}%
  \BibitemOpen
  \bibfield  {author} {\bibinfo {author} {\bibfnamefont {S.}~\bibnamefont
  {Kunisada}}, \bibinfo {author} {\bibfnamefont {S.}~\bibnamefont {Isono}},
  \bibinfo {author} {\bibfnamefont {Y.}~\bibnamefont {Kohama}}, \bibinfo
  {author} {\bibfnamefont {S.}~\bibnamefont {Sakai}}, \bibinfo {author}
  {\bibfnamefont {C.}~\bibnamefont {Bareille}}, \bibinfo {author}
  {\bibfnamefont {S.}~\bibnamefont {Sakuragi}}, \bibinfo {author}
  {\bibfnamefont {R.}~\bibnamefont {Noguchi}}, \bibinfo {author} {\bibfnamefont
  {K.}~\bibnamefont {Kurokawa}}, \bibinfo {author} {\bibfnamefont
  {K.}~\bibnamefont {Kuroda}}, \bibinfo {author} {\bibfnamefont
  {Y.}~\bibnamefont {Ishida}}, \bibinfo {author} {\bibfnamefont
  {S.}~\bibnamefont {Adachi}}, \bibinfo {author} {\bibfnamefont
  {R.}~\bibnamefont {Sekine}}, \bibinfo {author} {\bibfnamefont {T.~K.}\
  \bibnamefont {Kim}}, \bibinfo {author} {\bibfnamefont {C.}~\bibnamefont
  {Cacho}}, \bibinfo {author} {\bibfnamefont {S.}~\bibnamefont {Shin}},
  \bibinfo {author} {\bibfnamefont {T.}~\bibnamefont {Tohyama}}, \bibinfo
  {author} {\bibfnamefont {K.}~\bibnamefont {Tokiwa}},\ and\ \bibinfo {author}
  {\bibfnamefont {T.}~\bibnamefont {Kondo}},\ }\bibfield  {title} {\bibinfo
  {title} {Observation of small {{Fermi}} pockets protected by clean
  {{CuO$_2$}} sheets of a high-{{$T_C$}} superconductor},\ }\href
  {https://doi.org/10.1126/science.aay7311} {\bibfield  {journal} {\bibinfo
  {journal} {Science}\ }\textbf {\bibinfo {volume} {369}},\ \bibinfo {pages}
  {833} (\bibinfo {year} {2020})}\BibitemShut {NoStop}%
\bibitem [{\citenamefont {Kurokawa}\ \emph {et~al.}(2023)\citenamefont
  {Kurokawa}, \citenamefont {Isono}, \citenamefont {Kohama}, \citenamefont
  {Kunisada}, \citenamefont {Sakai}, \citenamefont {Sekine}, \citenamefont
  {Okubo}, \citenamefont {Watson}, \citenamefont {Kim}, \citenamefont {Cacho},
  \citenamefont {Shin}, \citenamefont {Tohyama}, \citenamefont {Tokiwa},\ and\
  \citenamefont {Kondo}}]{kurokawa2023}%
  \BibitemOpen
  \bibfield  {author} {\bibinfo {author} {\bibfnamefont {K.}~\bibnamefont
  {Kurokawa}}, \bibinfo {author} {\bibfnamefont {S.}~\bibnamefont {Isono}},
  \bibinfo {author} {\bibfnamefont {Y.}~\bibnamefont {Kohama}}, \bibinfo
  {author} {\bibfnamefont {S.}~\bibnamefont {Kunisada}}, \bibinfo {author}
  {\bibfnamefont {S.}~\bibnamefont {Sakai}}, \bibinfo {author} {\bibfnamefont
  {R.}~\bibnamefont {Sekine}}, \bibinfo {author} {\bibfnamefont
  {M.}~\bibnamefont {Okubo}}, \bibinfo {author} {\bibfnamefont {M.~D.}\
  \bibnamefont {Watson}}, \bibinfo {author} {\bibfnamefont {T.~K.}\
  \bibnamefont {Kim}}, \bibinfo {author} {\bibfnamefont {C.}~\bibnamefont
  {Cacho}}, \bibinfo {author} {\bibfnamefont {S.}~\bibnamefont {Shin}},
  \bibinfo {author} {\bibfnamefont {T.}~\bibnamefont {Tohyama}}, \bibinfo
  {author} {\bibfnamefont {K.}~\bibnamefont {Tokiwa}},\ and\ \bibinfo {author}
  {\bibfnamefont {T.}~\bibnamefont {Kondo}},\ }\bibfield  {title} {\bibinfo
  {title} {Unveiling phase diagram of the lightly doped high-{{Tc}} cuprate
  superconductors with disorder removed},\ }\href
  {https://doi.org/10.1038/s41467-023-39457-7} {\bibfield  {journal} {\bibinfo
  {journal} {Nat Commun}\ }\textbf {\bibinfo {volume} {14}},\ \bibinfo {pages}
  {4064} (\bibinfo {year} {2023})}\BibitemShut {NoStop}%
\bibitem [{\citenamefont {Oliviero}\ \emph {et~al.}(2022)\citenamefont
  {Oliviero}, \citenamefont {Benhabib}, \citenamefont {Gilmutdinov},
  \citenamefont {Vignolle}, \citenamefont {Drigo}, \citenamefont
  {Massoudzadegan}, \citenamefont {Leroux}, \citenamefont {Rikken},
  \citenamefont {Forget}, \citenamefont {Colson}, \citenamefont {Vignolles},\
  and\ \citenamefont {Proust}}]{oliviero2022a}%
  \BibitemOpen
  \bibfield  {author} {\bibinfo {author} {\bibfnamefont {V.}~\bibnamefont
  {Oliviero}}, \bibinfo {author} {\bibfnamefont {S.}~\bibnamefont {Benhabib}},
  \bibinfo {author} {\bibfnamefont {I.}~\bibnamefont {Gilmutdinov}}, \bibinfo
  {author} {\bibfnamefont {B.}~\bibnamefont {Vignolle}}, \bibinfo {author}
  {\bibfnamefont {L.}~\bibnamefont {Drigo}}, \bibinfo {author} {\bibfnamefont
  {M.}~\bibnamefont {Massoudzadegan}}, \bibinfo {author} {\bibfnamefont
  {M.}~\bibnamefont {Leroux}}, \bibinfo {author} {\bibfnamefont {G.~L. J.~A.}\
  \bibnamefont {Rikken}}, \bibinfo {author} {\bibfnamefont {A.}~\bibnamefont
  {Forget}}, \bibinfo {author} {\bibfnamefont {D.}~\bibnamefont {Colson}},
  \bibinfo {author} {\bibfnamefont {D.}~\bibnamefont {Vignolles}},\ and\
  \bibinfo {author} {\bibfnamefont {C.}~\bibnamefont {Proust}},\ }\bibfield
  {title} {\bibinfo {title} {Magnetotransport signatures of antiferromagnetism
  coexisting with charge order in the trilayer cuprate
  {{HgBa$_2$Ca$_2$Cu$_3$O$_{8+\delta}$}}},\ }\href
  {https://doi.org/10.1038/s41467-022-29134-6} {\bibfield  {journal} {\bibinfo
  {journal} {Nat Commun}\ }\textbf {\bibinfo {volume} {13}},\ \bibinfo {pages}
  {1568} (\bibinfo {year} {2022})}\BibitemShut {NoStop}%
\bibitem [{\citenamefont {Bellaiche}\ and\ \citenamefont
  {Vanderbilt}(2000)}]{bellaiche2000}%
  \BibitemOpen
  \bibfield  {author} {\bibinfo {author} {\bibfnamefont {L.}~\bibnamefont
  {Bellaiche}}\ and\ \bibinfo {author} {\bibfnamefont {D.}~\bibnamefont
  {Vanderbilt}},\ }\bibfield  {title} {\bibinfo {title} {Virtual crystal
  approximation revisited: Application to dielectric and piezoelectric
  properties of perovskites},\ }\href
  {https://doi.org/10.1103/PhysRevB.61.7877} {\bibfield  {journal} {\bibinfo
  {journal} {Phys. Rev. B}\ }\textbf {\bibinfo {volume} {61}},\ \bibinfo
  {pages} {7877} (\bibinfo {year} {2000})}\BibitemShut {NoStop}%
\bibitem [{\citenamefont {{van Veenendaal}}\ \emph {et~al.}(1994)\citenamefont
  {{van Veenendaal}}, \citenamefont {Sawatzky},\ and\ \citenamefont
  {Groen}}]{vanveenendaal1994}%
  \BibitemOpen
  \bibfield  {author} {\bibinfo {author} {\bibfnamefont {M.~A.}\ \bibnamefont
  {{van Veenendaal}}}, \bibinfo {author} {\bibfnamefont {G.~A.}\ \bibnamefont
  {Sawatzky}},\ and\ \bibinfo {author} {\bibfnamefont {W.~A.}\ \bibnamefont
  {Groen}},\ }\bibfield  {title} {\bibinfo {title} {Electronic structure of
  {{${\mathrm{Bi}}_{2}$${\mathrm{Sr}}_{2}$${\mathrm{Ca}}_{1\mathrm{\ensuremath{-}}\mathit{x}}$${\mathrm{Y}}_{\mathit{x}}$${\mathrm{Cu}}_{2}$${\mathrm{O}}_{8+\mathrm{\ensuremath{\delta}}}$}}:
  {{Cu}} 2p x-ray-photoelectron spectra and occupied and unoccupied low-energy
  states},\ }\href {https://doi.org/10.1103/PhysRevB.49.1407} {\bibfield
  {journal} {\bibinfo  {journal} {Phys. Rev. B}\ }\textbf {\bibinfo {volume}
  {49}},\ \bibinfo {pages} {1407} (\bibinfo {year} {1994})}\BibitemShut
  {NoStop}%
\bibitem [{\citenamefont {Damascelli}\ \emph {et~al.}(2003)\citenamefont
  {Damascelli}, \citenamefont {Hussain},\ and\ \citenamefont
  {Shen}}]{damascelli2003}%
  \BibitemOpen
  \bibfield  {author} {\bibinfo {author} {\bibfnamefont {A.}~\bibnamefont
  {Damascelli}}, \bibinfo {author} {\bibfnamefont {Z.}~\bibnamefont
  {Hussain}},\ and\ \bibinfo {author} {\bibfnamefont {Z.-X.}\ \bibnamefont
  {Shen}},\ }\bibfield  {title} {\bibinfo {title} {Angle-resolved photoemission
  studies of the cuprate superconductors},\ }\href
  {https://doi.org/10.1103/RevModPhys.75.473} {\bibfield  {journal} {\bibinfo
  {journal} {Rev. Mod. Phys.}\ }\textbf {\bibinfo {volume} {75}},\ \bibinfo
  {pages} {473} (\bibinfo {year} {2003})}\BibitemShut {NoStop}%
\bibitem [{\citenamefont {Hu}\ \emph {et~al.}(2021)\citenamefont {Hu},
  \citenamefont {Zhao}, \citenamefont {Gao}, \citenamefont {Yan}, \citenamefont
  {Rong}, \citenamefont {Huang}, \citenamefont {Liu}, \citenamefont {Cai},
  \citenamefont {Li}, \citenamefont {Chen}, \citenamefont {Zhao}, \citenamefont
  {Liu}, \citenamefont {Jin}, \citenamefont {Xu}, \citenamefont {Xiang},\ and\
  \citenamefont {Zhou}}]{hu2021}%
  \BibitemOpen
  \bibfield  {author} {\bibinfo {author} {\bibfnamefont {C.}~\bibnamefont
  {Hu}}, \bibinfo {author} {\bibfnamefont {J.}~\bibnamefont {Zhao}}, \bibinfo
  {author} {\bibfnamefont {Q.}~\bibnamefont {Gao}}, \bibinfo {author}
  {\bibfnamefont {H.}~\bibnamefont {Yan}}, \bibinfo {author} {\bibfnamefont
  {H.}~\bibnamefont {Rong}}, \bibinfo {author} {\bibfnamefont {J.}~\bibnamefont
  {Huang}}, \bibinfo {author} {\bibfnamefont {J.}~\bibnamefont {Liu}}, \bibinfo
  {author} {\bibfnamefont {Y.}~\bibnamefont {Cai}}, \bibinfo {author}
  {\bibfnamefont {C.}~\bibnamefont {Li}}, \bibinfo {author} {\bibfnamefont
  {H.}~\bibnamefont {Chen}}, \bibinfo {author} {\bibfnamefont {L.}~\bibnamefont
  {Zhao}}, \bibinfo {author} {\bibfnamefont {G.}~\bibnamefont {Liu}}, \bibinfo
  {author} {\bibfnamefont {C.}~\bibnamefont {Jin}}, \bibinfo {author}
  {\bibfnamefont {Z.}~\bibnamefont {Xu}}, \bibinfo {author} {\bibfnamefont
  {T.}~\bibnamefont {Xiang}},\ and\ \bibinfo {author} {\bibfnamefont {X.~J.}\
  \bibnamefont {Zhou}},\ }\bibfield  {title} {\bibinfo {title}
  {Momentum-resolved visualization of electronic evolution in doping a {{Mott}}
  insulator},\ }\href {https://doi.org/10.1038/s41467-021-21605-6} {\bibfield
  {journal} {\bibinfo  {journal} {Nat Commun}\ }\textbf {\bibinfo {volume}
  {12}},\ \bibinfo {pages} {1356} (\bibinfo {year} {2021})}\BibitemShut
  {NoStop}%
\bibitem [{\citenamefont {Hunter}\ \emph {et~al.}(1994)\citenamefont {Hunter},
  \citenamefont {Jorgensen}, \citenamefont {Wagner}, \citenamefont {Radaelli},
  \citenamefont {Hinks}, \citenamefont {Shaked}, \citenamefont {Hitterman},\
  and\ \citenamefont {Von~Dreele}}]{hunter1994}%
  \BibitemOpen
  \bibfield  {author} {\bibinfo {author} {\bibfnamefont {B.~A.}\ \bibnamefont
  {Hunter}}, \bibinfo {author} {\bibfnamefont {J.~D.}\ \bibnamefont
  {Jorgensen}}, \bibinfo {author} {\bibfnamefont {J.~L.}\ \bibnamefont
  {Wagner}}, \bibinfo {author} {\bibfnamefont {P.~G.}\ \bibnamefont
  {Radaelli}}, \bibinfo {author} {\bibfnamefont {D.~G.}\ \bibnamefont {Hinks}},
  \bibinfo {author} {\bibfnamefont {H.}~\bibnamefont {Shaked}}, \bibinfo
  {author} {\bibfnamefont {R.~L.}\ \bibnamefont {Hitterman}},\ and\ \bibinfo
  {author} {\bibfnamefont {R.~B.}\ \bibnamefont {Von~Dreele}},\ }\bibfield
  {title} {\bibinfo {title} {Pressure-induced structural changes in
  superconducting {{HgBa$_2$Ca$_{n-1}$Cu$_{n}$O$_{2n+2+\delta}$}} (n = 1, 2, 3)
  compounds},\ }\href {https://doi.org/10.1016/0921-4534(94)90659-9} {\bibfield
   {journal} {\bibinfo  {journal} {Physica C: Superconductivity}\ }\textbf
  {\bibinfo {volume} {221}},\ \bibinfo {pages} {1} (\bibinfo {year}
  {1994})}\BibitemShut {NoStop}%
\bibitem [{\citenamefont {Antipov}\ \emph {et~al.}(2002)\citenamefont
  {Antipov}, \citenamefont {Abakumov},\ and\ \citenamefont
  {Putilin}}]{antipov2002}%
  \BibitemOpen
  \bibfield  {author} {\bibinfo {author} {\bibfnamefont {E.~V.}\ \bibnamefont
  {Antipov}}, \bibinfo {author} {\bibfnamefont {A.~M.}\ \bibnamefont
  {Abakumov}},\ and\ \bibinfo {author} {\bibfnamefont {S.~N.}\ \bibnamefont
  {Putilin}},\ }\bibfield  {title} {\bibinfo {title} {Chemistry and structure
  of {{Hg-based}} superconducting {{Cu}} mixed oxides},\ }\href
  {https://doi.org/10.1088/0953-2048/15/7/201} {\bibfield  {journal} {\bibinfo
  {journal} {Supercond. Sci. Technol.}\ }\textbf {\bibinfo {volume} {15}},\
  \bibinfo {pages} {R31} (\bibinfo {year} {2002})}\BibitemShut {NoStop}%
\bibitem [{\citenamefont {Iyo}\ \emph {et~al.}(2006)\citenamefont {Iyo},
  \citenamefont {Tanaka}, \citenamefont {Kodama}, \citenamefont {Kito},
  \citenamefont {Tokiwa},\ and\ \citenamefont {Watanabe}}]{iyo2006}%
  \BibitemOpen
  \bibfield  {author} {\bibinfo {author} {\bibfnamefont {A.}~\bibnamefont
  {Iyo}}, \bibinfo {author} {\bibfnamefont {Y.}~\bibnamefont {Tanaka}},
  \bibinfo {author} {\bibfnamefont {Y.}~\bibnamefont {Kodama}}, \bibinfo
  {author} {\bibfnamefont {H.}~\bibnamefont {Kito}}, \bibinfo {author}
  {\bibfnamefont {K.}~\bibnamefont {Tokiwa}},\ and\ \bibinfo {author}
  {\bibfnamefont {T.}~\bibnamefont {Watanabe}},\ }\bibfield  {title} {\bibinfo
  {title} {Synthesis and physical properties of multilayered cuprates},\ }\href
  {https://doi.org/10.1016/j.physc.2006.03.067} {\bibfield  {journal} {\bibinfo
   {journal} {Physica C: Superconductivity and its Applications}\ }\bibinfo
  {series} {Proceedings of the 18th {{International Symposium}} on
  {{Superconductivity}} ({{ISS}} 2005)},\ \textbf {\bibinfo {volume}
  {445--448}},\ \bibinfo {pages} {17} (\bibinfo {year} {2006})}\BibitemShut
  {NoStop}%
\bibitem [{\citenamefont {Iyo}\ \emph {et~al.}(2007)\citenamefont {Iyo},
  \citenamefont {Tanaka}, \citenamefont {Kito}, \citenamefont {Kodama},
  \citenamefont {M.~Shirage}, \citenamefont {D.~Shivagan}, \citenamefont
  {Matsuhata}, \citenamefont {Tokiwa},\ and\ \citenamefont
  {Watanabe}}]{iyo2007}%
  \BibitemOpen
  \bibfield  {author} {\bibinfo {author} {\bibfnamefont {A.}~\bibnamefont
  {Iyo}}, \bibinfo {author} {\bibfnamefont {Y.}~\bibnamefont {Tanaka}},
  \bibinfo {author} {\bibfnamefont {H.}~\bibnamefont {Kito}}, \bibinfo {author}
  {\bibfnamefont {Y.}~\bibnamefont {Kodama}}, \bibinfo {author} {\bibfnamefont
  {P.}~\bibnamefont {M.~Shirage}}, \bibinfo {author} {\bibfnamefont
  {D.}~\bibnamefont {D.~Shivagan}}, \bibinfo {author} {\bibfnamefont
  {H.}~\bibnamefont {Matsuhata}}, \bibinfo {author} {\bibfnamefont
  {K.}~\bibnamefont {Tokiwa}},\ and\ \bibinfo {author} {\bibfnamefont
  {T.}~\bibnamefont {Watanabe}},\ }\bibfield  {title} {\bibinfo {title}
  {{{$T_C$}} vs {{$n$}} relationship for multilayered {{High-$T_C$}}
  superconductors},\ }\href {https://doi.org/10.1143/JPSJ.76.094711} {\bibfield
   {journal} {\bibinfo  {journal} {J. Phys. Soc. Jpn.}\ }\textbf {\bibinfo
  {volume} {76}},\ \bibinfo {pages} {094711} (\bibinfo {year}
  {2007})}\BibitemShut {NoStop}%
\bibitem [{\citenamefont {Pavarini}\ \emph {et~al.}(2001)\citenamefont
  {Pavarini}, \citenamefont {Dasgupta}, \citenamefont {{Saha-Dasgupta}},
  \citenamefont {Jepsen},\ and\ \citenamefont {Andersen}}]{pavarini2001}%
  \BibitemOpen
  \bibfield  {author} {\bibinfo {author} {\bibfnamefont {E.}~\bibnamefont
  {Pavarini}}, \bibinfo {author} {\bibfnamefont {I.}~\bibnamefont {Dasgupta}},
  \bibinfo {author} {\bibfnamefont {T.}~\bibnamefont {{Saha-Dasgupta}}},
  \bibinfo {author} {\bibfnamefont {O.}~\bibnamefont {Jepsen}},\ and\ \bibinfo
  {author} {\bibfnamefont {O.~K.}\ \bibnamefont {Andersen}},\ }\bibfield
  {title} {\bibinfo {title} {Band-structure trend in hole-doped cuprates and
  correlation with {{$T_{Cmax}$}}},\ }\href
  {https://doi.org/10.1103/PhysRevLett.87.047003} {\bibfield  {journal}
  {\bibinfo  {journal} {Phys. Rev. Lett.}\ }\textbf {\bibinfo {volume} {87}},\
  \bibinfo {pages} {047003} (\bibinfo {year} {2001})}\BibitemShut {NoStop}%
\bibitem [{\citenamefont {Hiroi}\ \emph {et~al.}(1994)\citenamefont {Hiroi},
  \citenamefont {Kobayashi},\ and\ \citenamefont {Takano}}]{hiroi1994}%
  \BibitemOpen
  \bibfield  {author} {\bibinfo {author} {\bibfnamefont {Z.}~\bibnamefont
  {Hiroi}}, \bibinfo {author} {\bibfnamefont {N.}~\bibnamefont {Kobayashi}},\
  and\ \bibinfo {author} {\bibfnamefont {M.}~\bibnamefont {Takano}},\
  }\bibfield  {title} {\bibinfo {title} {Probable hole-doped superconductivity
  without apical oxygens in {{(Ca, Na)$_2$CuO$_2$Cl$_2$}}},\ }\href
  {https://doi.org/10.1038/371139a0} {\bibfield  {journal} {\bibinfo  {journal}
  {Nature}\ }\textbf {\bibinfo {volume} {371}},\ \bibinfo {pages} {139}
  (\bibinfo {year} {1994})}\BibitemShut {NoStop}%
\bibitem [{\citenamefont {Kohsaka}\ \emph {et~al.}(2002)\citenamefont
  {Kohsaka}, \citenamefont {Azuma}, \citenamefont {Yamada}, \citenamefont
  {Sasagawa}, \citenamefont {Hanaguri}, \citenamefont {Takano},\ and\
  \citenamefont {Takagi}}]{kohsaka2002}%
  \BibitemOpen
  \bibfield  {author} {\bibinfo {author} {\bibfnamefont {Y.}~\bibnamefont
  {Kohsaka}}, \bibinfo {author} {\bibfnamefont {M.}~\bibnamefont {Azuma}},
  \bibinfo {author} {\bibfnamefont {I.}~\bibnamefont {Yamada}}, \bibinfo
  {author} {\bibfnamefont {T.}~\bibnamefont {Sasagawa}}, \bibinfo {author}
  {\bibfnamefont {T.}~\bibnamefont {Hanaguri}}, \bibinfo {author}
  {\bibfnamefont {M.}~\bibnamefont {Takano}},\ and\ \bibinfo {author}
  {\bibfnamefont {H.}~\bibnamefont {Takagi}},\ }\bibfield  {title} {\bibinfo
  {title} {Growth of {{Na}}-doped {{Ca$_2$CuO$_2$Cl$_2$}} single crystals under
  high pressures of several {{GPa}}},\ }\href
  {https://doi.org/10.1021/ja026680i} {\bibfield  {journal} {\bibinfo
  {journal} {Journal of the American Chemical Society}\ }\textbf {\bibinfo
  {volume} {124}},\ \bibinfo {pages} {12275} (\bibinfo {year}
  {2002})}\BibitemShut {NoStop}%
\bibitem [{\citenamefont {Yamada}\ \emph {et~al.}(2005)\citenamefont {Yamada},
  \citenamefont {Belik}, \citenamefont {Azuma}, \citenamefont {Harjo},
  \citenamefont {Kamiyama}, \citenamefont {Shimakawa},\ and\ \citenamefont
  {Takano}}]{yamada2005}%
  \BibitemOpen
  \bibfield  {author} {\bibinfo {author} {\bibfnamefont {I.}~\bibnamefont
  {Yamada}}, \bibinfo {author} {\bibfnamefont {A.~A.}\ \bibnamefont {Belik}},
  \bibinfo {author} {\bibfnamefont {M.}~\bibnamefont {Azuma}}, \bibinfo
  {author} {\bibfnamefont {S.}~\bibnamefont {Harjo}}, \bibinfo {author}
  {\bibfnamefont {T.}~\bibnamefont {Kamiyama}}, \bibinfo {author}
  {\bibfnamefont {Y.}~\bibnamefont {Shimakawa}},\ and\ \bibinfo {author}
  {\bibfnamefont {M.}~\bibnamefont {Takano}},\ }\bibfield  {title} {\bibinfo
  {title} {Single-layer oxychloride superconductor
  {{${\mathrm{Ca}}_{2\ensuremath{-}x}{\mathrm{CuO}}_{2}{\mathrm{Cl}}_{2}$}}
  with {{$A$}}-site cation deficiency},\ }\href
  {https://doi.org/10.1103/PhysRevB.72.224503} {\bibfield  {journal} {\bibinfo
  {journal} {Phys. Rev. B}\ }\textbf {\bibinfo {volume} {72}},\ \bibinfo
  {pages} {224503} (\bibinfo {year} {2005})}\BibitemShut {NoStop}%
\bibitem [{\citenamefont {Sowa}\ \emph {et~al.}(1990)\citenamefont {Sowa},
  \citenamefont {Hiratani},\ and\ \citenamefont {Miyauchi}}]{sowa1990}%
  \BibitemOpen
  \bibfield  {author} {\bibinfo {author} {\bibfnamefont {T.}~\bibnamefont
  {Sowa}}, \bibinfo {author} {\bibfnamefont {M.}~\bibnamefont {Hiratani}},\
  and\ \bibinfo {author} {\bibfnamefont {K.}~\bibnamefont {Miyauchi}},\
  }\bibfield  {title} {\bibinfo {title} {A new chlorooxocuprate,
  {{Ca$_3$Cu$_2$O$_4$Cl$_2$}}, with an oxygen defect intergrowth},\ }\href
  {https://doi.org/10.1016/0022-4596(90)90197-6} {\bibfield  {journal}
  {\bibinfo  {journal} {Journal of Solid State Chemistry}\ }\textbf {\bibinfo
  {volume} {84}},\ \bibinfo {pages} {178} (\bibinfo {year} {1990})}\BibitemShut
  {NoStop}%
\bibitem [{\citenamefont {Ruan}\ \emph {et~al.}(2016)\citenamefont {Ruan},
  \citenamefont {Hu}, \citenamefont {Zhao}, \citenamefont {Cai}, \citenamefont
  {Peng}, \citenamefont {Ye}, \citenamefont {Yu}, \citenamefont {Li},
  \citenamefont {Hao}, \citenamefont {Jin}, \citenamefont {Zhou}, \citenamefont
  {Weng},\ and\ \citenamefont {Wang}}]{ruan2016a}%
  \BibitemOpen
  \bibfield  {author} {\bibinfo {author} {\bibfnamefont {W.}~\bibnamefont
  {Ruan}}, \bibinfo {author} {\bibfnamefont {C.}~\bibnamefont {Hu}}, \bibinfo
  {author} {\bibfnamefont {J.}~\bibnamefont {Zhao}}, \bibinfo {author}
  {\bibfnamefont {P.}~\bibnamefont {Cai}}, \bibinfo {author} {\bibfnamefont
  {Y.}~\bibnamefont {Peng}}, \bibinfo {author} {\bibfnamefont {C.}~\bibnamefont
  {Ye}}, \bibinfo {author} {\bibfnamefont {R.}~\bibnamefont {Yu}}, \bibinfo
  {author} {\bibfnamefont {X.}~\bibnamefont {Li}}, \bibinfo {author}
  {\bibfnamefont {Z.}~\bibnamefont {Hao}}, \bibinfo {author} {\bibfnamefont
  {C.}~\bibnamefont {Jin}}, \bibinfo {author} {\bibfnamefont {X.}~\bibnamefont
  {Zhou}}, \bibinfo {author} {\bibfnamefont {Z.-Y.}\ \bibnamefont {Weng}},\
  and\ \bibinfo {author} {\bibfnamefont {Y.}~\bibnamefont {Wang}},\ }\bibfield
  {title} {\bibinfo {title} {Relationship between the parent charge transfer
  gap and maximum transition temperature in cuprates},\ }\href
  {https://doi.org/10.1007/s11434-016-1204-x} {\bibfield  {journal} {\bibinfo
  {journal} {Science Bulletin}\ }\textbf {\bibinfo {volume} {61}},\ \bibinfo
  {pages} {1826} (\bibinfo {year} {2016})}\BibitemShut {NoStop}%
\bibitem [{\citenamefont {Hohenberg}\ and\ \citenamefont
  {Kohn}(1964)}]{hohenberg1964}%
  \BibitemOpen
  \bibfield  {author} {\bibinfo {author} {\bibfnamefont {P.}~\bibnamefont
  {Hohenberg}}\ and\ \bibinfo {author} {\bibfnamefont {W.}~\bibnamefont
  {Kohn}},\ }\bibfield  {title} {\bibinfo {title} {Inhomogeneous electron
  gas},\ }\href {https://doi.org/10.1103/PhysRev.136.B864} {\bibfield
  {journal} {\bibinfo  {journal} {Phys. Rev.}\ }\textbf {\bibinfo {volume}
  {136}},\ \bibinfo {pages} {B864} (\bibinfo {year} {1964})}\BibitemShut
  {NoStop}%
\bibitem [{\citenamefont {Kohn}\ and\ \citenamefont {Sham}(1965)}]{kohn1965}%
  \BibitemOpen
  \bibfield  {author} {\bibinfo {author} {\bibfnamefont {W.}~\bibnamefont
  {Kohn}}\ and\ \bibinfo {author} {\bibfnamefont {L.~J.}\ \bibnamefont
  {Sham}},\ }\bibfield  {title} {\bibinfo {title} {Self-consistent equations
  including exchange and correlation effects},\ }\href
  {https://doi.org/10.1103/PhysRev.140.A1133} {\bibfield  {journal} {\bibinfo
  {journal} {Phys. Rev.}\ }\textbf {\bibinfo {volume} {140}},\ \bibinfo {pages}
  {A1133} (\bibinfo {year} {1965})}\BibitemShut {NoStop}%
\bibitem [{\citenamefont {Perdew}\ \emph {et~al.}(1996)\citenamefont {Perdew},
  \citenamefont {Burke},\ and\ \citenamefont {Ernzerhof}}]{perdew1996}%
  \BibitemOpen
  \bibfield  {author} {\bibinfo {author} {\bibfnamefont {J.~P.}\ \bibnamefont
  {Perdew}}, \bibinfo {author} {\bibfnamefont {K.}~\bibnamefont {Burke}},\ and\
  \bibinfo {author} {\bibfnamefont {M.}~\bibnamefont {Ernzerhof}},\ }\bibfield
  {title} {\bibinfo {title} {Generalized gradient approximation made simple},\
  }\href {https://doi.org/10.1103/PhysRevLett.77.3865} {\bibfield  {journal}
  {\bibinfo  {journal} {Phys. Rev. Lett.}\ }\textbf {\bibinfo {volume} {77}},\
  \bibinfo {pages} {3865} (\bibinfo {year} {1996})}\BibitemShut {NoStop}%
\bibitem [{\citenamefont {Lichtenstein}\ and\ \citenamefont
  {Katsnelson}(1998)}]{lichtenstein1998}%
  \BibitemOpen
  \bibfield  {author} {\bibinfo {author} {\bibfnamefont {A.~I.}\ \bibnamefont
  {Lichtenstein}}\ and\ \bibinfo {author} {\bibfnamefont {M.~I.}\ \bibnamefont
  {Katsnelson}},\ }\bibfield  {title} {\bibinfo {title} {Ab initio calculations
  of quasiparticle band structure in correlated systems: Lda++ approach},\
  }\href {https://doi.org/10.1103/PhysRevB.57.6884} {\bibfield  {journal}
  {\bibinfo  {journal} {Phys. Rev. B}\ }\textbf {\bibinfo {volume} {57}},\
  \bibinfo {pages} {6884} (\bibinfo {year} {1998})}\BibitemShut {NoStop}%
\bibitem [{\citenamefont {Kotliar}\ \emph {et~al.}(2006)\citenamefont
  {Kotliar}, \citenamefont {Savrasov}, \citenamefont {Haule}, \citenamefont
  {Oudovenko}, \citenamefont {Parcollet},\ and\ \citenamefont
  {Marianetti}}]{kotliar2006a}%
  \BibitemOpen
  \bibfield  {author} {\bibinfo {author} {\bibfnamefont {G.}~\bibnamefont
  {Kotliar}}, \bibinfo {author} {\bibfnamefont {S.~Y.}\ \bibnamefont
  {Savrasov}}, \bibinfo {author} {\bibfnamefont {K.}~\bibnamefont {Haule}},
  \bibinfo {author} {\bibfnamefont {V.~S.}\ \bibnamefont {Oudovenko}}, \bibinfo
  {author} {\bibfnamefont {O.}~\bibnamefont {Parcollet}},\ and\ \bibinfo
  {author} {\bibfnamefont {C.~A.}\ \bibnamefont {Marianetti}},\ }\bibfield
  {title} {\bibinfo {title} {Electronic structure calculations with dynamical
  mean-field theory},\ }\href {https://doi.org/10.1103/RevModPhys.78.865}
  {\bibfield  {journal} {\bibinfo  {journal} {Rev. Mod. Phys.}\ }\textbf
  {\bibinfo {volume} {78}},\ \bibinfo {pages} {865} (\bibinfo {year}
  {2006})}\BibitemShut {NoStop}%
\bibitem [{\citenamefont {Held}(2007)}]{held2007}%
  \BibitemOpen
  \bibfield  {author} {\bibinfo {author} {\bibfnamefont {K.}~\bibnamefont
  {Held}},\ }\bibfield  {title} {\bibinfo {title} {Electronic structure
  calculations using dynamical mean field theory},\ }\href
  {https://doi.org/10.1080/00018730701619647} {\bibfield  {journal} {\bibinfo
  {journal} {Advances in Physics}\ }\textbf {\bibinfo {volume} {56}},\ \bibinfo
  {pages} {829} (\bibinfo {year} {2007})}\BibitemShut {NoStop}%
\bibitem [{\citenamefont {Paul}\ and\ \citenamefont {Birol}(2019)}]{paul2019}%
  \BibitemOpen
  \bibfield  {author} {\bibinfo {author} {\bibfnamefont {A.}~\bibnamefont
  {Paul}}\ and\ \bibinfo {author} {\bibfnamefont {T.}~\bibnamefont {Birol}},\
  }\bibfield  {title} {\bibinfo {title} {Applications of {{DFT+DMFT}} in
  materials science},\ }\href
  {https://doi.org/https://doi.org/10.1146/annurev-matsci-070218-121825}
  {\bibfield  {journal} {\bibinfo  {journal} {Annual Review of Materials
  Research}\ }\textbf {\bibinfo {volume} {49}},\ \bibinfo {pages} {31}
  (\bibinfo {year} {2019})}\BibitemShut {NoStop}%
\bibitem [{\citenamefont {Shim}\ \emph {et~al.}(2007)\citenamefont {Shim},
  \citenamefont {Haule},\ and\ \citenamefont {Kotliar}}]{shim2007}%
  \BibitemOpen
  \bibfield  {author} {\bibinfo {author} {\bibfnamefont {J.~H.}\ \bibnamefont
  {Shim}}, \bibinfo {author} {\bibfnamefont {K.}~\bibnamefont {Haule}},\ and\
  \bibinfo {author} {\bibfnamefont {G.}~\bibnamefont {Kotliar}},\ }\bibfield
  {title} {\bibinfo {title} {Modeling the localized-to-itinerant electronic
  transition in the heavy fermion system {{CeIrIn$_5$}}},\ }\href
  {https://doi.org/10.1126/science.1149064} {\bibfield  {journal} {\bibinfo
  {journal} {Science}\ }\textbf {\bibinfo {volume} {318}},\ \bibinfo {pages}
  {1615} (\bibinfo {year} {2007})}\BibitemShut {NoStop}%
\bibitem [{\citenamefont {Haule}\ \emph {et~al.}(2014)\citenamefont {Haule},
  \citenamefont {Birol},\ and\ \citenamefont {Kotliar}}]{haule2014}%
  \BibitemOpen
  \bibfield  {author} {\bibinfo {author} {\bibfnamefont {K.}~\bibnamefont
  {Haule}}, \bibinfo {author} {\bibfnamefont {T.}~\bibnamefont {Birol}},\ and\
  \bibinfo {author} {\bibfnamefont {G.}~\bibnamefont {Kotliar}},\ }\bibfield
  {title} {\bibinfo {title} {Covalency in transition-metal oxides within
  all-electron dynamical mean-field theory},\ }\href
  {https://doi.org/10.1103/PhysRevB.90.075136} {\bibfield  {journal} {\bibinfo
  {journal} {Phys. Rev. B}\ }\textbf {\bibinfo {volume} {90}},\ \bibinfo
  {pages} {075136} (\bibinfo {year} {2014})}\BibitemShut {NoStop}%
\bibitem [{\citenamefont {Haule}(2018{\natexlab{b}})}]{haule2018}%
  \BibitemOpen
  \bibfield  {author} {\bibinfo {author} {\bibfnamefont {K.}~\bibnamefont
  {Haule}},\ }\bibfield  {title} {\bibinfo {title} {Structural predictions for
  correlated electron materials using the functional dynamical mean field
  theory approach},\ }\href {https://doi.org/10.7566/JPSJ.87.041005} {\bibfield
   {journal} {\bibinfo  {journal} {J. Phys. Soc. Jpn.}\ }\textbf {\bibinfo
  {volume} {87}},\ \bibinfo {pages} {041005} (\bibinfo {year}
  {2018}{\natexlab{b}})}\BibitemShut {NoStop}%
\bibitem [{\citenamefont {Mandal}\ \emph {et~al.}(2019)\citenamefont {Mandal},
  \citenamefont {Haule}, \citenamefont {Rabe},\ and\ \citenamefont
  {Vanderbilt}}]{mandal2019}%
  \BibitemOpen
  \bibfield  {author} {\bibinfo {author} {\bibfnamefont {S.}~\bibnamefont
  {Mandal}}, \bibinfo {author} {\bibfnamefont {K.}~\bibnamefont {Haule}},
  \bibinfo {author} {\bibfnamefont {K.~M.}\ \bibnamefont {Rabe}},\ and\
  \bibinfo {author} {\bibfnamefont {D.}~\bibnamefont {Vanderbilt}},\ }\bibfield
   {title} {\bibinfo {title} {Systematic beyond-{{DFT}} study of binary
  transition metal oxides},\ }\href
  {https://doi.org/https://doi.org/10.1038/s41524-019-0251-7} {\bibfield
  {journal} {\bibinfo  {journal} {npj Computational Materials}\ }\textbf
  {\bibinfo {volume} {5}},\ \bibinfo {pages} {115} (\bibinfo {year}
  {2019})}\BibitemShut {NoStop}%
\bibitem [{\citenamefont {Haule}(2015)}]{haule2015}%
  \BibitemOpen
  \bibfield  {author} {\bibinfo {author} {\bibfnamefont {K.}~\bibnamefont
  {Haule}},\ }\bibfield  {title} {\bibinfo {title} {Exact {{Double Counting}}
  in combining the dynamical mean field theory and the density functional
  theory},\ }\href {https://doi.org/10.1103/PhysRevLett.115.196403} {\bibfield
  {journal} {\bibinfo  {journal} {Phys. Rev. Lett.}\ }\textbf {\bibinfo
  {volume} {115}},\ \bibinfo {pages} {196403} (\bibinfo {year}
  {2015})}\BibitemShut {NoStop}%
\bibitem [{\citenamefont {Anisimov}\ \emph {et~al.}(1993)\citenamefont
  {Anisimov}, \citenamefont {Solovyev}, \citenamefont {Korotin}, \citenamefont
  {Czy{\.z}yk},\ and\ \citenamefont {Sawatzky}}]{anisimov1993}%
  \BibitemOpen
  \bibfield  {author} {\bibinfo {author} {\bibfnamefont {V.~I.}\ \bibnamefont
  {Anisimov}}, \bibinfo {author} {\bibfnamefont {I.~V.}\ \bibnamefont
  {Solovyev}}, \bibinfo {author} {\bibfnamefont {M.~A.}\ \bibnamefont
  {Korotin}}, \bibinfo {author} {\bibfnamefont {M.~T.}\ \bibnamefont
  {Czy{\.z}yk}},\ and\ \bibinfo {author} {\bibfnamefont {G.~A.}\ \bibnamefont
  {Sawatzky}},\ }\bibfield  {title} {\bibinfo {title} {Density-functional
  theory and {{NiO}} photoemission spectra},\ }\href
  {https://doi.org/10.1103/PhysRevB.48.16929} {\bibfield  {journal} {\bibinfo
  {journal} {Phys. Rev. B}\ }\textbf {\bibinfo {volume} {48}},\ \bibinfo
  {pages} {16929} (\bibinfo {year} {1993})}\BibitemShut {NoStop}%
\bibitem [{\citenamefont {Lichtenstein}\ and\ \citenamefont
  {Katsnelson}(2000)}]{lichtenstein2000}%
  \BibitemOpen
  \bibfield  {author} {\bibinfo {author} {\bibfnamefont {A.~I.}\ \bibnamefont
  {Lichtenstein}}\ and\ \bibinfo {author} {\bibfnamefont {M.~I.}\ \bibnamefont
  {Katsnelson}},\ }\bibfield  {title} {\bibinfo {title} {Antiferromagnetism and
  d-wave superconductivity in cuprates: A cluster dynamical mean-field
  theory},\ }\href {https://doi.org/10.1103/PhysRevB.62.R9283} {\bibfield
  {journal} {\bibinfo  {journal} {Phys. Rev. B}\ }\textbf {\bibinfo {volume}
  {62}},\ \bibinfo {pages} {R9283} (\bibinfo {year} {2000})}\BibitemShut
  {NoStop}%
\bibitem [{\citenamefont {Kotliar}\ \emph {et~al.}(2001)\citenamefont
  {Kotliar}, \citenamefont {Savrasov}, \citenamefont {P\'alsson},\ and\
  \citenamefont {Biroli}}]{kotliar2001}%
  \BibitemOpen
  \bibfield  {author} {\bibinfo {author} {\bibfnamefont {G.}~\bibnamefont
  {Kotliar}}, \bibinfo {author} {\bibfnamefont {S.~Y.}\ \bibnamefont
  {Savrasov}}, \bibinfo {author} {\bibfnamefont {G.}~\bibnamefont
  {P\'alsson}},\ and\ \bibinfo {author} {\bibfnamefont {G.}~\bibnamefont
  {Biroli}},\ }\bibfield  {title} {\bibinfo {title} {Cellular dynamical mean
  field approach to strongly correlated systems},\ }\href
  {https://doi.org/10.1103/PhysRevLett.87.186401} {\bibfield  {journal}
  {\bibinfo  {journal} {Phys. Rev. Lett.}\ }\textbf {\bibinfo {volume} {87}},\
  \bibinfo {pages} {186401} (\bibinfo {year} {2001})}\BibitemShut {NoStop}%
\bibitem [{\citenamefont {Maier}\ \emph {et~al.}(2005)\citenamefont {Maier},
  \citenamefont {Jarrell}, \citenamefont {Schulthess}, \citenamefont {Kent},\
  and\ \citenamefont {White}}]{maier2005a}%
  \BibitemOpen
  \bibfield  {author} {\bibinfo {author} {\bibfnamefont {T.~A.}\ \bibnamefont
  {Maier}}, \bibinfo {author} {\bibfnamefont {M.}~\bibnamefont {Jarrell}},
  \bibinfo {author} {\bibfnamefont {T.~C.}\ \bibnamefont {Schulthess}},
  \bibinfo {author} {\bibfnamefont {P.~R.~C.}\ \bibnamefont {Kent}},\ and\
  \bibinfo {author} {\bibfnamefont {J.~B.}\ \bibnamefont {White}},\ }\bibfield
  {title} {\bibinfo {title} {Systematic study of {{$d$}}-wave superconductivity
  in the {{2D}} repulsive {{Hubbard}} model},\ }\href
  {https://doi.org/10.1103/PhysRevLett.95.237001} {\bibfield  {journal}
  {\bibinfo  {journal} {Phys. Rev. Lett.}\ }\textbf {\bibinfo {volume} {95}},\
  \bibinfo {pages} {237001} (\bibinfo {year} {2005})}\BibitemShut {NoStop}%
\bibitem [{\citenamefont {Tremblay}\ \emph {et~al.}(2006)\citenamefont
  {Tremblay}, \citenamefont {Kyung},\ and\ \citenamefont
  {S\'en\'echal}}]{LTP:2006}%
  \BibitemOpen
  \bibfield  {author} {\bibinfo {author} {\bibfnamefont {A.~M.~S.}\
  \bibnamefont {Tremblay}}, \bibinfo {author} {\bibfnamefont {B.}~\bibnamefont
  {Kyung}},\ and\ \bibinfo {author} {\bibfnamefont {D.}~\bibnamefont
  {S\'en\'echal}},\ }\bibfield  {title} {\bibinfo {title} {Pseudogap and
  high-temperature superconductivity from weak to strong coupling. {{Towards}}
  a quantitative theory},\ }\href {http://dx.doi.org/10.1063/1.2199446}
  {\bibfield  {journal} {\bibinfo  {journal} {Low Temp. Phys.}\ }\textbf
  {\bibinfo {volume} {32}},\ \bibinfo {pages} {424} (\bibinfo {year}
  {2006})}\BibitemShut {NoStop}%
\bibitem [{\citenamefont {Kim}\ \emph {et~al.}(2020)\citenamefont {Kim},
  \citenamefont {Haule},\ and\ \citenamefont
  {Vanderbilt}}]{PhysRevB.102.081105}%
  \BibitemOpen
  \bibfield  {author} {\bibinfo {author} {\bibfnamefont {H.-S.}\ \bibnamefont
  {Kim}}, \bibinfo {author} {\bibfnamefont {K.}~\bibnamefont {Haule}},\ and\
  \bibinfo {author} {\bibfnamefont {D.}~\bibnamefont {Vanderbilt}},\ }\bibfield
   {title} {\bibinfo {title} {Molecular {{Mott}} state in the deficient spinel
  {{${\mathrm{GaV}}_{4}{\mathrm{S}}_{8}$}}},\ }\href
  {https://doi.org/10.1103/PhysRevB.102.081105} {\bibfield  {journal} {\bibinfo
   {journal} {Phys. Rev. B}\ }\textbf {\bibinfo {volume} {102}},\ \bibinfo
  {pages} {081105} (\bibinfo {year} {2020})}\BibitemShut {NoStop}%
\bibitem [{\citenamefont {Rubel}\ \emph {et~al.}(2014)\citenamefont {Rubel},
  \citenamefont {Bokhanchuk}, \citenamefont {Ahmed},\ and\ \citenamefont
  {Assmann}}]{rubel2014}%
  \BibitemOpen
  \bibfield  {author} {\bibinfo {author} {\bibfnamefont {O.}~\bibnamefont
  {Rubel}}, \bibinfo {author} {\bibfnamefont {A.}~\bibnamefont {Bokhanchuk}},
  \bibinfo {author} {\bibfnamefont {S.~J.}\ \bibnamefont {Ahmed}},\ and\
  \bibinfo {author} {\bibfnamefont {E.}~\bibnamefont {Assmann}},\ }\bibfield
  {title} {\bibinfo {title} {Unfolding the band structure of disordered solids:
  {{From}} bound states to high-mobility {{Kane}} fermions},\ }\href
  {https://doi.org/10.1103/PhysRevB.90.115202} {\bibfield  {journal} {\bibinfo
  {journal} {Phys. Rev. B}\ }\textbf {\bibinfo {volume} {90}},\ \bibinfo
  {pages} {115202} (\bibinfo {year} {2014})}\BibitemShut {NoStop}%
\bibitem [{\citenamefont {S\'en\'echal}\ \emph {et~al.}(2000)\citenamefont
  {S\'en\'echal}, \citenamefont {Perez},\ and\ \citenamefont
  {Pioro-Ladri\`ere}}]{senechal2000}%
  \BibitemOpen
  \bibfield  {author} {\bibinfo {author} {\bibfnamefont {D.}~\bibnamefont
  {S\'en\'echal}}, \bibinfo {author} {\bibfnamefont {D.}~\bibnamefont
  {Perez}},\ and\ \bibinfo {author} {\bibfnamefont {M.}~\bibnamefont
  {Pioro-Ladri\`ere}},\ }\bibfield  {title} {\bibinfo {title} {Spectral weight
  of the {{Hubbard}} model through cluster perturbation theory},\ }\href
  {https://doi.org/10.1103/PhysRevLett.84.522} {\bibfield  {journal} {\bibinfo
  {journal} {Phys. Rev. Lett.}\ }\textbf {\bibinfo {volume} {84}},\ \bibinfo
  {pages} {522} (\bibinfo {year} {2000})}\BibitemShut {NoStop}%
\bibitem [{\citenamefont {Stanescu}\ and\ \citenamefont
  {Kotliar}(2006)}]{stanescu2006}%
  \BibitemOpen
  \bibfield  {author} {\bibinfo {author} {\bibfnamefont {T.~D.}\ \bibnamefont
  {Stanescu}}\ and\ \bibinfo {author} {\bibfnamefont {G.}~\bibnamefont
  {Kotliar}},\ }\bibfield  {title} {\bibinfo {title} {Fermi arcs and hidden
  zeros of the {{Green}} function in the pseudogap state},\ }\href
  {https://doi.org/10.1103/PhysRevB.74.125110} {\bibfield  {journal} {\bibinfo
  {journal} {Phys. Rev. B}\ }\textbf {\bibinfo {volume} {74}},\ \bibinfo
  {pages} {125110} (\bibinfo {year} {2006})}\BibitemShut {NoStop}%
\bibitem [{\citenamefont {Verret}\ \emph {et~al.}(2019)\citenamefont {Verret},
  \citenamefont {Roy}, \citenamefont {Foley}, \citenamefont {Charlebois},
  \citenamefont {S\'en\'echal},\ and\ \citenamefont {Tremblay}}]{verret2019}%
  \BibitemOpen
  \bibfield  {author} {\bibinfo {author} {\bibfnamefont {S.}~\bibnamefont
  {Verret}}, \bibinfo {author} {\bibfnamefont {J.}~\bibnamefont {Roy}},
  \bibinfo {author} {\bibfnamefont {A.}~\bibnamefont {Foley}}, \bibinfo
  {author} {\bibfnamefont {M.}~\bibnamefont {Charlebois}}, \bibinfo {author}
  {\bibfnamefont {D.}~\bibnamefont {S\'en\'echal}},\ and\ \bibinfo {author}
  {\bibfnamefont {A.-M.~S.}\ \bibnamefont {Tremblay}},\ }\bibfield  {title}
  {\bibinfo {title} {Intrinsic cluster-shaped density waves in cellular
  dynamical mean-field theory},\ }\href
  {https://doi.org/10.1103/PhysRevB.100.224520} {\bibfield  {journal} {\bibinfo
   {journal} {Phys. Rev. B}\ }\textbf {\bibinfo {volume} {100}},\ \bibinfo
  {pages} {224520} (\bibinfo {year} {2019})}\BibitemShut {NoStop}%
\bibitem [{\citenamefont {Bolech}\ \emph {et~al.}(2003)\citenamefont {Bolech},
  \citenamefont {Kancharla},\ and\ \citenamefont {Kotliar}}]{bolech2003}%
  \BibitemOpen
  \bibfield  {author} {\bibinfo {author} {\bibfnamefont {C.~J.}\ \bibnamefont
  {Bolech}}, \bibinfo {author} {\bibfnamefont {S.~S.}\ \bibnamefont
  {Kancharla}},\ and\ \bibinfo {author} {\bibfnamefont {G.}~\bibnamefont
  {Kotliar}},\ }\bibfield  {title} {\bibinfo {title} {Cellular dynamical
  mean-field theory for the one-dimensional extended {{Hubbard}} model},\
  }\href {https://doi.org/10.1103/PhysRevB.67.075110} {\bibfield  {journal}
  {\bibinfo  {journal} {Phys. Rev. B}\ }\textbf {\bibinfo {volume} {67}},\
  \bibinfo {pages} {075110} (\bibinfo {year} {2003})}\BibitemShut {NoStop}%
\bibitem [{\citenamefont {Kancharla}\ \emph {et~al.}(2008)\citenamefont
  {Kancharla}, \citenamefont {Kyung}, \citenamefont {S{\'e}n{\'e}chal},
  \citenamefont {Civelli}, \citenamefont {Capone}, \citenamefont {Kotliar},\
  and\ \citenamefont {Tremblay}}]{kancharla2008b}%
  \BibitemOpen
  \bibfield  {author} {\bibinfo {author} {\bibfnamefont {S.~S.}\ \bibnamefont
  {Kancharla}}, \bibinfo {author} {\bibfnamefont {B.}~\bibnamefont {Kyung}},
  \bibinfo {author} {\bibfnamefont {D.}~\bibnamefont {S{\'e}n{\'e}chal}},
  \bibinfo {author} {\bibfnamefont {M.}~\bibnamefont {Civelli}}, \bibinfo
  {author} {\bibfnamefont {M.}~\bibnamefont {Capone}}, \bibinfo {author}
  {\bibfnamefont {G.}~\bibnamefont {Kotliar}},\ and\ \bibinfo {author}
  {\bibfnamefont {A.-M.~S.}\ \bibnamefont {Tremblay}},\ }\bibfield  {title}
  {\bibinfo {title} {Anomalous superconductivity and its competition with
  antiferromagnetism in doped {{Mott}} insulators},\ }\href
  {https://doi.org/10.1103/PhysRevB.77.184516} {\bibfield  {journal} {\bibinfo
  {journal} {Phys. Rev. B}\ }\textbf {\bibinfo {volume} {77}},\ \bibinfo
  {pages} {184516} (\bibinfo {year} {2008})}\BibitemShut {NoStop}%
\bibitem [{\citenamefont {Foley}\ \emph {et~al.}(2019)\citenamefont {Foley},
  \citenamefont {Verret}, \citenamefont {Tremblay},\ and\ \citenamefont
  {S{\'e}n{\'e}chal}}]{foley2019a}%
  \BibitemOpen
  \bibfield  {author} {\bibinfo {author} {\bibfnamefont {A.}~\bibnamefont
  {Foley}}, \bibinfo {author} {\bibfnamefont {S.}~\bibnamefont {Verret}},
  \bibinfo {author} {\bibfnamefont {A.-M.~S.}\ \bibnamefont {Tremblay}},\ and\
  \bibinfo {author} {\bibfnamefont {D.}~\bibnamefont {S{\'e}n{\'e}chal}},\
  }\bibfield  {title} {\bibinfo {title} {Coexistence of superconductivity and
  antiferromagnetism in the {{Hubbard}} model for cuprates},\ }\href
  {https://doi.org/10.1103/PhysRevB.99.184510} {\bibfield  {journal} {\bibinfo
  {journal} {Phys. Rev. B}\ }\textbf {\bibinfo {volume} {99}},\ \bibinfo
  {pages} {184510} (\bibinfo {year} {2019})}\BibitemShut {NoStop}%
\bibitem [{\citenamefont {Dash}\ and\ \citenamefont
  {S{\'e}n{\'e}chal}(2019)}]{dash2019}%
  \BibitemOpen
  \bibfield  {author} {\bibinfo {author} {\bibfnamefont {S.~S.}\ \bibnamefont
  {Dash}}\ and\ \bibinfo {author} {\bibfnamefont {D.}~\bibnamefont
  {S{\'e}n{\'e}chal}},\ }\bibfield  {title} {\bibinfo {title} {Pseudogap
  transition within the superconducting phase in the three-band {{Hubbard}}
  model},\ }\href {https://doi.org/10.1103/PhysRevB.100.214509} {\bibfield
  {journal} {\bibinfo  {journal} {Phys. Rev. B}\ }\textbf {\bibinfo {volume}
  {100}},\ \bibinfo {pages} {214509} (\bibinfo {year} {2019})}\BibitemShut
  {NoStop}%
\bibitem [{\citenamefont {Bai}\ \emph {et~al.}(2000)\citenamefont {Bai},
  \citenamefont {Demmel}, \citenamefont {Dongarra}, \citenamefont {Ruhe},\ and\
  \citenamefont {van~der Vorst}}]{bai2000}%
  \BibitemOpen
  \bibfield  {author} {\bibinfo {author} {\bibfnamefont {Z.}~\bibnamefont
  {Bai}}, \bibinfo {author} {\bibfnamefont {J.}~\bibnamefont {Demmel}},
  \bibinfo {author} {\bibfnamefont {J.}~\bibnamefont {Dongarra}}, \bibinfo
  {author} {\bibfnamefont {A.}~\bibnamefont {Ruhe}},\ and\ \bibinfo {author}
  {\bibfnamefont {H.}~\bibnamefont {van~der Vorst}},\ }\href
  {https://doi.org/10.1137/1.9780898719581} {\emph {\bibinfo {title} {Templates
  for the Solution of Algebraic Eigenvalue Problems}}},\ edited by\ \bibinfo
  {editor} {\bibfnamefont {Z.}~\bibnamefont {Bai}}, \bibinfo {editor}
  {\bibfnamefont {J.}~\bibnamefont {Demmel}}, \bibinfo {editor} {\bibfnamefont
  {J.}~\bibnamefont {Dongarra}}, \bibinfo {editor} {\bibfnamefont
  {A.}~\bibnamefont {Ruhe}},\ and\ \bibinfo {editor} {\bibfnamefont
  {H.}~\bibnamefont {van~der Vorst}}\ (\bibinfo  {publisher} {Society for
  Industrial and Applied Mathematics},\ \bibinfo {year} {2000})\BibitemShut
  {NoStop}%
\bibitem [{\citenamefont {Koch}\ \emph {et~al.}(2008)\citenamefont {Koch},
  \citenamefont {Sangiovanni},\ and\ \citenamefont {Gunnarsson}}]{koch2008}%
  \BibitemOpen
  \bibfield  {author} {\bibinfo {author} {\bibfnamefont {E.}~\bibnamefont
  {Koch}}, \bibinfo {author} {\bibfnamefont {G.}~\bibnamefont {Sangiovanni}},\
  and\ \bibinfo {author} {\bibfnamefont {O.}~\bibnamefont {Gunnarsson}},\
  }\bibfield  {title} {\bibinfo {title} {Sum rules and bath parametrization for
  quantum cluster theories},\ }\href
  {https://doi.org/10.1103/PhysRevB.78.115102} {\bibfield  {journal} {\bibinfo
  {journal} {Phys. Rev. B}\ }\textbf {\bibinfo {volume} {78}},\ \bibinfo
  {pages} {115102} (\bibinfo {year} {2008})}\BibitemShut {NoStop}%
\bibitem [{\citenamefont {Haule}(2007)}]{haule2007}%
  \BibitemOpen
  \bibfield  {author} {\bibinfo {author} {\bibfnamefont {K.}~\bibnamefont
  {Haule}},\ }\bibfield  {title} {\bibinfo {title} {Quantum monte carlo
  impurity solver for cluster dynamical mean-field theory and electronic
  structure calculations with adjustable cluster base},\ }\href
  {https://doi.org/10.1103/PhysRevB.75.155113} {\bibfield  {journal} {\bibinfo
  {journal} {Phys. Rev. B}\ }\textbf {\bibinfo {volume} {75}},\ \bibinfo
  {pages} {155113} (\bibinfo {year} {2007})}\BibitemShut {NoStop}%
\bibitem [{\citenamefont {Kaye}\ \emph {et~al.}(2022)\citenamefont {Kaye},
  \citenamefont {Chen},\ and\ \citenamefont {Parcollet}}]{kaye2022}%
  \BibitemOpen
  \bibfield  {author} {\bibinfo {author} {\bibfnamefont {J.}~\bibnamefont
  {Kaye}}, \bibinfo {author} {\bibfnamefont {K.}~\bibnamefont {Chen}},\ and\
  \bibinfo {author} {\bibfnamefont {O.}~\bibnamefont {Parcollet}},\ }\bibfield
  {title} {\bibinfo {title} {Discrete {{Lehmann}} representation of imaginary
  time {{Green's}} functions},\ }\href
  {https://doi.org/10.1103/PhysRevB.105.235115} {\bibfield  {journal} {\bibinfo
   {journal} {Phys. Rev. B}\ }\textbf {\bibinfo {volume} {105}},\ \bibinfo
  {pages} {235115} (\bibinfo {year} {2022})}\BibitemShut {NoStop}%
\bibitem [{\citenamefont {Gull}\ \emph {et~al.}(2011)\citenamefont {Gull},
  \citenamefont {Millis}, \citenamefont {Lichtenstein}, \citenamefont
  {Rubtsov}, \citenamefont {Troyer},\ and\ \citenamefont {Werner}}]{gull2011}%
  \BibitemOpen
  \bibfield  {author} {\bibinfo {author} {\bibfnamefont {E.}~\bibnamefont
  {Gull}}, \bibinfo {author} {\bibfnamefont {A.~J.}\ \bibnamefont {Millis}},
  \bibinfo {author} {\bibfnamefont {A.~I.}\ \bibnamefont {Lichtenstein}},
  \bibinfo {author} {\bibfnamefont {A.~N.}\ \bibnamefont {Rubtsov}}, \bibinfo
  {author} {\bibfnamefont {M.}~\bibnamefont {Troyer}},\ and\ \bibinfo {author}
  {\bibfnamefont {P.}~\bibnamefont {Werner}},\ }\bibfield  {title} {\bibinfo
  {title} {Continuous-time monte carlo methods for quantum impurity models},\
  }\href {https://doi.org/10.1103/RevModPhys.83.349} {\bibfield  {journal}
  {\bibinfo  {journal} {Rev. Mod. Phys.}\ }\textbf {\bibinfo {volume} {83}},\
  \bibinfo {pages} {349} (\bibinfo {year} {2011})}\BibitemShut {NoStop}%
\bibitem [{\citenamefont {Luo}\ \emph {et~al.}(2023{\natexlab{a}})\citenamefont
  {Luo}, \citenamefont {Chen}, \citenamefont {Li}, \citenamefont {Gao},
  \citenamefont {Yin}, \citenamefont {Yan}, \citenamefont {Miao}, \citenamefont
  {Luo}, \citenamefont {Shu}, \citenamefont {Chen}, \citenamefont {Lin},
  \citenamefont {Zhang}, \citenamefont {Wang}, \citenamefont {Zhang},
  \citenamefont {Yang}, \citenamefont {Peng}, \citenamefont {Liu},
  \citenamefont {Zhao}, \citenamefont {Xu}, \citenamefont {Xiang},\ and\
  \citenamefont {Zhou}}]{luo2023a}%
  \BibitemOpen
  \bibfield  {author} {\bibinfo {author} {\bibfnamefont {X.}~\bibnamefont
  {Luo}}, \bibinfo {author} {\bibfnamefont {H.}~\bibnamefont {Chen}}, \bibinfo
  {author} {\bibfnamefont {Y.}~\bibnamefont {Li}}, \bibinfo {author}
  {\bibfnamefont {Q.}~\bibnamefont {Gao}}, \bibinfo {author} {\bibfnamefont
  {C.}~\bibnamefont {Yin}}, \bibinfo {author} {\bibfnamefont {H.}~\bibnamefont
  {Yan}}, \bibinfo {author} {\bibfnamefont {T.}~\bibnamefont {Miao}}, \bibinfo
  {author} {\bibfnamefont {H.}~\bibnamefont {Luo}}, \bibinfo {author}
  {\bibfnamefont {Y.}~\bibnamefont {Shu}}, \bibinfo {author} {\bibfnamefont
  {Y.}~\bibnamefont {Chen}}, \bibinfo {author} {\bibfnamefont {C.}~\bibnamefont
  {Lin}}, \bibinfo {author} {\bibfnamefont {S.}~\bibnamefont {Zhang}}, \bibinfo
  {author} {\bibfnamefont {Z.}~\bibnamefont {Wang}}, \bibinfo {author}
  {\bibfnamefont {F.}~\bibnamefont {Zhang}}, \bibinfo {author} {\bibfnamefont
  {F.}~\bibnamefont {Yang}}, \bibinfo {author} {\bibfnamefont {Q.}~\bibnamefont
  {Peng}}, \bibinfo {author} {\bibfnamefont {G.}~\bibnamefont {Liu}}, \bibinfo
  {author} {\bibfnamefont {L.}~\bibnamefont {Zhao}}, \bibinfo {author}
  {\bibfnamefont {Z.}~\bibnamefont {Xu}}, \bibinfo {author} {\bibfnamefont
  {T.}~\bibnamefont {Xiang}},\ and\ \bibinfo {author} {\bibfnamefont {X.~J.}\
  \bibnamefont {Zhou}},\ }\bibfield  {title} {\bibinfo {title} {Electronic
  origin of high superconducting critical temperature in trilayer cuprates},\
  }\href {https://doi.org/10.1038/s41567-023-02206-0} {\bibfield  {journal}
  {\bibinfo  {journal} {Nat. Phys.}\ ,\ \bibinfo {pages} {1}} (\bibinfo {year}
  {2023}{\natexlab{a}})}\BibitemShut {NoStop}%
\bibitem [{\citenamefont {Blaha}\ \emph {et~al.}(2020)\citenamefont {Blaha},
  \citenamefont {Schwarz}, \citenamefont {Tran}, \citenamefont {Laskowski},
  \citenamefont {Madsen},\ and\ \citenamefont {Marks}}]{blaha2020}%
  \BibitemOpen
  \bibfield  {author} {\bibinfo {author} {\bibfnamefont {P.}~\bibnamefont
  {Blaha}}, \bibinfo {author} {\bibfnamefont {K.}~\bibnamefont {Schwarz}},
  \bibinfo {author} {\bibfnamefont {F.}~\bibnamefont {Tran}}, \bibinfo {author}
  {\bibfnamefont {R.}~\bibnamefont {Laskowski}}, \bibinfo {author}
  {\bibfnamefont {G.~K.~H.}\ \bibnamefont {Madsen}},\ and\ \bibinfo {author}
  {\bibfnamefont {L.~D.}\ \bibnamefont {Marks}},\ }\bibfield  {title} {\bibinfo
  {title} {{{WIEN2k}}: {{An APW}}+lo program for calculating the properties of
  solids},\ }\href {https://doi.org/10.1063/1.5143061} {\bibfield  {journal}
  {\bibinfo  {journal} {J. Chem. Phys.}\ }\textbf {\bibinfo {volume} {152}},\
  \bibinfo {pages} {074101} (\bibinfo {year} {2020})}\BibitemShut {NoStop}%
\bibitem [{\citenamefont {Momma}\ and\ \citenamefont
  {Izumi}(2011)}]{momma2011}%
  \BibitemOpen
  \bibfield  {author} {\bibinfo {author} {\bibfnamefont {K.}~\bibnamefont
  {Momma}}\ and\ \bibinfo {author} {\bibfnamefont {F.}~\bibnamefont {Izumi}},\
  }\bibfield  {title} {\bibinfo {title} {Vesta 3 for three-dimensional
  visualization of crystal, volumetric and morphology data},\ }\href
  {https://doi.org/https://doi.org/10.1107/S0021889811038970} {\bibfield
  {journal} {\bibinfo  {journal} {Journal of applied crystallography}\ }\textbf
  {\bibinfo {volume} {44}},\ \bibinfo {pages} {1272} (\bibinfo {year}
  {2011})}\BibitemShut {NoStop}%
\bibitem [{\citenamefont {Ronning}\ \emph {et~al.}(2005)\citenamefont
  {Ronning}, \citenamefont {Shen}, \citenamefont {Armitage}, \citenamefont
  {Damascelli}, \citenamefont {Lu}, \citenamefont {Shen}, \citenamefont
  {Miller},\ and\ \citenamefont {Kim}}]{ronning2005}%
  \BibitemOpen
  \bibfield  {author} {\bibinfo {author} {\bibfnamefont {F.}~\bibnamefont
  {Ronning}}, \bibinfo {author} {\bibfnamefont {K.~M.}\ \bibnamefont {Shen}},
  \bibinfo {author} {\bibfnamefont {N.~P.}\ \bibnamefont {Armitage}}, \bibinfo
  {author} {\bibfnamefont {A.}~\bibnamefont {Damascelli}}, \bibinfo {author}
  {\bibfnamefont {D.~H.}\ \bibnamefont {Lu}}, \bibinfo {author} {\bibfnamefont
  {Z.-X.}\ \bibnamefont {Shen}}, \bibinfo {author} {\bibfnamefont {L.~L.}\
  \bibnamefont {Miller}},\ and\ \bibinfo {author} {\bibfnamefont
  {C.}~\bibnamefont {Kim}},\ }\bibfield  {title} {\bibinfo {title} {Anomalous
  high-energy dispersion in angle-resolved photoemission spectra from the
  insulating cuprate {{Ca}} 2 {{CuO}} 2 {{Cl}} 2},\ }\href
  {https://doi.org/10.1103/PhysRevB.71.094518} {\bibfield  {journal} {\bibinfo
  {journal} {Phys. Rev. B}\ }\textbf {\bibinfo {volume} {71}},\ \bibinfo
  {pages} {094518} (\bibinfo {year} {2005})}\BibitemShut {NoStop}%
\bibitem [{\citenamefont {Perkins}\ \emph {et~al.}(1993)\citenamefont
  {Perkins}, \citenamefont {Graybeal}, \citenamefont {Kastner}, \citenamefont
  {Birgeneau}, \citenamefont {Falck},\ and\ \citenamefont
  {Greven}}]{perkins1993}%
  \BibitemOpen
  \bibfield  {author} {\bibinfo {author} {\bibfnamefont {J.~D.}\ \bibnamefont
  {Perkins}}, \bibinfo {author} {\bibfnamefont {J.~M.}\ \bibnamefont
  {Graybeal}}, \bibinfo {author} {\bibfnamefont {M.~A.}\ \bibnamefont
  {Kastner}}, \bibinfo {author} {\bibfnamefont {R.~J.}\ \bibnamefont
  {Birgeneau}}, \bibinfo {author} {\bibfnamefont {J.~P.}\ \bibnamefont
  {Falck}},\ and\ \bibinfo {author} {\bibfnamefont {M.}~\bibnamefont
  {Greven}},\ }\bibfield  {title} {\bibinfo {title} {Mid-infrared optical
  absorption in undoped lamellar copper oxides},\ }\href
  {https://doi.org/10.1103/PhysRevLett.71.1621} {\bibfield  {journal} {\bibinfo
   {journal} {Phys. Rev. Lett.}\ }\textbf {\bibinfo {volume} {71}},\ \bibinfo
  {pages} {1621} (\bibinfo {year} {1993})}\BibitemShut {NoStop}%
\bibitem [{\citenamefont {Falck}\ \emph {et~al.}(1994)\citenamefont {Falck},
  \citenamefont {Perkins}, \citenamefont {Levy}, \citenamefont {Kastner},
  \citenamefont {Graybeal},\ and\ \citenamefont {Birgeneau}}]{falck1994}%
  \BibitemOpen
  \bibfield  {author} {\bibinfo {author} {\bibfnamefont {J.~P.}\ \bibnamefont
  {Falck}}, \bibinfo {author} {\bibfnamefont {J.~D.}\ \bibnamefont {Perkins}},
  \bibinfo {author} {\bibfnamefont {A.}~\bibnamefont {Levy}}, \bibinfo {author}
  {\bibfnamefont {M.~A.}\ \bibnamefont {Kastner}}, \bibinfo {author}
  {\bibfnamefont {J.~M.}\ \bibnamefont {Graybeal}},\ and\ \bibinfo {author}
  {\bibfnamefont {R.~J.}\ \bibnamefont {Birgeneau}},\ }\bibfield  {title}
  {\bibinfo {title} {Midinfrared electroreflectance in
  {{${\mathrm{La}}_{2}$${\mathrm{CuO}}_{4+\mathit{y}}$}}},\ }\href
  {https://doi.org/10.1103/PhysRevB.49.6246} {\bibfield  {journal} {\bibinfo
  {journal} {Phys. Rev. B}\ }\textbf {\bibinfo {volume} {49}},\ \bibinfo
  {pages} {6246} (\bibinfo {year} {1994})}\BibitemShut {NoStop}%
\bibitem [{\citenamefont {Perkins}\ \emph {et~al.}(1998)\citenamefont
  {Perkins}, \citenamefont {Birgeneau}, \citenamefont {Graybeal}, \citenamefont
  {Kastner},\ and\ \citenamefont {Kleinberg}}]{perkins1998}%
  \BibitemOpen
  \bibfield  {author} {\bibinfo {author} {\bibfnamefont {J.~D.}\ \bibnamefont
  {Perkins}}, \bibinfo {author} {\bibfnamefont {R.~J.}\ \bibnamefont
  {Birgeneau}}, \bibinfo {author} {\bibfnamefont {J.~M.}\ \bibnamefont
  {Graybeal}}, \bibinfo {author} {\bibfnamefont {M.~A.}\ \bibnamefont
  {Kastner}},\ and\ \bibinfo {author} {\bibfnamefont {D.~S.}\ \bibnamefont
  {Kleinberg}},\ }\bibfield  {title} {\bibinfo {title} {Midinfrared optical
  excitations in undoped lamellar copper oxides},\ }\href
  {https://doi.org/10.1103/PhysRevB.58.9390} {\bibfield  {journal} {\bibinfo
  {journal} {Phys. Rev. B}\ }\textbf {\bibinfo {volume} {58}},\ \bibinfo
  {pages} {9390} (\bibinfo {year} {1998})}\BibitemShut {NoStop}%
\bibitem [{\citenamefont {Waku}\ \emph {et~al.}(2004)\citenamefont {Waku},
  \citenamefont {Katsufuji}, \citenamefont {Kohsaka}, \citenamefont {Sasagawa},
  \citenamefont {Takagi}, \citenamefont {Kishida}, \citenamefont {Okamoto},
  \citenamefont {Azuma},\ and\ \citenamefont {Takano}}]{waku2004}%
  \BibitemOpen
  \bibfield  {author} {\bibinfo {author} {\bibfnamefont {K.}~\bibnamefont
  {Waku}}, \bibinfo {author} {\bibfnamefont {T.}~\bibnamefont {Katsufuji}},
  \bibinfo {author} {\bibfnamefont {Y.}~\bibnamefont {Kohsaka}}, \bibinfo
  {author} {\bibfnamefont {T.}~\bibnamefont {Sasagawa}}, \bibinfo {author}
  {\bibfnamefont {H.}~\bibnamefont {Takagi}}, \bibinfo {author} {\bibfnamefont
  {H.}~\bibnamefont {Kishida}}, \bibinfo {author} {\bibfnamefont
  {H.}~\bibnamefont {Okamoto}}, \bibinfo {author} {\bibfnamefont
  {M.}~\bibnamefont {Azuma}},\ and\ \bibinfo {author} {\bibfnamefont
  {M.}~\bibnamefont {Takano}},\ }\bibfield  {title} {\bibinfo {title} {Charge
  dynamics of
  {{${\mathrm{Ca}}_{2\ensuremath{-}x}{\mathrm{Na}}_{x}{\mathrm{CuO}}_{2}{\mathrm{Cl}}_{2}$}}
  as a correlated electron system with the ideal tetragonal lattice},\ }\href
  {https://doi.org/10.1103/PhysRevB.70.134501} {\bibfield  {journal} {\bibinfo
  {journal} {Phys. Rev. B}\ }\textbf {\bibinfo {volume} {70}},\ \bibinfo
  {pages} {134501} (\bibinfo {year} {2004})}\BibitemShut {NoStop}%
\bibitem [{\citenamefont {Schuster}\ \emph {et~al.}(2009)\citenamefont
  {Schuster}, \citenamefont {Pyon}, \citenamefont {Knupfer}, \citenamefont
  {Fink}, \citenamefont {Azuma}, \citenamefont {Takano}, \citenamefont
  {Takagi},\ and\ \citenamefont {B{\"u}chner}}]{schuster2009}%
  \BibitemOpen
  \bibfield  {author} {\bibinfo {author} {\bibfnamefont {R.}~\bibnamefont
  {Schuster}}, \bibinfo {author} {\bibfnamefont {S.}~\bibnamefont {Pyon}},
  \bibinfo {author} {\bibfnamefont {M.}~\bibnamefont {Knupfer}}, \bibinfo
  {author} {\bibfnamefont {J.}~\bibnamefont {Fink}}, \bibinfo {author}
  {\bibfnamefont {M.}~\bibnamefont {Azuma}}, \bibinfo {author} {\bibfnamefont
  {M.}~\bibnamefont {Takano}}, \bibinfo {author} {\bibfnamefont
  {H.}~\bibnamefont {Takagi}},\ and\ \bibinfo {author} {\bibfnamefont
  {B.}~\bibnamefont {B{\"u}chner}},\ }\bibfield  {title} {\bibinfo {title}
  {Charge-transfer excitons in underdoped
  {{${\text{Ca}}_{2\ensuremath{-}x}{\text{Na}}_{x}{\text{CuO}}_{2}{\text{Cl}}_{2}$}}
  studied by electron energy-loss spectroscopy},\ }\href
  {https://doi.org/10.1103/PhysRevB.79.214517} {\bibfield  {journal} {\bibinfo
  {journal} {Phys. Rev. B}\ }\textbf {\bibinfo {volume} {79}},\ \bibinfo
  {pages} {214517} (\bibinfo {year} {2009})}\BibitemShut {NoStop}%
\bibitem [{\citenamefont {Feenstra}\ and\ \citenamefont
  {Stroscio}(1987)}]{feenstra1987}%
  \BibitemOpen
  \bibfield  {author} {\bibinfo {author} {\bibfnamefont {R.~M.}\ \bibnamefont
  {Feenstra}}\ and\ \bibinfo {author} {\bibfnamefont {J.~A.}\ \bibnamefont
  {Stroscio}},\ }\bibfield  {title} {\bibinfo {title} {Tunneling spectroscopy
  of the {{GaAs}}(110) surface},\ }\href {https://doi.org/10.1116/1.583691}
  {\bibfield  {journal} {\bibinfo  {journal} {Journal of Vacuum Science \&
  Technology B: Microelectronics Processing and Phenomena}\ }\textbf {\bibinfo
  {volume} {5}},\ \bibinfo {pages} {923} (\bibinfo {year} {1987})}\BibitemShut
  {NoStop}%
\bibitem [{\citenamefont {Weimer}\ \emph {et~al.}(1989)\citenamefont {Weimer},
  \citenamefont {Kramar},\ and\ \citenamefont {Baldeschwieler}}]{weimer1989}%
  \BibitemOpen
  \bibfield  {author} {\bibinfo {author} {\bibfnamefont {M.}~\bibnamefont
  {Weimer}}, \bibinfo {author} {\bibfnamefont {J.}~\bibnamefont {Kramar}},\
  and\ \bibinfo {author} {\bibfnamefont {J.~D.}\ \bibnamefont
  {Baldeschwieler}},\ }\bibfield  {title} {\bibinfo {title} {Band bending and
  the apparent barrier height in scanning tunneling microscopy},\ }\href
  {https://doi.org/10.1103/PhysRevB.39.5572} {\bibfield  {journal} {\bibinfo
  {journal} {Phys. Rev. B}\ }\textbf {\bibinfo {volume} {39}},\ \bibinfo
  {pages} {5572} (\bibinfo {year} {1989})}\BibitemShut {NoStop}%
\bibitem [{\citenamefont {Kyung}\ \emph {et~al.}(2009)\citenamefont {Kyung},
  \citenamefont {S{\'e}n{\'e}chal},\ and\ \citenamefont
  {Tremblay}}]{kyung2009}%
  \BibitemOpen
  \bibfield  {author} {\bibinfo {author} {\bibfnamefont {B.}~\bibnamefont
  {Kyung}}, \bibinfo {author} {\bibfnamefont {D.}~\bibnamefont
  {S{\'e}n{\'e}chal}},\ and\ \bibinfo {author} {\bibfnamefont {A.-M.~S.}\
  \bibnamefont {Tremblay}},\ }\bibfield  {title} {\bibinfo {title} {Pairing
  dynamics in strongly correlated superconductivity},\ }\href
  {https://doi.org/10.1103/PhysRevB.80.205109} {\bibfield  {journal} {\bibinfo
  {journal} {Phys. Rev. B}\ }\textbf {\bibinfo {volume} {80}},\ \bibinfo
  {pages} {205109} (\bibinfo {year} {2009})}\BibitemShut {NoStop}%
\bibitem [{\citenamefont {Mu{\ss}hoff}\ \emph {et~al.}(2021)\citenamefont
  {Mu{\ss}hoff}, \citenamefont {Kiani},\ and\ \citenamefont
  {Pavarini}}]{musshoff2021a}%
  \BibitemOpen
  \bibfield  {author} {\bibinfo {author} {\bibfnamefont {J.}~\bibnamefont
  {Mu{\ss}hoff}}, \bibinfo {author} {\bibfnamefont {A.}~\bibnamefont {Kiani}},\
  and\ \bibinfo {author} {\bibfnamefont {E.}~\bibnamefont {Pavarini}},\
  }\bibfield  {title} {\bibinfo {title} {Magnetic response trends in cuprates
  and the {{$t-t'$ Hubbard}} model},\ }\href
  {https://doi.org/10.1103/PhysRevB.103.075136} {\bibfield  {journal} {\bibinfo
   {journal} {Phys. Rev. B}\ }\textbf {\bibinfo {volume} {103}},\ \bibinfo
  {pages} {075136} (\bibinfo {year} {2021})}\BibitemShut {NoStop}%
\bibitem [{\citenamefont {Lebert}\ \emph {et~al.}(2017)\citenamefont {Lebert},
  \citenamefont {Dean}, \citenamefont {Nicolaou}, \citenamefont {Pelliciari},
  \citenamefont {Dantz}, \citenamefont {Schmitt}, \citenamefont {Yu},
  \citenamefont {Azuma}, \citenamefont {Castellan}, \citenamefont {Miao},
  \citenamefont {Gauzzi}, \citenamefont {Baptiste},\ and\ \citenamefont
  {{d'Astuto}}}]{lebert2017a}%
  \BibitemOpen
  \bibfield  {author} {\bibinfo {author} {\bibfnamefont {B.~W.}\ \bibnamefont
  {Lebert}}, \bibinfo {author} {\bibfnamefont {M.~P.~M.}\ \bibnamefont {Dean}},
  \bibinfo {author} {\bibfnamefont {A.}~\bibnamefont {Nicolaou}}, \bibinfo
  {author} {\bibfnamefont {J.}~\bibnamefont {Pelliciari}}, \bibinfo {author}
  {\bibfnamefont {M.}~\bibnamefont {Dantz}}, \bibinfo {author} {\bibfnamefont
  {T.}~\bibnamefont {Schmitt}}, \bibinfo {author} {\bibfnamefont
  {R.}~\bibnamefont {Yu}}, \bibinfo {author} {\bibfnamefont {M.}~\bibnamefont
  {Azuma}}, \bibinfo {author} {\bibfnamefont {J.-P.}\ \bibnamefont
  {Castellan}}, \bibinfo {author} {\bibfnamefont {H.}~\bibnamefont {Miao}},
  \bibinfo {author} {\bibfnamefont {A.}~\bibnamefont {Gauzzi}}, \bibinfo
  {author} {\bibfnamefont {B.}~\bibnamefont {Baptiste}},\ and\ \bibinfo
  {author} {\bibfnamefont {M.}~\bibnamefont {{d'Astuto}}},\ }\bibfield  {title}
  {\bibinfo {title} {Resonant inelastic x-ray scattering study of spin-wave
  excitations in the cuprate parent compound {{Ca$_2$CuO$_2$Cl$_2$}}},\ }\href
  {https://doi.org/10.1103/PhysRevB.95.155110} {\bibfield  {journal} {\bibinfo
  {journal} {Phys. Rev. B}\ }\textbf {\bibinfo {volume} {95}},\ \bibinfo
  {pages} {155110} (\bibinfo {year} {2017})}\BibitemShut {NoStop}%
\bibitem [{\citenamefont {Lebert}\ \emph {et~al.}(2023)\citenamefont {Lebert},
  \citenamefont {{Bacq-Labreuil}}, \citenamefont {Dean}, \citenamefont
  {Ruotsalainen}, \citenamefont {Nicolaou}, \citenamefont {Huotari},
  \citenamefont {Yamada}, \citenamefont {Yamamoto}, \citenamefont {Azuma},
  \citenamefont {Brookes}, \citenamefont {Yakhou}, \citenamefont {Miao},
  \citenamefont {{Santos-Cottin}}, \citenamefont {Lenz}, \citenamefont
  {Biermann},\ and\ \citenamefont {{d'Astuto}}}]{lebert2023a}%
  \BibitemOpen
  \bibfield  {author} {\bibinfo {author} {\bibfnamefont {B.~W.}\ \bibnamefont
  {Lebert}}, \bibinfo {author} {\bibfnamefont {B.}~\bibnamefont
  {{Bacq-Labreuil}}}, \bibinfo {author} {\bibfnamefont {M.~P.~M.}\ \bibnamefont
  {Dean}}, \bibinfo {author} {\bibfnamefont {K.}~\bibnamefont {Ruotsalainen}},
  \bibinfo {author} {\bibfnamefont {A.}~\bibnamefont {Nicolaou}}, \bibinfo
  {author} {\bibfnamefont {S.}~\bibnamefont {Huotari}}, \bibinfo {author}
  {\bibfnamefont {I.}~\bibnamefont {Yamada}}, \bibinfo {author} {\bibfnamefont
  {H.}~\bibnamefont {Yamamoto}}, \bibinfo {author} {\bibfnamefont
  {M.}~\bibnamefont {Azuma}}, \bibinfo {author} {\bibfnamefont {N.~B.}\
  \bibnamefont {Brookes}}, \bibinfo {author} {\bibfnamefont {F.}~\bibnamefont
  {Yakhou}}, \bibinfo {author} {\bibfnamefont {H.}~\bibnamefont {Miao}},
  \bibinfo {author} {\bibfnamefont {D.}~\bibnamefont {{Santos-Cottin}}},
  \bibinfo {author} {\bibfnamefont {B.}~\bibnamefont {Lenz}}, \bibinfo {author}
  {\bibfnamefont {S.}~\bibnamefont {Biermann}},\ and\ \bibinfo {author}
  {\bibfnamefont {M.}~\bibnamefont {{d'Astuto}}},\ }\bibfield  {title}
  {\bibinfo {title} {Paramagnon dispersion and damping in doped
  {{Na$_x$Ca$_{2-x}$CuO$_2$Cl$_2$}}},\ }\href
  {https://doi.org/10.1103/PhysRevB.108.024506} {\bibfield  {journal} {\bibinfo
   {journal} {Phys. Rev. B}\ }\textbf {\bibinfo {volume} {108}},\ \bibinfo
  {pages} {024506} (\bibinfo {year} {2023})}\BibitemShut {NoStop}%
\bibitem [{\citenamefont {Kunisada}\ \emph {et~al.}(2017)\citenamefont
  {Kunisada}, \citenamefont {Adachi}, \citenamefont {Sakai}, \citenamefont
  {Sasaki}, \citenamefont {Nakayama}, \citenamefont {Akebi}, \citenamefont
  {Kuroda}, \citenamefont {Sasagawa}, \citenamefont {Watanabe}, \citenamefont
  {Shin},\ and\ \citenamefont {Kondo}}]{kunisada2017}%
  \BibitemOpen
  \bibfield  {author} {\bibinfo {author} {\bibfnamefont {S.}~\bibnamefont
  {Kunisada}}, \bibinfo {author} {\bibfnamefont {S.}~\bibnamefont {Adachi}},
  \bibinfo {author} {\bibfnamefont {S.}~\bibnamefont {Sakai}}, \bibinfo
  {author} {\bibfnamefont {N.}~\bibnamefont {Sasaki}}, \bibinfo {author}
  {\bibfnamefont {M.}~\bibnamefont {Nakayama}}, \bibinfo {author}
  {\bibfnamefont {S.}~\bibnamefont {Akebi}}, \bibinfo {author} {\bibfnamefont
  {K.}~\bibnamefont {Kuroda}}, \bibinfo {author} {\bibfnamefont
  {T.}~\bibnamefont {Sasagawa}}, \bibinfo {author} {\bibfnamefont
  {T.}~\bibnamefont {Watanabe}}, \bibinfo {author} {\bibfnamefont
  {S.}~\bibnamefont {Shin}},\ and\ \bibinfo {author} {\bibfnamefont
  {T.}~\bibnamefont {Kondo}},\ }\bibfield  {title} {\bibinfo {title}
  {Observation of {{Bogoliubov}} band hybridization in the optimally doped
  trilayer {{Bi$_2$Sr$_2$Ca$_2$Cu$_3$O$_{10+\delta}$}}},\ }\href
  {https://doi.org/10.1103/PhysRevLett.119.217001} {\bibfield  {journal}
  {\bibinfo  {journal} {Phys. Rev. Lett.}\ }\textbf {\bibinfo {volume} {119}},\
  \bibinfo {pages} {217001} (\bibinfo {year} {2017})}\BibitemShut {NoStop}%
\bibitem [{\citenamefont {Mostofi}\ \emph {et~al.}(2008)\citenamefont
  {Mostofi}, \citenamefont {Yates}, \citenamefont {Lee}, \citenamefont {Souza},
  \citenamefont {Vanderbilt},\ and\ \citenamefont {Marzari}}]{mostofi2008}%
  \BibitemOpen
  \bibfield  {author} {\bibinfo {author} {\bibfnamefont {A.~A.}\ \bibnamefont
  {Mostofi}}, \bibinfo {author} {\bibfnamefont {J.~R.}\ \bibnamefont {Yates}},
  \bibinfo {author} {\bibfnamefont {Y.-S.}\ \bibnamefont {Lee}}, \bibinfo
  {author} {\bibfnamefont {I.}~\bibnamefont {Souza}}, \bibinfo {author}
  {\bibfnamefont {D.}~\bibnamefont {Vanderbilt}},\ and\ \bibinfo {author}
  {\bibfnamefont {N.}~\bibnamefont {Marzari}},\ }\bibfield  {title} {\bibinfo
  {title} {Wannier90: {{A}} tool for obtaining maximally-localised {{Wannier}}
  functions},\ }\href {https://doi.org/10.1016/j.cpc.2007.11.016} {\bibfield
  {journal} {\bibinfo  {journal} {Computer Physics Communications}\ }\textbf
  {\bibinfo {volume} {178}},\ \bibinfo {pages} {685} (\bibinfo {year}
  {2008})}\BibitemShut {NoStop}%
\bibitem [{\citenamefont {Mostofi}\ \emph {et~al.}(2014)\citenamefont
  {Mostofi}, \citenamefont {Yates}, \citenamefont {Pizzi}, \citenamefont {Lee},
  \citenamefont {Souza}, \citenamefont {Vanderbilt},\ and\ \citenamefont
  {Marzari}}]{mostofi2014}%
  \BibitemOpen
  \bibfield  {author} {\bibinfo {author} {\bibfnamefont {A.~A.}\ \bibnamefont
  {Mostofi}}, \bibinfo {author} {\bibfnamefont {J.~R.}\ \bibnamefont {Yates}},
  \bibinfo {author} {\bibfnamefont {G.}~\bibnamefont {Pizzi}}, \bibinfo
  {author} {\bibfnamefont {Y.-S.}\ \bibnamefont {Lee}}, \bibinfo {author}
  {\bibfnamefont {I.}~\bibnamefont {Souza}}, \bibinfo {author} {\bibfnamefont
  {D.}~\bibnamefont {Vanderbilt}},\ and\ \bibinfo {author} {\bibfnamefont
  {N.}~\bibnamefont {Marzari}},\ }\bibfield  {title} {\bibinfo {title} {An
  updated version of wannier90: {{A}} tool for obtaining maximally-localised
  {{Wannier}} functions},\ }\href {https://doi.org/10.1016/j.cpc.2014.05.003}
  {\bibfield  {journal} {\bibinfo  {journal} {Computer Physics Communications}\
  }\textbf {\bibinfo {volume} {185}},\ \bibinfo {pages} {2309} (\bibinfo {year}
  {2014})}\BibitemShut {NoStop}%
\bibitem [{\citenamefont {Gauthier}\ \emph {et~al.}(2020)\citenamefont
  {Gauthier}, \citenamefont {Sobota}, \citenamefont {Hashimoto}, \citenamefont
  {Pfau}, \citenamefont {Lu}, \citenamefont {Bauer}, \citenamefont {Ronning},
  \citenamefont {Kirchmann},\ and\ \citenamefont {Shen}}]{gauthier2020}%
  \BibitemOpen
  \bibfield  {author} {\bibinfo {author} {\bibfnamefont {N.}~\bibnamefont
  {Gauthier}}, \bibinfo {author} {\bibfnamefont {J.~A.}\ \bibnamefont
  {Sobota}}, \bibinfo {author} {\bibfnamefont {M.}~\bibnamefont {Hashimoto}},
  \bibinfo {author} {\bibfnamefont {H.}~\bibnamefont {Pfau}}, \bibinfo {author}
  {\bibfnamefont {D.-H.}\ \bibnamefont {Lu}}, \bibinfo {author} {\bibfnamefont
  {E.~D.}\ \bibnamefont {Bauer}}, \bibinfo {author} {\bibfnamefont
  {F.}~\bibnamefont {Ronning}}, \bibinfo {author} {\bibfnamefont {P.~S.}\
  \bibnamefont {Kirchmann}},\ and\ \bibinfo {author} {\bibfnamefont {Z.-X.}\
  \bibnamefont {Shen}},\ }\bibfield  {title} {\bibinfo {title} {Quantum-well
  states in fractured crystals of the heavy-fermion material {{CeCoIn$_5$}}},\
  }\href {https://doi.org/10.1103/PhysRevB.102.125111} {\bibfield  {journal}
  {\bibinfo  {journal} {Phys. Rev. B}\ }\textbf {\bibinfo {volume} {102}},\
  \bibinfo {pages} {125111} (\bibinfo {year} {2020})}\BibitemShut {NoStop}%
\bibitem [{\citenamefont {Chiang}(2000)}]{chiang2000}%
  \BibitemOpen
  \bibfield  {author} {\bibinfo {author} {\bibfnamefont {T.~C.}\ \bibnamefont
  {Chiang}},\ }\bibfield  {title} {\bibinfo {title} {Photoemission studies of
  quantum well states in thin films},\ }\href
  {https://doi.org/10.1016/S0167-5729(00)00006-6} {\bibfield  {journal}
  {\bibinfo  {journal} {Surface Science Reports}\ }\textbf {\bibinfo {volume}
  {39}},\ \bibinfo {pages} {181} (\bibinfo {year} {2000})}\BibitemShut
  {NoStop}%
\bibitem [{\citenamefont {Kawakami}\ \emph {et~al.}(1999)\citenamefont
  {Kawakami}, \citenamefont {Rotenberg}, \citenamefont {Choi}, \citenamefont
  {{Escorcia-Aparicio}}, \citenamefont {Bowen}, \citenamefont {Wolfe},
  \citenamefont {Arenholz}, \citenamefont {Zhang}, \citenamefont {Smith},\ and\
  \citenamefont {Qiu}}]{kawakami1999}%
  \BibitemOpen
  \bibfield  {author} {\bibinfo {author} {\bibfnamefont {R.~K.}\ \bibnamefont
  {Kawakami}}, \bibinfo {author} {\bibfnamefont {E.}~\bibnamefont {Rotenberg}},
  \bibinfo {author} {\bibfnamefont {H.~J.}\ \bibnamefont {Choi}}, \bibinfo
  {author} {\bibfnamefont {E.~J.}\ \bibnamefont {{Escorcia-Aparicio}}},
  \bibinfo {author} {\bibfnamefont {M.~O.}\ \bibnamefont {Bowen}}, \bibinfo
  {author} {\bibfnamefont {J.~H.}\ \bibnamefont {Wolfe}}, \bibinfo {author}
  {\bibfnamefont {E.}~\bibnamefont {Arenholz}}, \bibinfo {author}
  {\bibfnamefont {Z.~D.}\ \bibnamefont {Zhang}}, \bibinfo {author}
  {\bibfnamefont {N.~V.}\ \bibnamefont {Smith}},\ and\ \bibinfo {author}
  {\bibfnamefont {Z.~Q.}\ \bibnamefont {Qiu}},\ }\bibfield  {title} {\bibinfo
  {title} {Quantum-well states in copper thin films},\ }\href
  {https://doi.org/10.1038/18178} {\bibfield  {journal} {\bibinfo  {journal}
  {Nature}\ }\textbf {\bibinfo {volume} {398}},\ \bibinfo {pages} {132}
  (\bibinfo {year} {1999})}\BibitemShut {NoStop}%
\bibitem [{\citenamefont {Paggel}\ \emph {et~al.}(1999)\citenamefont {Paggel},
  \citenamefont {Miller},\ and\ \citenamefont {Chiang}}]{paggel1999}%
  \BibitemOpen
  \bibfield  {author} {\bibinfo {author} {\bibfnamefont {J.~J.}\ \bibnamefont
  {Paggel}}, \bibinfo {author} {\bibfnamefont {T.}~\bibnamefont {Miller}},\
  and\ \bibinfo {author} {\bibfnamefont {T.-C.}\ \bibnamefont {Chiang}},\
  }\bibfield  {title} {\bibinfo {title} {Quantum-well states as
  {{Fabry-P{\'e}rot}} modes in a thin-film electron interferometer},\ }\href
  {https://doi.org/10.1126/science.283.5408.1709} {\bibfield  {journal}
  {\bibinfo  {journal} {Science}\ }\textbf {\bibinfo {volume} {283}},\ \bibinfo
  {pages} {1709} (\bibinfo {year} {1999})}\BibitemShut {NoStop}%
\bibitem [{\citenamefont {LaBollita}\ and\ \citenamefont
  {Botana}(2022)}]{labollita2022}%
  \BibitemOpen
  \bibfield  {author} {\bibinfo {author} {\bibfnamefont {H.}~\bibnamefont
  {LaBollita}}\ and\ \bibinfo {author} {\bibfnamefont {A.~S.}\ \bibnamefont
  {Botana}},\ }\bibfield  {title} {\bibinfo {title} {Correlated electronic
  structure of a quintuple-layer nickelate},\ }\href
  {https://doi.org/10.1103/PhysRevB.105.085118} {\bibfield  {journal} {\bibinfo
   {journal} {Phys. Rev. B}\ }\textbf {\bibinfo {volume} {105}},\ \bibinfo
  {pages} {085118} (\bibinfo {year} {2022})}\BibitemShut {NoStop}%
\bibitem [{\citenamefont {Singh}(1993)}]{singh1993}%
  \BibitemOpen
  \bibfield  {author} {\bibinfo {author} {\bibfnamefont {D.~J.}\ \bibnamefont
  {Singh}},\ }\bibfield  {title} {\bibinfo {title} {Electronic structure of
  {{${\mathrm{HgBa}}_{2}$${\mathrm{Ca}}_{2}$${\mathrm{Cu}}_{3}$${\mathrm{O}}_{8}$}}:
  The role of mercury},\ }\href {https://doi.org/10.1103/PhysRevB.48.3571}
  {\bibfield  {journal} {\bibinfo  {journal} {Phys. Rev. B}\ }\textbf {\bibinfo
  {volume} {48}},\ \bibinfo {pages} {3571} (\bibinfo {year}
  {1993})}\BibitemShut {NoStop}%
\bibitem [{\citenamefont {Singh}\ and\ \citenamefont
  {Pickett}(1994)}]{singh1994}%
  \BibitemOpen
  \bibfield  {author} {\bibinfo {author} {\bibfnamefont {D.~J.}\ \bibnamefont
  {Singh}}\ and\ \bibinfo {author} {\bibfnamefont {W.~E.}\ \bibnamefont
  {Pickett}},\ }\bibfield  {title} {\bibinfo {title} {Unconventional oxygen
  doping behavior in
  {{$\mathrm{Hg}{\mathrm{Ba}}_{2}{\mathrm{Ca}}_{2}{\mathrm{Cu}}_{3}{\mathrm{O}}_{8+\ensuremath{\delta}}$}}},\
  }\href {https://doi.org/10.1103/PhysRevLett.73.476} {\bibfield  {journal}
  {\bibinfo  {journal} {Phys. Rev. Lett.}\ }\textbf {\bibinfo {volume} {73}},\
  \bibinfo {pages} {476} (\bibinfo {year} {1994})}\BibitemShut {NoStop}%
\bibitem [{\citenamefont {Chan}\ \emph {et~al.}(1991)\citenamefont {Chan},
  \citenamefont {Harshman}, \citenamefont {Lynn}, \citenamefont {Massidda},\
  and\ \citenamefont {Mitzi}}]{chan1991}%
  \BibitemOpen
  \bibfield  {author} {\bibinfo {author} {\bibfnamefont {L.~P.}\ \bibnamefont
  {Chan}}, \bibinfo {author} {\bibfnamefont {D.~R.}\ \bibnamefont {Harshman}},
  \bibinfo {author} {\bibfnamefont {K.~G.}\ \bibnamefont {Lynn}}, \bibinfo
  {author} {\bibfnamefont {S.}~\bibnamefont {Massidda}},\ and\ \bibinfo
  {author} {\bibfnamefont {D.~B.}\ \bibnamefont {Mitzi}},\ }\bibfield  {title}
  {\bibinfo {title} {Pair momentum distribution in
  {{${\mathrm{Bi}}_{2}$${\mathrm{Sr}}_{2}$${\mathrm{CaCu}}_{2}$${\mathrm{O}}_{8+\mathrm{\ensuremath{\delta}}}$}}
  measured by positron annihilation: Existence and nature of the fermi
  surface},\ }\href {https://doi.org/10.1103/PhysRevLett.67.1350} {\bibfield
  {journal} {\bibinfo  {journal} {Phys. Rev. Lett.}\ }\textbf {\bibinfo
  {volume} {67}},\ \bibinfo {pages} {1350} (\bibinfo {year}
  {1991})}\BibitemShut {NoStop}%
\bibitem [{\citenamefont {Lin}\ \emph {et~al.}(2006)\citenamefont {Lin},
  \citenamefont {Sahrakorpi}, \citenamefont {Markiewicz},\ and\ \citenamefont
  {Bansil}}]{lin2006}%
  \BibitemOpen
  \bibfield  {author} {\bibinfo {author} {\bibfnamefont {H.}~\bibnamefont
  {Lin}}, \bibinfo {author} {\bibfnamefont {S.}~\bibnamefont {Sahrakorpi}},
  \bibinfo {author} {\bibfnamefont {R.~S.}\ \bibnamefont {Markiewicz}},\ and\
  \bibinfo {author} {\bibfnamefont {A.}~\bibnamefont {Bansil}},\ }\bibfield
  {title} {\bibinfo {title} {Raising {{Bi-O}} bands above the fermi energy
  level of hole-doped
  {{${\mathrm{Bi}}_{2}{\mathrm{Sr}}_{2}{\mathrm{CaCu}}_{2}{\mathrm{O}}_{8+\ensuremath{\delta}}$}}
  and other cuprate superconductors},\ }\href
  {https://doi.org/10.1103/PhysRevLett.96.097001} {\bibfield  {journal}
  {\bibinfo  {journal} {Phys. Rev. Lett.}\ }\textbf {\bibinfo {volume} {96}},\
  \bibinfo {pages} {097001} (\bibinfo {year} {2006})}\BibitemShut {NoStop}%
\bibitem [{\citenamefont {Nokelainen}\ \emph {et~al.}(2020)\citenamefont
  {Nokelainen}, \citenamefont {Lane}, \citenamefont {Markiewicz}, \citenamefont
  {Barbiellini}, \citenamefont {Pulkkinen}, \citenamefont {Singh},
  \citenamefont {Sun}, \citenamefont {Pussi},\ and\ \citenamefont
  {Bansil}}]{nokelainen2020}%
  \BibitemOpen
  \bibfield  {author} {\bibinfo {author} {\bibfnamefont {J.}~\bibnamefont
  {Nokelainen}}, \bibinfo {author} {\bibfnamefont {C.}~\bibnamefont {Lane}},
  \bibinfo {author} {\bibfnamefont {R.~S.}\ \bibnamefont {Markiewicz}},
  \bibinfo {author} {\bibfnamefont {B.}~\bibnamefont {Barbiellini}}, \bibinfo
  {author} {\bibfnamefont {A.}~\bibnamefont {Pulkkinen}}, \bibinfo {author}
  {\bibfnamefont {B.}~\bibnamefont {Singh}}, \bibinfo {author} {\bibfnamefont
  {J.}~\bibnamefont {Sun}}, \bibinfo {author} {\bibfnamefont {K.}~\bibnamefont
  {Pussi}},\ and\ \bibinfo {author} {\bibfnamefont {A.}~\bibnamefont
  {Bansil}},\ }\bibfield  {title} {\bibinfo {title} {Ab initio description of
  the {{Bi$_2$Sr$_2$CaCu$_2$O$_{8+\delta}$ }} electronic structure},\ }\href
  {https://doi.org/10.1103/PhysRevB.101.214523} {\bibfield  {journal} {\bibinfo
   {journal} {Phys. Rev. B}\ }\textbf {\bibinfo {volume} {101}},\ \bibinfo
  {pages} {214523} (\bibinfo {year} {2020})}\BibitemShut {NoStop}%
\bibitem [{\citenamefont {Lee}\ \emph {et~al.}(2006)\citenamefont {Lee},
  \citenamefont {Nagaosa},\ and\ \citenamefont {Wen}}]{lee2006}%
  \BibitemOpen
  \bibfield  {author} {\bibinfo {author} {\bibfnamefont {P.~A.}\ \bibnamefont
  {Lee}}, \bibinfo {author} {\bibfnamefont {N.}~\bibnamefont {Nagaosa}},\ and\
  \bibinfo {author} {\bibfnamefont {X.-G.}\ \bibnamefont {Wen}},\ }\bibfield
  {title} {\bibinfo {title} {Doping a {{Mott}} insulator: Physics of
  high-temperature superconductivity},\ }\href
  {https://doi.org/10.1103/RevModPhys.78.17} {\bibfield  {journal} {\bibinfo
  {journal} {Rev. Mod. Phys.}\ }\textbf {\bibinfo {volume} {78}},\ \bibinfo
  {pages} {17} (\bibinfo {year} {2006})}\BibitemShut {NoStop}%
\bibitem [{\citenamefont {Sordi}\ \emph {et~al.}(2010)\citenamefont {Sordi},
  \citenamefont {Haule},\ and\ \citenamefont {Tremblay}}]{sordi2010}%
  \BibitemOpen
  \bibfield  {author} {\bibinfo {author} {\bibfnamefont {G.}~\bibnamefont
  {Sordi}}, \bibinfo {author} {\bibfnamefont {K.}~\bibnamefont {Haule}},\ and\
  \bibinfo {author} {\bibfnamefont {A.-M.~S.}\ \bibnamefont {Tremblay}},\
  }\bibfield  {title} {\bibinfo {title} {Finite doping signatures of the
  {{Mott}} transition in the two-dimensional {{Hubbard}} model},\ }\href
  {https://doi.org/10.1103/PhysRevLett.104.226402} {\bibfield  {journal}
  {\bibinfo  {journal} {Phys. Rev. Lett.}\ }\textbf {\bibinfo {volume} {104}},\
  \bibinfo {pages} {226402} (\bibinfo {year} {2010})}\BibitemShut {NoStop}%
\bibitem [{\citenamefont {Alloul}\ \emph {et~al.}(1989)\citenamefont {Alloul},
  \citenamefont {Ohno},\ and\ \citenamefont {Mendels}}]{alloul1989}%
  \BibitemOpen
  \bibfield  {author} {\bibinfo {author} {\bibfnamefont {H.}~\bibnamefont
  {Alloul}}, \bibinfo {author} {\bibfnamefont {T.}~\bibnamefont {Ohno}},\ and\
  \bibinfo {author} {\bibfnamefont {P.}~\bibnamefont {Mendels}},\ }\bibfield
  {title} {\bibinfo {title} {{{$^{89}$Y NMR}} evidence for a {{Fermi}}-liquid
  behavior in {{YBa$_2$Cu$_3$O$_{6+x}$}}},\ }\href
  {https://doi.org/10.1103/PhysRevLett.63.1700} {\bibfield  {journal} {\bibinfo
   {journal} {Phys. Rev. Lett.}\ }\textbf {\bibinfo {volume} {63}},\ \bibinfo
  {pages} {1700} (\bibinfo {year} {1989})}\BibitemShut {NoStop}%
\bibitem [{\citenamefont {Warren}\ \emph {et~al.}(1989)\citenamefont {Warren},
  \citenamefont {Walstedt}, \citenamefont {Brennert}, \citenamefont {Cava},
  \citenamefont {Tycko}, \citenamefont {Bell},\ and\ \citenamefont
  {Dabbagh}}]{warren1989}%
  \BibitemOpen
  \bibfield  {author} {\bibinfo {author} {\bibfnamefont {W.~W.}\ \bibnamefont
  {Warren}}, \bibinfo {author} {\bibfnamefont {R.~E.}\ \bibnamefont
  {Walstedt}}, \bibinfo {author} {\bibfnamefont {G.~F.}\ \bibnamefont
  {Brennert}}, \bibinfo {author} {\bibfnamefont {R.~J.}\ \bibnamefont {Cava}},
  \bibinfo {author} {\bibfnamefont {R.}~\bibnamefont {Tycko}}, \bibinfo
  {author} {\bibfnamefont {R.~F.}\ \bibnamefont {Bell}},\ and\ \bibinfo
  {author} {\bibfnamefont {G.}~\bibnamefont {Dabbagh}},\ }\bibfield  {title}
  {\bibinfo {title} {Cu spin dynamics and superconducting precursor effects in
  planes above {{$T_C$}} in {{YBa$_2$Cu$_3$O$_{6.7}$}}},\ }\href
  {https://doi.org/10.1103/PhysRevLett.62.1193} {\bibfield  {journal} {\bibinfo
   {journal} {Phys. Rev. Lett.}\ }\textbf {\bibinfo {volume} {62}},\ \bibinfo
  {pages} {1193} (\bibinfo {year} {1989})}\BibitemShut {NoStop}%
\bibitem [{\citenamefont {Ito}\ \emph {et~al.}(1993)\citenamefont {Ito},
  \citenamefont {Takenaka},\ and\ \citenamefont {Uchida}}]{ito1993}%
  \BibitemOpen
  \bibfield  {author} {\bibinfo {author} {\bibfnamefont {T.}~\bibnamefont
  {Ito}}, \bibinfo {author} {\bibfnamefont {K.}~\bibnamefont {Takenaka}},\ and\
  \bibinfo {author} {\bibfnamefont {S.}~\bibnamefont {Uchida}},\ }\bibfield
  {title} {\bibinfo {title} {Systematic deviation from {{T-linear}} behavior in
  the in-plane resistivity of {{YBa$_2$Cu$_3$O$_{7-y}$}}: {{Evidence}} for
  dominant spin scattering},\ }\href
  {https://doi.org/10.1103/PhysRevLett.70.3995} {\bibfield  {journal} {\bibinfo
   {journal} {Phys. Rev. Lett.}\ }\textbf {\bibinfo {volume} {70}},\ \bibinfo
  {pages} {3995} (\bibinfo {year} {1993})}\BibitemShut {NoStop}%
\bibitem [{\citenamefont {Timusk}\ and\ \citenamefont
  {Statt}(1999)}]{timusk1999}%
  \BibitemOpen
  \bibfield  {author} {\bibinfo {author} {\bibfnamefont {T.}~\bibnamefont
  {Timusk}}\ and\ \bibinfo {author} {\bibfnamefont {B.}~\bibnamefont {Statt}},\
  }\bibfield  {title} {\bibinfo {title} {The pseudogap in high-temperature
  superconductors: An experimental survey},\ }\href
  {https://doi.org/10.1088/0034-4885/62/1/002} {\bibfield  {journal} {\bibinfo
  {journal} {Rep. Prog. Phys.}\ }\textbf {\bibinfo {volume} {62}},\ \bibinfo
  {pages} {61} (\bibinfo {year} {1999})}\BibitemShut {NoStop}%
\bibitem [{\citenamefont {Huscroft}\ \emph {et~al.}(2001)\citenamefont
  {Huscroft}, \citenamefont {Jarrell}, \citenamefont {Maier}, \citenamefont
  {Moukouri},\ and\ \citenamefont {Tahvildarzadeh}}]{Huscroft2001}%
  \BibitemOpen
  \bibfield  {author} {\bibinfo {author} {\bibfnamefont {C.}~\bibnamefont
  {Huscroft}}, \bibinfo {author} {\bibfnamefont {M.}~\bibnamefont {Jarrell}},
  \bibinfo {author} {\bibfnamefont {{\relax Th}.}~\bibnamefont {Maier}},
  \bibinfo {author} {\bibfnamefont {S.}~\bibnamefont {Moukouri}},\ and\
  \bibinfo {author} {\bibfnamefont {A.~N.}\ \bibnamefont {Tahvildarzadeh}},\
  }\bibfield  {title} {\bibinfo {title} {Pseudogaps in the {{2D Hubbard}}
  model},\ }\href {https://doi.org/10.1103/PhysRevLett.86.139} {\bibfield
  {journal} {\bibinfo  {journal} {Phys. Rev. Lett.}\ }\textbf {\bibinfo
  {volume} {86}},\ \bibinfo {pages} {139} (\bibinfo {year} {2001})}\BibitemShut
  {NoStop}%
\bibitem [{\citenamefont {Kohsaka}\ \emph {et~al.}(2003)\citenamefont
  {Kohsaka}, \citenamefont {Sasagawa}, \citenamefont {Ronning}, \citenamefont
  {Yoshida}, \citenamefont {Kim}, \citenamefont {Hanaguri}, \citenamefont
  {Azuma}, \citenamefont {Takano}, \citenamefont {Xun~Shen},\ and\
  \citenamefont {Takagi}}]{kohsaka2003}%
  \BibitemOpen
  \bibfield  {author} {\bibinfo {author} {\bibfnamefont {Y.}~\bibnamefont
  {Kohsaka}}, \bibinfo {author} {\bibfnamefont {T.}~\bibnamefont {Sasagawa}},
  \bibinfo {author} {\bibfnamefont {F.}~\bibnamefont {Ronning}}, \bibinfo
  {author} {\bibfnamefont {T.}~\bibnamefont {Yoshida}}, \bibinfo {author}
  {\bibfnamefont {C.}~\bibnamefont {Kim}}, \bibinfo {author} {\bibfnamefont
  {T.}~\bibnamefont {Hanaguri}}, \bibinfo {author} {\bibfnamefont
  {M.}~\bibnamefont {Azuma}}, \bibinfo {author} {\bibfnamefont
  {M.}~\bibnamefont {Takano}}, \bibinfo {author} {\bibfnamefont
  {Z.}~\bibnamefont {Xun~Shen}},\ and\ \bibinfo {author} {\bibfnamefont
  {H.}~\bibnamefont {Takagi}},\ }\bibfield  {title} {\bibinfo {title}
  {Angle-{{Resolved Photoemission Spectroscopy}} of
  ({{Ca}},{{Na}}){\textsubscript{2}}{{CuO}}{\textsubscript{2}}{{Cl}}{\textsubscript{2}}
  {{Crystals}}: {{Fingerprints}} of a {{Magnetic Insulator}} in a {{Heavily
  Underdoped Superconductor}}},\ }\href {https://doi.org/10.1143/JPSJ.72.1018}
  {\bibfield  {journal} {\bibinfo  {journal} {J. Phys. Soc. Jpn.}\ }\textbf
  {\bibinfo {volume} {72}},\ \bibinfo {pages} {1018} (\bibinfo {year}
  {2003})}\BibitemShut {NoStop}%
\bibitem [{\citenamefont {Graf}\ \emph
  {et~al.}(2007{\natexlab{a}})\citenamefont {Graf}, \citenamefont {Gweon},\
  and\ \citenamefont {Lanzara}}]{graf2007a}%
  \BibitemOpen
  \bibfield  {author} {\bibinfo {author} {\bibfnamefont {J.}~\bibnamefont
  {Graf}}, \bibinfo {author} {\bibfnamefont {G.-H.}\ \bibnamefont {Gweon}},\
  and\ \bibinfo {author} {\bibfnamefont {A.}~\bibnamefont {Lanzara}},\
  }\bibfield  {title} {\bibinfo {title} {Universal waterfall-like feature in
  the spectral function of high temperature superconductors},\ }\href
  {https://doi.org/https://doi.org/10.1016/j.physc.2007.03.005} {\bibfield
  {journal} {\bibinfo  {journal} {Physica C: Superconductivity and its
  applications}\ }\textbf {\bibinfo {volume} {460}},\ \bibinfo {pages} {194}
  (\bibinfo {year} {2007}{\natexlab{a}})}\BibitemShut {NoStop}%
\bibitem [{\citenamefont {Graf}\ \emph
  {et~al.}(2007{\natexlab{b}})\citenamefont {Graf}, \citenamefont {Gweon},
  \citenamefont {McElroy}, \citenamefont {Zhou}, \citenamefont {Jozwiak},
  \citenamefont {Rotenberg}, \citenamefont {Bill}, \citenamefont {Sasagawa},
  \citenamefont {Eisaki}, \citenamefont {Uchida}, \citenamefont {Takagi},
  \citenamefont {Lee},\ and\ \citenamefont {Lanzara}}]{graf2007}%
  \BibitemOpen
  \bibfield  {author} {\bibinfo {author} {\bibfnamefont {J.}~\bibnamefont
  {Graf}}, \bibinfo {author} {\bibfnamefont {G.-H.}\ \bibnamefont {Gweon}},
  \bibinfo {author} {\bibfnamefont {K.}~\bibnamefont {McElroy}}, \bibinfo
  {author} {\bibfnamefont {S.~Y.}\ \bibnamefont {Zhou}}, \bibinfo {author}
  {\bibfnamefont {C.}~\bibnamefont {Jozwiak}}, \bibinfo {author} {\bibfnamefont
  {E.}~\bibnamefont {Rotenberg}}, \bibinfo {author} {\bibfnamefont
  {A.}~\bibnamefont {Bill}}, \bibinfo {author} {\bibfnamefont {T.}~\bibnamefont
  {Sasagawa}}, \bibinfo {author} {\bibfnamefont {H.}~\bibnamefont {Eisaki}},
  \bibinfo {author} {\bibfnamefont {S.}~\bibnamefont {Uchida}}, \bibinfo
  {author} {\bibfnamefont {H.}~\bibnamefont {Takagi}}, \bibinfo {author}
  {\bibfnamefont {D.-H.}\ \bibnamefont {Lee}},\ and\ \bibinfo {author}
  {\bibfnamefont {A.}~\bibnamefont {Lanzara}},\ }\bibfield  {title} {\bibinfo
  {title} {Universal high energy anomaly in the angle-resolved photoemission
  spectra of high temperature superconductors: Possible evidence of spinon and
  holon branches},\ }\href {https://doi.org/10.1103/PhysRevLett.98.067004}
  {\bibfield  {journal} {\bibinfo  {journal} {Phys. Rev. Lett.}\ }\textbf
  {\bibinfo {volume} {98}},\ \bibinfo {pages} {067004} (\bibinfo {year}
  {2007}{\natexlab{b}})}\BibitemShut {NoStop}%
\bibitem [{\citenamefont {Kordyuk}\ \emph {et~al.}(2006)\citenamefont
  {Kordyuk}, \citenamefont {Borisenko}, \citenamefont {Zabolotnyy},
  \citenamefont {Geck}, \citenamefont {Knupfer}, \citenamefont {Fink},
  \citenamefont {B\"uchner}, \citenamefont {Lin}, \citenamefont {Keimer},
  \citenamefont {Berger}, \citenamefont {Pan}, \citenamefont {Komiya},\ and\
  \citenamefont {Ando}}]{kordyuk2006}%
  \BibitemOpen
  \bibfield  {author} {\bibinfo {author} {\bibfnamefont {A.~A.}\ \bibnamefont
  {Kordyuk}}, \bibinfo {author} {\bibfnamefont {S.~V.}\ \bibnamefont
  {Borisenko}}, \bibinfo {author} {\bibfnamefont {V.~B.}\ \bibnamefont
  {Zabolotnyy}}, \bibinfo {author} {\bibfnamefont {J.}~\bibnamefont {Geck}},
  \bibinfo {author} {\bibfnamefont {M.}~\bibnamefont {Knupfer}}, \bibinfo
  {author} {\bibfnamefont {J.}~\bibnamefont {Fink}}, \bibinfo {author}
  {\bibfnamefont {B.}~\bibnamefont {B\"uchner}}, \bibinfo {author}
  {\bibfnamefont {C.~T.}\ \bibnamefont {Lin}}, \bibinfo {author} {\bibfnamefont
  {B.}~\bibnamefont {Keimer}}, \bibinfo {author} {\bibfnamefont
  {H.}~\bibnamefont {Berger}}, \bibinfo {author} {\bibfnamefont {A.~V.}\
  \bibnamefont {Pan}}, \bibinfo {author} {\bibfnamefont {S.}~\bibnamefont
  {Komiya}},\ and\ \bibinfo {author} {\bibfnamefont {Y.}~\bibnamefont {Ando}},\
  }\bibfield  {title} {\bibinfo {title} {Constituents of the quasiparticle
  spectrum along the nodal direction of high-${T}_{c}$ cuprates},\ }\href
  {https://doi.org/10.1103/PhysRevLett.97.017002} {\bibfield  {journal}
  {\bibinfo  {journal} {Phys. Rev. Lett.}\ }\textbf {\bibinfo {volume} {97}},\
  \bibinfo {pages} {017002} (\bibinfo {year} {2006})}\BibitemShut {NoStop}%
\bibitem [{\citenamefont {Borisenko}\ \emph
  {et~al.}(2006{\natexlab{a}})\citenamefont {Borisenko}, \citenamefont
  {Kordyuk}, \citenamefont {Zabolotnyy}, \citenamefont {Geck}, \citenamefont
  {Inosov}, \citenamefont {Koitzsch}, \citenamefont {Fink}, \citenamefont
  {Knupfer}, \citenamefont {B\"uchner}, \citenamefont {Hinkov}, \citenamefont
  {Lin}, \citenamefont {Keimer}, \citenamefont {Wolf}, \citenamefont
  {Chiuzb\ifmmode~\u{a}\else \u{a}\fi{}ian}, \citenamefont {Patthey},\ and\
  \citenamefont {Follath}}]{borisenko2006a}%
  \BibitemOpen
  \bibfield  {author} {\bibinfo {author} {\bibfnamefont {S.~V.}\ \bibnamefont
  {Borisenko}}, \bibinfo {author} {\bibfnamefont {A.~A.}\ \bibnamefont
  {Kordyuk}}, \bibinfo {author} {\bibfnamefont {V.}~\bibnamefont {Zabolotnyy}},
  \bibinfo {author} {\bibfnamefont {J.}~\bibnamefont {Geck}}, \bibinfo {author}
  {\bibfnamefont {D.}~\bibnamefont {Inosov}}, \bibinfo {author} {\bibfnamefont
  {A.}~\bibnamefont {Koitzsch}}, \bibinfo {author} {\bibfnamefont
  {J.}~\bibnamefont {Fink}}, \bibinfo {author} {\bibfnamefont {M.}~\bibnamefont
  {Knupfer}}, \bibinfo {author} {\bibfnamefont {B.}~\bibnamefont {B\"uchner}},
  \bibinfo {author} {\bibfnamefont {V.}~\bibnamefont {Hinkov}}, \bibinfo
  {author} {\bibfnamefont {C.~T.}\ \bibnamefont {Lin}}, \bibinfo {author}
  {\bibfnamefont {B.}~\bibnamefont {Keimer}}, \bibinfo {author} {\bibfnamefont
  {T.}~\bibnamefont {Wolf}}, \bibinfo {author} {\bibfnamefont {S.~G.}\
  \bibnamefont {Chiuzb\ifmmode~\u{a}\else \u{a}\fi{}ian}}, \bibinfo {author}
  {\bibfnamefont {L.}~\bibnamefont {Patthey}},\ and\ \bibinfo {author}
  {\bibfnamefont {R.}~\bibnamefont {Follath}},\ }\bibfield  {title} {\bibinfo
  {title} {Kinks, nodal bilayer splitting, and interband scattering in
  {{${\mathrm{YBa}}_{2}{\mathrm{Cu}}_{3}{\mathrm{O}}_{6+x}$}}},\ }\href
  {https://doi.org/10.1103/PhysRevLett.96.117004} {\bibfield  {journal}
  {\bibinfo  {journal} {Phys. Rev. Lett.}\ }\textbf {\bibinfo {volume} {96}},\
  \bibinfo {pages} {117004} (\bibinfo {year} {2006}{\natexlab{a}})}\BibitemShut
  {NoStop}%
\bibitem [{\citenamefont {Borisenko}\ \emph
  {et~al.}(2006{\natexlab{b}})\citenamefont {Borisenko}, \citenamefont
  {Kordyuk}, \citenamefont {Koitzsch}, \citenamefont {Fink}, \citenamefont
  {Geck}, \citenamefont {Zabolotnyy}, \citenamefont {Knupfer}, \citenamefont
  {B\"uchner}, \citenamefont {Berger}, \citenamefont {Falub}, \citenamefont
  {Shi}, \citenamefont {Krempasky},\ and\ \citenamefont
  {Patthey}}]{borisenko2006b}%
  \BibitemOpen
  \bibfield  {author} {\bibinfo {author} {\bibfnamefont {S.~V.}\ \bibnamefont
  {Borisenko}}, \bibinfo {author} {\bibfnamefont {A.~A.}\ \bibnamefont
  {Kordyuk}}, \bibinfo {author} {\bibfnamefont {A.}~\bibnamefont {Koitzsch}},
  \bibinfo {author} {\bibfnamefont {J.}~\bibnamefont {Fink}}, \bibinfo {author}
  {\bibfnamefont {J.}~\bibnamefont {Geck}}, \bibinfo {author} {\bibfnamefont
  {V.}~\bibnamefont {Zabolotnyy}}, \bibinfo {author} {\bibfnamefont
  {M.}~\bibnamefont {Knupfer}}, \bibinfo {author} {\bibfnamefont
  {B.}~\bibnamefont {B\"uchner}}, \bibinfo {author} {\bibfnamefont
  {H.}~\bibnamefont {Berger}}, \bibinfo {author} {\bibfnamefont
  {M.}~\bibnamefont {Falub}}, \bibinfo {author} {\bibfnamefont
  {M.}~\bibnamefont {Shi}}, \bibinfo {author} {\bibfnamefont {J.}~\bibnamefont
  {Krempasky}},\ and\ \bibinfo {author} {\bibfnamefont {L.}~\bibnamefont
  {Patthey}},\ }\bibfield  {title} {\bibinfo {title} {Parity of the pairing
  bosons in a high-temperature
  {{$\mathrm{Pb}\mathrm{\text{\ensuremath{-}}}{\mathrm{Bi}}_{2}{\mathrm{Sr}}_{2}{\mathrm{CaCu}}_{2}{\mathrm{O}}_{8}$}}
  bilayer superconductor by angle-resolved photoemission spectroscopy},\ }\href
  {https://doi.org/10.1103/PhysRevLett.96.067001} {\bibfield  {journal}
  {\bibinfo  {journal} {Phys. Rev. Lett.}\ }\textbf {\bibinfo {volume} {96}},\
  \bibinfo {pages} {067001} (\bibinfo {year} {2006}{\natexlab{b}})}\BibitemShut
  {NoStop}%
\bibitem [{\citenamefont {Valla}\ \emph {et~al.}(2007)\citenamefont {Valla},
  \citenamefont {Kidd}, \citenamefont {Yin}, \citenamefont {Gu}, \citenamefont
  {Johnson}, \citenamefont {Pan},\ and\ \citenamefont {Fedorov}}]{valla2007}%
  \BibitemOpen
  \bibfield  {author} {\bibinfo {author} {\bibfnamefont {T.}~\bibnamefont
  {Valla}}, \bibinfo {author} {\bibfnamefont {T.~E.}\ \bibnamefont {Kidd}},
  \bibinfo {author} {\bibfnamefont {W.-G.}\ \bibnamefont {Yin}}, \bibinfo
  {author} {\bibfnamefont {G.~D.}\ \bibnamefont {Gu}}, \bibinfo {author}
  {\bibfnamefont {P.~D.}\ \bibnamefont {Johnson}}, \bibinfo {author}
  {\bibfnamefont {Z.-H.}\ \bibnamefont {Pan}},\ and\ \bibinfo {author}
  {\bibfnamefont {A.~V.}\ \bibnamefont {Fedorov}},\ }\bibfield  {title}
  {\bibinfo {title} {High-energy kink observed in the electron dispersion of
  high-temperature cuprate superconductors},\ }\href
  {https://doi.org/10.1103/PhysRevLett.98.167003} {\bibfield  {journal}
  {\bibinfo  {journal} {Phys. Rev. Lett.}\ }\textbf {\bibinfo {volume} {98}},\
  \bibinfo {pages} {167003} (\bibinfo {year} {2007})}\BibitemShut {NoStop}%
\bibitem [{\citenamefont {Xie}\ \emph {et~al.}(2007)\citenamefont {Xie},
  \citenamefont {Yang}, \citenamefont {Shen}, \citenamefont {Zhao},
  \citenamefont {Ou}, \citenamefont {Weil}, \citenamefont {Gu}, \citenamefont
  {Arita}, \citenamefont {Qiao}, \citenamefont {Namatame}, \citenamefont
  {Taniguchi}, \citenamefont {Kaneko}, \citenamefont {Eisaki}, \citenamefont
  {Tsuei}, \citenamefont {Cheng}, \citenamefont {Vobornik}, \citenamefont
  {Fujii}, \citenamefont {Rossi}, \citenamefont {Yang},\ and\ \citenamefont
  {Feng}}]{xie2007}%
  \BibitemOpen
  \bibfield  {author} {\bibinfo {author} {\bibfnamefont {B.~P.}\ \bibnamefont
  {Xie}}, \bibinfo {author} {\bibfnamefont {K.}~\bibnamefont {Yang}}, \bibinfo
  {author} {\bibfnamefont {D.~W.}\ \bibnamefont {Shen}}, \bibinfo {author}
  {\bibfnamefont {J.~F.}\ \bibnamefont {Zhao}}, \bibinfo {author}
  {\bibfnamefont {H.~W.}\ \bibnamefont {Ou}}, \bibinfo {author} {\bibfnamefont
  {J.}~\bibnamefont {Weil}}, \bibinfo {author} {\bibfnamefont {S.~Y.}\
  \bibnamefont {Gu}}, \bibinfo {author} {\bibfnamefont {M.}~\bibnamefont
  {Arita}}, \bibinfo {author} {\bibfnamefont {S.}~\bibnamefont {Qiao}},
  \bibinfo {author} {\bibfnamefont {H.}~\bibnamefont {Namatame}}, \bibinfo
  {author} {\bibfnamefont {M.}~\bibnamefont {Taniguchi}}, \bibinfo {author}
  {\bibfnamefont {N.}~\bibnamefont {Kaneko}}, \bibinfo {author} {\bibfnamefont
  {H.}~\bibnamefont {Eisaki}}, \bibinfo {author} {\bibfnamefont {K.~D.}\
  \bibnamefont {Tsuei}}, \bibinfo {author} {\bibfnamefont {C.~M.}\ \bibnamefont
  {Cheng}}, \bibinfo {author} {\bibfnamefont {I.}~\bibnamefont {Vobornik}},
  \bibinfo {author} {\bibfnamefont {J.}~\bibnamefont {Fujii}}, \bibinfo
  {author} {\bibfnamefont {G.}~\bibnamefont {Rossi}}, \bibinfo {author}
  {\bibfnamefont {Z.~Q.}\ \bibnamefont {Yang}},\ and\ \bibinfo {author}
  {\bibfnamefont {D.~L.}\ \bibnamefont {Feng}},\ }\bibfield  {title} {\bibinfo
  {title} {High-energy scale revival and giant kink in the dispersion of a
  cuprate superconductor},\ }\href
  {https://doi.org/10.1103/PhysRevLett.98.147001} {\bibfield  {journal}
  {\bibinfo  {journal} {Phys. Rev. Lett.}\ }\textbf {\bibinfo {volume} {98}},\
  \bibinfo {pages} {147001} (\bibinfo {year} {2007})}\BibitemShut {NoStop}%
\bibitem [{\citenamefont {Moritz}\ \emph {et~al.}(2009)\citenamefont {Moritz},
  \citenamefont {Schmitt}, \citenamefont {Johnston}, \citenamefont {Motoyama},
  \citenamefont {Greven}, \citenamefont {Lu}, \citenamefont {Kim},
  \citenamefont {Scalettar}, \citenamefont {Shen} \emph {et~al.}}]{moritz2009}%
  \BibitemOpen
  \bibfield  {author} {\bibinfo {author} {\bibfnamefont {B.}~\bibnamefont
  {Moritz}}, \bibinfo {author} {\bibfnamefont {W.}~\bibnamefont {Schmitt},
  \bibfnamefont {Fand~Meevasana}}, \bibinfo {author} {\bibfnamefont
  {S.}~\bibnamefont {Johnston}}, \bibinfo {author} {\bibfnamefont
  {E.}~\bibnamefont {Motoyama}}, \bibinfo {author} {\bibfnamefont
  {M.}~\bibnamefont {Greven}}, \bibinfo {author} {\bibfnamefont
  {D.}~\bibnamefont {Lu}}, \bibinfo {author} {\bibfnamefont {C.}~\bibnamefont
  {Kim}}, \bibinfo {author} {\bibfnamefont {R.}~\bibnamefont {Scalettar}},
  \bibinfo {author} {\bibfnamefont {Z.}~\bibnamefont {Shen}}, \emph {et~al.},\
  }\bibfield  {title} {\bibinfo {title} {Effect of strong correlations on the
  high energy anomaly in hole-and electron-doped high-{{$T_C$}}
  superconductors},\ }\href {https://doi.org/10.1088/1367-2630/11/9/093020}
  {\bibfield  {journal} {\bibinfo  {journal} {New Journal of Physics}\ }\textbf
  {\bibinfo {volume} {11}},\ \bibinfo {pages} {093020} (\bibinfo {year}
  {2009})}\BibitemShut {NoStop}%
\bibitem [{\citenamefont {Zhang}\ \emph {et~al.}(2008)\citenamefont {Zhang},
  \citenamefont {Liu}, \citenamefont {Meng}, \citenamefont {Zhao},
  \citenamefont {Liu}, \citenamefont {Dong}, \citenamefont {Lu}, \citenamefont
  {Wen}, \citenamefont {Xu}, \citenamefont {Gu}, \citenamefont {Sasagawa},
  \citenamefont {Wang}, \citenamefont {Zhu}, \citenamefont {Zhang},
  \citenamefont {Zhou}, \citenamefont {Wang}, \citenamefont {Zhao},
  \citenamefont {Chen}, \citenamefont {Xu},\ and\ \citenamefont
  {Zhou}}]{zhang2008}%
  \BibitemOpen
  \bibfield  {author} {\bibinfo {author} {\bibfnamefont {W.}~\bibnamefont
  {Zhang}}, \bibinfo {author} {\bibfnamefont {G.}~\bibnamefont {Liu}}, \bibinfo
  {author} {\bibfnamefont {J.}~\bibnamefont {Meng}}, \bibinfo {author}
  {\bibfnamefont {L.}~\bibnamefont {Zhao}}, \bibinfo {author} {\bibfnamefont
  {H.}~\bibnamefont {Liu}}, \bibinfo {author} {\bibfnamefont {X.}~\bibnamefont
  {Dong}}, \bibinfo {author} {\bibfnamefont {W.}~\bibnamefont {Lu}}, \bibinfo
  {author} {\bibfnamefont {J.~S.}\ \bibnamefont {Wen}}, \bibinfo {author}
  {\bibfnamefont {Z.~J.}\ \bibnamefont {Xu}}, \bibinfo {author} {\bibfnamefont
  {G.~D.}\ \bibnamefont {Gu}}, \bibinfo {author} {\bibfnamefont
  {T.}~\bibnamefont {Sasagawa}}, \bibinfo {author} {\bibfnamefont
  {G.}~\bibnamefont {Wang}}, \bibinfo {author} {\bibfnamefont {Y.}~\bibnamefont
  {Zhu}}, \bibinfo {author} {\bibfnamefont {H.}~\bibnamefont {Zhang}}, \bibinfo
  {author} {\bibfnamefont {Y.}~\bibnamefont {Zhou}}, \bibinfo {author}
  {\bibfnamefont {X.}~\bibnamefont {Wang}}, \bibinfo {author} {\bibfnamefont
  {Z.}~\bibnamefont {Zhao}}, \bibinfo {author} {\bibfnamefont {C.}~\bibnamefont
  {Chen}}, \bibinfo {author} {\bibfnamefont {Z.}~\bibnamefont {Xu}},\ and\
  \bibinfo {author} {\bibfnamefont {X.~J.}\ \bibnamefont {Zhou}},\ }\bibfield
  {title} {\bibinfo {title} {High energy dispersion relations for the high
  temperature
  {{${\mathrm{Bi}}_{2}{\mathrm{Sr}}_{2}{\mathrm{CaCu}}_{2}{\mathrm{O}}_{8}$}}
  superconductor from laser-based angle-resolved photoemission spectroscopy},\
  }\href {https://doi.org/10.1103/PhysRevLett.101.017002} {\bibfield  {journal}
  {\bibinfo  {journal} {Phys. Rev. Lett.}\ }\textbf {\bibinfo {volume} {101}},\
  \bibinfo {pages} {017002} (\bibinfo {year} {2008})}\BibitemShut {NoStop}%
\bibitem [{\citenamefont {Basak}\ \emph {et~al.}(2009)\citenamefont {Basak},
  \citenamefont {Das}, \citenamefont {Lin}, \citenamefont {Nieminen},
  \citenamefont {Lindroos}, \citenamefont {Markiewicz},\ and\ \citenamefont
  {Bansil}}]{basak2009}%
  \BibitemOpen
  \bibfield  {author} {\bibinfo {author} {\bibfnamefont {S.}~\bibnamefont
  {Basak}}, \bibinfo {author} {\bibfnamefont {T.}~\bibnamefont {Das}}, \bibinfo
  {author} {\bibfnamefont {H.}~\bibnamefont {Lin}}, \bibinfo {author}
  {\bibfnamefont {J.}~\bibnamefont {Nieminen}}, \bibinfo {author}
  {\bibfnamefont {M.}~\bibnamefont {Lindroos}}, \bibinfo {author}
  {\bibfnamefont {R.~S.}\ \bibnamefont {Markiewicz}},\ and\ \bibinfo {author}
  {\bibfnamefont {A.}~\bibnamefont {Bansil}},\ }\bibfield  {title} {\bibinfo
  {title} {Origin of the high-energy kink in the photoemission spectrum of the
  high-temperature superconductor
  {{${\text{Bi}}_{2}{\text{Sr}}_{2}{\text{CaCu}}_{2}{\text{O}}_{8}$}}},\ }\href
  {https://doi.org/10.1103/PhysRevB.80.214520} {\bibfield  {journal} {\bibinfo
  {journal} {Phys. Rev. B}\ }\textbf {\bibinfo {volume} {80}},\ \bibinfo
  {pages} {214520} (\bibinfo {year} {2009})}\BibitemShut {NoStop}%
\bibitem [{\citenamefont {Zhou}\ \emph {et~al.}(2010)\citenamefont {Zhou},
  \citenamefont {Liu}, \citenamefont {Meng}, \citenamefont {Zhang},
  \citenamefont {Liu}, \citenamefont {Zhao},\ and\ \citenamefont
  {Jia}}]{zhou2010}%
  \BibitemOpen
  \bibfield  {author} {\bibinfo {author} {\bibfnamefont {X.}~\bibnamefont
  {Zhou}}, \bibinfo {author} {\bibfnamefont {G.}~\bibnamefont {Liu}}, \bibinfo
  {author} {\bibfnamefont {J.}~\bibnamefont {Meng}}, \bibinfo {author}
  {\bibfnamefont {W.}~\bibnamefont {Zhang}}, \bibinfo {author} {\bibfnamefont
  {H.}~\bibnamefont {Liu}}, \bibinfo {author} {\bibfnamefont {L.}~\bibnamefont
  {Zhao}},\ and\ \bibinfo {author} {\bibfnamefont {X.}~\bibnamefont {Jia}},\
  }\bibfield  {title} {\bibinfo {title} {High resolution angle-resolved
  photoemission spectroscopy on {{Cu}}-based and {{Fe}}-based high-{{$T_C$}}
  superconductors},\ }\href
  {https://doi.org/https://doi.org/10.1002/pssa.201026447} {\bibfield
  {journal} {\bibinfo  {journal} {physica status solidi (a)}\ }\textbf
  {\bibinfo {volume} {207}},\ \bibinfo {pages} {2674} (\bibinfo {year}
  {2010})}\BibitemShut {NoStop}%
\bibitem [{\citenamefont {Sakakibara}\ \emph {et~al.}(2010)\citenamefont
  {Sakakibara}, \citenamefont {Usui}, \citenamefont {Kuroki}, \citenamefont
  {Arita},\ and\ \citenamefont {Aoki}}]{sakakibara2010}%
  \BibitemOpen
  \bibfield  {author} {\bibinfo {author} {\bibfnamefont {H.}~\bibnamefont
  {Sakakibara}}, \bibinfo {author} {\bibfnamefont {H.}~\bibnamefont {Usui}},
  \bibinfo {author} {\bibfnamefont {K.}~\bibnamefont {Kuroki}}, \bibinfo
  {author} {\bibfnamefont {R.}~\bibnamefont {Arita}},\ and\ \bibinfo {author}
  {\bibfnamefont {H.}~\bibnamefont {Aoki}},\ }\bibfield  {title} {\bibinfo
  {title} {Two-orbital model explains the higher transition temperature of the
  single-layer {{Hg}}-cuprate superconductor compared to that of the
  {{La}}-cuprate superconductor},\ }\href
  {https://doi.org/10.1103/PhysRevLett.105.057003} {\bibfield  {journal}
  {\bibinfo  {journal} {Phys. Rev. Lett.}\ }\textbf {\bibinfo {volume} {105}},\
  \bibinfo {pages} {057003} (\bibinfo {year} {2010})}\BibitemShut {NoStop}%
\bibitem [{\citenamefont {Sakakibara}\ \emph {et~al.}(2012)\citenamefont
  {Sakakibara}, \citenamefont {Usui}, \citenamefont {Kuroki}, \citenamefont
  {Arita},\ and\ \citenamefont {Aoki}}]{sakakibara2012}%
  \BibitemOpen
  \bibfield  {author} {\bibinfo {author} {\bibfnamefont {H.}~\bibnamefont
  {Sakakibara}}, \bibinfo {author} {\bibfnamefont {H.}~\bibnamefont {Usui}},
  \bibinfo {author} {\bibfnamefont {K.}~\bibnamefont {Kuroki}}, \bibinfo
  {author} {\bibfnamefont {R.}~\bibnamefont {Arita}},\ and\ \bibinfo {author}
  {\bibfnamefont {H.}~\bibnamefont {Aoki}},\ }\bibfield  {title} {\bibinfo
  {title} {Origin of the material dependence of {{$T_C$}} in the single-layered
  cuprates},\ }\href {https://doi.org/10.1103/PhysRevB.85.064501} {\bibfield
  {journal} {\bibinfo  {journal} {Phys. Rev. B}\ }\textbf {\bibinfo {volume}
  {85}},\ \bibinfo {pages} {064501} (\bibinfo {year} {2012})}\BibitemShut
  {NoStop}%
\bibitem [{\citenamefont {Meinders}\ \emph {et~al.}(1993)\citenamefont
  {Meinders}, \citenamefont {Eskes},\ and\ \citenamefont
  {Sawatzky}}]{meinders1993}%
  \BibitemOpen
  \bibfield  {author} {\bibinfo {author} {\bibfnamefont {M.~B.~J.}\
  \bibnamefont {Meinders}}, \bibinfo {author} {\bibfnamefont {H.}~\bibnamefont
  {Eskes}},\ and\ \bibinfo {author} {\bibfnamefont {G.~A.}\ \bibnamefont
  {Sawatzky}},\ }\bibfield  {title} {\bibinfo {title} {Spectral-weight
  transfer: {{Breakdown}} of low-energy-scale sum rules in correlated
  systems},\ }\href {https://doi.org/10.1103/PhysRevB.48.3916} {\bibfield
  {journal} {\bibinfo  {journal} {Phys. Rev. B}\ }\textbf {\bibinfo {volume}
  {48}},\ \bibinfo {pages} {3916} (\bibinfo {year} {1993})}\BibitemShut
  {NoStop}%
\bibitem [{\citenamefont {Wang}\ \emph {et~al.}(2020)\citenamefont {Wang},
  \citenamefont {He}, \citenamefont {Wohlfeld}, \citenamefont {Hashimoto},
  \citenamefont {Huang}, \citenamefont {Lu}, \citenamefont {Mo}, \citenamefont
  {Komiya}, \citenamefont {Jia}, \citenamefont {Moritz}, \citenamefont {Shen},\
  and\ \citenamefont {Devereaux}}]{wang2020}%
  \BibitemOpen
  \bibfield  {author} {\bibinfo {author} {\bibfnamefont {Y.}~\bibnamefont
  {Wang}}, \bibinfo {author} {\bibfnamefont {Y.}~\bibnamefont {He}}, \bibinfo
  {author} {\bibfnamefont {K.}~\bibnamefont {Wohlfeld}}, \bibinfo {author}
  {\bibfnamefont {M.}~\bibnamefont {Hashimoto}}, \bibinfo {author}
  {\bibfnamefont {E.~W.}\ \bibnamefont {Huang}}, \bibinfo {author}
  {\bibfnamefont {D.}~\bibnamefont {Lu}}, \bibinfo {author} {\bibfnamefont
  {S.-K.}\ \bibnamefont {Mo}}, \bibinfo {author} {\bibfnamefont
  {S.}~\bibnamefont {Komiya}}, \bibinfo {author} {\bibfnamefont
  {C.}~\bibnamefont {Jia}}, \bibinfo {author} {\bibfnamefont {B.}~\bibnamefont
  {Moritz}}, \bibinfo {author} {\bibfnamefont {Z.-X.}\ \bibnamefont {Shen}},\
  and\ \bibinfo {author} {\bibfnamefont {T.~P.}\ \bibnamefont {Devereaux}},\
  }\bibfield  {title} {\bibinfo {title} {Emergence of quasiparticles in a doped
  {{Mott}} insulator},\ }\href {https://doi.org/10.1038/s42005-020-00480-5}
  {\bibfield  {journal} {\bibinfo  {journal} {Commun Phys}\ }\textbf {\bibinfo
  {volume} {3}},\ \bibinfo {pages} {1} (\bibinfo {year} {2020})}\BibitemShut
  {NoStop}%
\bibitem [{\citenamefont {Sakai}(2023)}]{sakai2023}%
  \BibitemOpen
  \bibfield  {author} {\bibinfo {author} {\bibfnamefont {S.}~\bibnamefont
  {Sakai}},\ }\bibfield  {title} {\bibinfo {title} {Nonperturbative
  calculations for spectroscopic properties of cuprate high-temperature
  superconductors},\ }\href {https://doi.org/10.7566/JPSJ.92.092001} {\bibfield
   {journal} {\bibinfo  {journal} {J. Phys. Soc. Jpn.}\ }\textbf {\bibinfo
  {volume} {92}},\ \bibinfo {pages} {092001} (\bibinfo {year}
  {2023})}\BibitemShut {NoStop}%
\bibitem [{\citenamefont {Luo}\ \emph {et~al.}(2023{\natexlab{b}})\citenamefont
  {Luo}, \citenamefont {Hu}, \citenamefont {Wang}, \citenamefont {W{\'u}},\
  and\ \citenamefont {Yao}}]{luo2023}%
  \BibitemOpen
  \bibfield  {author} {\bibinfo {author} {\bibfnamefont {Z.}~\bibnamefont
  {Luo}}, \bibinfo {author} {\bibfnamefont {X.}~\bibnamefont {Hu}}, \bibinfo
  {author} {\bibfnamefont {M.}~\bibnamefont {Wang}}, \bibinfo {author}
  {\bibfnamefont {W.}~\bibnamefont {W{\'u}}},\ and\ \bibinfo {author}
  {\bibfnamefont {D.-X.}\ \bibnamefont {Yao}},\ }\bibfield  {title} {\bibinfo
  {title} {Bilayer two-orbital model of {{La$_3$Ni$_2$O$_7$}} under pressure},\
  }\href {https://doi.org/10.1103/PhysRevLett.131.126001} {\bibfield  {journal}
  {\bibinfo  {journal} {Phys. Rev. Lett.}\ }\textbf {\bibinfo {volume} {131}},\
  \bibinfo {pages} {126001} (\bibinfo {year} {2023}{\natexlab{b}})}\BibitemShut
  {NoStop}%
\bibitem [{\citenamefont {Zunger}\ and\ \citenamefont
  {Malyi}(2021)}]{zunger2021}%
  \BibitemOpen
  \bibfield  {author} {\bibinfo {author} {\bibfnamefont {A.}~\bibnamefont
  {Zunger}}\ and\ \bibinfo {author} {\bibfnamefont {O.~I.}\ \bibnamefont
  {Malyi}},\ }\bibfield  {title} {\bibinfo {title} {Understanding doping of
  quantum materials},\ }\href {https://doi.org/10.1021/acs.chemrev.0c00608}
  {\bibfield  {journal} {\bibinfo  {journal} {Chem. Rev.}\ }\textbf {\bibinfo
  {volume} {121}},\ \bibinfo {pages} {3031} (\bibinfo {year}
  {2021})}\BibitemShut {NoStop}%
\bibitem [{\citenamefont {Lany}\ \emph {et~al.}(2007)\citenamefont {Lany},
  \citenamefont {{Osorio-Guill{\'e}n}},\ and\ \citenamefont
  {Zunger}}]{lany2007}%
  \BibitemOpen
  \bibfield  {author} {\bibinfo {author} {\bibfnamefont {S.}~\bibnamefont
  {Lany}}, \bibinfo {author} {\bibfnamefont {J.}~\bibnamefont
  {{Osorio-Guill{\'e}n}}},\ and\ \bibinfo {author} {\bibfnamefont
  {A.}~\bibnamefont {Zunger}},\ }\bibfield  {title} {\bibinfo {title} {Origins
  of the doping asymmetry in oxides: {{Hole}} doping in {{NiO}} versus electron
  doping in {{ZnO}}},\ }\href {https://doi.org/10.1103/PhysRevB.75.241203}
  {\bibfield  {journal} {\bibinfo  {journal} {Phys. Rev. B}\ }\textbf {\bibinfo
  {volume} {75}},\ \bibinfo {pages} {241203} (\bibinfo {year}
  {2007})}\BibitemShut {NoStop}%
\bibitem [{\citenamefont {Ideta}\ \emph {et~al.}(2010)\citenamefont {Ideta},
  \citenamefont {Takashima}, \citenamefont {Hashimoto}, \citenamefont
  {Yoshida}, \citenamefont {Fujimori}, \citenamefont {Anzai}, \citenamefont
  {Fujita}, \citenamefont {Nakashima}, \citenamefont {Ino}, \citenamefont
  {Arita}, \citenamefont {Namatame}, \citenamefont {Taniguchi}, \citenamefont
  {Ono}, \citenamefont {Kubota}, \citenamefont {Lu}, \citenamefont {Shen},
  \citenamefont {Kojima},\ and\ \citenamefont {Uchida}}]{ideta2010}%
  \BibitemOpen
  \bibfield  {author} {\bibinfo {author} {\bibfnamefont {S.}~\bibnamefont
  {Ideta}}, \bibinfo {author} {\bibfnamefont {K.}~\bibnamefont {Takashima}},
  \bibinfo {author} {\bibfnamefont {M.}~\bibnamefont {Hashimoto}}, \bibinfo
  {author} {\bibfnamefont {T.}~\bibnamefont {Yoshida}}, \bibinfo {author}
  {\bibfnamefont {A.}~\bibnamefont {Fujimori}}, \bibinfo {author}
  {\bibfnamefont {H.}~\bibnamefont {Anzai}}, \bibinfo {author} {\bibfnamefont
  {T.}~\bibnamefont {Fujita}}, \bibinfo {author} {\bibfnamefont
  {Y.}~\bibnamefont {Nakashima}}, \bibinfo {author} {\bibfnamefont
  {A.}~\bibnamefont {Ino}}, \bibinfo {author} {\bibfnamefont {M.}~\bibnamefont
  {Arita}}, \bibinfo {author} {\bibfnamefont {H.}~\bibnamefont {Namatame}},
  \bibinfo {author} {\bibfnamefont {M.}~\bibnamefont {Taniguchi}}, \bibinfo
  {author} {\bibfnamefont {K.}~\bibnamefont {Ono}}, \bibinfo {author}
  {\bibfnamefont {M.}~\bibnamefont {Kubota}}, \bibinfo {author} {\bibfnamefont
  {D.~H.}\ \bibnamefont {Lu}}, \bibinfo {author} {\bibfnamefont {Z.-X.}\
  \bibnamefont {Shen}}, \bibinfo {author} {\bibfnamefont {K.~M.}\ \bibnamefont
  {Kojima}},\ and\ \bibinfo {author} {\bibfnamefont {S.}~\bibnamefont
  {Uchida}},\ }\bibfield  {title} {\bibinfo {title} {Enhanced {{Superconducting
  Gaps}} in the {{Trilayer High-Temperature}}
  {{Bi$_2$Sr$_2$Ca$_2$Cu$_3$O$_{10+\delta}$}} {{Cuprate Superconductor}}},\
  }\href {https://doi.org/10.1103/PhysRevLett.104.227001} {\bibfield  {journal}
  {\bibinfo  {journal} {Phys. Rev. Lett.}\ }\textbf {\bibinfo {volume} {104}},\
  \bibinfo {pages} {227001} (\bibinfo {year} {2010})}\BibitemShut {NoStop}%
\bibitem [{\citenamefont {Stewart}(2017)}]{stewart2017}%
  \BibitemOpen
  \bibfield  {author} {\bibinfo {author} {\bibfnamefont {G.~R.}\ \bibnamefont
  {Stewart}},\ }\bibfield  {title} {\bibinfo {title} {Unconventional
  superconductivity},\ }\href {https://doi.org/10.1080/00018732.2017.1331615}
  {\bibfield  {journal} {\bibinfo  {journal} {Advances in Physics}\ }\textbf
  {\bibinfo {volume} {66}},\ \bibinfo {pages} {75} (\bibinfo {year}
  {2017})}\BibitemShut {NoStop}%
\bibitem [{\citenamefont {Alloul}(2024)}]{alloul2024}%
  \BibitemOpen
  \bibfield  {author} {\bibinfo {author} {\bibfnamefont {H.}~\bibnamefont
  {Alloul}},\ }\bibfield  {title} {\bibinfo {title} {What do we learn from
  impurities and disorder in high-{{$T_C$}} cuprates?},\ }\bibfield  {journal}
  {\bibinfo  {journal} {Frontiers in Physics}\ }\textbf {\bibinfo {volume}
  {12}},\ \href {https://doi.org/10.3389/fphy.2024.1406242}
  {10.3389/fphy.2024.1406242} (\bibinfo {year} {2024})\BibitemShut {NoStop}%
\bibitem [{\citenamefont {Aichhorn}\ \emph {et~al.}(2006)\citenamefont
  {Aichhorn}, \citenamefont {Arrigoni}, \citenamefont {Potthoff},\ and\
  \citenamefont {Hanke}}]{aichhorn2006}%
  \BibitemOpen
  \bibfield  {author} {\bibinfo {author} {\bibfnamefont {M.}~\bibnamefont
  {Aichhorn}}, \bibinfo {author} {\bibfnamefont {E.}~\bibnamefont {Arrigoni}},
  \bibinfo {author} {\bibfnamefont {M.}~\bibnamefont {Potthoff}},\ and\
  \bibinfo {author} {\bibfnamefont {W.}~\bibnamefont {Hanke}},\ }\bibfield
  {title} {\bibinfo {title} {Antiferromagnetic to superconducting phase
  transition in the hole- and electron-doped {{Hubbard}} model at zero
  temperature},\ }\href {https://doi.org/10.1103/PhysRevB.74.024508} {\bibfield
   {journal} {\bibinfo  {journal} {Phys. Rev. B}\ }\textbf {\bibinfo {volume}
  {74}},\ \bibinfo {pages} {024508} (\bibinfo {year} {2006})}\BibitemShut
  {NoStop}%
\bibitem [{\citenamefont {Civelli}(2009)}]{civelli2009}%
  \BibitemOpen
  \bibfield  {author} {\bibinfo {author} {\bibfnamefont {M.}~\bibnamefont
  {Civelli}},\ }\bibfield  {title} {\bibinfo {title} {Doping-driven evolution
  of the superconducting state from a doped {{Mott}} insulator: {{Cluster}}
  dynamical mean-field theory},\ }\href
  {https://doi.org/10.1103/PhysRevB.79.195113} {\bibfield  {journal} {\bibinfo
  {journal} {Phys. Rev. B}\ }\textbf {\bibinfo {volume} {79}},\ \bibinfo
  {pages} {195113} (\bibinfo {year} {2009})}\BibitemShut {NoStop}%
\bibitem [{Note1()}]{Note1}%
  \BibitemOpen
  \bibinfo {note} {We have not investigated the competition with
  antiferromagnetism, that may also be a factor explaining the finite doping
  threshold~\cite
  {senechal2005competition,capone2006competition,kancharla2008b,foley2019a}.}\BibitemShut
  {Stop}%
\bibitem [{\citenamefont {Weber}(2021)}]{weber2021}%
  \BibitemOpen
  \bibfield  {author} {\bibinfo {author} {\bibfnamefont {C.}~\bibnamefont
  {Weber}},\ }\bibfield  {title} {\bibinfo {title} {Unifying guiding principles
  for designing optimized superconductors},\ }\href
  {https://doi.org/10.1073/pnas.2115874118} {\bibfield  {journal} {\bibinfo
  {journal} {Proceedings of the National Academy of Sciences}\ }\textbf
  {\bibinfo {volume} {118}},\ \bibinfo {pages} {e2115874118} (\bibinfo {year}
  {2021})}\BibitemShut {NoStop}%
\bibitem [{\citenamefont {Honma}\ and\ \citenamefont
  {Hor}(2008)}]{Honma_Hor_2008}%
  \BibitemOpen
  \bibfield  {author} {\bibinfo {author} {\bibfnamefont {T.}~\bibnamefont
  {Honma}}\ and\ \bibinfo {author} {\bibfnamefont {P.~H.}\ \bibnamefont
  {Hor}},\ }\bibfield  {title} {\bibinfo {title} {Unified electronic phase
  diagram for hole-doped high-{{Tc}} cuprates},\ }\href
  {https://doi.org/10.1103/PhysRevB.77.184520} {\bibfield  {journal} {\bibinfo
  {journal} {Physical Review B}\ }\textbf {\bibinfo {volume} {77}},\ \bibinfo
  {pages} {184520} (\bibinfo {year} {2008})}\BibitemShut {NoStop}%
\bibitem [{\citenamefont {Chakraborty}\ \emph {et~al.}(2010)\citenamefont
  {Chakraborty}, \citenamefont {Galanakis},\ and\ \citenamefont
  {Phillips}}]{Chakraborty_Galanakis_Phillips_2010}%
  \BibitemOpen
  \bibfield  {author} {\bibinfo {author} {\bibfnamefont {S.}~\bibnamefont
  {Chakraborty}}, \bibinfo {author} {\bibfnamefont {D.}~\bibnamefont
  {Galanakis}},\ and\ \bibinfo {author} {\bibfnamefont {P.}~\bibnamefont
  {Phillips}},\ }\bibfield  {title} {\bibinfo {title} {Emergence of
  particle-hole symmetry near optimal doping in high-temperature copper oxide
  superconductors},\ }\href {https://doi.org/10.1103/PhysRevB.82.214503}
  {\bibfield  {journal} {\bibinfo  {journal} {Physical Review B}\ }\textbf
  {\bibinfo {volume} {82}},\ \bibinfo {pages} {214503} (\bibinfo {year}
  {2010})}\BibitemShut {NoStop}%
\bibitem [{\citenamefont {Haule}\ and\ \citenamefont
  {Kotliar}(2007)}]{Haule_Kotliar_Avoided_2007}%
  \BibitemOpen
  \bibfield  {author} {\bibinfo {author} {\bibfnamefont {K.}~\bibnamefont
  {Haule}}\ and\ \bibinfo {author} {\bibfnamefont {G.}~\bibnamefont
  {Kotliar}},\ }\bibfield  {title} {\bibinfo {title} {Avoided criticality in
  near-optimally doped high-temperature superconductors},\ }\href
  {https://doi.org/10.1103/PhysRevB.76.092503} {\bibfield  {journal} {\bibinfo
  {journal} {Phys. Rev. B}\ }\textbf {\bibinfo {volume} {76}},\ \bibinfo
  {pages} {092503} (\bibinfo {year} {2007})}\BibitemShut {NoStop}%
\bibitem [{\citenamefont {Vidhyadhiraja}\ \emph {et~al.}(2009)\citenamefont
  {Vidhyadhiraja}, \citenamefont {Macridin}, \citenamefont
  {\ifmmode~\mbox{\c{S}}\else \c{S}\fi{}en}, \citenamefont {Jarrell},\ and\
  \citenamefont {Ma}}]{Vidhyadhiraja:2009}%
  \BibitemOpen
  \bibfield  {author} {\bibinfo {author} {\bibfnamefont {N.~S.}\ \bibnamefont
  {Vidhyadhiraja}}, \bibinfo {author} {\bibfnamefont {A.}~\bibnamefont
  {Macridin}}, \bibinfo {author} {\bibfnamefont {C.}~\bibnamefont
  {\ifmmode~\mbox{\c{S}}\else \c{S}\fi{}en}}, \bibinfo {author} {\bibfnamefont
  {M.}~\bibnamefont {Jarrell}},\ and\ \bibinfo {author} {\bibfnamefont
  {M.}~\bibnamefont {Ma}},\ }\bibfield  {title} {\bibinfo {title} {Quantum
  critical point at finite doping in the 2d hubbard model: A dynamical cluster
  quantum monte carlo study},\ }\href
  {https://doi.org/10.1103/PhysRevLett.102.206407} {\bibfield  {journal}
  {\bibinfo  {journal} {Phys. Rev. Lett.}\ }\textbf {\bibinfo {volume} {102}},\
  \bibinfo {pages} {206407} (\bibinfo {year} {2009})}\BibitemShut {NoStop}%
\bibitem [{\citenamefont {Fratino}\ \emph
  {et~al.}(2016{\natexlab{a}})\citenamefont {Fratino}, \citenamefont {S\'emon},
  \citenamefont {Sordi},\ and\ \citenamefont {Tremblay}}]{fratino2016a}%
  \BibitemOpen
  \bibfield  {author} {\bibinfo {author} {\bibfnamefont {L.}~\bibnamefont
  {Fratino}}, \bibinfo {author} {\bibfnamefont {P.}~\bibnamefont {S\'emon}},
  \bibinfo {author} {\bibfnamefont {G.}~\bibnamefont {Sordi}},\ and\ \bibinfo
  {author} {\bibfnamefont {A.-M.~S.}\ \bibnamefont {Tremblay}},\ }\bibfield
  {title} {\bibinfo {title} {Pseudogap and superconductivity in two-dimensional
  doped charge-transfer insulators},\ }\href
  {https://doi.org/10.1103/PhysRevB.93.245147} {\bibfield  {journal} {\bibinfo
  {journal} {Phys. Rev. B}\ }\textbf {\bibinfo {volume} {93}},\ \bibinfo
  {pages} {245147} (\bibinfo {year} {2016}{\natexlab{a}})}\BibitemShut
  {NoStop}%
\bibitem [{\citenamefont {Emery}\ and\ \citenamefont
  {Kivelson}(1995)}]{Emery_Kivelson_1995}%
  \BibitemOpen
  \bibfield  {author} {\bibinfo {author} {\bibfnamefont {V.~J.}\ \bibnamefont
  {Emery}}\ and\ \bibinfo {author} {\bibfnamefont {S.~A.}\ \bibnamefont
  {Kivelson}},\ }\bibfield  {title} {\bibinfo {title} {Importance of phase
  fluctuations in superconductors with small superfluid density},\ }\href
  {https://doi.org/10.1038/374434a0} {\bibfield  {journal} {\bibinfo  {journal}
  {Nature}\ }\textbf {\bibinfo {volume} {374}},\ \bibinfo {pages} {434–437}
  (\bibinfo {year} {1995})}\BibitemShut {NoStop}%
\bibitem [{\citenamefont {Kowalski}(2021)}]{Kowalski_MSc}%
  \BibitemOpen
  \bibfield  {author} {\bibinfo {author} {\bibfnamefont {N.}~\bibnamefont
  {Kowalski}},\ }\href@noop {} {\bibinfo {title} {An analysis of example}}
  (\bibinfo {year} {2021}),\ \bibinfo {note} {available at
  \url{https://savoirs.usherbrooke.ca/bitstream/handle/11143/18438/kowalski_nicolas_MSc_2021.pdf?sequence=7&isAllowed=y}}\BibitemShut
  {NoStop}%
\bibitem [{\citenamefont {Maier}\ and\ \citenamefont
  {Scalapino}(2019)}]{Maier_Scalapino_2019}%
  \BibitemOpen
  \bibfield  {author} {\bibinfo {author} {\bibfnamefont {T.~A.}\ \bibnamefont
  {Maier}}\ and\ \bibinfo {author} {\bibfnamefont {D.~J.}\ \bibnamefont
  {Scalapino}},\ }\bibfield  {title} {\bibinfo {title} {Pairfield fluctuations
  of a 2d hubbard model},\ }\href {https://doi.org/10.1038/s41535-019-0169-9}
  {\bibfield  {journal} {\bibinfo  {journal} {npj Quantum Materials}\ }\textbf
  {\bibinfo {volume} {4}},\ \bibinfo {pages} {30} (\bibinfo {year}
  {2019})}\BibitemShut {NoStop}%
\bibitem [{\citenamefont {Uemura}\ \emph {et~al.}(1989)\citenamefont {Uemura},
  \citenamefont {Le}, \citenamefont {Luke}, \citenamefont {Sternlieb},
  \citenamefont {Brewer}, \citenamefont {Kadono}, \citenamefont {Kiefl},
  \citenamefont {Kreitzman},\ and\ \citenamefont
  {Riseman}}]{uemura1989probing}%
  \BibitemOpen
  \bibfield  {author} {\bibinfo {author} {\bibfnamefont {Y.}~\bibnamefont
  {Uemura}}, \bibinfo {author} {\bibfnamefont {L.}~\bibnamefont {Le}}, \bibinfo
  {author} {\bibfnamefont {G.}~\bibnamefont {Luke}}, \bibinfo {author}
  {\bibfnamefont {B.}~\bibnamefont {Sternlieb}}, \bibinfo {author}
  {\bibfnamefont {J.}~\bibnamefont {Brewer}}, \bibinfo {author} {\bibfnamefont
  {R.}~\bibnamefont {Kadono}}, \bibinfo {author} {\bibfnamefont
  {R.}~\bibnamefont {Kiefl}}, \bibinfo {author} {\bibfnamefont
  {S.}~\bibnamefont {Kreitzman}},\ and\ \bibinfo {author} {\bibfnamefont
  {T.}~\bibnamefont {Riseman}},\ }\bibfield  {title} {\bibinfo {title} {Probing
  magnetism and superconductivity of high-tc systems with positive muons},\
  }\href {https://doi.org/https://doi.org/10.1016/0921-4534(89)90493-0}
  {\bibfield  {journal} {\bibinfo  {journal} {Physica C: Superconductivity and
  its Applications}\ }\textbf {\bibinfo {volume} {162}},\ \bibinfo {pages}
  {857} (\bibinfo {year} {1989})}\BibitemShut {NoStop}%
\bibitem [{\citenamefont {Simard}\ \emph {et~al.}(2019)\citenamefont {Simard},
  \citenamefont {H{\'e}bert}, \citenamefont {Foley}, \citenamefont
  {S{\'e}n{\'e}chal},\ and\ \citenamefont {Tremblay}}]{simard2019}%
  \BibitemOpen
  \bibfield  {author} {\bibinfo {author} {\bibfnamefont {O.}~\bibnamefont
  {Simard}}, \bibinfo {author} {\bibfnamefont {C.-D.}\ \bibnamefont
  {H{\'e}bert}}, \bibinfo {author} {\bibfnamefont {A.}~\bibnamefont {Foley}},
  \bibinfo {author} {\bibfnamefont {D.}~\bibnamefont {S{\'e}n{\'e}chal}},\ and\
  \bibinfo {author} {\bibfnamefont {A.-M.~S.}\ \bibnamefont {Tremblay}},\
  }\bibfield  {title} {\bibinfo {title} {Superfluid stiffness in cuprates:
  {{Effect}} of {{Mott}} transition and phase competition},\ }\href
  {https://doi.org/10.1103/PhysRevB.100.094506} {\bibfield  {journal} {\bibinfo
   {journal} {Phys. Rev. B}\ }\textbf {\bibinfo {volume} {100}},\ \bibinfo
  {pages} {094506} (\bibinfo {year} {2019})}\BibitemShut {NoStop}%
\bibitem [{\citenamefont {Fratino}\ \emph
  {et~al.}(2016{\natexlab{b}})\citenamefont {Fratino}, \citenamefont
  {S{\'e}mon}, \citenamefont {Sordi},\ and\ \citenamefont
  {Tremblay}}]{fratino2016}%
  \BibitemOpen
  \bibfield  {author} {\bibinfo {author} {\bibfnamefont {L.}~\bibnamefont
  {Fratino}}, \bibinfo {author} {\bibfnamefont {P.}~\bibnamefont {S{\'e}mon}},
  \bibinfo {author} {\bibfnamefont {G.}~\bibnamefont {Sordi}},\ and\ \bibinfo
  {author} {\bibfnamefont {A.-M.~S.}\ \bibnamefont {Tremblay}},\ }\bibfield
  {title} {\bibinfo {title} {An organizing principle for two-dimensional
  strongly correlated superconductivity},\ }\href
  {https://doi.org/10.1038/srep22715} {\bibfield  {journal} {\bibinfo
  {journal} {Sci Rep}\ }\textbf {\bibinfo {volume} {6}},\ \bibinfo {pages}
  {22715} (\bibinfo {year} {2016}{\natexlab{b}})}\BibitemShut {NoStop}%
\bibitem [{\citenamefont {Zenitani}\ \emph {et~al.}(1995)\citenamefont
  {Zenitani}, \citenamefont {Inari}, \citenamefont {Sahoda}, \citenamefont
  {Uehara}, \citenamefont {Akimitsu}, \citenamefont {Kubota},\ and\
  \citenamefont {Ayabe}}]{zenitani1995}%
  \BibitemOpen
  \bibfield  {author} {\bibinfo {author} {\bibfnamefont {Y.}~\bibnamefont
  {Zenitani}}, \bibinfo {author} {\bibfnamefont {K.}~\bibnamefont {Inari}},
  \bibinfo {author} {\bibfnamefont {S.}~\bibnamefont {Sahoda}}, \bibinfo
  {author} {\bibfnamefont {M.}~\bibnamefont {Uehara}}, \bibinfo {author}
  {\bibfnamefont {J.}~\bibnamefont {Akimitsu}}, \bibinfo {author}
  {\bibfnamefont {N.}~\bibnamefont {Kubota}},\ and\ \bibinfo {author}
  {\bibfnamefont {M.}~\bibnamefont {Ayabe}},\ }\bibfield  {title} {\bibinfo
  {title} {Superconductivity in (ca, na) 2cacu2o4cl2 the new simplest
  double-layer cuprate with apical chlorine},\ }\href
  {https://doi.org/10.1016/0921-4534(95)00234-0} {\bibfield  {journal}
  {\bibinfo  {journal} {Physica C: Superconductivity}\ }\textbf {\bibinfo
  {volume} {248}},\ \bibinfo {pages} {167} (\bibinfo {year}
  {1995})}\BibitemShut {NoStop}%
\bibitem [{\citenamefont {Paeckel}\ \emph {et~al.}(2023)\citenamefont
  {Paeckel}, \citenamefont {Köhler}, \citenamefont {Manmana},\ and\
  \citenamefont {Lenz}}]{paeckel2023}%
  \BibitemOpen
  \bibfield  {author} {\bibinfo {author} {\bibfnamefont {S.}~\bibnamefont
  {Paeckel}}, \bibinfo {author} {\bibfnamefont {T.}~\bibnamefont {Köhler}},
  \bibinfo {author} {\bibfnamefont {S.~R.}\ \bibnamefont {Manmana}},\ and\
  \bibinfo {author} {\bibfnamefont {B.}~\bibnamefont {Lenz}},\ }\bibfield
  {title} {\bibinfo {title} {Matrix-product-state-based band-lanczos solver for
  quantum cluster approaches},\ }\bibfield  {journal} {\bibinfo  {journal}
  {arXiv.org}\ }\href {https://doi.org/https://arxiv.org/abs/2310.10799}
  {https://arxiv.org/abs/2310.10799} (\bibinfo {year} {2023})\BibitemShut
  {NoStop}%
\bibitem [{\citenamefont {N\'u\~nez Fern\'andez}\ \emph
  {et~al.}(2022)\citenamefont {N\'u\~nez Fern\'andez}, \citenamefont {Jeannin},
  \citenamefont {Dumitrescu}, \citenamefont {Kloss}, \citenamefont {Kaye},
  \citenamefont {Parcollet},\ and\ \citenamefont {Waintal}}]{fernandez2022}%
  \BibitemOpen
  \bibfield  {author} {\bibinfo {author} {\bibfnamefont {Y.}~\bibnamefont
  {N\'u\~nez Fern\'andez}}, \bibinfo {author} {\bibfnamefont {M.}~\bibnamefont
  {Jeannin}}, \bibinfo {author} {\bibfnamefont {P.~T.}\ \bibnamefont
  {Dumitrescu}}, \bibinfo {author} {\bibfnamefont {T.}~\bibnamefont {Kloss}},
  \bibinfo {author} {\bibfnamefont {J.}~\bibnamefont {Kaye}}, \bibinfo {author}
  {\bibfnamefont {O.}~\bibnamefont {Parcollet}},\ and\ \bibinfo {author}
  {\bibfnamefont {X.}~\bibnamefont {Waintal}},\ }\bibfield  {title} {\bibinfo
  {title} {Learning {{Feynman}} diagrams with tensor trains},\ }\href
  {https://doi.org/10.1103/PhysRevX.12.041018} {\bibfield  {journal} {\bibinfo
  {journal} {Phys. Rev. X}\ }\textbf {\bibinfo {volume} {12}},\ \bibinfo
  {pages} {041018} (\bibinfo {year} {2022})}\BibitemShut {NoStop}%
\bibitem [{\citenamefont {Erpenbeck}\ \emph {et~al.}(2023)\citenamefont
  {Erpenbeck}, \citenamefont {Lin}, \citenamefont {Blommel}, \citenamefont
  {Zhang}, \citenamefont {Iskakov}, \citenamefont {Bernheimer}, \citenamefont
  {N\'u\~nez Fern\'andez}, \citenamefont {Cohen}, \citenamefont {Parcollet},
  \citenamefont {Waintal},\ and\ \citenamefont {Gull}}]{erpenbeck2023}%
  \BibitemOpen
  \bibfield  {author} {\bibinfo {author} {\bibfnamefont {A.}~\bibnamefont
  {Erpenbeck}}, \bibinfo {author} {\bibfnamefont {W.-T.}\ \bibnamefont {Lin}},
  \bibinfo {author} {\bibfnamefont {T.}~\bibnamefont {Blommel}}, \bibinfo
  {author} {\bibfnamefont {L.}~\bibnamefont {Zhang}}, \bibinfo {author}
  {\bibfnamefont {S.}~\bibnamefont {Iskakov}}, \bibinfo {author} {\bibfnamefont
  {L.}~\bibnamefont {Bernheimer}}, \bibinfo {author} {\bibfnamefont
  {Y.}~\bibnamefont {N\'u\~nez Fern\'andez}}, \bibinfo {author} {\bibfnamefont
  {G.}~\bibnamefont {Cohen}}, \bibinfo {author} {\bibfnamefont
  {O.}~\bibnamefont {Parcollet}}, \bibinfo {author} {\bibfnamefont
  {X.}~\bibnamefont {Waintal}},\ and\ \bibinfo {author} {\bibfnamefont
  {E.}~\bibnamefont {Gull}},\ }\bibfield  {title} {\bibinfo {title} {Tensor
  train continuous time solver for quantum impurity models},\ }\href
  {https://doi.org/10.1103/PhysRevB.107.245135} {\bibfield  {journal} {\bibinfo
   {journal} {Phys. Rev. B}\ }\textbf {\bibinfo {volume} {107}},\ \bibinfo
  {pages} {245135} (\bibinfo {year} {2023})}\BibitemShut {NoStop}%
\bibitem [{\citenamefont
  {Haule}(2023)}]{haule2023strongcouplingquantumimpurity}%
  \BibitemOpen
  \bibfield  {author} {\bibinfo {author} {\bibfnamefont {K.}~\bibnamefont
  {Haule}},\ }\bibfield  {title} {\bibinfo {title} {Strong coupling quantum
  impurity solver on the real and imaginary axis},\ }\href
  {https://arxiv.org/abs/2311.09412} {\bibfield  {journal} {\bibinfo  {journal}
  {arXiv.org}\ } (\bibinfo {year} {2023})}\BibitemShut {NoStop}%
\bibitem [{\citenamefont {Tanaka}(2019)}]{tanaka2019}%
  \BibitemOpen
  \bibfield  {author} {\bibinfo {author} {\bibfnamefont {A.}~\bibnamefont
  {Tanaka}},\ }\bibfield  {title} {\bibinfo {title} {Metal-insulator transition
  in the two-dimensional {{Hubbard}} model: Dual fermion approach with
  {{Lanczos }}exact diagonalization},\ }\href
  {https://doi.org/10.1103/PhysRevB.99.205133} {\bibfield  {journal} {\bibinfo
  {journal} {Phys. Rev. B}\ }\textbf {\bibinfo {volume} {99}},\ \bibinfo
  {pages} {205133} (\bibinfo {year} {2019})}\BibitemShut {NoStop}%
\bibitem [{\citenamefont {Liang}\ \emph {et~al.}(2017)\citenamefont {Liang},
  \citenamefont {Vanhala}, \citenamefont {Peotta}, \citenamefont {Siro},
  \citenamefont {Harju},\ and\ \citenamefont {T{\"o}rm{\"a}}}]{liang2017}%
  \BibitemOpen
  \bibfield  {author} {\bibinfo {author} {\bibfnamefont {L.}~\bibnamefont
  {Liang}}, \bibinfo {author} {\bibfnamefont {T.~I.}\ \bibnamefont {Vanhala}},
  \bibinfo {author} {\bibfnamefont {S.}~\bibnamefont {Peotta}}, \bibinfo
  {author} {\bibfnamefont {T.}~\bibnamefont {Siro}}, \bibinfo {author}
  {\bibfnamefont {A.}~\bibnamefont {Harju}},\ and\ \bibinfo {author}
  {\bibfnamefont {P.}~\bibnamefont {T{\"o}rm{\"a}}},\ }\bibfield  {title}
  {\bibinfo {title} {Band geometry, {{Berry}} curvature, and superfluid
  weight},\ }\href {https://doi.org/10.1103/PhysRevB.95.024515} {\bibfield
  {journal} {\bibinfo  {journal} {Phys. Rev. B}\ }\textbf {\bibinfo {volume}
  {95}},\ \bibinfo {pages} {024515} (\bibinfo {year} {2017})}\BibitemShut
  {NoStop}%
\bibitem [{\citenamefont {Charlebois}\ \emph {et~al.}(2015)\citenamefont
  {Charlebois}, \citenamefont {S\'en\'echal}, \citenamefont {Gagnon},\ and\
  \citenamefont {Tremblay}}]{charlebois2015}%
  \BibitemOpen
  \bibfield  {author} {\bibinfo {author} {\bibfnamefont {M.}~\bibnamefont
  {Charlebois}}, \bibinfo {author} {\bibfnamefont {D.}~\bibnamefont
  {S\'en\'echal}}, \bibinfo {author} {\bibfnamefont {A.-M.}\ \bibnamefont
  {Gagnon}},\ and\ \bibinfo {author} {\bibfnamefont {A.-M.~S.}\ \bibnamefont
  {Tremblay}},\ }\bibfield  {title} {\bibinfo {title} {Impurity-induced
  magnetic moments on the graphene-lattice {{Hubbard}} model: {{An}}
  inhomogeneous cluster dynamical mean-field theory study},\ }\href
  {https://doi.org/10.1103/PhysRevB.91.035132} {\bibfield  {journal} {\bibinfo
  {journal} {Phys. Rev. B}\ }\textbf {\bibinfo {volume} {91}},\ \bibinfo
  {pages} {035132} (\bibinfo {year} {2015})}\BibitemShut {NoStop}%
\bibitem [{\citenamefont {Pahlevanzadeh}\ \emph {et~al.}(2021)\citenamefont
  {Pahlevanzadeh}, \citenamefont {Sahebsara},\ and\ \citenamefont
  {Sénéchal}}]{pahlevanzadeh2021}%
  \BibitemOpen
  \bibfield  {author} {\bibinfo {author} {\bibfnamefont {B.}~\bibnamefont
  {Pahlevanzadeh}}, \bibinfo {author} {\bibfnamefont {P.}~\bibnamefont
  {Sahebsara}},\ and\ \bibinfo {author} {\bibfnamefont {D.}~\bibnamefont
  {Sénéchal}},\ }\bibfield  {title} {\bibinfo {title} {{Chiral $p$-wave
  superconductivity in twisted bilayer graphene from dynamical mean field
  theory}},\ }\href {https://doi.org/10.21468/SciPostPhys.11.1.017} {\bibfield
  {journal} {\bibinfo  {journal} {SciPost Phys.}\ }\textbf {\bibinfo {volume}
  {11}},\ \bibinfo {pages} {017} (\bibinfo {year} {2021})}\BibitemShut
  {NoStop}%
\bibitem [{\citenamefont {S{\'e}n{\'e}chal}\ \emph {et~al.}(2005)\citenamefont
  {S{\'e}n{\'e}chal}, \citenamefont {Lavertu}, \citenamefont {Marois},\ and\
  \citenamefont {Tremblay}}]{senechal2005competition}%
  \BibitemOpen
  \bibfield  {author} {\bibinfo {author} {\bibfnamefont {D.}~\bibnamefont
  {S{\'e}n{\'e}chal}}, \bibinfo {author} {\bibfnamefont {P.-L.}\ \bibnamefont
  {Lavertu}}, \bibinfo {author} {\bibfnamefont {M.-A.}\ \bibnamefont
  {Marois}},\ and\ \bibinfo {author} {\bibfnamefont {A.-M.}\ \bibnamefont
  {Tremblay}},\ }\bibfield  {title} {\bibinfo {title} {Competition between
  antiferromagnetism and superconductivity in high-{{$T_C$}} cuprates},\ }\href
  {https://doi.org/10.1103/PhysRevLett.94.156404} {\bibfield  {journal}
  {\bibinfo  {journal} {Physical review letters}\ }\textbf {\bibinfo {volume}
  {94}},\ \bibinfo {pages} {156404} (\bibinfo {year} {2005})}\BibitemShut
  {NoStop}%
\bibitem [{\citenamefont {Capone}\ and\ \citenamefont
  {Kotliar}(2006)}]{capone2006competition}%
  \BibitemOpen
  \bibfield  {author} {\bibinfo {author} {\bibfnamefont {M.}~\bibnamefont
  {Capone}}\ and\ \bibinfo {author} {\bibfnamefont {G.}~\bibnamefont
  {Kotliar}},\ }\bibfield  {title} {\bibinfo {title} {Competition between
  {{$d$}}-wave superconductivity and antiferromagnetism in the two-dimensional
  {{Hubbard}} model},\ }\href {https://doi.org/10.1103/PhysRevB.74.054513}
  {\bibfield  {journal} {\bibinfo  {journal} {Phys. Rev. B}\ }\textbf {\bibinfo
  {volume} {74}},\ \bibinfo {pages} {054513} (\bibinfo {year}
  {2006})}\BibitemShut {NoStop}%
\end{thebibliography}%

\end{document}